\documentclass[12pt,twoside,vi]{mitthesis}
\usepackage{lgrind}
\usepackage{cmap}
\usepackage{amsmath}
\usepackage{amssymb}
\usepackage{crimson}
\usepackage[T1]{fontenc}
\usepackage{listings}
\usepackage{xcolor}
\usepackage{tikz}
\usepackage{enumitem}
\usepackage{sourcecodepro}
\usepackage{subcaption}
\usepackage{hyperref}
\usepackage{titlesec}
\usepackage{blindtext}
\usepackage{color}
\usepackage{float}
\usepackage[font={small}]{caption}
\usepackage{titleps}
\usepackage{tabulary}
\usepackage{booktabs}
\usepackage{overpic}
\usepackage{pgfplots}
\usepackage{makecell}
\usepackage{appendix}
\usepackage{multirow}
\usepackage{bm}
\usepackage{mathtools}
\usepackage{algorithm}
\usepackage{algpseudocode}
\usepackage{setspace}
\usepackage{bibentry}
\usepackage{etoolbox}
\usepackage{microtype}

\apptocmd{\thebibliography}{\raggedright}{}{}

\nobibliography*

\definecolor{grey}{rgb}{0.5,0.5,0.5}
\definecolor{blue}{rgb}{0.0,0.2,0.7}
\definecolor{gray75}{gray}{0.75}

\newcommand{\hsp}{\hspace{20pt}}
\titleformat{\chapter}[hang]{\LARGE\bfseries}{\thechapter\hsp\textcolor{gray75}{|}\hsp}{0pt}{\LARGE\bfseries}

\usetikzlibrary{arrows}
\usetikzlibrary{quotes}

\DeclareMathOperator*{\argmin}{arg\,min}

\renewcommand*\copyright{{\textcopyright}}

\begin{document}

\pdfsuppresswarningpagegroup=1

% -*-latex-*-
% 
% For questions, comments, concerns or complaints:
% thesis@mit.edu
% 
%
% $Log: cover.tex,v $
% Revision 1.8  2008/05/13 15:02:15  jdreed
% Degree month is June, not May.  Added note about prevdegrees.
% Arthur Smith's title updated
%
% Revision 1.7  2001/02/08 18:53:16  boojum
% changed some \newpages to \cleardoublepages
%
% Revision 1.6  1999/10/21 14:49:31  boojum
% changed comment referring to documentstyle
%
% Revision 1.5  1999/10/21 14:39:04  boojum
% *** empty log message ***
%
% Revision 1.4  1997/04/18  17:54:10  othomas
% added page numbers on abstract and cover, and made 1 abstract
% page the default rather than 2.  (anne hunter tells me this
% is the new institute standard.)
%
% Revision 1.4  1997/04/18  17:54:10  othomas
% added page numbers on abstract and cover, and made 1 abstract
% page the default rather than 2.  (anne hunter tells me this
% is the new institute standard.)
%
% Revision 1.3  93/05/17  17:06:29  starflt
% Added acknowledgements section (suggested by tompalka)
% 
% Revision 1.2  92/04/22  13:13:13  epeisach
% Fixes for 1991 course 6 requirements
% Phrase "and to grant others the right to do so" has been added to 
% permission clause
% Second copy of abstract is not counted as separate pages so numbering works
% out
% 
% Revision 1.1  92/04/22  13:08:20  epeisach

% NOTE:
% These templates make an effort to conform to the MIT Thesis specifications,
% however the specifications can change.  We recommend that you verify the
% layout of your title page with your thesis advisor and/or the MIT 
% Libraries before printing your final copy.
\title{Differentiable Visual Computing}

\author{Tzu-Mao Li}
% If you wish to list your previous degrees on the cover page, use the 
% previous degrees command:
%       \prevdegrees{A.A., Harvard University (1985)}
% You can use the \\ command to list multiple previous degrees
%       \prevdegrees{B.S., University of California (1978) \\
%                    S.M., Massachusetts Institute of Technology (1981)}
\department{Department of Electrical Engineering and Computer Science}

% If the thesis is for two degrees simultaneously, list them both
% separated by \and like this:
\degree{Doctor of Philosophy}
% \degree{Doctor of Philosophy \and Master of Science}
% \degree{Bachelor of Science in Computer Science and Engineering}

% As of the 2007-08 academic year, valid degree months are September, 
% February, or June.  The default is June.
\degreemonth{June}
\degreeyear{2019}
\thesisdate{May 10, 2019}

%% By default, the thesis will be copyrighted to MIT.  If you need to copyright
%% the thesis to yourself, just specify the `vi' documentclass option.  If for
%% some reason you want to exactly specify the copyright notice text, you can
%% use the \copyrightnoticetext command.  
%\copyrightnoticetext{\copyright IBM, 1990.  Do not open till Xmas.}

% If there is more than one supervisor, use the \supervisor command
% once for each.
\supervisor{Fr\'edo Durand}{Professor of Electrical Engineering and Computer Science}

% This is the department committee chairman, not the thesis committee
% chairman.  You should replace this with your Department's Committee
% Chairman.
\chairman{Leslie A. Kolodziejski}{Professor of Electrical Engineering and Computer Science \\ Chair, Department Committee on Graduate Students}

% Make the titlepage based on the above information.  If you need
% something special and can't use the standard form, you can specify
% the exact text of the titlepage yourself.  Put it in a titlepage
% environment and leave blank lines where you want vertical space.
% The spaces will be adjusted to fill the entire page.  The dotted
% lines for the signatures are made with the \signature command.
\maketitle

% The abstractpage environment sets up everything on the page except
% the text itself.  The title and other header material are put at the
% top of the page, and the supervisors are listed at the bottom.  A
% new page is begun both before and after.  Of course, an abstract may
% be more than one page itself.  If you need more control over the
% format of the page, you can use the abstract environment, which puts
% the word "Abstract" at the beginning and single spaces its text.

%% You can either \input (*not* \include) your abstract file, or you can put
%% the text of the abstract directly between the \begin{abstractpage} and
%% \end{abstractpage} commands.

% First copy: start a new page, and save the page number.
\cleardoublepage
% Uncomment the next line if you do NOT want a page number on your
% abstract and acknowledgments pages.
% \pagestyle{empty}
\setcounter{savepage}{\thepage}
\begin{abstractpage}
Derivatives of computer graphics, image processing, and deep learning algorithms have tremendous use in guiding parameter space searches, or solving inverse problems. As the algorithms become more sophisticated, we no longer only need to differentiate simple mathematical functions, but have to deal with general programs which encode complex transformations of data. This dissertation introduces three tools, for addressing the challenges that arise when obtaining and applying the derivatives for complex graphics algorithms.

Traditionally, practitioners have been constrained to composing programs with a limited set of coarse-grained operators, or hand-deriving derivatives. We extend the image processing language Halide with reverse-mode automatic differentiation, and the ability to automatically optimize the gradient computations. This enables automatic generation of the gradients of arbitrary Halide programs, at high performance, with little programmer effort. We demonstrate several applications, including how our system enables quality improvements of even traditional, feed-forward image processing algorithms, blurring the distinction between classical and deep learning methods.

In 3D rendering, the gradient is required with respect to variables such as camera parameters, light sources, geometry, and appearance. However, computing the gradient is challenging because the rendering integral includes visibility terms that are not differentiable. We introduce, to our knowledge, the first general-purpose differentiable ray tracer that solves the full rendering equation, while correctly taking the geometric discontinuities into account. We show prototype applications in inverse rendering and the generation of adversarial examples for neural networks.

Finally, we demonstrate that the derivatives of light path throughput, especially the second-order ones, can also be useful for guiding sampling in forward rendering. Simulating light transport in the presence of multi-bounce glossy effects and motion in 3D rendering is challenging due to the high-dimensional integrand and narrow  high-contribution areas. We extend the Metropolis Light Transport algorithm by adapting to the local shape of the integrand, thereby increasing sampling efficiency. In particular, the Hessian is able to capture the strong anisotropy of the integrand. We use ideas from Hamiltonian Monte Carlo and simulate physics in Taylor expansion to draw samples from high-contribution region.

\end{abstractpage}

% Additional copy: start a new page, and reset the page number.  This way,
% the second copy of the abstract is not counted as separate pages.
% Uncomment the next 6 lines if you need two copies of the abstract
% page.
% \setcounter{page}{\thesavepage}
% \begin{abstractpage}
% \input{abstract}
% \end{abstractpage}

\cleardoublepage

\section*{Acknowledgments}

I would like to acknowledge by briefly reflecting how I ended up writing this dissertation.

I was fascinated by computer science since I knew the existence of computers. I love the idea of automating tedious computation and the intellectual challenges involved in the process of automation. Naturally, I enrolled in a computer science program during my undergraduate education. At there I was introduced to the enchanting world of computer graphics, where people are able to generate beautiful images with equations and code, instead of pen and paper. During my undergraduate and master's studies at National Taiwan University, I worked with Yung-Yu Chuang, the professor who brought me into computer graphics research. I started to read academic papers, and was mesmerized by people who contribute their own ideas to improve image generation. In the end I was also able to contribute my own little idea, by publishing a paper related to denoising Monte Carlo rendering images during my Master's study.

I hoped to do more, and decided to pursue a Ph.D. next. Among all the academic literature I studied, a few names caught my eyes. My Ph.D. advisor Fr{\'e}do Durand is one of them. His work on frequency analysis for light transport simulation provides a rigorous and insightful theoretical foundation for sampling and reconstruction in light transport. I am also amazed by his versatility in research. At the point this dissertation was written, he has already worked on physically-based rendering, non-photorealistic rendering, computational photography, computer vision, geometric and material modeling, human-computer interaction, medical imaging, and programming systems. After I first met him, I also found out that he has a great sense of humor and is in general a very likable and good-tempered person.

I joined the MIT graphics group as a rendering person. Fr{\'e}do and I decided that research on Metropolis light transport~\cite{Veach:1997:MLT} aligns our interests the most. Metropolis light transport is a classical rendering algorithm that was recently revitalized thanks to Wenzel Jakob's reimplementation of this notoriously difficult-to-implement algorithm. We found the classical literature of Langevin Monte Carlo~\cite{Roberts:1996:ECL} and Hamiltonian Monte Carlo~\cite{Duane:1987:HMC}, and thought that the derivatives information these algorithms use can also help Metropolis light transport. I also learned about the field of automatic differentiation, which later became the most essential tool of this thesis. Fr{\'e}do, Luke Anderson, and Shuang Zhao had a lot of discussions with me on this topic, which helped to shape my thoughts. Fr{\'e}do also brought Ravi Ramamoorthi, Jaakko Lehtinen, and Wenzel Jakob into this project. Ravi helped the most on this particular project. When we were collaborating, every week he would monitor my progress, and try to understand the current obstacles and provide advice. While the discussions with Fr{\'e}do and others are usually higher-level, Ravi tried to understand every single detail of what I do. Explaining my thought process to them clarified my thinking. Ravi and Fr{\'e}do also have strong influences on my paper writing style. Fr{\'e}do taught me to focus on the high-level pictures and Ravi taught me to explain everything in a clear manner.

Like many other computer science research projects, this project ended up to be a huge undertaking of software engineering. Our algorithm requires the Hessian and gradient of the light transport contribution. Implementing this efficiently requires complex metaprogramming and is very difficult with existing tools. This also motivates my later work on extending the Halide programming language~\cite{Ragan-Kelley:2012:DAS} for generating gradients. Nevertheless, I was still able to implement the first differentiable bidirectional path tracer that is able to generate gradient and Hessian of path contribution. We published a paper in 2015.

The results from our first project on improving Metropolis light transport were quite encouraging. This motivated us to look deeper into this subfield. Our derivative-based algorithm helped local exploration, but the bigger issue of these Markov chain Monte Carlo methods lies in the global exploration. As light transport contribution is inherently multi-modal due to discontinuities, the Markov chains in Metropolis light transport algorithms usually have bad mixing. This typically manifests as blotchy artifacts on the images. We were hoping to have a better understanding of the global structure of light transport path space, in order to resolve the exploration issue. Unfortunately, this turns out to be more difficult than we imagined, due to the curse of dimensionality. As the dimensionality of the path space increases, the difficulty to capture the structure increases exponentially. We were able to get decent results by fitting Gaussians on low-dimensional cases (say, 4D), but I got stuck as soon as I proceeded to higher-dimensional space.\footnote{Recently, Reibold et al.~\cite{Reibold:2018:SGS} published a similar idea. A key feature that makes their idea works, in my opinion, is that they focus on fitting block-tridiagonal covariance matrices for their Gaussian mixtures, instead of the full covariance as we tried. This makes their problem significantly more tractable.}

I stuck on the global Metropolis light transport project for nearly two years. In the meantime, I also explored a few other directions. For example, I tried to generalize gradient domain rendering~\cite{Lehtinen:2013:GDM} to the wavelet domain. None of these attempts were very successful. As most researchers already knew, when working on a research project, it is very difficult to know whether the researcher is missing something, or the project simply will go nowhere in the first place. Still, I gained a lot of useful knowledge in these years. During this period, I helped Luke Anderson on his programming system for rendering, Aether~\cite{Anderson:2017:AED}. The system stores the Monte Carlo sampling process symbolically, and automatically produces the probability density function of this process. Like my first Metropolis rendering project, this process also heavily involves metaprogramming and huge engineering efforts. This increased my interests in systems research -- many computer graphics researches are so engineering heavy, that we need better tools to help researchers and engineers for fast prototyping. I also did an internship at Weta digital to work with the Manuka rendering team, including Marc Droske, Jir{\'\i} Vorba, Jorge Schwarzhaupt, Luca Fascione. I also met Lingqi Yan, who was also an intern there working on appearance modeling. We often chatted about research and video games together. The internship at Weta taught me a lot about production rendering, visual effects practices, and the beauty of New Zealand.

After I stagnated on the projects for a while, to avoid sunk cost fallacy, Fr{\'e}do and I decided to temporarily move on. Inspired by the recent success of deep learning, and my frustration on the lack of tools for general and efficient automatic differentiation, we chose to work on automatic differentiating Halide code. We picked Halide because it was developed by our group, so we are sufficiently familiar with it. Halide also strikes the balance between having a more general computation model than most deep learning frameworks, and focused enough for us to optimize the gradient code generation. We contacted Jonathan Ragan-Kelley and Andrew Adams, the parents of Halide, who also had the idea of adding automatic differentiation to Halide for a long time. We also brought in my labmate Micha{\"e}l Gharbi, who is one of the most knowledgable and likable people in the world, to work on this. I learned a lot about deep learning and data-driven computing from Micha{\"e}l and a lot about parallel programming systems from Jonathan and Andrew. It was super fun working with these people. I was also happy that my knowledge of automatic differentiation became useful in fields outside of rendering. 

At the summer of 2017, I did an internship at Nvidia research at Seattle. Since we published the 2015 paper on improving Metropolis light transport using derivative information, we were always thinking about using it also for inverse rendering. Previous work focused on either simplified model~\cite{Loper:2014:OAD} or volumetric scattering~\cite{Gkioulekas:2013:IVR}, and we think there are interesting cases on inverse surface light transport where derivatives can be useful. My collaborator and friend Jaakko saw my internship as an opportunity for pushing me to work in this direction. He also pointed out that the main technical challenge in the surface light transport case would be the non-differentiability. Under the help of Jaakko, Marco Salvi, and Aaron Lefohn, we experimented several different approximation algorithms there. They work well in many cases, but there were always some edge cases that would break the algorithms. The time at Nvidia was a fresh break from graduate school. I met Lingqi again there, and Christoph Peters, another intern who has an unusual passion for the theory of moment reconstruction problems, and always have only apples for his lunch. I also met two other interns Qi Sun, Tri Nguyen and maintained friendships with them since then.

After I wrapped up my internship at Nvidia and returned to MIT, I and Jaakko talked with Fr\'edo and Miika Aittala about the inverse rendering project. Fr\'edo pointed out the relation between the mesh discontinuities and silhouette rasterization~\cite{Sander:2001:DEO}. After more discussions, we realized that this is highly related to Ravi's first-order analysis work back in 2007~\cite{Ramamoorthi:2007:FAL}. I worked out the math and generalized Ravi's theory to arbitrary parameters, primary visibility, and global illumination. Based on my experience on the differentiable bidirectional path tracer from 2015, I was able to quickly come up with a prototype renderer and to write a paper about this.

This is how this dissertation was written.

Since I mostly focused on the research part of the story, many people were left out. I am grateful to the labmates in MIT graphics group. Lukas Murmann and Alexandre Kaspar were the students who joined the group at the same time as me. Naturally, we hung out a lot together (by my standard). YiChang Shih, Abe Davis, and Valentina Shin are the senior students who guided us when we were lost. Zoya Bylinski made sure we always have enough snacks and coffee in the office. I enjoyed all the trash-talks with Tiam Jaroensri. Prafull Sharma's jokes are sometimes funny, sometimes not too much. Camille Biscarrat and Caroline Chan brought energy and fresh air to our offices. I had a lot of nerdy programming languages chat with Yuanming Hu. It was also fun to chat about research with Gaurav Chaurasia, Aleksandar Zlateski, and David Levin. Luke Anderson proofread every single paper I have written, including this dissertation. Nathaniel Jones broadened my view on rendering's application in architectural visualization.

The reviewer \#1 of the inverse rendering paper, who I suspect to be Ravi, provided an extremely detailed and helpful review. My thesis committee Justin Solomon and Wojciech Matusik also provided useful comments. Justin pointed out the relation between the Reynold transport theorem and our edge sampling method in the inverse rendering project. I should also thank NSF and Toyota Research Institute for the funding support, so I don't freeze to death in Boston.

Anton Kaplanyan invited me to intern at Facebook Reality Lab during 2018. I met Thomas Leimk\"{u}hler, Steve Bako, Christoph Schied, and Michael Doggett there. FRL was a very competitive and vigorous environment. I loved the free food. They had smoked salmon for the breakfast! I worked with Dejan Azinovi\'c, Matthias, Nie{\ss}ner and Anton there on a material and lighting reconstruction project.

I met my girlfriend Ailin Deng at the end of 2017. She has since then become the oasis that shelters me when I am tired of programming and research. I thank I-Chao Shen and Sheng-Chieh Chin for being patient for listening to my rants and non-sensical research ideas. I am grateful that my parents remain supportive throughout my studies on computer science.

% This is the acknowledgements section.  You should replace this with your
% own acknowledgements.

%%%%%%%%%%%%%%%%%%%%%%%%%%%%%%%%%%%%%%%%%%%%%%%%%%%%%%%%%%%%%%%%%%%%%%
% -*-latex-*-

\newpagestyle{mypagestyle}{
  \setheadrule{.4pt}%
  \sethead[\textit {\chaptertitle}]%
    []%
    [\thepage]%
    {\thepage}%
    {}%
    {\textit {\chaptertitle}}%
}
\pagestyle{mypagestyle}
  % -*- Mode:TeX -*-
%% This file simply contains the commands that actually generate the table of
%% contents and lists of figures and tables.  You can omit any or all of
%% these files by simply taking out the appropriate command.  For more
%% information on these files, see appendix C.3.3 of the LaTeX manual. 
\tableofcontents
%\newpage
%\listoffigures
%\newpage
%\listoftables

\chapter{Introduction}

\begin{figure}[t]
  \centering
  \includegraphics[width=\linewidth]{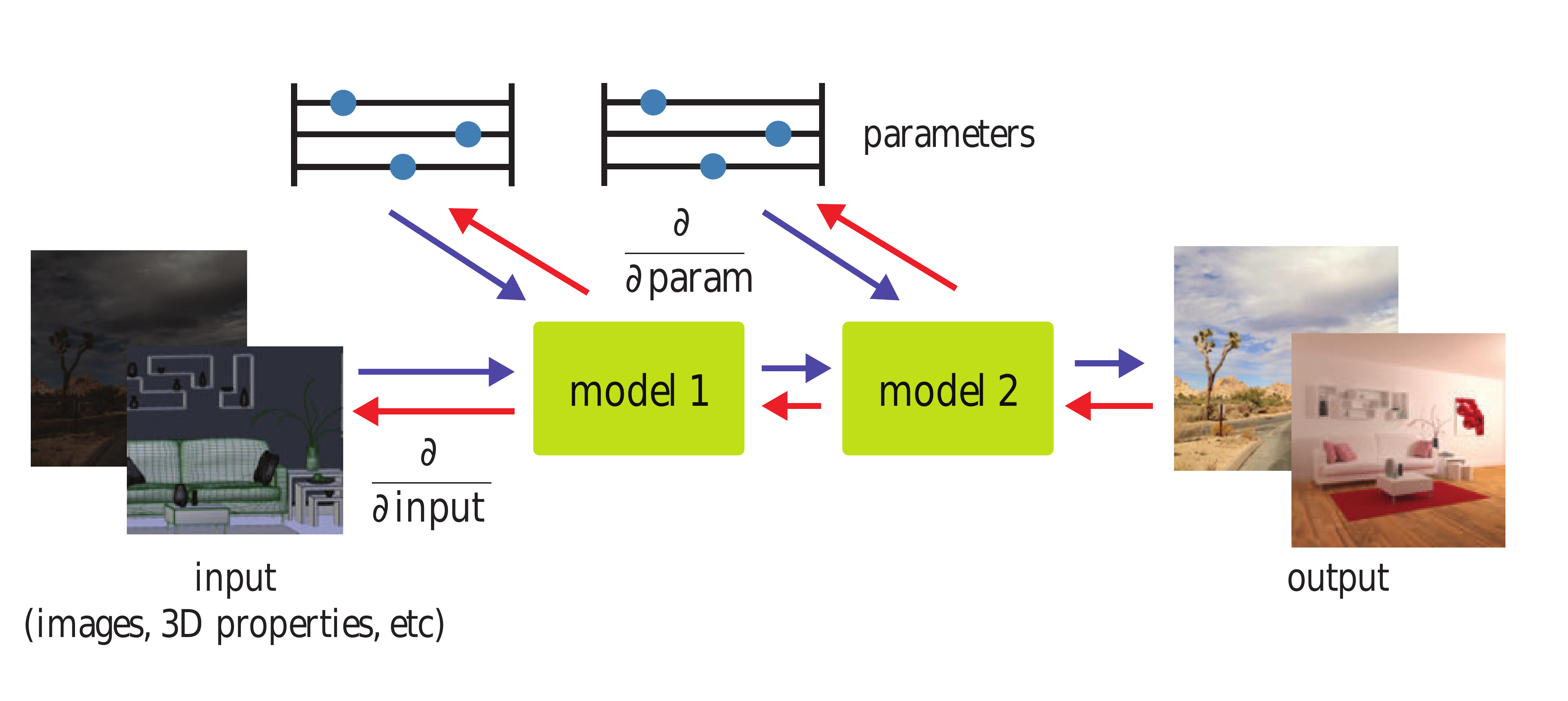}
  \caption{\textbf{Differentiable visual computing.} Derivatives enable us to make smart decisions in our models. The derivatives of a model's output can be taken with respect to the parameters or the inputs of the model. This makes it possible to find corresponding inputs for a given output, solving an inverse problem, or we can find model parameters that map inputs to outputs. This is useful in both data-driven and non-data-driven applications.}
  \label{fig:differentiable_programming}
\end{figure}

Differential calculus seeks to characterize the local geometry of a function. By definition, the derivatives at a point of a function tell us what happens to the outputs if we slightly move the point. This property enables us to make smarter decisions, and makes derivatives a fundamental tool for various tasks including parameter tuning, solving inverse problems, and sampling.

As computer graphics and image processing algorithms become more sophisticated, computing derivatives for functions defined in these algorithms becomes more important. Derivatives are useful in both data-driven and non data-driven scenarios. For one thing, as the number of parameters of the algorithm increases, it becomes infeasible to manually adjust them to achieve the desired behavior. Data-driven approaches allow us to automatically tune the parameters of our model. For another thing, these sophisticated \textit{forward} models can be used for solving \textit{inverse} problems. For example, the computer graphics community has developed mature models of how photons interact with scenes and cameras, and it is desirable to incorporate this knowledge, instead of learning it from scratch using a data-driven approach. Finally, differentiable algorithms are \textit{composable}, which means we can piece different differentiable algorithms together, and have an end-to-end differentiable system as a whole. This enables us to compose novel algorithms by adding other differentiable components to the pipeline, such as deep learning architectures. Figure~\ref{fig:differentiable_programming} illustrates the use of derivatives.

% Deep learning is not enough
While deep learning has popularized the use of gradient-based optimization over highly-parameterized functions, the current building blocks used in deep learning methods are very limited. A typical deep learning architecture is usually composed of convolution filters, linear combinations of elements (``fully connected'' layer), subsampling by an integer factor (``pooling'', usually by a factor of 2), and element-wise nonlinearities. Most visual computing algorithms are far more sophisticated than these. They often combine neighboring pixels using non-linear kernels (e.g.~\cite{Tomasi:1998:BFG, Buades:2005:NAI}), downsample a signal prefiltered by some antialiasing filters (e.g.~\cite{Mitchell:1988:RFC}), use heavy-tailed non-linear functions to model the reflectance of surfaces (e.g.~\cite{Cook:1982:RMC}), or traverse trees for finding intersections between objects (e.g.~\cite{Clark:1976:HGM}).

% We should do differentiable programming
I argue that most numerical algorithms in computer graphics and image processing should be implemented in a differentiable manner. This is beneficial for both data-driven and non-data-driven applications. Comparing to deep learning approaches, this allows better control and interpretability by integrating the domain knowledge into the model. It makes debugging models a lot easier since we have a better idea of how data should interact with the model. It is often more efficient both in time and memory, and more accurate, since the model is more tailored to the applications.

% Challenges of differential programming
Efficiently evaluating derivatives from algorithms that perform complex transformations on 3D data or 2D images presents challenges in both systems and algorithms. Firstly, existing deep learning frameworks (e.g.~\cite{Abadi:2015:TLM,Paszke:2017:ADP}) only have limited expressiveness. While automatic differentiation methods (e.g.~\cite{Griewank:2008:EDP}) can generate derivatives from almost arbitrary algorithms, generating efficient derivative code while taking parallelism and locality into consideration is still difficult. Secondly, the algorithms can introduce discontinuities. For example, in 3D rendering, the visibility term is discontinuous, which prevents direct application of automatic differentiation. Finally, designing algorithms that efficiently utilize the obtained derivatives is also important.

% Contribution
The contributions of this dissertation are three novel tools for addressing the challenges and for investigating the use of derivatives in the context of visual computing.

\paragraph{Efficient Automatic Differentiation for Image Processing and Deep Learning}

In Chapter~\ref{chap:gradient_halide}, we address the systems challenges for efficiently generating derivatives code from image processing algorithms. Existing tools for automatically generating derivatives have at least one of the following two issues:
\begin{itemize}
	\item General automatic differentiation systems (e.g. ~\cite{Bischof:1992:AAD,Griewank:1996:AAP,Hascoet:2013:TAD,Hogan:2014:FRA,Wiltschko:2017:TAD}) are \emph{inefficient} because they do not take parallelism and locality into consideration.
	\item Deep learning frameworks (e.g. ~\cite{Abadi:2015:TLM,Paszke:2017:ADP}) are \emph{inflexible} because they are composed of coarse-grained, specialized operators, such as convolutions or element-wise operations. For many applications, it is often difficult to assemble these operators to build the desired algorithm. Even when done successfully, the resulting code is often both slow and memory-inefficient, saving and reloading entire arrays of intermediate results between each step, causing costly cache misses.
\end{itemize}
These limitations are one of the main obstacles preventing researchers and developers from inventing novel differentiable algorithms, since they are often required to manually derive and implement the derivatives in lower-level languages such as C++ or CUDA.

In this chapter, we focus on image processing and deep learning. We build on the image processing language Halide~\cite{Ragan-Kelley:2012:DAS, Ragan-Kelley:2013:HLC}, and extend it with the ability to generate gradient code. Halide provides a concise and natural language for expressing image processing algorithms, while allowing the separation between high-level algorithm and low-level scheduling for achieving high-performance across platforms. To generate efficient gradient code, we develop a compiler transformation for generating gradient code automatically from Halide algorithms. Keys to making the transformation work are a scatter-to-gather conversion algorithm which preserves parallelism, and a simple automatic scheduling algorithm which specializes in the patterns in gradient code and provides a GPU backend.

% Building on Halide brings us several benefits. Firstly, it provides a concise and natural language for expressing image processing algorithms. Secondly, the language allows the separation between high-level algorithm and low-level scheduling, such as parallelization and memory trade-offs. This enables performance portability across different target architectures such as mobile CPUs, specialized digital signal processors, and data center GPUs, without the need to rewrite high-level algorithms. Finally, Halide's existing scheduling constructs naturally express and generalize ''checkpointing``, an essential optimization from traditional automatic differentiation literature for controlling memory usage.

% While Halide allows the users to write high-performance image processing code, it still requires manual derivation and implementation of the gradients. It also requires the user to write the low-level scheduling code for high-performance. We develop a compiler transformation for generating gradient code automatically from Halide algorithms. Key to make the transformation work are a scatter-to-gather conversion algorithm which preserves parallelism, and a simple automatic scheduling algorithm specializes to the patterns in gradient code and supports GPU backend.

Using this new extension of Halide, we show first that we can concisely and efficiently implement existing custom deep learning operators, which previously required implementation in low-level CUDA. Our generated code is as fast or even faster than the corresponding high-performance hand-written code, with less than $1/10$ of the lines of code. Secondly, we show that gradient-based parameter optimization is useful outside of traditional deep learning approaches. We significantly improve the accuracy of two traditional image processing algorithms by augmenting their parameters and automatically optimizing them. Thirdly, we show that the system is also useful for inverse problems. We implement a novel joint burst demosaicking and superresolution algorithm by building a forward image formation model. Finally, we demonstrate our extension's versatility by implementing two applications outside of image processing -- lens design optimization through a ray tracer and a classical fluid simulator in computer graphics~\cite{Stam:1999:SF}.

\paragraph{Differentiable Monte Carlo Ray Tracing through Edge Sampling}

While automatic differentiation generates derivatives, it does not handle non-differentiability in individual code paths. In particular, for computer graphics, we are interested in the gradients of the 3D rendering operation with respect to variables such as camera parameters, light sources, scene geometry, and appearance. While the rendering integral is differentiable, the \emph{integrand} is discontinuous due to visibility. Previous works on differentiable rendering (e.g.~\cite{Loper:2014:OAD, Kato:2018:N3M}) focused on fast approximate solutions, and do not handle secondary effects such as shadows or global illumination.

In Chapter~\ref{chap:redner}, we introduce a general-purpose differentiable ray tracer, which, to our knowledge, is the first comprehensive solution that is able to compute the gradients of the rendering integral with respect to scene parameters, while correctly taking geometric discontinuities into consideration. We observe that the discontinuities in the rendering integral become Dirac delta functions when taking the gradient. Therefore we develop a novel method for explicit sampling of the triangle edges that introduce the discontinuities. This requires new spatial acceleration techniques and importance sampling for efficiently selecting edges.

We integrate our differentiable ray tracer with the automatic differentiation library PyTorch~\cite{Paszke:2017:ADP}, and demonstrate prototype applications for inverse rendering and finding adversarial examples for neural networks. 

\paragraph{Hessian-Hamiltonian Monte Carlo Rendering}

Finally, we show that derivatives, especially the second-order ones, can also be used for accelerating forward rendering by guiding light path sampling. In Chapter~\ref{chap:h2mc}, we present a Markov chain Monte Carlo rendering algorithm that automatically and explicitly adapts to the local shape of the integrand using the second-order Taylor expansion, thereby increasing sampling efficiency. In particular, the Hessian is able to capture the strong anisotropy caused by challenging effects such as multi-bounce glossy effects and motion.

Using derivatives in the context of sampling instead of optimization requires more care. The second-order Taylor expansion does not define a proper distribution, and therefore cannot be directly importance sampled. We use ideas from Hamiltonian Monte Carlo~\cite{Duane:1987:HMC} that simulates Hamiltonian dynamics in a flipped version of the Taylor expansion where gravity pulls particles towards the high-contribution region. The quadratic landscape leads to a closed-form anisotropic Gaussian distribution, and results in a standard Metropolis-Hastings algorithm~\cite{Hastings:1970:MCS}.

Unlike previous works that derive the sampling procedures manually and only consider specific effects, our resulting algorithm is general thanks to automatic differentiation. In particular, our method is the first Markov chain Monte Carlo rendering algorithm that is able to resolve the anisotropy in the time dimension and render difficult moving caustics.

\section{Background and Target Audience}

Chapter~\ref{chap:autodiff} and Chapter~\ref{chap:optimization_sampling} review the background of automatic differentiation, optimization, and sampling, and their relationship. These are not novel components of this dissertation, but they represent important components that are glossed over in the individual publications. Moreover, they connect the central themes of this dissertation: differentiating algorithms and making use of the resulting derivatives.

I imagine the majority of readers of this dissertation to be researchers in the fields of computer graphics, image processing, systems or machine learning, who are interested in using the individual tools and want to know the details better, or people who are building their own differentiable systems. For both groups of people, I hope the examples in this dissertation can improve your intuition on building differentiable systems in the future.

\section{Publications}

The content of this dissertation has appeared in the following publications:
\begin{itemize}
	\item Chapter~\ref{chap:gradient_halide}: \bibentry{Li:2018:DPI}
	\item Chapter~\ref{chap:redner} \bibentry{Li:2018:DMC}
	\item Chapter~\ref{chap:h2mc} \bibentry{Li:2015:AGM}
\end{itemize}

The source code of the projects can be downloaded from the corresponding project sites:
\begin{itemize}
	\item http://gradients.halide-lang.org/
	\item https://people.csail.mit.edu/tzumao/diffrt/
	\item https://people.csail.mit.edu/tzumao/h2mc/
\end{itemize}

For Chapter~\ref{chap:gradient_halide}, I added a hindsight on differentiating scan operations (Chapter~\ref{sec:partial_update}) and an example of fluid simulation (Chapter~\ref{sub:non_image_processing}) since the publication.

For Chapter~\ref{chap:redner}, I added more discussions about the pathelogical parallel edges condition where our method can produce incorrect result (Chapter~\ref{sec:redner_method}). There is also some discussions regarding non-linear camera models (Chapter~\ref{sec:redner_method}). I added some discussion related to Reynolds transport theorem and shape optimization (Chapter~\ref{sec:redner_reynolds}). I also revised the edge selection algorithm (Chapter~\ref{sec:edge_tree}), added some discussions about a GPU implementation (Chapter~\ref{sec:redner_results}), and added discussions about differentiable geometry buffer rendering (Chapter~\ref{sec:redner_g_buffer}).

For Chapter~\ref{chap:h2mc}, I added a description of an improved large step mutation method (Chapter~\ref{sec:h2mc_implementation}), and some discussions to recent works (Chapter~\ref{sec:h2mc_limitation}).

\chapter{Automatic Differentiation}
\label{chap:autodiff}

\lstset{
    language=C++,
    basicstyle=\footnotesize\fontfamily{SourceCodePro-TLF}\selectfont,
    breaklines=true,
    showstringspaces=false,
    keywordstyle=\color{green!40!black},
    commentstyle=\itshape\color{gray},
    numberstyle=\color{blue},
    morekeywords={function, to, not},
    moredelim=**[is][\color{red}]{@}{@},
    keepspaces=true
}

Evaluating derivatives for computer graphics and image processing algorithms is the key to this dissertation. We will use them to minimize cost functions, solve inverse problems, and guide sampling procedures. Intuitively speaking, the derivatives of a function characterize the local behavior at a given point, e.g. if I move the point to this direction, will the output values become larger or smaller? This allows us to find points that result in certain function values, such as maximizing a utility function, or minimizing the difference between the output and a target.

In this chapter, we review the methods for generating derivatives from numerical programs. The chapter serves as an introductory article to the theory and practice of automatic differentiation. The reader is encouraged to read Griewank and Walther's textbook~\cite{Griewank:2008:EDP} for a comprehensive treatment of the topic.

Given a computer program containing control flow, loops, and/or recursion, with some real number inputs and some real number outputs, our goal is to compute the derivatives between the outputs and the inputs. Sometimes there is only a scalar output but more than one input, in which case we are interested in the gradient vector. Sometimes there are multiple outputs as well, and we are interested in the Jacobian matrix. Sometimes we are interested in the higher-order derivatives such as the Hessian matrix.

While the title of this chapter is \emph{automatic} differentiation, we will also talk about how to differentiate a program \emph{manually}, which is less difficult than one might imagine. We show how to systematically write down the derivative code just by looking at a program, without lengthy and convoluted mathematical notation. While this is still more tedious and error-prone than an automatic compiler transformation (which is why we develop the tool in Chapter~\ref{chap:gradient_halide}), it is a useful practice for understanding the structure of derivative code, and is even practical sometimes if it is difficult to parse and transform the code.

\section{Finite Differences and Symbolic Derivatives}
Before discussing automatic differentiation algorithms, it is useful to review other ways of generating derivatives, and compare them to automatic differentiation.

A common approximation for derivatives are finite differences, sometimes also called numerical derivatives. Given a function $f(x)$ and an input $x$, we approximate the derivative by perturbing $x$ by a small amount $h$:
\begin{align}
\frac{d f(x)}{d x} &\approx \frac{f(x + h) - f(x)}{h} \text{\qquad or} \\
\frac{d f(x)}{d x} &\approx \frac{f(x + h) - f(x - h)}{2h}.
\label{eq:finite_difference}
\end{align}
The problem with this approximation is two-fold. First, the optimal choice of the step size $h$ in a computer system is problem dependent. If the step size is too small, the rounding error of the floating point representation becomes too large. On the other hand, if the step size is too large, the result becomes a poor approximation to the true derivative. Second, the method is inefficient for multivariate functions. For a function with $100$ variables and a scalar output, computing the full gradient vector requires at least $101$ evaluations of the original function.

Another alternative is to treat the content of the function $f$ as a sequence of mathematical operations, and symbolically differentiate the function. Indeed, most of the rules for differentiation are mechanical, and we can apply the rules to generate $f'(x)$. However, in our case, $f(x)$ is usually an \emph{algorithm}, and symbolic differentiation does not scale well with the number of symbols. Consider the following code:

\begin{figure}[ht]
\centering
\begin{tabular}{c}
\begin{lstlisting}
function f(x):
    result = x
    for i = 1 to 8:
        result = exp(result)
    return result
\end{lstlisting}
\end{tabular}
\caption{A code example that iteratively computes a nested exponential for demonstrating the difference between symbolic differentiation and automatic differentiation.}
\label{fig:exp_example_code}
\end{figure}

Using the symbolic differentiation tools from mathematical software such as Mathematica~\cite{Wolfram:2019:Mathematica} would result in the following expression:
\begin{equation}
\frac{d f(x)}{d x} = e^{x + e^{e^{e^{e^{e^{e^{e^{x}}}}}}} + e^{e^{e^{e^{e^{e^{x}}}}}} + e^{e^{e^{e^{e^{x}}}}} + e^{e^{e^{e^{x}}}} + e^{e^{e^{x}}} + e^{e^{x}} + e^{x}}.
\label{eq:symbolic_derivative}
\end{equation} 

The size of derivative expression will become intractable when the size of the loop grows much larger. Using forward-mode automatic differentiation, which will be introduced later, we can generate the following code for computing derivatives:
\begin{center}
\begin{tabular}{c}
\begin{lstlisting}
function d_f(x):
    result = x
    d_result = 1
    for i = 1 to 8:
        result = exp(result)
        d_result = d_result * result
    return d_result
\end{lstlisting}
\end{tabular}
\end{center}

The code above outputs the exact same values as the symbolic derivative (Equation~\ref{eq:symbolic_derivative}), but is significantly more efficient ($8$ v.s. $37$ exponentials). This is due to automatic differentiation's better use of the intermediate values and the careful factorization of common subexpressions.

\section{Algorithms for Generating Derivatives}
\label{sec:autodiff_algo}

For a better understanding of automatic differentiation, before introducing the fully automatic solution, we will first discuss how to manually differentiate a code example. We start from programs with only function calls and elementary operations such as addition and multiplication. In particular, we do not allow recursive or circular function calls. Later in Chapter~\ref{sec:control_flow}, we generalize the idea to handle control flow such as loops and branches, and handle recursion. Throughout the chapter, we assume all function calls are side-effect free. To the author's knowledge, there are no known automatic differentiation algorithms for transforming arbitrary functions with side effects.

The key to automatic differentiation is the chain rule. Consider the following code with input \lstinline{x} and output \lstinline{z}:

\begin{center}
\begin{tabular}{c}
\begin{lstlisting}
y = f(x)
z = g(y)
\end{lstlisting}
\end{tabular}
\end{center}

Assume we already know the derivative functions $\frac{df(x)}{dx}$ and $\frac{dg(y)}{dy}$, and we are interested in the derivative of the output \lstinline{z} with respect to input \lstinline{x}. We can compute the derivative by applying the chain rule:

\begin{center}
\begin{tabular}{c}
\begin{lstlisting}
dydx = dfdx(x)
dzdy = dgdy(y)
dzdx = dzdy * dydx
\end{lstlisting}
\end{tabular}
\end{center}

We can recursively apply the rule to generate derivative functions, until the function is an elementary function for which we know the analytical derivatives, such as addition, multiplication, \lstinline{sin()}, or \lstinline{exp()}.

A useful mental model for automatic differentiation is the \emph{computational graph}. It can be used for representing dependencies between variables. The nodes of the graph are the variables and the edges are the derivatives between the adjacent vertices. In the case above the graph is linear:

\begin{center}
\begin{tikzpicture}
\tikzset{vertex/.style = {shape=circle,draw,minimum size=1.5em}}
\tikzset{edge/.style = {->,> = latex'}}
\node[vertex] (x) at (0,0) {\lstinline{x}};
\node[vertex] (y) at (2,0) {\lstinline{y}};
\node[vertex] (z) at (4,0) {\lstinline{z}};
\draw[edge] (x) to["$\frac{dy}{dx}$" {midway}] (y);
\draw[edge] (y) to["$\frac{dz}{dy}$" {midway}] (z);
\end{tikzpicture}
\end{center}

Computing derivatives from a computational graph involves traversal of the graph, and gathering of different paths that connect inputs and outputs.

In practice, most functions are multivariate, and often times we want to have multiple derivatives such as for the gradient vector. In this case, different derivatives may have common paths in the computational graph that can be factored out, which can greatly impact efficiency. Consider the following code example and its computational graph:

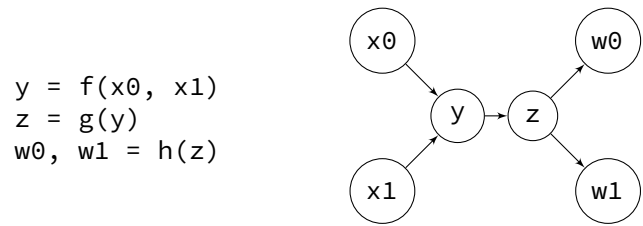
\begin{figure}[ht]
\centering
\begin{minipage}{.3\textwidth}
\begin{tabular}{c}
\begin{lstlisting}
y = f(x0, x1)
z = g(y)
w0, w1 = h(z)
\end{lstlisting}
\end{tabular}
\end{minipage}
\begin{minipage}{.3\textwidth}
\begin{tikzpicture}
\tikzset{vertex/.style = {shape=circle,draw,minimum size=1.5em}}
\tikzset{edge/.style = {->,> = latex'}}
\node[vertex] (x0) at  (0,2) {\lstinline{x0}};
\node[vertex] (x1) at  (0,0) {\lstinline{x1}};
\node[vertex] (y) at  (1,1) {\lstinline{y}};
\node[vertex] (z) at  (2,1) {\lstinline{z}};
\node[vertex] (w0) at  (3,2) {\lstinline{w0}};
\node[vertex] (w1) at  (3,0) {\lstinline{w1}};
\draw[edge] (x0) to (y);
\draw[edge] (x1) to (y);
\draw[edge] (y) to (z);
\draw[edge] (z) to (w0);
\draw[edge] (z) to (w1);
\end{tikzpicture}
\end{minipage}
\caption{Code example and computational graph with two inputs \lstinline{x0, x1} and two outputs \lstinline{w0, w1}}
\label{fig:two_by_two_code_example}
\end{figure}

There are four derivatives between the two outputs and two inputs. We can obtain them by traversing the four corresponding paths in the computational graph:

\begin{figure}[ht]
\centering
\begin{subfigure}[t]{0.22\textwidth}
    \begin{tikzpicture}[scale = 0.72]
    \tikzset{vertex/.style = {shape=circle,draw,minimum size=1.5em}}
    \tikzset{edge/.style = {->,> = latex'}}
    \node[vertex] (x0) at  (0,2) {\lstinline{x0}};
    \node[vertex] (x1) at  (0,0) {\lstinline{x1}};
    \node[vertex] (y) at  (1,1) {\lstinline{y}};
    \node[vertex] (z) at  (2.3,1) {\lstinline{z}};
    \node[vertex] (w0) at  (3.3,2) {\lstinline{w0}};
    \node[vertex] (w1) at  (3.3,0) {\lstinline{w1}};
    \draw[edge, red, thick] (x0) to (y);
    \draw[edge] (x1) to (y);
    \draw[edge, red, thick] (y) to (z);
    \draw[edge, red, thick] (z) to (w0);
    \draw[edge] (z) to (w1);
    \end{tikzpicture}
    \caption{$\frac{\partial w_0}{ \partial x_0}$}
\end{subfigure} \quad
\begin{subfigure}[t]{0.22\textwidth}
    \begin{tikzpicture}[scale = 0.72]
    \tikzset{vertex/.style = {shape=circle,draw,minimum size=1.5em}}
    \tikzset{edge/.style = {->,> = latex'}}
    \node[vertex] (x0) at  (0,2) {\lstinline{x0}};
    \node[vertex] (x1) at  (0,0) {\lstinline{x1}};
    \node[vertex] (y) at  (1,1) {\lstinline{y}};
    \node[vertex] (z) at  (2.3,1) {\lstinline{z}};
    \node[vertex] (w0) at  (3.3,2) {\lstinline{w0}};
    \node[vertex] (w1) at  (3.3,0) {\lstinline{w1}};
    \draw[edge] (x0) to (y);
    \draw[edge, red, thick] (x1) to (y);
    \draw[edge, red, thick] (y) to (z);
    \draw[edge, red, thick] (z) to (w0);
    \draw[edge] (z) to (w1);
    \end{tikzpicture}
    \caption{$\frac{\partial w_0}{ \partial x_1}$}
\end{subfigure} \quad
\begin{subfigure}[t]{0.22\textwidth}
    \begin{tikzpicture}[scale = 0.72]
    \tikzset{vertex/.style = {shape=circle,draw,minimum size=1.5em}}
    \tikzset{edge/.style = {->,> = latex'}}
    \node[vertex] (x0) at  (0,2) {\lstinline{x0}};
    \node[vertex] (x1) at  (0,0) {\lstinline{x1}};
    \node[vertex] (y) at  (1,1) {\lstinline{y}};
    \node[vertex] (z) at  (2.3,1) {\lstinline{z}};
    \node[vertex] (w0) at  (3.3,2) {\lstinline{w0}};
    \node[vertex] (w1) at  (3.3,0) {\lstinline{w1}};
    \draw[edge, red, thick] (x0) to (y);
    \draw[edge] (x1) to (y);
    \draw[edge, red, thick] (y) to (z);
    \draw[edge] (z) to (w0);
    \draw[edge, red, thick] (z) to (w1);
    \end{tikzpicture}
    \caption{$\frac{\partial w_1}{ \partial x_0}$}
\end{subfigure} \quad
\begin{subfigure}[t]{0.22\textwidth}
    \begin{tikzpicture}[scale = 0.72]
    \tikzset{vertex/.style = {shape=circle,draw,minimum size=1.5em}}
    \tikzset{edge/.style = {->,> = latex'}}
    \node[vertex] (x0) at  (0,2) {\lstinline{x0}};
    \node[vertex] (x1) at  (0,0) {\lstinline{x1}};
    \node[vertex] (y) at  (1,1) {\lstinline{y}};
    \node[vertex] (z) at  (2.3,1) {\lstinline{z}};
    \node[vertex] (w0) at  (3.3,2) {\lstinline{w0}};
    \node[vertex] (w1) at  (3.3,0) {\lstinline{w1}};
    \draw[edge] (x0) to (y);
    \draw[edge, red, thick] (x1) to (y);
    \draw[edge, red, thick] (y) to (z);
    \draw[edge] (z) to (w0);
    \draw[edge, red, thick] (z) to (w1);
    \end{tikzpicture}
    \caption{$\frac{\partial w_1}{ \partial x_1}$}
\end{subfigure}
\end{figure}

For example, in (a), the derivative of \lstinline{w0} with respect to \lstinline{x0} 
is the product of the three red edges:
\begin{equation}
\frac{\partial w_0}{\partial x_0} = \frac{\partial w_0}{\partial z} \frac{\partial z}{\partial y} \frac{\partial y}{\partial x_0},
\end{equation}
and in (b), the derivative of \lstinline{w0} with respect to \lstinline{x1} is
\begin{equation}
\frac{\partial w_0}{\partial x_1} = \frac{\partial w_0}{\partial z} \frac{\partial z}{\partial y} \frac{\partial y}{\partial x_1}.
\end{equation}

We can observe that some of the derivatives share common subpaths in the computational graph. For example the two derivatives above $\frac{\partial w_0}{\partial x_0}$ and $\frac{\partial w_0}{\partial x_1}$ share the same subpath \lstinline{y, z, w0}. We can therefore factor this subpath out and premultiply $\frac{\partial w_0}{\partial y} = \frac{\partial w_0}{\partial z}\frac{\partial z}{\partial y}$ for the two derivatives. In a larger computational graph, this factorization can have enormous impact on the performance of the derivative code, even affecting the time complexity in terms of the number of inputs or outputs.

Different automatic differentiation algorithms find common factors in the computational graph in different ways. In the most general case, finding a factorization that results in minimal operations is NP-hard~\cite{Naumann:2008:OJA}. Fortunately, in many common cases, such as factorization for the gradient vector, there are efficient solutions.\footnote{However, this does not take parallelism and memory efficiency into consideration. We show in Chapter~\ref{chap:gradient_halide} how we address this issue.}

If the input is a scalar variable, no matter how many variables there are in the output, \emph{forward-mode} automatic differentiation generates derivative code that has the same time complexity as the original algorithm. On the other hand, if the output is a scalar variable, no matter how many input variables there are, \emph{reverse-mode} automatic differentiation generates derivative code that has the same time complexity as the original algorithm. The latter case is particularly interesting, since it means that we can compute the gradient with the same time complexity (the ``cheap gradient principle''), which can be useful for various optimization and sampling algorithms.

Next, we demonstrate several algorithms for computing the derivatives while carefully taking the common subexpressions into consideration. We show how to transform a numerical algorithm with control flow, loops, or recursion to code that generates the derivatives.

\subsection{Forward-mode}
\label{sec:forward_mode}
We start with the simplest algorithm, usually called forward-mode automatic differentiation, and sometimes also called dual number (see Chapter~\ref{sec:historical_remarks} for historical remarks). Forward-mode traverses the computational graph from the inputs to outputs, computing derivatives of the intermediate nodes with respect to all input variables along the way. Forward-mode is efficient when the input dimension is low and the output dimension is high, since for each node in the computational graph, we need to compute the derivatives with respect to every single input variable.

In computer graphics, forward-mode has been used for computing screen-space derivatives for texture prefiltering in 3D rendering~\cite{Igehy:1999:TRD, Kulla:2018:SPI}, for computing derivatives in differential equations for physical simulation~\cite{Grinspun:2003:DS}, and for estimating motion in specular objects~\cite{Zimmer:2015:PME}. Forward-mode is also useful for computing the Hessian, where one can first apply forward-mode then apply reverse-mode on each output to obtain the full Hessian matrix.

We will describe forward-mode using the previous example in Figure~\ref{fig:two_by_two_code_example}. Starting from the inputs, the goal is to propagate the derivatives with respect to the inputs using the chain rule. To handle function calls, for every function \lstinline{f(x)} referenced by the output variables, we generate a derivative function \lstinline{df(x, dx)}, where \lstinline{dx} is the derivative of \lstinline{x} with respect to the input variables.

We start from the inputs \lstinline{x0, x1} and generate $\frac{\partial x_0}{\partial x_0} = 1$ and $\frac{\partial x_1}{\partial x_1} = 1$. We use a 2D vector \lstinline{dx0dx} to represent the derivatives of \lstinline{x0} with respect to \lstinline{x0} and \lstinline{x1}.

\begin{center}
\begin{minipage}{.4\textwidth}
\begin{tabular}{c}
\begin{lstlisting}
@dx0dx = {1, 0}@
@dx1dx = {0, 1}@
y = f(x0, x1)
z = g(y)
w0, w1 = h(z)
\end{lstlisting}
\end{tabular}
\end{minipage}
\begin{minipage}{.3\textwidth}
\begin{tikzpicture}
\tikzset{vertex/.style = {shape=circle,draw,minimum size=1.5em}}
\tikzset{edge/.style = {->,> = latex'}}
\node[vertex, red] (x0) at  (0,2) {\lstinline{x0}};
\node[vertex, red] (x1) at  (0,0) {\lstinline{x1}};
\node[vertex] (y) at  (1,1) {\lstinline{y}};
\node[vertex] (z) at  (2,1) {\lstinline{z}};
\node[vertex] (w0) at  (3,2) {\lstinline{w0}};
\node[vertex] (w1) at  (3,0) {\lstinline{w1}};
\draw[edge] (x0) to (y);
\draw[edge] (x1) to (y);
\draw[edge] (y) to (z);
\draw[edge] (z) to (w0);
\draw[edge] (z) to (w1);
\end{tikzpicture}
\end{minipage}
\end{center}

We then obtain the derivatives for \lstinline{y} with respect to the inputs. We assume we already applied forward-mode automatic differentiation for \lstinline{f}, so we have a derivative function \lstinline{df(x0, dx0dx, x1, dx1dx)}.
\begin{center}
\begin{minipage}{.4\textwidth}
\begin{tabular}{c}
\begin{lstlisting}
dx0dx = {1, 0}
dx1dx = {0, 1}
y = f(x0, x1)
@dydx = df(x0, dx0dx,@
@          x1, dx1dx)@
z = g(y)
w0, w1 = h(z)
\end{lstlisting}
\end{tabular}
\end{minipage}
\begin{minipage}{.3\textwidth}
\begin{tikzpicture}
\tikzset{vertex/.style = {shape=circle,draw,minimum size=1.5em}}
\tikzset{edge/.style = {->,> = latex'}}
\node[vertex] (x0) at  (0,2) {\lstinline{x0}};
\node[vertex] (x1) at  (0,0) {\lstinline{x1}};
\node[vertex, red] (y) at  (1,1) {\lstinline{y}};
\node[vertex] (z) at  (2,1) {\lstinline{z}};
\node[vertex] (w0) at  (3,2) {\lstinline{w0}};
\node[vertex] (w1) at  (3,0) {\lstinline{w1}};
\draw[edge, red] (x0) to (y);
\draw[edge, red] (x1) to (y);
\draw[edge] (y) to (z);
\draw[edge] (z) to (w0);
\draw[edge] (z) to (w1);
\end{tikzpicture}
\end{minipage}
\end{center}

We then propagate the derivative to \lstinline{z}:
\begin{center}
\begin{minipage}{.4\textwidth}
\begin{tabular}{c}
\begin{lstlisting}
dx0dx = {1, 0}
dx1dx = {0, 1}
y = f(x0, x1)
dydx = df(x0, dx0dx,
          x1, dx1dx)
z = g(y)
@dzdx = dg(y, dydx)@
w0, w1 = h(z)
\end{lstlisting}
\end{tabular}
\end{minipage}
\begin{minipage}{.3\textwidth}
\begin{tikzpicture}
\tikzset{vertex/.style = {shape=circle,draw,minimum size=1.5em}}
\tikzset{edge/.style = {->,> = latex'}}
\node[vertex] (x0) at  (0,2) {\lstinline{x0}};
\node[vertex] (x1) at  (0,0) {\lstinline{x1}};
\node[vertex] (y) at  (1,1) {\lstinline{y}};
\node[vertex, red] (z) at  (2,1) {\lstinline{z}};
\node[vertex] (w0) at  (3,2) {\lstinline{w0}};
\node[vertex] (w1) at  (3,0) {\lstinline{w1}};
\draw[edge] (x0) to (y);
\draw[edge] (x1) to (y);
\draw[edge, red] (y) to (z);
\draw[edge] (z) to (w0);
\draw[edge] (z) to (w1);
\end{tikzpicture}
\end{minipage}
\end{center}

Finally, we propagate the derivatives from \lstinline{z} to the outputs \lstinline{w0, w1}.
\begin{center}
\begin{minipage}{.4\textwidth}
\begin{tabular}{c}
\begin{lstlisting}
dx0dx = {1, 0}
dx1dx = {0, 1}
y = f(x0, x1)
dydx = df(x0, dx0dx,
          x1, dx1dx)
z = g(y)
dzdx = dg(y, dydx)
w0, w1 = h(z)
@dw0dx, dw1dx = dh(z, dzdx)@
\end{lstlisting}
\end{tabular}
\end{minipage}
\begin{minipage}{.3\textwidth}
\begin{tikzpicture}
\tikzset{vertex/.style = {shape=circle,draw,minimum size=1.5em}}
\tikzset{edge/.style = {->,> = latex'}}
\node[vertex] (x0) at  (0,2) {\lstinline{x0}};
\node[vertex] (x1) at  (0,0) {\lstinline{x1}};
\node[vertex] (y) at  (1,1) {\lstinline{y}};
\node[vertex] (z) at  (2,1) {\lstinline{z}};
\node[vertex, red] (w0) at  (3,2) {\lstinline{w0}};
\node[vertex, red] (w1) at  (3,0) {\lstinline{w1}};
\draw[edge] (x0) to (y);
\draw[edge] (x1) to (y);
\draw[edge] (y) to (z);
\draw[edge, red] (z) to (w0);
\draw[edge, red] (z) to (w1);
\end{tikzpicture}
\end{minipage}
\end{center}

The time complexity of the code generated by forward-mode automatic differentiation is $O(d)$ times the time complexity of the original algorithm, where $d$ is the number of input variables. It is efficient for functions with few input variables.

However, for many applications of derivatives, we need to differentiate functions with thousands or even millions of input variables. Using forward-mode for this would be infeasible, as we need to compute the derivatives with respect to \emph{all} input variables for every output in the computational graph. Fortunately, there is another algorithm called reverse-mode automatic differentiation that can generate derivative code that has the same time complexity as the original algorithm when there is only a single output, regardless of the number of input variables.

\subsection{Reverse-mode}
\label{sec:reverse_mode}

Reverse-mode propagates the derivatives from outputs to inputs, unlike forward-mode, which propagates the derivatives from inputs to outputs. For each node in the computational graph, we compute the derivatives of all outputs with respect to the variable at that node. Therefore reverse-mode is much more efficient when the input dimension is large and the output dimension is low. However, reverse-mode is also more complicated to implement since it needs to run the original algorithm backward to propagate the derivatives.

We again use the same previous example in Figure~\ref{fig:two_by_two_code_example} to illustrate how reverse-mode works. Similar to forward-mode, we need to handle function calls. For every function \lstinline{y = f(x)} referenced by the output variables, we generate a derivative function \lstinline{df(x, dy)}, where $dy$ is a vector of derivatives of the final output with respect to the function's output \lstinline{y} (in contrast, in forward-mode, the derivative functions take the input derivatives as arguments). Handling control flow and recursion in reverse-mode is more complicated. We discuss them in Chapter~\ref{sec:control_flow}.

We start from the outputs \lstinline{w0, w1} using $\frac{\partial w_0}{\partial w_0} = 1$ and $\frac{\partial w_1}{\partial w_1} = 1$. We use a 2D vector \lstinline{dwdw0} to represent the derivatives of \lstinline{w0, w1} with respect to \lstinline{w0}.

\begin{center}
\begin{minipage}{.41\textwidth}
\begin{tabular}{c}
\begin{lstlisting}
y = f(x0, x1)
z = g(y)
w0, w1 = h(z)

@dwdw0 = {1, 0}@
@dwdw1 = {0, 1}@
\end{lstlisting}
\end{tabular}
\end{minipage}
\begin{minipage}{.3\textwidth}
\begin{tikzpicture}
\tikzset{vertex/.style = {shape=circle,draw,minimum size=1.5em}}
\tikzset{edge/.style = {->,> = latex'}}
\node[vertex] (x0) at  (0,2) {\lstinline{x0}};
\node[vertex] (x1) at  (0,0) {\lstinline{x1}};
\node[vertex] (y) at  (1,1) {\lstinline{y}};
\node[vertex] (z) at  (2,1) {\lstinline{z}};
\node[vertex, red] (w0) at  (3,2) {\lstinline{w0}};
\node[vertex, red] (w1) at  (3,0) {\lstinline{w1}};
\draw[edge] (x0) to (y);
\draw[edge] (x1) to (y);
\draw[edge] (y) to (z);
\draw[edge] (z) to (w0);
\draw[edge] (z) to (w1);
\end{tikzpicture}
\end{minipage}
\end{center}

Next, we propagate the derivatives to variable \lstinline{z} on which the two outputs depend. We assume we already applied reverse-mode to the function \lstinline{h} and have \lstinline{dh(z, dwdw0, dwdw1)}.

\begin{center}
\begin{minipage}{.41\textwidth}
\begin{tabular}{c}
\begin{lstlisting}
y = f(x0, x1)
z = g(y)
w0, w1 = h(z)

dwdw0 = {1, 0}
dwdw1 = {0, 1}
@dwdz = dh(z, dwdw0, dwdw1)@
\end{lstlisting}
\end{tabular}
\end{minipage}
\begin{minipage}{.3\textwidth}
\begin{tikzpicture}
\tikzset{vertex/.style = {shape=circle,draw,minimum size=1.5em}}
\tikzset{edge/.style = {->,> = latex'}}
\node[vertex] (x0) at  (0,2) {\lstinline{x0}};
\node[vertex] (x1) at  (0,0) {\lstinline{x1}};
\node[vertex] (y) at  (1,1) {\lstinline{y}};
\node[vertex, red] (z) at  (2,1) {\lstinline{z}};
\node[vertex] (w0) at  (3,2) {\lstinline{w0}};
\node[vertex] (w1) at  (3,0) {\lstinline{w1}};
\draw[edge] (x0) to (y);
\draw[edge] (x1) to (y);
\draw[edge] (y) to (z);
\draw[edge, red] (z) to (w0);
\draw[edge, red] (z) to (w1);
\end{tikzpicture}
\end{minipage}
\end{center}

Similarly, we propagate to \lstinline{y} from \lstinline{z}.

\begin{center}
\begin{minipage}{.41\textwidth}
\begin{tabular}{c}
\begin{lstlisting}
y = f(x0, x1)
z = g(y)
w0, w1 = h(z)

dwdw0 = {1, 0}
dwdw1 = {0, 1}
dwdz = dh(z, dwdw0, dwdw1)
@dwdy = dg(y, dwdz)@
\end{lstlisting}
\end{tabular}
\end{minipage}
\begin{minipage}{.3\textwidth}
\begin{tikzpicture}
\tikzset{vertex/.style = {shape=circle,draw,minimum size=1.5em}}
\tikzset{edge/.style = {->,> = latex'}}
\node[vertex] (x0) at  (0,2) {\lstinline{x0}};
\node[vertex] (x1) at  (0,0) {\lstinline{x1}};
\node[vertex, red] (y) at  (1,1) {\lstinline{y}};
\node[vertex] (z) at  (2,1) {\lstinline{z}};
\node[vertex] (w0) at  (3,2) {\lstinline{w0}};
\node[vertex] (w1) at  (3,0) {\lstinline{w1}};
\draw[edge] (x0) to (y);
\draw[edge] (x1) to (y);
\draw[edge, red] (y) to (z);
\draw[edge] (z) to (w0);
\draw[edge] (z) to (w1);
\end{tikzpicture}
\end{minipage}
\end{center}

Finally we obtain the derivatives of the outputs \lstinline{w} with respect to the two inputs.

\begin{center}
\begin{minipage}{.41\textwidth}
\begin{tabular}{c}
\begin{lstlisting}
y = f(x0, x1)
z = g(y)
w0, w1 = h(z)

dwdw0 = {1, 0}
dwdw1 = {0, 1}
dwdz = dh(z, dwdw0, dwdw1)
dwdy = dg(y, dwdz)
@dwx0, dwx1 = df(x0, x1, dwdy)@
\end{lstlisting}
\end{tabular}
\end{minipage}
\begin{minipage}{.3\textwidth}
\begin{tikzpicture}
\tikzset{vertex/.style = {shape=circle,draw,minimum size=1.5em}}
\tikzset{edge/.style = {->,> = latex'}}
\node[vertex, red] (x0) at  (0,2) {\lstinline{x0}};
\node[vertex, red] (x1) at  (0,0) {\lstinline{x1}};
\node[vertex] (y) at  (1,1) {\lstinline{y}};
\node[vertex] (z) at  (2,1) {\lstinline{z}};
\node[vertex] (w0) at  (3,2) {\lstinline{w0}};
\node[vertex] (w1) at  (3,0) {\lstinline{w1}};
\draw[edge, red] (x0) to (y);
\draw[edge, red] (x1) to (y);
\draw[edge] (y) to (z);
\draw[edge] (z) to (w0);
\draw[edge] (z) to (w1);
\end{tikzpicture}
\end{minipage}
\end{center}

A major difference between reverse-mode and forward-mode that makes the implementation of reverse-mode much more complicated, is that we can only start the differentiation after the final output is computed. This makes it impossible to interleave the derivative code with the original code like in forward-mode. This issue has the most impact when differentiating programs with control flow or recursion. We discuss them in Chapter~\ref{sec:control_flow}.

\subsection{Beyond Forward and Reverse Modes}
\label{sec:beyond}

As we have discussed, forward-mode is efficient when the number of inputs is small, while reverse-mode is efficient when the number of outputs is small. When both the number of inputs and the number of outputs are large, and we are interested in the Jacobian or its subset, both forward and reverse modes can be inefficient.

In general, we can think of derivative computation as a pathfinding problem on the computational graph: We want to find all the paths that connect between inputs and outputs. Many of the paths share common subpaths and it is more computationally efficient to factor out the common subpaths. Forward-mode and reverse-mode are two different greedy approaches that factor out the common subpaths either from the input node or output node, and they can deliver suboptimal results that do not have the minimal amount of computation.

For the general Jacobian, finding the factorization that results in the minimal amount of computation is called the ``Jacobian accumulation problem'' and is proven to be NP-Hard~\cite{Naumann:2008:OJA}. However, there exist several heuristics (e.g.~\cite{Griewank:1991:OCJ, Naumann:1999:ECJ, Guenter:2007:ESD}). Usually, the heuristics use some form of a greedy approach to factor the node that is reused by the most paths. These heuristics can also be used for higher-order derivatives such as Hessian matrices~\cite{Gower:2012:NFC, Wang:2016:EPE}, since the Hessian is the Jacobian of the gradient vector with respect to the input dimensions.

\section{Automatic Differentiation as Program Transformation}

In this section, we discuss the practical implementation of automatic differentiation. Typically the implementation of automatic differentiation systems can be categorized as a point in a spectrum, depending on how much is done at compile-time. At one end of the spectrum, the \emph{tracing} approach, or sometimes called the \emph{taping} approach, re-compiles the derivatives whenever we evaluate the function. At the other end of the spectrum, the \emph{source transformation} approach does as much at compile-time as possible by compiling the derivative code only once. The tracing approach has the benefit of simpler implementation, and is easier to incorporate into existing code, while the source transformation approach has better performance, but usually can only handle a subset of a general-purpose language and is much more difficult to implement.

\paragraph{Tracing}
The tracing approach bears similarity to the tracing just-in-time compilation technique used by various interpreters. Tracing automatic differentiation usually records a linear sequence of the computation at run-time (usually called a \emph{tape} or Wengert list~\cite{Wengert:1964:SAD}). Typically, all the control flows will be flattened in the trace. The system then ``compiles'' the derivatives just-in-time by traversing the linear sequence. A typical implementation is to use operator overloading on a special floating point type, replacing all the elementary operations by the overloaded functions. The user is then required to replace all the floating point type occurences with the special type in their program, and call a compile function to start the differentiation.

Tracing is the most popular method for implementing general automatic differentiation systems. Most of the popular automatic differentiation systems use tracing (e.g. CppAD~\cite{Bell:2003:CPPAD}, ADOL-C~\cite{Griewank:1996:AAP}, Adept~\cite{Hogan:2014:FRA}, and Stan~\cite{Stan:2015}). However, tracing is inefficient due to the limited amount of work that can be done during the just-in-time differentiation. For example, if a function is linear, all of the derivatives of it are constant, however, tracing approaches often fail to perform constant folding optimization, since folding the constant at run-time is often more costly than just computing the constant. Metaprogramming techniques such as expression templates can help mitigate this issue~\cite{Hogan:2014:FRA, Sagebaum:2018:ETP}, but they cannot optimize across functions or even statements.

\paragraph{Source Transformation}
Another approach is to take the source code of some numerical program, and generate the code for the derivatives. It is also possible to build an abstract syntax tree using operator overloading, then generate derivative code from the tree (the systems in Chapter~\ref{chap:gradient_halide} and Chapter~\ref{chap:redner} used precisely this approach). This approach is much more efficient compared to tracing due to the number of optimizations that can be done at compile-time (constant folding, copy elision, common subexpression elimination, etc). However, it is more difficult to integrate into existing languages, and often can only handle a subset of the language features. For example, none of the existing source transformation methods is able to handle functions with arbitrary side effects.

In Chapter~\ref{sec:autodiff_algo} we already discussed general rules for handling elementary operations and function calls. A straightforward line to line syntax tree transformation should do the job. In the subsection below, we briefly discuss how source transformation can be done for programs with control flow including for loops and while loops, and how to handle recursion or cyclic function calls.

\subsection{Control Flow and Recursion}
\label{sec:control_flow}

Handling control flow and recursion in forward-mode is trivial. We do not need to modify the flow at all. Since forward-mode propagates from the inputs, for each statement, we can compute its derivative immediately after like we did in Chapter~\ref{sec:forward_mode}.

In reverse-mode, however, control flow and recursion introduce challenges, since we need to \emph{revert} the flow. Consider the iterative exponential example from Figure~\ref{fig:exp_example_code}. To apply reverse-mode, we need to revert the for loop. We observe an issue here: we need the intermediate \lstinline{exp(result)} values for the derivatives. To resolve this, we will need to record the intermediate values during the first pass of the loop:

\begin{center}
\begin{tabular}{c}
\begin{lstlisting}
function d_f(x):
    result = x
    @results = []@
    for i = 1 to 8:
        @results.push(result)@
        result = exp(result)
    
    @d_result = 1@
    for i = 8 to 1:
        // one-based indexing
        @d_result = d_result * exp(results[i])@
    return d_result 
\end{lstlisting}
\end{tabular}
\end{center}

The general strategy for transforming loops in reverse-mode is to push intermediate variables into a stack for each loop~\cite{Villard:1999:AIA}, then pop the items during the reverse loop. Nested loops can be handled in the same way. For efficient code generation, dependency analysis is often required to push only variables that will be used later to the stack (e.g.~\cite{Strout:2006:LAA}).

The same strategy of storing intermediate variables in a stack also works for loop continuations, early exits, and conditioned while loops. We can use the size of the stack as the termination criteria. For example, we modify the previous example to a while loop and highlight the derivative code in red:
\begin{center}
\begin{tabular}{c}
\begin{lstlisting}
function d_f(x):
    result = x
    @results = []@
    while result > 0.1 and result < 10:
        @results.push(result)@
        result = exp(result)

    @d_result = 1@
    for i = len(results) to 1:
        @d_result = d_result * exp(results[i])@
    return d_result
\end{lstlisting}
\end{tabular}
\end{center}

Recursion is equally or even more troublesome compared to control flow for reverse-mode. Consider the following tail recursion that represents the same function:
\begin{center}
\begin{tabular}{c}
\begin{lstlisting}
function f(x):
    if x <= 0.1 or x >= 10:
        return x
    result = f(exp(x))
    return result
\end{lstlisting}
\end{tabular}
\end{center}
It is tempting to use the reverse-mode rules we developed in Chapter~\ref{sec:reverse_mode} to differentiate the function like the following:
\begin{center}
\begin{tabular}{c}
\begin{lstlisting}
function d_f(x, d_result):
    if x <= 0.1 or x >= 10:
        return 1
    result = f(exp(x))
    return d_f(result, d_result) * exp(x)
\end{lstlisting}
\end{tabular}
\end{center}
However, a close inspection reveals that the generated derivative function \lstinline{d_f} has higher time complexity compared to the original function ($O(N^2)$ v.s. $O(N)$), since every time we call \lstinline{d_f} we will recompute \lstinline{f(exp(x))}, resulting in redundant computation.

A solution to this, similar to the case of loops, is to use the technique of memoization. We can cache the result of recursive function calls in a stack, and traverse the recursion tree in reverse by traversing the stack:
\begin{center}
\begin{tabular}{c}
\begin{lstlisting}
function d_f(x, d_result):
    if x <= 0.1 or x >= 10:
        return 1
    results = []
    @result = f(exp(x), results)@

    @d_result = 1@
    for i = len(results) to 1:
        @d_result = d_result * exp(results[i])@
    return d_result
\end{lstlisting}
\end{tabular}
\end{center}
This also works in the case where \lstinline{f} recursively calls itself several times. A possible implementation is to use a tree instead of a stack to store the intermediate results.

The transformations above reveal an issue with the reverse-mode approach. While for scalar output, reverse-mode is efficient in time complexity, it is not efficient in memory complexity, since the memory usage depends on the number of instructions, or the length of the loops. A classical optimization to reduce memory usage is called ``checkpointing''. The key idea is to only push to, or to \emph{checkpoint}, the intermediate variable stack sporadically, and recompute the loop from the closest checkpoint every time. Griewank~\cite{Griewank:1992:ALG} showed that by checkpointing only $O\left(\log\left(N\right)\right)$ times for a loop with length $O(N)$, we can achieve memory complexity of $O\left(\log\left(N\right)\right)$, and time complexity of $O\left(N\log\left(N\right)\right)$ for reverse-mode.

Higher-order derivatives can be obtained by successive applications of forward- and reverse-modes. Applying reverse-mode more than once can be difficult since the stack introduces side-effects (see Chapter~\ref{sec:further_readings} for more discussions). Furthermore, in the case of the Jacobian computation, it is difficult to devise transformation rules for control flow for methods beyond forward- and reverse-modes.

\section{Historical Remarks}
\label{sec:historical_remarks}

Automatic differentiation is perhaps one of the most rediscovered ideas in the scientific literature. Forward-mode is equivalent to the dual number algebra introduced in 1871~\cite{Clifford:1871:PSB}. The idea of reverse-mode was floating around in the 1960s (e.g.~\cite{Kelley:1960:GTO}), and most likely materialized first in 1970~\cite{Linnainmaa:1970:RCR} for estimating the rounding error of an algorithm, and was later applied to neural networks and rebranded as backpropagation~\cite{Werbos:1982:AAN, Rumelhart:1986:LIR}. In computer graphics, the field of animation control has a long history of using automatic differentiation. Witkin and Kass developed a Lisp-based system that can automatically generate derivatives for optimizing character animation~\cite{Witkin:1988:SC}. The field of optimal control theory, which is highly related to animation control, is also an early user of automatic differentiation. They take the differential equation perspective and usually call forward-mode ``tangent'' or ``sensitivity'' while calling reverse-mode ``adjoint''. One of the earlier large-scale usages of automatic differentiation is oceanography (e.g.~\cite{Marotzke:1999:CAM}), where the derivatives of fluid simulators are used for sensitivity and optimization studies. Due to the strong interest from the science and engineering communities, many early automatic differentiation tools are developed in Fortran (e.g. ADIFOR~\cite{Bischof:1992:AAD}, TAMC~\cite{Giering:1998:RAC}, OpenAD~\cite{Utke:2008:OMO}). See Griewank and Schmidhuber's articles~\cite{Griewank:2012:WIR, Schmidhuber:2015:WIB} for more remarks.

\section{Further Readings}
\label{sec:further_readings}

\paragraph{Deep learning frameworks} The core of deep learning is backpropagation, or equivalently reverse-mode automatic differentiation. There are several recent deep learning frameworks for implementing neural network architectures. Some of them are closer to the tracing approach~\cite{Paszke:2017:ADP}, while some of them are closer to the source transformation approach~\cite{Bergstra:2010:TCG, Dong:2014:ICN, Abadi:2015:TLM}. However, all of them only differentiate the code at a coarse-level of operators, while the operators (e.g. convolution, element-wise operations, pooling) and their gradients are implemented by experts. When the desired operation is easy to express by a few of these operators, these frameworks deliver efficient performance. However, for many novel operators, it is either inefficient or impossible to implement on top of these frameworks, and practitioners often end up implementing their own custom operators in C++ or CUDA, and derive the derivatives by hand. In Chapter~\ref{chap:gradient_halide} we discuss this in the context of image processing and deep learning.

\paragraph{Stochastic approximation of derivatives} In addition to finite differences and symbolic differentiation, one can also employ stochastic approximation to gradients or higher-order derivatives. Simultaneous perturbation stochastic approximation (SPSA)~\cite{Bhatnagar:2012:SRA} and evolution strategy~\cite{Beyer:2002:ESN} are two examples of this. Curvature propagation~\cite{Martens:2012:EHB} takes a similar idea to stochastically approximate Hessian matrix using exact gradients. These methods sidestep the time complexity of finite differences, at the cost of having variance on the derivatives depending on the local dimensionality of the function.

\paragraph{Nested applications of reverse-mode} An issue with the approach for handling control flow and recursion we introduced in Chapter~\ref{sec:control_flow} is that it does not form a \emph{closure}, that is, the derivative code that uses the stack cannot be differentiated again, since the stack introduces side-effects. Pearlmutter and Siskind~\cite{Pearlmutter:2008:RAF} propose a solution for this using Lambda calculus, by developing proper transformation rules in a side-effect free functional language, which produces closure. The generated code has similar performance to the stack approach, but has the benefit of supporting nested applications of reverse-mode. The resulting transformation is non-local (in contrast, the one we describe in Chapter~\ref{sec:autodiff_algo} is local), in the sense that the functions generated can be vastly different from the original ones. Recently, Shaikhha et al.~\cite{Shaikhha:2018:EDP} generalize Pearlmutter and Siskind's idea to handle array inputs in a functional language. Their current implementation does not generate vectorized code, but it is possible to further generalize their approach for better code generation.

\paragraph{Higher-order derivatives} For Hessian computation, Gower and Mello develop a reverse-mode-like algorithm that utilizes the symmetry and sparsity~\cite{Gower:2012:NFC}. It was later shown to be equivalent to one of the heursitics for computing Jacobian accumulation~\cite{Wang:2016:EPE}. Betancourt~\cite{Betancourt:2018:GTH} explores the connections between automatic differentiation and differential geometry, and develops algorithms for higher-order derivatives similar to Gower and Mello's method.

% \paragraph{Subderivatives}

% Write something about vector computation
\chapter[Derivative-based Optimization and Markov Chain Monte Carlo Sampling]{Derivative-based Optimization and \\ Markov Chain Monte Carlo Sampling}
\label{chap:optimization_sampling}
Most of the uses of derivatives in this dissertation are for optimizing or sampling a function. For optimization, we are interested in the \emph{mode} of a function, whereas for sampling we are interested in the \emph{statistics}, such as mean or variance. In this chapter, we briefly introduce classical methods that use derivatives for optimization and Markov chain sampling. This is a massive topic and it deserves multiple university courses. Therefore, this chapter is by no means a comprehensive introduction. I only discuss methods more relevant to the dissertation. Readers are encouraged to read textbooks from Boyd and Vandenberghe~\cite{Boyd:2004:CO}, Nocedal and Wright~\cite{Nocedal:2006:NO} (both for optimization), Brooks et al.~\cite{Brooks:2011:HMC} (for sampling, focus on Markov chain Monte Carlo methods), and Owen~\cite{Owen:2013:MCT} (for sampling, introduces various Monte Carlo integration methods).

Optimization and sampling have myriad applications across all fields of computational science. Optimization can be used for finding the parameters of a model given training input and output pairs, or solving inverse problems, where we want to find inputs that map to certain outputs. Markov chain sampling can be used for integrating light path contribution in physically-based rendering, characterizing posterior distributions in Bayesian statistics, or generating molecular structures for computational chemistry. We also discuss the relationship between optimization and sampling in Chapter~\ref{sec:relation_optimization_sampling}.

\section{Optimization}

Given a function $f(\mathbf{x}) : \mathbb{R}^n \rightarrow \mathbb{R}$, we are interested in finding an input $\mathbf{x}^{*}$ that minimizes the function. The function $f$ we want to minimize is often call the \emph{cost} function, \emph{loss} function, or \emph{energy} function, where the last term is borrowed from molecular dynamics. Formally, this is usually written as:
\begin{equation}
    \mathbf{x}^{*} = \argmin_{\mathbf{x}} f(\mathbf{x}).
\end{equation}

For example, we may want to recover an unknown pose $\mathbf{p}$ of a camera, such that when we pass it to a rendering function $r\left(\mathbf{p}\right)$, the output matches an observation image $\mathbf{I}$. We can define the loss function as the squared difference between the rendering output and the observed image:
\begin{equation}
    \mathbf{p}^* = \argmin_{\mathbf{p}} \sum_{i} \left\| r\left(\mathbf{p}\right) - \mathbf{I} \right\|^2.
    \label{eq:inverse_problem}
\end{equation}
The goal of optimization is then to find a camera pose $\mathbf{p}$ that renders an image similar to the observation $\mathbf{I}$. These problems are usually called \emph{inverse problems}, since we have a forward model $r$, and we are interested in inverting the model.

Another use case is when we have a sequence of example inputs $\mathbf{a}_i$ and outputs $\mathbf{b}_i$, and we want to learn a mapping between them. We can define the mapping as $g\left(\mathbf{a}_i; \Theta\right)$, where $\Theta$ is some set of parameters. We can then define the loss function as the difference between the mapped outputs and the example outputs:
\begin{equation}
    \Theta^* = \argmin_{\Theta} \sum_{i} \left\|g\left(\mathbf{a}_i; \Theta\right) - \mathbf{b}_i \right\|^2,
\end{equation}
and optimize the mapping parameters $\Theta$. In statistics, this is often called regression, while in machine learning this is called supervised learning, or empirical risk minimization.

Blindly searching for inputs or parameters that minimize the loss function is inefficient, especially when the dimension $n$ is high. Intuitively speaking, the space to search grows exponentially with respect to the dimensionality. Therefore, it is important to guide the search towards a direction that lowers the cost function. This is precisely what a gradient vector does. The gradient points in the direction where the function increases the most in the infinitesimal neighborhood. If we move along the negative gradient direction, we expect the cost function to decrease. This motivates our first optimization algorithm, gradient descent.

\subsection{Gradient Descent}

\begin{figure}[t]
  \centering
  \includegraphics[width=0.5\linewidth]{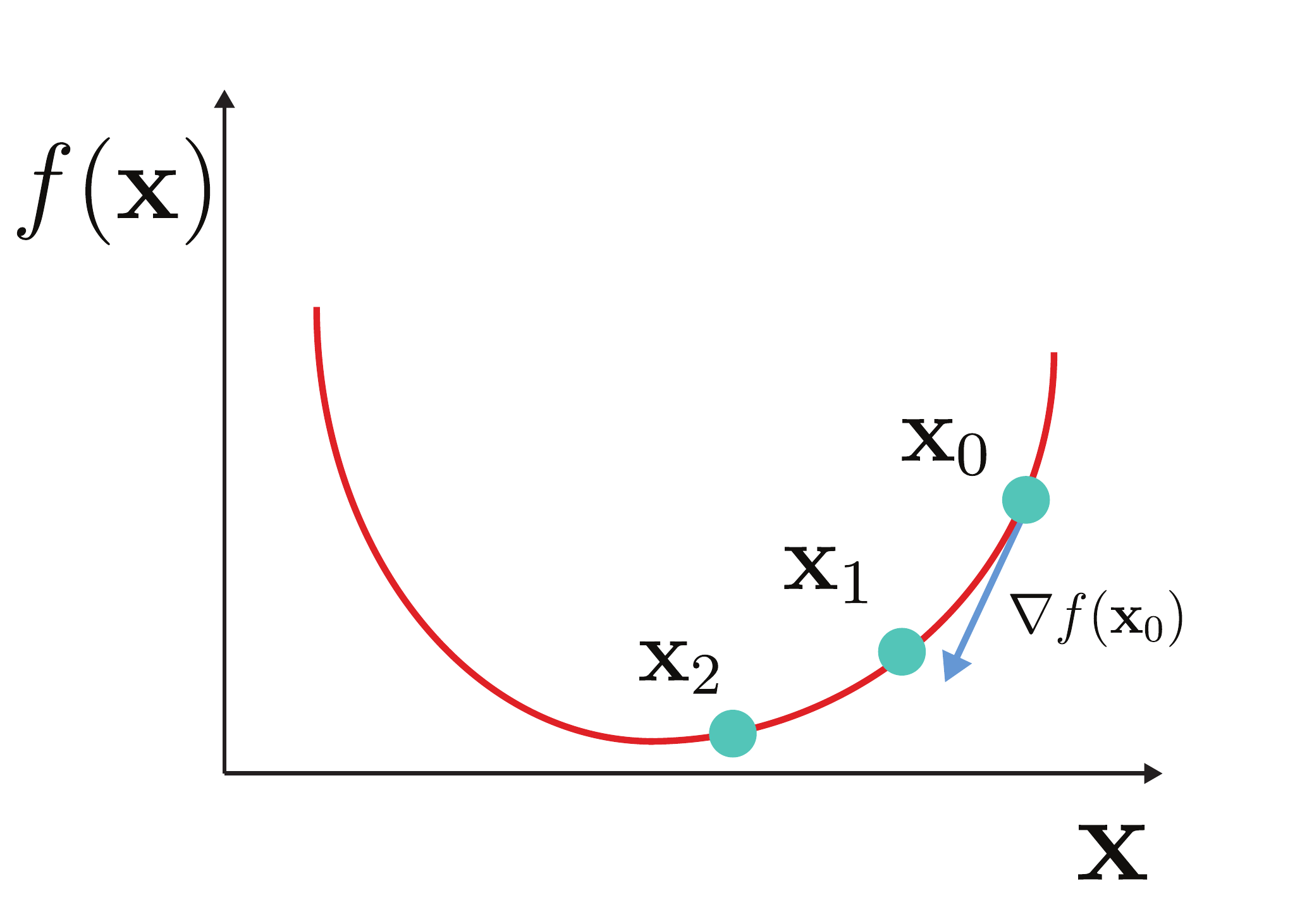}
  
  \caption{\label{fig:gradient_descent} Gradient descent minimizes a function by iteratively following the negative gradient direction.}
\end{figure}

The idea of gradient descent dates back to Cauchy~\cite{Cauchy:1847:MGP}. Figure~\ref{fig:gradient_descent} illustrates the process. For a loss function $f(\mathbf{x})$, starting from an initial guess $\mathbf{x}_0$, we iteratively refine the guess using the gradient $\nabla f(\mathbf{x})$:
\begin{equation}
\mathbf{x}_{i+1} = \mathbf{x}_{i} - \gamma \nabla f(\mathbf{x}_i),
\end{equation}
where $\gamma$ is the \emph{step size} parameter, sometimes called the \emph{learning rate}, which determines how far we move along the negative gradient direction. Choosing the right step size is difficult, as it usually depends on the smoothness of the cost function, and typically the best step size is different for each dimension.

Gradient descent and all optimization methods we introduce in this chapter are \emph{local} search methods. This means that they only find a local minimum of the function, while there can be a global minimum that is lower than the minimum they find.
% Maybe say something about the "eliminating bad local minima thing"

Without any assumption on the function $f$, there is no guarantee that gradient descent will converge even to a local minimum. For example, if we reach a saddle point, the gradient would be zero and the iteration would stop. The convergence rate of gradient descent depends on how \emph{convex} the function is (is it a ``bowl shape'' so that it only has a single minimum?), and whether it is \emph{Lipschitz continuous} (is there a bound on how fast the function is changing?). Curious readers can consult textbooks (e.g. ~\cite{Boyd:2004:CO}) for more convergence proofs.

\subsection{Stochastic Gradient Descent}
\label{sec:sgd}

In many applications, the gradient $\nabla f(\mathbf{x})$ we compute may not be fully accurate. For example, in regression, our cost function is a sum over example input-output pairs. If we have a huge database of example pairs, say one million, doing one step of gradient descent would require inefficiently enumerating all pairs of inputs and outputs. It would be desirable to randomly select a \emph{mini-batch} each time we perform a gradient descent step (say, four from the one million). Furthermore, sometimes in an inverse problem, our forward model itself is a stochastic approximation to an integral (e.g. the rendering function in Chapter~\ref{chap:redner}), and so is our loss function and gradients.

Fortunately, if our gradient approximation is \emph{unbiased} (the expectation is the same as the true gradient) or \emph{consistent} (the expectation converges to the true gradient if we use more samples), gradient descent can still converge to a local minimum~\cite{Robbins:1951:SAM, Chen:2018:SGD}. The condition for convergence is a gradually reducing step size $\gamma$ over iterations, or equivalently, an increasing number of samples for gradient approximation. Intuitively, the noise we introduce in the gradient approximation brings some randomness to the steps in the gradient descent iterations, but on average, they still go in the right direction. When we are closer to the optimum, the noise makes it harder to hit the exact optimum, so we either need to take smaller steps, or reduce the noise by increasing the number of samples.

In addition to computational efficiency, it is observed that the randomness can help stochastic gradient descent escapes from saddle points~\cite{Ge:2015:ESP}. The noise also acts as an \emph{early stopping} mechanism~\cite{Prechelt:1998:ESW, Hardt:2016:TFG}, which helps regression to generalize better to data not in the examples, thereby avoiding \emph{overfitting}. See Chapter~\ref{sec:relation_optimization_sampling} for more discussions on this, and the relationship between stochastic gradient descent and other sampling-based methods.

\subsection{Newton's Method}
Choosing the right step size for gradient descent methods is difficult and problem-dependent. Intuitively, for flat regions of cost functions, we want to choose a larger step size, while sharp regions require a smaller step size. Second-order derivatives are a good measure of how flat a function is: if the magnitude of the second-order derivatives is large, then the gradient is changing fast, so we should not take a large step.

In the 1D case, assuming the loss function always has positive second derivatives (which means it has a bowl shape or is convex), the update step of Newton's method is
\begin{equation}
x_{i+1} = x_{i} - \frac{f'(x_i)}{f''(x_i)},
\label{eq:one_d_newton}
\end{equation}
where we replace the step size $\gamma$ with the inverse of the second derivative.

To derive Newton's method for the multivariate case, let us expand the loss function using the second-order Taylor expansion around $\mathbf{x}_i$:
\begin{equation}
f(\mathbf{x}_i + \Delta_{\mathbf{x}}) \approx f(\mathbf{x}_i) + \nabla f(\mathbf{x}_i) \Delta_{\mathbf{x}} + \frac{1}{2}\Delta_{\mathbf{x}}^T \mathbf{H}(\mathbf{x}_i) \Delta_{\mathbf{x}},
\end{equation}
where $\mathbf{H}(\mathbf{x}_i)$ is the Hessian matrix. If we solve for the critical point of this approximation by taking the gradient of $\Delta_{\mathbf{x}}$ and setting it to zero, we arrive at an update rule:
\begin{equation}
\mathbf{x}_{i+1} = \mathbf{x}_{i} - \mathbf{H}(\mathbf{x}_i)^{-1} \nabla f(\mathbf{x}_i).
\end{equation}
Essentially we replace the division of the second derivative in Equation~\ref{eq:one_d_newton} by multiplication by the inverse of the Hessian matrix.

Newton's method can also be modified to work in a stochastic setting, where both the gradient and Hessian are an approximation to the true ones (e.g.~\cite{Roosta:2016:SNM, Roosta:2016:SNM2}).

Newton's method eliminates the need for choosing the step size, at the cost of several disadvantages. First, the critical point of the Taylor expansion is not necessarily the minimum: it is only the minimum when the Hessian matrix is positive definite (all eigenvalues are positive). Second, computing and inverting the Hessian is expensive in high-dimensional cases. Various methods address these issues. Quasi-Newton methods or Gauss-Newton methods approximate the Hessian using first-order derivatives. Hessian-free methods (e.g.~\cite{Martens:2010:DLH}) use the Hessian-vector product (much cheaper than full Hessian computation) to obtain second-order information. Some methods approximate the Hessian using its diagonal (e.g.~\cite{Martens:2012:EHB}). Adaptive gradient methods adjust the learning rate per dimension using the statistics of gradients from previous iterations.

Next, we will briefly introduce adaptive gradient methods, as we use them extensively in the following chapters. We will skip the discussions on Quasi-Newton, Gauss-Newton methods and others, since they are less relevant to this dissertation. 

\subsection{Adaptive Gradient Methods}
\label{sec:adaptive_gradient_method}

How do we assess the flatness of a function, or how fast the gradients are changing, without looking at the second-order derivatives? The idea is to look at previous gradient descent iterations. The magnitude of the gradients is often a good indicator: if the magnitude is large, the function is changing fast. \emph{Adagrad}~\cite{Duchi:2011:ASM} builds on this idea and uses the inverse of average gradient magnitude per dimension as the step size:
\begin{equation}
\mathbf{x}_{i+1} = \mathbf{x}_{i} - \frac{\gamma}{\sqrt{G_{i}^2 + \epsilon}} \circ \nabla f(\mathbf{x}_i),
\end{equation}
where $G_{i}^2$ is a vector of the \emph{sum} of the squared gradients at or before iteration $i$ (the \emph{second moment} of the gradient), the division and the $\circ$ here denote element-wise division and multiplication, and $\epsilon$ is a small number (say, $10^{-8}$) to prevent division by zero.

Adagrad tends to reduce the learning rate quite aggressively, since it keeps the sum of squared gradient instead of average. Also, the smoothness of a function may be significantly different during the course of optimization. A possible modification is to only keep track of recent squared gradients. This can be done by an exponential moving average update (sometimes called an infinite impulse response filter):
\begin{equation}
{G'_{i}}^2 = \alpha {G'_{i - 1}}^2 + (1 - \alpha) \nabla f(\mathbf{x}_i)^2,
\end{equation}
where $\alpha$ is the weight update parameter. This leads us to the \emph{RMSProp} method~\cite{Tieleman:2012:LDG}, which replaces the second moment $G^2$ with the exponential moving average ${G'}^2$:
\begin{equation}
\mathbf{x}_{i+1} = \mathbf{x}_{i} - \frac{\gamma}{\sqrt{{G'_{i}}^2 + \epsilon}} \circ \nabla f(\mathbf{x}_i).
\end{equation}
Finally, in the stochastic setting, the gradient can be noisy, and the exponential moving average can filter out the noise. Therefore, we can also apply the moving average to the gradient in addition to the second moment, maintaining its first moment $m_i$:
\begin{equation}
m_i = \beta m_{i - 1} + (1 - \beta) \nabla f(\mathbf{x}_i),
\end{equation}
where $\beta$ is another weight update parameter. We can then use this smoothed gradient for the update. This results in the most popular gradient-based optimization algorithm as of the time this dissertation is written, Adam~\cite{Kingma:2015:AMS}\footnote{I omit the bias correction for the moving average here for simplicity. See the original paper for more details.}:
\begin{equation}
\mathbf{x}_{i+1} = \mathbf{x}_{i} - \frac{\gamma}{\sqrt{{G'_{i}}^2 + \epsilon}} \circ m_i.
\end{equation}

In regression, an optimizer that achieves low error in the example pairs mapping is not necessarily going to be considered a \emph{good} optimizer. What matters more is generalization, that is, how does the mapping perform for pairs that are not in the examples. Per discussion in Chapter~\ref{sec:sgd} and ~\ref{sec:relation_optimization_sampling}, the noise in stochastic gradient descent sometimes acts as an early termination mechanism, making the loss function higher, but also making the mapping generalize better. This might also explain why most efforts on improving Adam recently are not replacing it for regression tasks. There are theories explaining the generalization behavior of stochastic gradient descent~\cite{Hardt:2016:TFG, Zhang:2017:TDL}. However, to the author's knowledge, so far no theory explains the differences in the generalization ability between different adaptive gradient methods.

There are many other variants of adaptive gradient methods, and they are still being actively developed. ADADELTA~\cite{Zeiler:2012:AAL} is an alternative that also keeps track of the second moment of the updates (in addition to just the gradient second moment). The moving average for the gradients in Adam is essentially the same as a popular acceleration method for gradient descent called momentum~\cite{Goh:2017:WMR}. Nesterov~\cite{Nesterov:1983:MUC} proposes an acceleration by extrapolating the momentum, achieving the same convergence rate as Newton's method in the convex and non-stochastic setting. It is also possible to incorporate Nesterov's method in Adam~\cite{Dozat:2016:INM}. Reddi et al.~\cite{Reddi:2018:OCA} study the convergence of Adam and find counterexamples in the convex setting where Adam does not converge. They propose a fix by using the maximum second moment. Maclaurin et al.~\cite{Maclaurin:2015:GHO} show that it is possible to optimize the hyperparameters of adaptive gradient descent methods by performing reverse-mode automatic differentiation on top of gradient descent. Stochastic Average Gradient~\cite{Schmidt:2017:MFS} and Stochastic Variance Reduced Gradient~\cite{Johnson:2013:ASG} focus on the mini-batch setting and perform variance reduction on the gradients by reusing previous mini-batches.

\section{Markov Chain Monte Carlo Sampling}
\label{sec:sampling}

\begin{figure}[t]
  \centering
  \begin{subfigure}[t]{0.32\linewidth}
    \includegraphics[width=\linewidth]{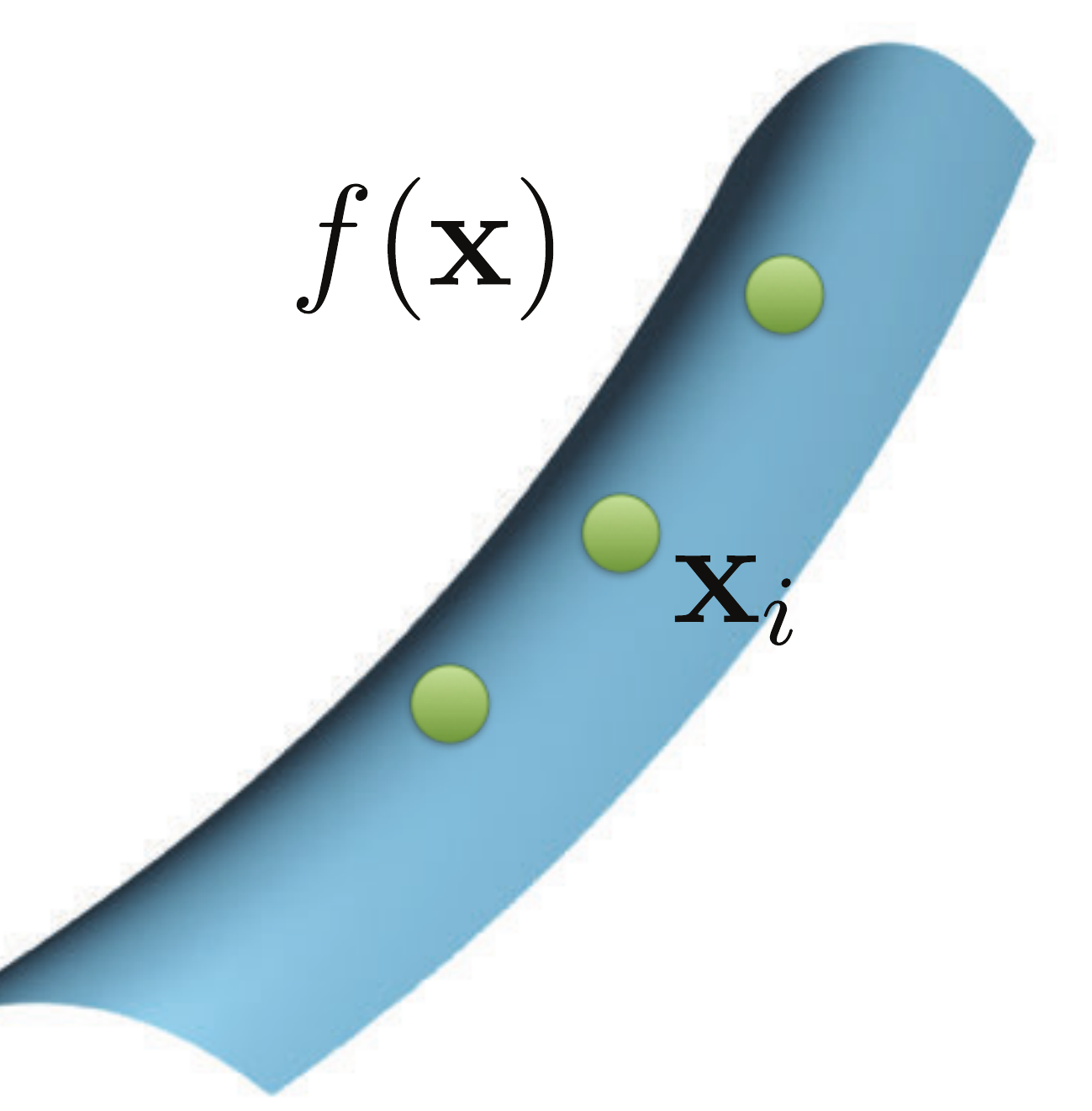}
    \caption{\label{fig:sampling} sampling}
  \end{subfigure}
  \begin{subfigure}[t]{0.32\linewidth}
    \includegraphics[width=\linewidth]{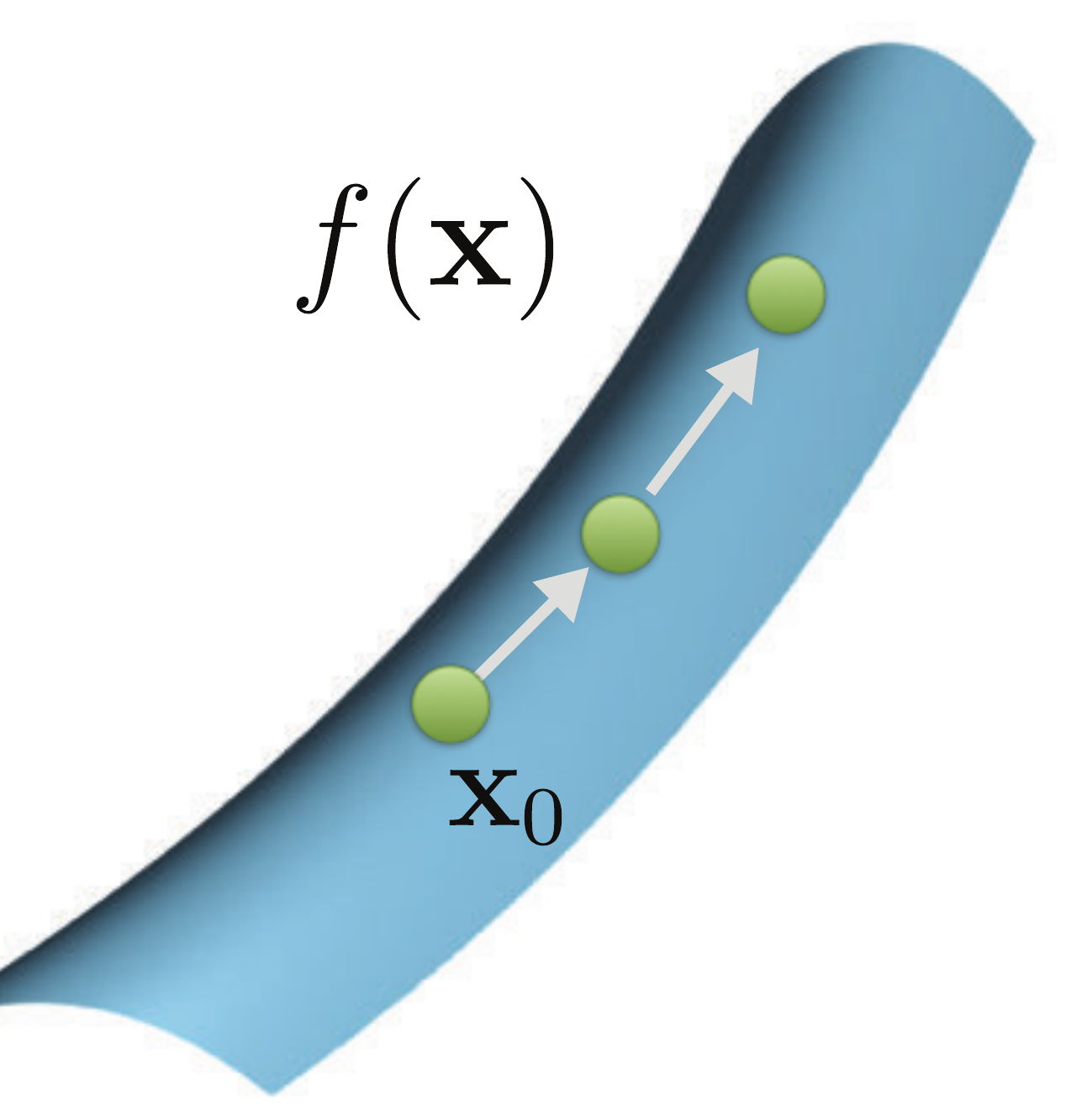}
    \caption{\label{fig:mcmc} Markov chain Monte Carlo}
  \end{subfigure}
  \begin{subfigure}[t]{0.32\linewidth}
    \includegraphics[width=\linewidth]{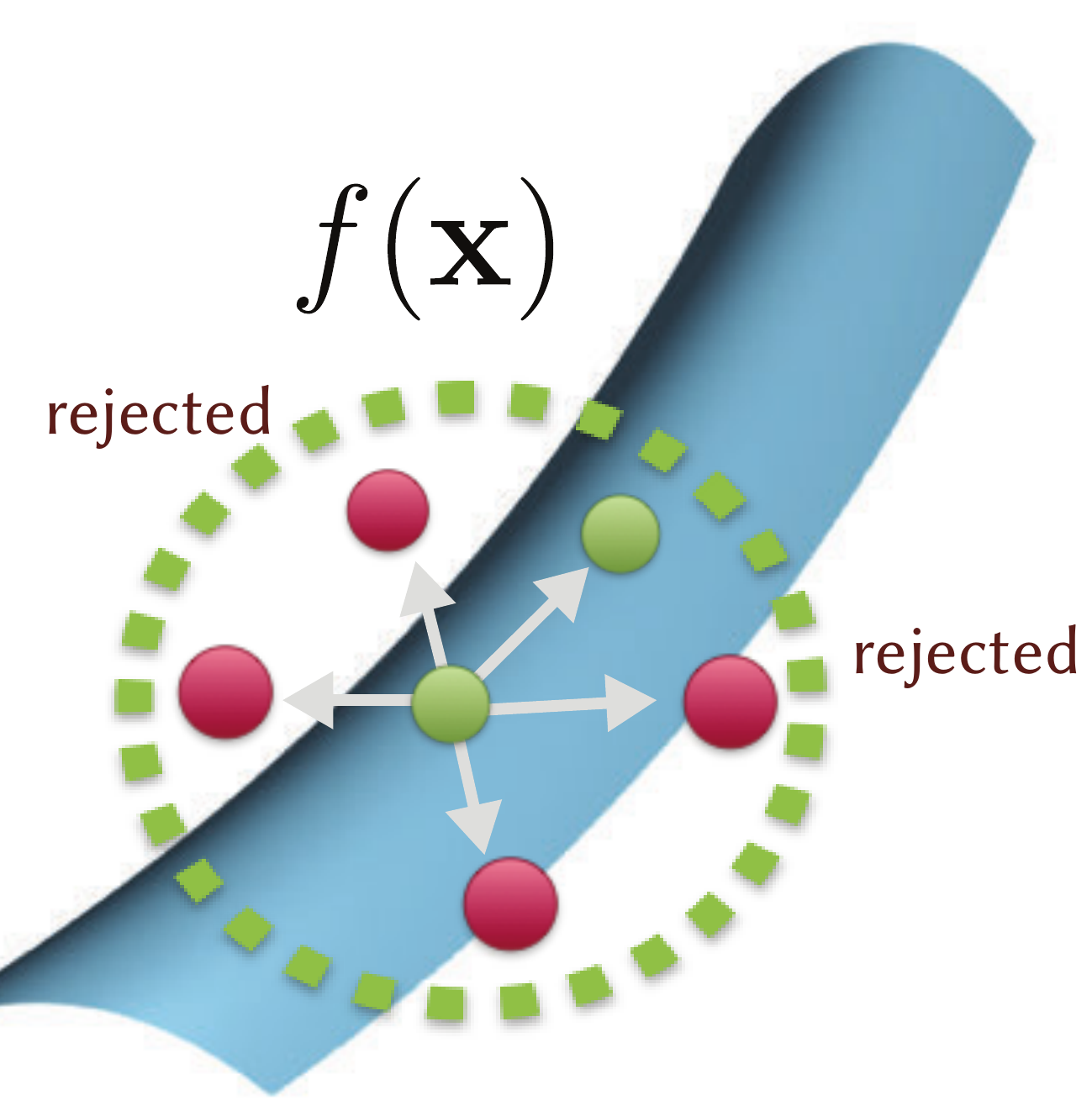}
    \caption{\label{fig:mcmc_proposal} proposal}
  \end{subfigure}

  \caption{\label{fig:sampling_goal} \textbf{Sampling.} (\subref{fig:sampling}) Given a function $f$ (2D in this case), the goal of sampling is to produce a set of samples $\mathbf{x}_i$ such that their distribution is proportional to $f$. (\subref{fig:mcmc}) Markov chain Monte Carlo samples from a function by generating a sequence of samples through a \emph{local} random walk. (\subref{fig:mcmc_proposal}) Each sample is generated from the previous one, and probabilistically rejected if the contribution is low, so that we have a higher probability of staying in high contribution regions.}
\end{figure}

In this section, we discuss sampling, an operation related to optimization. In contrast to optimization which is finding the \emph{mode} of a function, we are interested in the \emph{statistics} such as mean or variance. I focus most of the discussion on Markov chain Monte Carlo methods, since they are more related to the derivative-based scenario. For Monte Carlo integration in general, the reader can consult Owen's textbook for more information~\cite{Owen:2013:MCT}. For Monte Carlo integration for light transport, Veach's thesis~\cite{Veach:1998:RMC} and the textbook by Pharr et al.~\cite{Pharr:2016:PBR} are both excellent references.

Figure~\ref{fig:sampling_goal} illustrates our goal: Given a positive function $f(\mathbf{x}) : \mathbb{R}^n \rightarrow \mathbb{R}$, we want to generate a set of random samples $\mathbf{x}_i$, such that their probability densities $p(\mathbf{x}_i)$ are proportional to $f$:
\begin{align}
\mathbf{x}_i &\sim p(\mathbf{x}_i) \\
p(\mathbf{x}_i) &\propto f(\mathbf{x}_i).
\end{align}

This is highly related to the optimization problems: given a function $g$ to minimize, we can set $f = e^{-g}$ and sample from $f$, and pick the sample with highest $f$. Sampling has many uses in statistics, machine learning, and computer graphics. It is useful for estimating uncertainty: for example, every sample in Figure~\ref{fig:sampling_goal} achieves a high score, indicating that the problem is \emph{ill-posed}, in the sense that many inputs have an equally good loss. For both inverse problems and regression, sampling is also a more natural solution in high-dimensional space from a probabilistic viewpoint (Chapter~\ref{sec:relation_optimization_sampling}). Finally, in physically-based rendering, we estimate the total energy passing through each pixel by sampling light paths connecting light sources to the eye.

There are many ways to sample from a function, but most of them do not generalize to arbitrary functions. Inverse transform sampling requires us to integrate $f$ to obtain the probability density and cumulative density function, and then invert the cumulative density functions. Rejection sampling is typically inefficient in high-dimensional space and requires us to have an upper bound on $f$.

We focus on a specific method for sampling from a function -- the Markov chain Monte Carlo method. It does not assume much on the function it samples from, and more importantly, there are extensions of Markov chain Monte Carlo methods that make use of derivatives of the function, making the sampling more directed in high-dimensional space. The downside is it generates correlated samples that reduce sampling efficiency. % There is another promising alternative named Sequential Monte Carlo, or particle filters, and we will also briefly discuss it.

\subsection{Metropolis-Hastings Algorithm}

Markov chain Monte Carlo generates samples in a sequence, forming a \emph{Markov chain} (Figure~\ref{fig:mcmc}). That is, the generation of each sample only depends on the previous sample. This allows us to employ a local random walk strategy like the ones we used for optimization. To generate a new sample from the current one, we define a proposal distribution $Q(\mathbf{a} \rightarrow \mathbf{b})$, we then probabilistically accept or reject the proposal based on its contribution $f$ (Figure~\ref{fig:mcmc_proposal}). Overall the algorithm generates a sequence of samples $\mathbf{x}_t$ as follows:
\begin{enumerate}
    \item Propose a new sample $\mathbf{x}'$ from the current sample $\mathbf{x}_t$ according to the proposal distribution: $\mathbf{x}' \sim Q(\mathbf{x}_t \rightarrow \mathbf{x}')$.
    \item Compute acceptance probability $a(\mathbf{x}', \mathbf{x}_t) = min \left( 1, \frac{f\left(\mathbf{x}'\right)}{f\left(\mathbf{x}_t\right)} \frac{Q(\mathbf{x}' \rightarrow \mathbf{x}_t)}{Q(\mathbf{x}_i \rightarrow \mathbf{x}')} \right)$
    \item Set $\mathbf{x}_{t+1} = \mathbf{x}'$ if accepted, otherwise $\mathbf{x}_{t+1} = \mathbf{x}_{t}$.
\end{enumerate}

The algorithm was developed by Metropolis~\cite{Metropolis:1953:ESC} for symmetric proposal distributions, later extended by Hastings~\cite{Hastings:1970:MCS} for handling asymmetric proposals, and extended again by Green~\cite{Green:1995:RJM} for handling spaces of varying dimensions. Intuitively speaking, this algorithm allows us to put more samples in the high contribution regions, while having non zero probability of visiting all of $f$'s domain. Below we provide a sketch of proof explaining why the sequence $\mathbf{x}_t$ is distributed proportionally to $f$.

It is easier to explain Markov chains in the discrete state space. Let us for now assume $\mathbf{x}_t$ represents a positive integer and $f$ maps from $\mathbb{N}$ to $\mathbb{R}$. Our transition distribution $Q$ becomes a matrix $Q_{ij}$ representing the probability to transition from $i$ to $j$, and the acceptance probability is also a matrix $a_{ij}$. All the statements below naturally generalize to continuous state space.

First, we need to define the concept of a \emph{stationary distribution}. We represent our current sample distribution as a probability mass function vector $\pi^t$. Each iteration in the Markov chain, is essentially transforming the probability mass function:
\begin{equation}
K \pi^{t}  = \pi^{t+1},
\end{equation}
where $K$ is the \emph{kernel} matrix of the Markov chain. We say $\pi$ is a stationary distribution of the kernel $K$ if $K\pi = \pi$. In other words, $\pi$ is a fixed point of the kernel $K$, or $\pi$ is the eigenvector corresponding to eigenvalue $1$. If a Markov chain is \emph{ergodic}, that is, after enough transitions, a state has a non-zero probability of reaching all states, then it has a unique stationary distribution. This means that, in the limit, any distribution will converge to the stationary distribution after enough iterations.

Next, we define the \emph{detailed balance} condition. A Markov chain with kernel $K$ and a distribution $\pi$ is said to satisfy the detailed balance condition if:
\begin{equation}
K_{ij} \pi_{i} = K_{ji} \pi_{j} \text{ } \forall i, j,
\label{eq:detailed_balance}
\end{equation}
where the kernel matrix $K_{ij}$ describes the probability of state $i$ transitioning to state $j$. Intuitively, this means that the probability of transitioning from $i$ to $j$ is the same as from $j$ to $i$. A kernel satisfying detailed balance implies that it has a stationary distribution, but a kernel with a stationary distribution does not necessarily satisfy detailed balance. This can be observed by summing over $j$ in Equation~\ref{eq:detailed_balance}:
\begin{equation}
\sum_{j} K_{ij} \pi_{i} = \pi_{i} = \sum_{j} K_{ji} \pi_{j},
\end{equation}
where the first equation comes from $\sum_{j} K_{ij} = 1$ since the probabilities of state transition sum to one.

Finally, we show that the kernel specified by the Metropolis-Hastings algorithm satisfies detailed balance for the distribution proportional to $f$. Therefore, as long as the transition distribution $T$ is ergodic, it will converge to the right solution. We observe that state $i$ transitions to state $j$ with probability $T_{ij} a_{ij}$ (accept), and stays in $i$ with $\sum_{j} T_{ij} (1 - a_{ij})$ (reject). Hence the kernel $K$ is:
\begin{equation}
K_{ij} = \begin{cases}
    T_{ij} a_{ij},& \text{if } i \neq j \\
    T_{ii} a_{ii} + \sum_{j} T_{ij} (1 - a_{ij}), & \text{i = j}
\end{cases}.
\end{equation}
By substituting $a_{ij} = min\left(1, \frac{f_j T_{ji}}{f_i T_{ij}} \right)$ into the kernel $K_{ij}$, and applying some algebra, it can be shown that $K_{ij} f_i = K_{ji} f_j$.

The same proof also applies to the continuous state space by replacing all sums with integrals.

While Metropolis-Hastings generates a correct distribution in the limit, the rate it reaches that limit can vary (usually called the \emph{mixing rate}). The success of Markov chain Monte Carlo methods depends on the transition kernels. If most proposals are rejected, we stay in the same state and waste many samples. On the other hand, even if all the proposals are accepted, if we do not move away enough from the starting position to explore the state space, we still get a bad mixing rate.

Similarly, in the optimization case, blindly moving samples around (say, using an isotropic Gaussian distribution as proposal distribution) can be inefficient, especially in the high-dimensional case and when the contribution function is sparse. Below we discuss two variants of Markov chain Monte Carlo methods that use derivatives to improve the mixing rate -- Langevin Monte Carlo and Hamiltonian Monte Carlo.

\subsection{Langevin Monte Carlo}

\begin{figure}[t]
  \centering
  \includegraphics[width=0.32\linewidth]{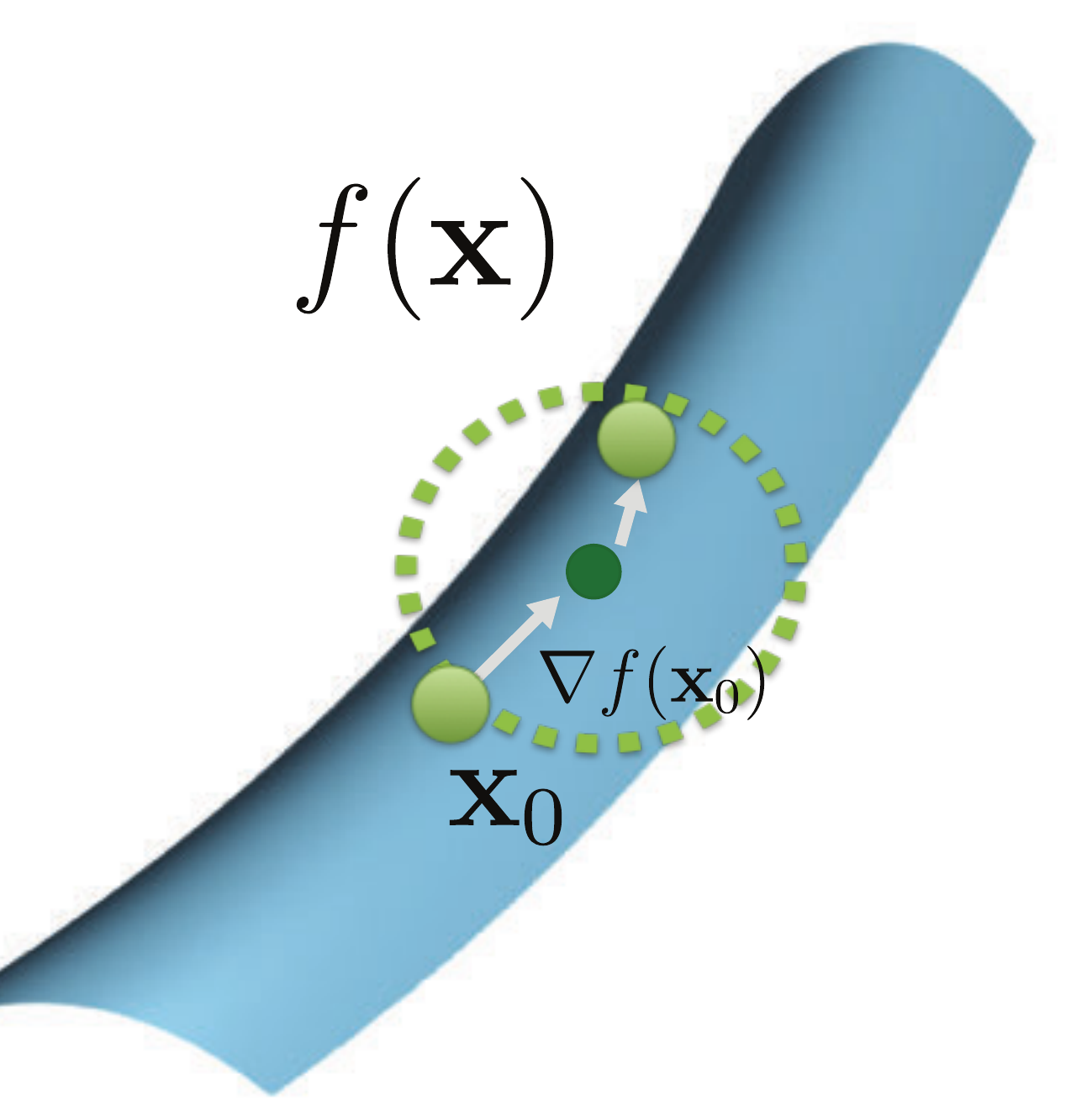}

  \caption{\label{fig:langevin} Langevin Monte Carlo, or the Metropolis-adjusted Langevin Algorithm~\cite{Roberts:1996:ECL} follows the gradient field by generating proposals from an isotropic Gaussian distribution, whose mean is shifted by the gradient of the sampling function.}
\end{figure}

Langevin Monte Carlo, or the Metropolis-adjusted Langevin Algorithm~\cite{Roberts:1996:ECL}, while derived from Langevin dynamics for describing the behavior of molecules, has a pretty simple intuition: it follows the gradient flow by using a proposal distribution whose center is shifted by the gradient (Figure~\ref{fig:langevin}). Formally the transition from state $\mathbf{x}$ to state $\mathbf{y}$ is:
\begin{equation}
    T(\mathbf{x} \rightarrow \mathbf{y}) \sim \mathcal{N}(-\gamma \nabla f(\mathbf{x}), \sigma^2 I),
\end{equation}
where $\mathcal{N}$ is the normal distribution, $\gamma$ is the step size or learning rate, $\sigma$ is a scalar standard deviation, and $I$ is the identity matrix.

This simple modification already brings significant benefits. Roberts and Rosenthal~\cite{Roberts:1998:OSD} show that if $f(\mathbf{x})$ is high-dimensional (say, more than $5$) and \emph{separable} (the dimensions of the input $\mathbf{x}$ are independent to each other), then the \emph{optimal} acceptance rate of Langevin Monte Carlo is around $57\%$, while the optimal acceptance rate of the Metropolis algorithm using isotropic Gaussian is around $23\%$. This means that the sampling efficiency of Langevin Monte Carlo is much better than zero-mean isotropic Gaussian in this case. Langevin Monte Carlo also produces less correlated samples. For $d$-dimensional separable functions, the expected number of samples needed to reach a nearly independent point grows as $d^{\frac{4}{3}}$, where when using isotropic Gaussian the number grows as $d^2$~\cite{Neal:2010:MUH}.

\subsection{Hamiltonian Monte Carlo}
\label{sec:hmc_background}

\begin{figure}[t]
  \centering
  \begin{subfigure}[t]{0.32\linewidth}
    \includegraphics[width=\linewidth]{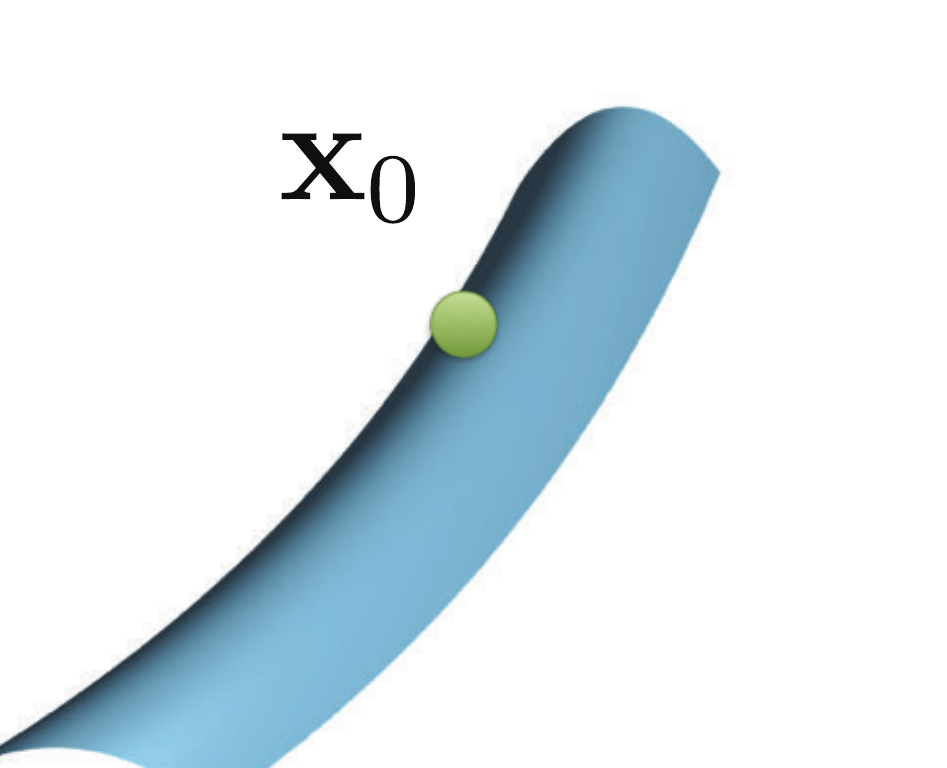}
    \caption{\label{fig:hmc_orig} original landscape}
  \end{subfigure}
  \begin{subfigure}[t]{0.32\linewidth}
    \includegraphics[width=\linewidth]{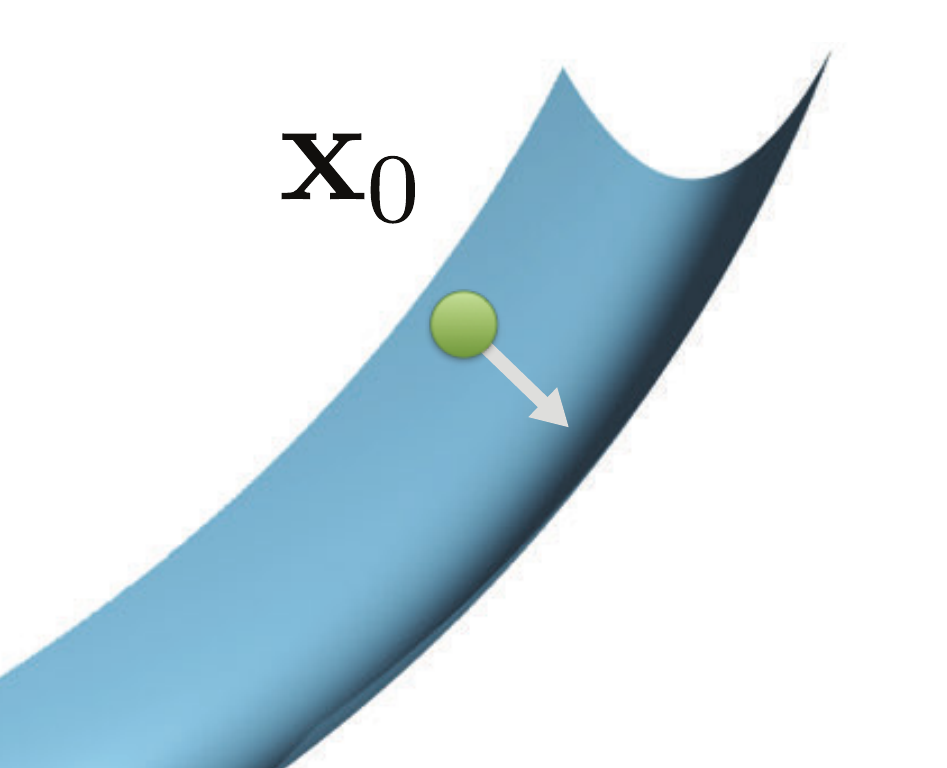}
    \caption{\label{fig:hmc_init} flip, random velocity}
  \end{subfigure}
  \begin{subfigure}[t]{0.32\linewidth}
    \includegraphics[width=\linewidth]{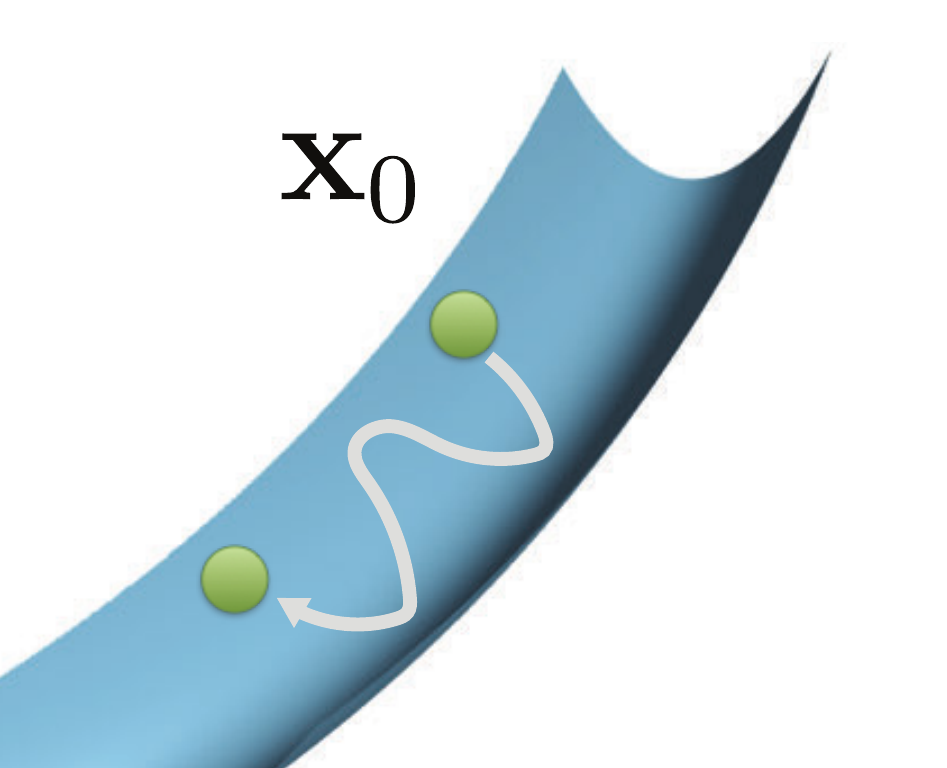}
    \caption{\label{fig:hmc_traj} simulate dynamics}
  \end{subfigure}

  \caption{\label{fig:hmc} Hamiltonian Monte Carlo~\cite{Duane:1987:HMC} takes the sampling function (\subref{fig:hmc_orig}) and flips it upside down (\subref{fig:hmc_init}). It then assigns a random velocity to the sample and simulates physics to let gravity pull the balls and stop at some fixed time (\subref{fig:hmc_traj}).}
\end{figure}

Hamiltonian Monte Carlo~\cite{Duane:1987:HMC} takes the idea of following gradient flow further, blurring the distinction between sampling and optimization. Figure~\ref{fig:hmc} illustrates a mental model for Hamiltonian Monte Carlo: we first flip the landscape of $f$ upside down, moving high contribution regions to the lower ground. Assuming the current sample is a rigid ball, we assign a random initial velocity to the ball, and simulate physics to let gravity pull the ball towards the lower ground, which are the higher contribution regions since we flipped $f$. In Chapter~\ref{chap:h2mc} we build on this idea to develop a Markov chain Monte Carlo rendering algorithm. We will provide a more formal introduction of Hamiltonian Monte Carlo and related work there.

The result of this physics simulation is that we follow the gradient field of the function guided by a momentum (similar to the gradient descent momentum we discussed in Chapter~\ref{sec:adaptive_gradient_method}). Comparing to Langevin Monte Carlo, Hamiltonian Monte Carlo reduces the randomness by only accepting or rejecting a sample after a fixed time. It scales even better with dimensionality ($d^{\frac{5}{4}}$ for $d$-dimensional separable functions) and also has a better optimal acceptance rate ($65\%$ for high-dimensional separable functions). The downside is it needs to discretize the physics simulation into multiple timesteps, making the cost of generating a sample very high. In fact, Langevin Monte Carlo is a special case of Hamiltonian Monte Carlo with only a single time step. The method we develop in Chapter~\ref{chap:h2mc} combines the benefits of Hamiltonian Monte Carlo and Lanvegin Monte Carlo by simulating Hamiltonian dynamics in a local neighborhood, while potentially sacrificing some benefits of the reduced randomness.

\subsection{Stochastic Langevin or Hamiltonian Monte Carlo}

Gradient descent works when the gradient is a stochastic approximation. Fortunately, Langevin Monte Carlo and Hamiltonian Monte Carlo also work when gradients are stochastic~\cite{Neal:2010:MUH, Welling:2011:BLS, Chen:2014:SGH}. It may seem trivial, since any proposal distribution that satisfies detailed balance should converge (the gradients do not even need to be unbiased or consistent). However, especially in the case of Hamiltonian Monte Carlo, to ensure a good convergence rate, care has to be taken to balance the dynamics to counter the noise injected in the trajectory~\cite{Chen:2014:SGH}.

\section{Relation between Optimization and Sampling}
\label{sec:relation_optimization_sampling}

From a probabilistic viewpoint, we can view optimization as trying to find a point that is maximizing the probability density distribution. For example, if we use a squared loss $f(\mathbf{x}) = \left| g(\mathbf{x}) - \mathbf{y} \right|^2$, we can see this as finding the mode of a normal distribution centered around $\mathbf{y}$ (more precisely, the density is $p\left(g\left(\mathbf{x}\right)\right) \propto e^{\frac{-f(\mathbf{x})}{\sigma^2}}$ for some standard deviation $\sigma$ representing the uncertainty). If we use the absolute difference, often called $L^1$ loss, it corresponds to the Laplace distribution. In contrast to optimization, sampling algorithms try to \emph{sample} from the Gaussian distribution, so points with higher density are more likely to be sampled.

Recently, researchers have started to explore the relationship between Markov chain Monte Carlo sampling and various gradient descent methods. In certain non-convex settings for optimization, Markov chain Monte Carlo methods, when used for optimization, have a faster convergence rate than gradient-based optimization algorithms~\cite{Ma:2018:SCF}. On the other hand, some variants of Langevin Monte Carlo always accept proposals~\cite{Welling:2011:BLS}, making the methods resemble gradient-based optimization more. There are many similar parallel developments between the sampling and optimization literature. Hamiltonian Monte Carlo's introduction of momentum to Langevin Monte Carlo is similar to the momentum in gradient descent. Gibbs' sampling~\cite{Geman:1984:SRG} is similar to coordinate descent by treating only a subset of input variables at a time. Riemannian Manifold Langevin and Hamiltonian Monte Carlo~\cite{Girolami:2011:RML} introduces the second-order derivatives similar to Newton's method or natural gradient method~\cite{Amari:1997:NLS}. It is fair to expect sampling and optimization algorithms to converge in the future, making it unnecessary to distinguish between them.
% https://arxiv.org/abs/1609.04388?
 
The connection between sampling and the generalizing effect of stochastic gradient descent~\cite{Hardt:2016:TFG, Mandt:2017:SGD} is also worth noting. In a high-dimensional space, most of the mass of distribution does not distribute around its mode~\cite{Carpenter:2017:TSC}. The intuition is that it is very unlikely that a person has the average height, average weight, average size of eyes and mouth, average length of arms and legs. Therefore, probabilistically, it makes little sense to find the exact minimum in high-dimensional space, since the minimum is not representative of the distribution. When we sample from a high-dimensional distribution, most of the samples would not be around the mode, but have a small distance to it. This is exactly what the noise in stochastic gradient descent is doing: in a practical number of iterations, it makes the optimization miss the exact minimum, but end up in a position having a small distance to the minimum. In effect this allows stochastic gradient descent to achieve better generalization, since they find a more typical instance of the probability distribution.

\newcommand{\figref}[1]{Figure~\ref{#1}}
\newcommand{\secref}[1]{Chapter~\ref{#1}}

\lstdefinestyle{Halide}{
    language=C++,
    basicstyle=\scriptsize\fontfamily{SourceCodePro-TLF}\selectfont,
    breaklines=true,
    showstringspaces=false,
    keywordstyle=\color{green!40!black},
    commentstyle=\itshape\color{grey},
    numberstyle=\color{blue},
    morekeywords={Func,Expr,Var,RDom,Param,Buffer},
    keepspaces=true
}

\newcommand{\notation}[1]{\ensuremath{#1}\xspace}
\newcommand{\Loss}{\notation{\mathcal{L}}}

\newfloat{lstfloat}{ht}{lop}
\floatname{lstfloat}{Listing}
\def\lstfloatautorefname{Listing}

\chapter[Differentiable Image Processing and Deep Learning in Halide]{Differentiable Image Processing and \\ Deep Learning in Halide}

\label{chap:gradient_halide}
\begin{figure}[h]
  \centering
  \vspace{-2.5em}
  \captionsetup[subfigure]{justification=centering}
  \begin{subfigure}[t]{0.3\linewidth}
    \includegraphics[width=\linewidth]{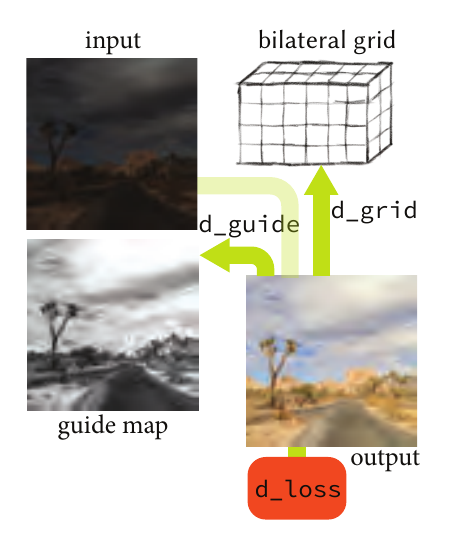}
    \caption{\label{fig:custom_nn} \footnotesize neural network operator: \\ bilateral slicing}
  \end{subfigure}
  \begin{subfigure}[t]{0.3\linewidth}
    \includegraphics[width=\linewidth]{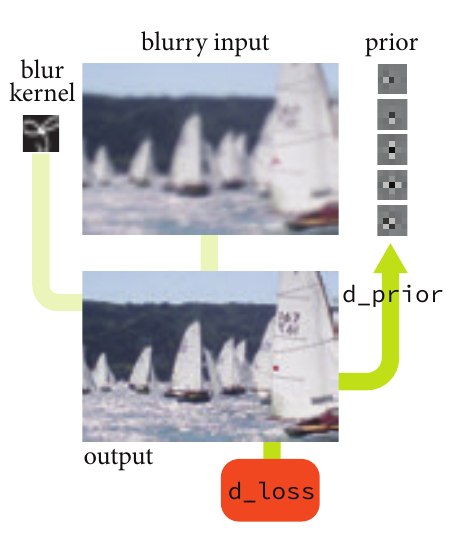}
    \caption{\label{fig:pipeline_tuning} \footnotesize optimizing the parameters of a \emph{forward} image processing pipeline}
  \end{subfigure}
  \begin{subfigure}[t]{0.36\linewidth}
    \includegraphics[width=\linewidth]{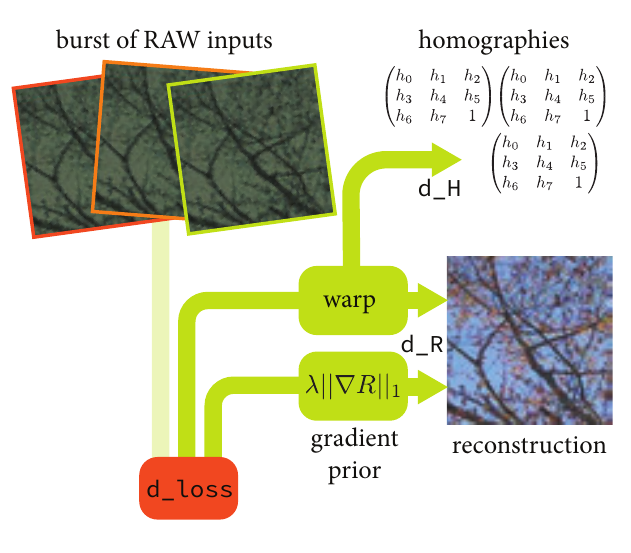}
    \caption{\label{fig:inverse_problem} \footnotesize optimizing the reconstruction and warping parameters of an \emph{inverse} problem}
  \end{subfigure}
  \caption{\label{fig:gradient_halide_teaser} \textbf{Differentiable image processing.} Our system automatically derives and optimizes gradient code for general image processing pipelines, and yields state-of-the-art performance on both CPUs and GPUs. This enables a variety of imaging applications, from training custom neural network layers (a), to optimizing the parameters of traditional image processing pipelines (b), to solving inverse problems (c). To support these applications, we extend the Halide language to automatically and efficiently compute gradients. We also introduce a new automatic performance optimization that can handle the specific computation patterns of the gradient. Using our system, a user can easily write high-level image processing algorithms, and then automatically derive high-performance gradient code for CPUs, GPUs, and other architectures. Images from left to right are from MIT5k dataset~\protect\cite{Bychkovsky:2011:LPG}, ImageNet~\protect\cite{Deng:2009:ILS}, and deep demosaicking dataset~\protect\cite{Gharbi:2016:DJD}, respectively.}
\end{figure}

Optimization and end-to-end learning are driving rapid progress in graphics and imaging, by viewing either the output image or large sets of pipeline parameters as unknowns, e.g.~\cite{Heide:2014:FFC, Jaderberg:2015:STN, Barron:2016:FBS, Gharbi:2017:DBL}. Key to this progress is the surprising power of gradient-based optimization methods to find solutions to nonlinear objectives over large sets of unknowns. Unfortunately, the computation of gradients remains a challenge in the general case, especially when performance is paramount such as for training neural networks or when solving for images via optimization. In Chapter~\ref{chap:autodiff}, we discussed methods for generating derivative code from programs, but they are not designed for image processing programs, and they do not take parallelism and locality into consideration. Typically, practitioners have to either manually derive gradients or they are limited to the composition of building blocks offered by deep learning libraries. The result is often inefficient, and when users decide to stray from existing operators, the implementation of fast GPU derivative code is a major undertaking.

At first glance, modern machine learning frameworks like PyTorch~\cite{Paszke:2017:ADP}, TensorFlow~\cite{Abadi:2015:TLM} or CNTK~\cite{Dong:2014:ICN} seem like appealing environments for new gradient-based graphics algorithms. When limited to their walled-gardens of pre-made, coarse-grained operations, these frameworks provide high-performance kernel implementations and automatic differentiation through chains of operations. As general programming languages, however, they are a poor fit for many imaging applications. Building new algorithms requires contorting a problem into complex and tangled compositions of existing building blocks. Even when done successfully, the resulting implementation is often both slow and memory-inefficient, saving and reloading entire arrays of intermediate results between each step, causing costly cache misses.

\begin{figure}[t]
  \includegraphics[width=\linewidth]{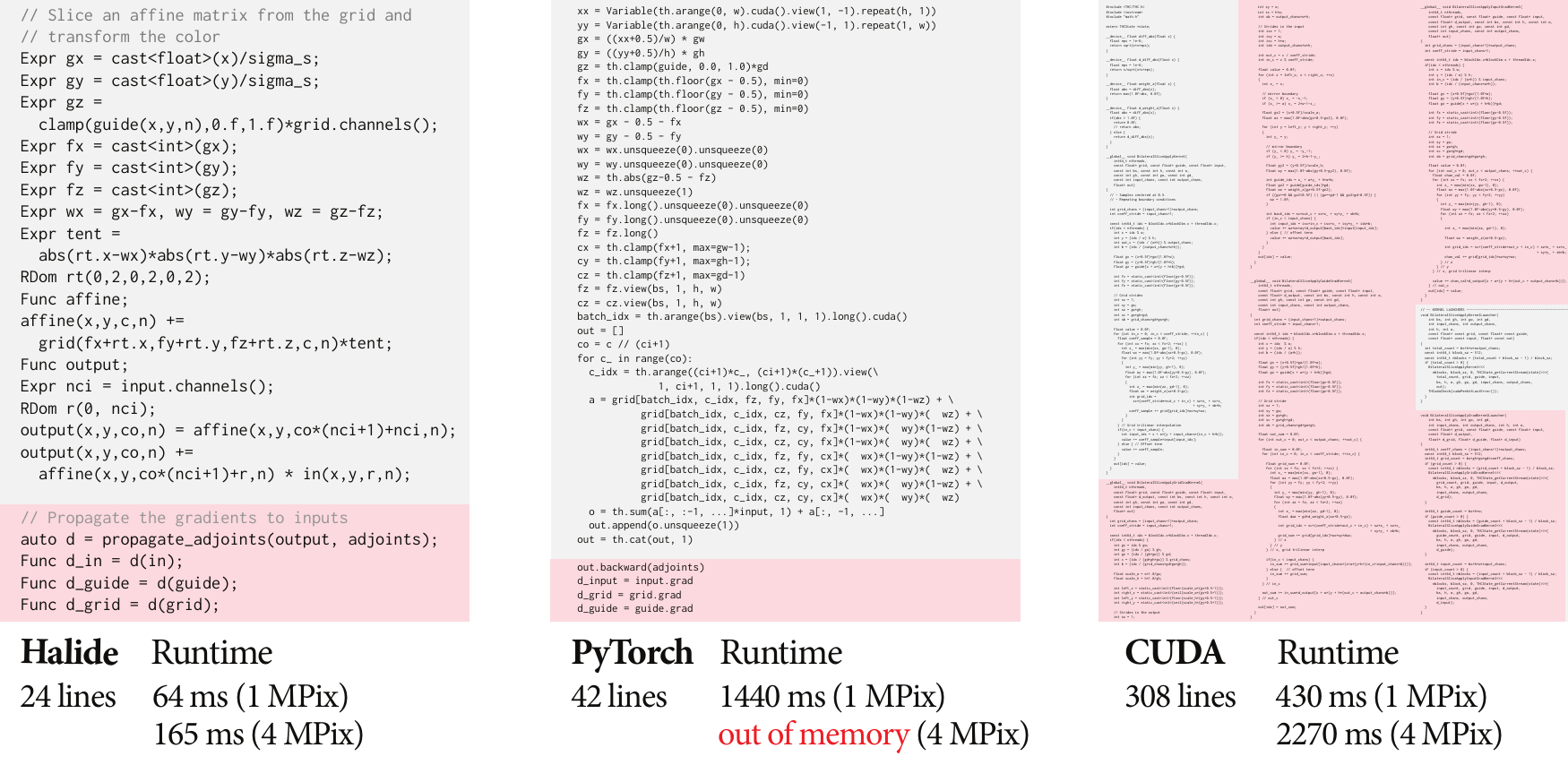}
  \caption{\textbf{Code comparison.} Implementations of the forward and gradient computations of the bilateral slicing layer~\cite{Gharbi:2017:DBL} in Halide, PyTorch, and CUDA\@. Using our automatic differentiation and scheduling extensions, the Halide implementation is clear, concise, and fast. The PyTorch implementation is modestly more complex, but runs $20\times$ slower on a $1k\times1k$ input, fails to complete (out of memory on a 12GB NVIDIA Titan Xp) on a $2k\times2k$ input, and is only possible thanks to new operators added to PyTorch since the original publication. The CUDA implementation, developed by the original authors, is not only complex (an order of magnitude larger than either Halide or PyTorch), but is dominated by hand-derived gradient computations. It is faster than PyTorch and scales to larger inputs, but is still about $10\times$ slower than the Halide version. Note: code size includes a few lines beyond the core logic shown for both Halide and PyTorch.
  }
  \label{fig:code_comparison}
\end{figure}

Consider the following example. A recent neural network-based operator for approximating image processing algorithms was built around a new ``bilateral slicing'' layer based on the bilateral grid~\cite{Gharbi:2017:DBL, Chen:2007:REI}. At the time it was published, neither PyTorch nor TensorFlow was even capable of practically expressing this computation.\footnote{Technically, TensorFlow graphs are Turing-complete, thanks to their inclusion of a while loop node. However, implementing the algorithm at this level would be both incredibly complex and run at least thousands of times slower.} As a result, the authors had to define an entirely new operator, written by hand in about 100 lines of CUDA for the forward pass and 200 lines more for its manually-derived gradient (\figref{fig:code_comparison}, right). This was a sizeable programming task which took significant time and expertise. While new operations now make it possible to implement this operation in 42 lines of PyTorch, this yields less than 1/3rd the performance on small inputs and runs out of memory on realistically-sized images (\figref{fig:code_comparison}, middle). The challenge of efficiently deriving and computing gradients for custom nodes remains a serious obstacle to deep learning.

This pattern is ubiquitous. New custom nodes require major effort to implement correctly and efficiently, making it hard to experiment. Similarly, general image processing pipelines often do not map well to deep learning toolboxes. As a result, most researchers limit themselves to consider only operations which are already well-supported by existing frameworks, while NVIDIA and the framework developers must constantly expand the set of native operations. The only alternative is to invest orders of magnitude more effort in developing custom operations, hand-deriving, reimplementing, and debugging gradient code for every change during the development of a new algorithm.

Recently, the Halide domain-specific language~\cite{Ragan-Kelley:2012:DAS, Ragan-Kelley:2013:HLC} has enabled the implementation of high-performance image-processing pipelines. It is an effective solution to implementing custom nodes and general image processing pipelines, but it still requires the manual derivation of gradients. Furthermore, our experience shows that the computation pattern of derivatives differs from that of forward code, which causes existing automatic performance optimizations in Halide to fail. Critically, the current built-in Halide autoscheduler does not support GPU schedules. 

We extend Halide with methods to automatically and efficiently compute the gradients of arbitrary Halide programs using reverse-mode automatic differentiation (\secref{sec:gradient_halide_method}). This transformation supports most existing features in the language, except for a few cases where side-effects are introduced (Section~\ref{sec:partial_update}). 

Building atop Halide has several advantages. It provides a concise, natural language in which to express image processing computations, and for which there is already a library of existing algorithms. The Halide compiler portably targets numerous processor and accelerator architectures, from mobile CPUs, to image processing DSPs, to data center GPUs, and supports compilation to very high-performance code. Finally, Halide's existing language and scheduling constructs compose with reverse-mode automatic differentiation to naturally express and generalize essential optimizations from the traditional automatic differentiation literature (\secref{sec:checkpointing}). Keys to making our compiler transformation work are a scatter-to-gather conversion algorithm which preserves parallelism (\secref{sec:scatter_gather}), and a simple automatic scheduling algorithm specialized to the patterns that appear in generated gradient code (\secref{sec:autoscheduler}). Halide's existing system of powerful dependence analyses is essential for both. In contrast to traditional Halide, automatic scheduling is critical given the complexity of the automatically-generated gradient code.

Using our new automatic gradient computation and automatic scheduler, we show how we can easily implement three recently-proposed neural network layers using code that is both faster and significantly simpler than the authors' original custom nodes written in C++ and CUDA (\secref{sub:nnlayer}). For example, the aforementioned bilateral slicing layer is expressed in 24 lines of Halide (\figref{fig:code_comparison}, left), including just four lines to compute and extract its gradients, while compiling automatically to an implementation about \(10\times\) faster than the authors' original handwritten CUDA, and \(20\times\) faster than a more limited version in PyTorch. We believe that this ease of implementation and performance tuning will dramatically facilitate prototyping, by delivering both automatic gradients and high performance at the outset of experimentation, not after-the-fact once the usefulness of a node has been established.

We also argue that this approach of gradient-based optimization through arbitrary programs is useful outside the traditional deep learning applications which have popularized it. Our vision is that any image-processing pipelines can benefit from an automatic tuning of internal parameters. Currently, this step is usually done by hand through user trial-and-error. The availability of automatic derivatives makes it possible to systematically optimize any internal parameter of an image processing pipeline, given some output objectives. This is especially appealing when gradients are available in the same language used for high-performance code deployment. We show how to significantly improve the performance of two traditional image processing algorithms by automatically optimizing their key parameters and filters (\secref{sub:param-opt}). We also develop a novel joint burst demosaicking and superresolution algorithm by inverting a forward image formation model including warps by unknown homographies, solving for the image and homographies simultaneously (\secref{sec:inverse_imaging}). Finally, we show the versatility of our approach and implement a lens design optimization by differentiating an optical simulator and fluid simulator (\secref{sub:non_image_processing}).

\section{Related Work}

\subsection{Automatic Differentiation and Deep Learning Frameworks}

% Autodiff frameworks
Following the methods in Chapter~\ref{chap:autodiff}, it is possible to generate derivative code from a given program. Many automatic differentiation frameworks have been developed for general programming languages~\cite{Bischof:1992:AAD, Griewank:1996:AAP, Hascoet:2013:TAD, Hogan:2014:FRA, Wiltschko:2017:TAD}, but general programming languages can be cumbersome for image processing applications. Writing efficient image processing code requires enormous effort to take parallelism, locality, and memory consumption/bandwidth into account~\cite{Ragan-Kelley:2012:DAS}. These difficulties are compounded when we also want to compute derivatives. In particular, none of the existing automatic differentiation compilers or libraries can handle automatic differentiation of vectorized code.

Recent deep learning packages provide higher level, highly optimized differentiable building blocks for users to assemble their program~\cite{Bergstra:2010:TCG, Dong:2014:ICN, Abadi:2015:TLM, Paszke:2017:ADP}. These packages are efficient when the algorithm to be implemented can be conveniently expressed by combining these building blocks. But it is quite common for users to write their own custom operators in low-level C++ or CUDA to extend a package's functionalities.

Using our approach, one can simply write the forward program. Our algorithm generates the derivatives and, thanks to Halide's decoupling of algorithm and schedule and our automatic scheduler, provides convenient handles to easily produce efficient code. 

\subsection{Image Processing Languages}

Our work builds on the Halide~\cite{Ragan-Kelley:2012:DAS} image processing language, which we briefly introduce in \secref{sec:halide_background}.

The Opt language~\cite{Devito:2017:ODS} focuses on nonlinear least squares problems. It provides language constructs to describe the least squares cost and automatically generates solvers. It uses the D* algorithm~\cite{Guenter:2007:ESD} to generate derivatives for the Jacobian.
The ProxImaL~\cite{Heide:2016:PEI} language, on the other hand, focuses on solving inverse problems using proximal gradient algorithms. The language provides a set of functions and their corresponding proximal operators. It then generates Halide code for optimization. Our system can be used to generate the adjoints required by new ProxImaL operators.

These languages focus on a specific set of solvers, namely nonlinear least squares and proximal methods, and provide high-level interfaces to them. On the other hand, we deal with any problem that requires the gradient of a program. Our system can also be used to solve for unknowns other than images, such as optimizing the hyperparameters of an algorithm or jointly optimizing images and parameters. \secref{sec:inverse_imaging} demonstrates this with some examples.

Recently, there have been attempts to automatically speed-up image processing pipelines~\cite{Mullapudi:2015:PAO, Yang:2016:VAV, Mullapudi:2016:ASH, Sioutas:2018:LTL, Baghdadi:2019:TPC}. We developed a new automatic scheduler in Halide with specialized mechanisms for parallel reductions~\cite{Suriana:2017:PAR}, which often occur in the gradients of image processing code. Our system could further benefit from future developments in automatic code optimization.

\subsection{Learning and Optimizing with Images}

Gradient-based optimization is commonly used in image processing. It has been used for image restoration~\cite{Rudin:1992:NTV}, image registration~\cite{Zitova:2003:IRM}, optical flow estimation~\cite{Horn:1981:DOF}, stereo vision~\cite{Barron:2016:FBS}, learning image priors~\cite{Roth:2005:FEF, Ulyanov:2017:DIP} and solving complex inverse problems~\cite{Heide:2014:FFC}. Our work alleviates the need to manually derive the gradient in such applications, which enables faster experimentation.

Deep learning has revitalized interest in building differentiable forward image processing pipelines whose parameters can be tuned by stochastic gradient descent. Successful instances include image restoration~\cite{Gharbi:2016:DJD, Zhang:2017:BGD}, photographic enhancement~\cite{Xu:2015:DEA}, and applications such as colorization~\cite{Iizuka:2016:LTC, Zhang:2016:CIC}, and style transfer~\cite{Gatys:2016:IST, Luan:2017:DPS}.
Some of these methods call for custom operators~\cite{Jaderberg:2015:STN, Ilg:2017:FEO, Gharbi:2017:DBL}, typically not available in mainstream frameworks. For these custom operators, forward and gradient operations are implemented manually. Our work provides a convenient way to explore new custom computations.

% Multigrid preconditioners Szeliski 
% RED denoising 

\section{The Halide Programming Language}
\label{sec:halide_background}

Our system extends the Halide programming language. We give a brief overview of the constructs in Halide that are relevant to our system. For more detail on Halide, see the original papers~\cite{Ragan-Kelley:2012:DAS, Ragan-Kelley:2013:HLC} and documentation.\footnote{\url{http://halide-lang.org/}}

Halide is a language designed to make it easy to write high-performance image- and array-processing code. The key idea in Halide is the separation of a program into the \textit{algorithm}, which specifies \emph{what} is computed, and the \textit{schedule}, which dictates the \emph{order} of computation and storage. The algorithm is expressed as a pure functional, feed-forward pipeline of arithmetic operations on multidimensional grids. The \emph{schedule} addresses concerns such as tiling, vectorization, parallelization, mapping to a GPU, etc. The language guarantees that the output of a program depends only on the \textit{algorithm} and not on the \textit{schedule}. This frees the user from worrying about low-level optimizations while writing the high-level algorithm. They can then explore optimization strategies without unintentionally altering the output.

By adding automatic differentiation to Halide, we build on this philosophy. To create a differentiable pipeline, the user no longer needs to worry about the correctness and efficiency of the gradient code. With the sole specification of a forward algorithm, our system synthesizes the gradient algorithm. Optimization strategies can then be explored for both, either manually or with an auto-scheduler.

The following code shows an example Halide program that performs gamma correction on an image and computes the $L^2$ norm between the output and a target image:
\begin{lstlisting}[style=Halide]
    Param<float> g; // Gamma parameter
    Buffer<float> im, tgt; // 2-D input and target buffers
    Var x, y; // Integer variables for the pixel coordinates
    Func f; // Halide function declarations
    // Halide function definition
    f(x, y) = pow(im(x, y), g); 
    // Reduction variables to loop over target's domain
    RDom r(tgt);
    Func loss; // We compute the MSE loss between f and tgt
    loss() = 0.f; // Initialize the sum to 0
    Expr diff = f(r.x, r.y) - tgt(r.x, r.y);
    loss() += diff * diff; // Update definition
\end{lstlisting}

Halide is embedded in C++. Halide pipeline stages are called \textit{functions} and represented in code by the C++ class \lstinline{Func}. Each Halide function is defined over an n-dimensional grid. The definition of a function comprises:
\begin{itemize}
    \item an \textit{initial value} that specifies a value for each grid point.
    \item optional \textit{recursive updates} that modify these values in-place.
\end{itemize}
The function definitions are specified as Halide \textit{expressions} (objects of type \lstinline{Expr}). Halide expressions are side-effect-free, including arithmetic, logical expressions, conditionals, and calls to other Halide functions, input buffers, or external code (such as \lstinline{sin} or \lstinline{exp}).

Reduction operators, such as summation or general convolution, are implemented through recursive updates of a Halide function. The domain of a reduction is represented in code as an \lstinline{RDom}, which implies a loop over that domain. All loops in Halide are implicit, whether over the domain of a function or a reduction.

Scheduling is expressed through methods exposed on \lstinline{Func}. There are many scheduling operators, which transform the computation to trade off between memory bandwidth, parallelism, and redundant computation. Halide lowers the schedule and algorithm into a set of loop nests and kernels. These are then compiled to machine code for various architectures. We use the CUDA and x86 backends for the applications demonstrated in this chapter.

\begin{figure}[t]
  \centering
  \includegraphics[width=\linewidth]{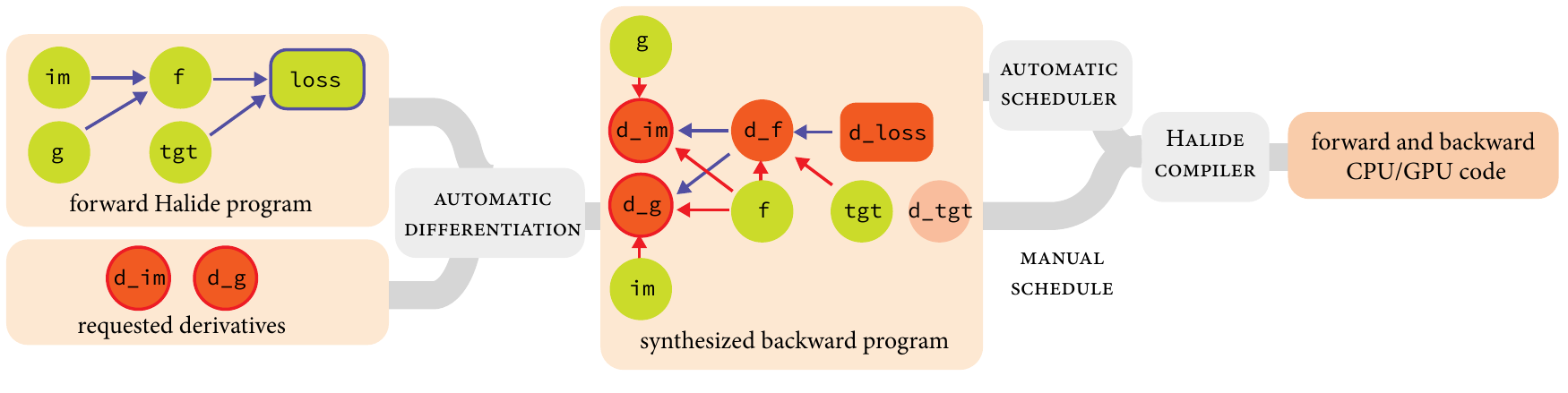}
    \caption{\textbf{Overview of our compiler.} The user writes a forward Halide program as they would normally. Then, they specify the set of outputs and gradients the system should produce. Our automatic differentiation generates new Halide functions that implement the requested gradients. The user can either manually schedule the pipeline or use our automatic scheduler. Finally, the Halide compiler generates machine code for the scheduled forward and backward algorithms.}
\label{fig:gradient_halide_overview}
\end{figure}

\section{Method}
\label{sec:gradient_halide_method}

To use our system, a programmer first writes a forward Halide algorithm. They then request the gradient of some scalar loss with respect to any Halide function, image buffer, or parameter in the pipeline. Our automatic differentiation system visits the graph of functions that describes the forward algorithm and synthesizes new Halide functions that implement the gradient computation (\secref{sec:autodiff_halide}).
The programmer can either specify the schedule for these new functions manually or use our automatic scheduler (\secref{sec:autoscheduler}). Unlike Halide's built-in auto-scheduler~\cite{Mullapudi:2016:ASH}, ours recognizes patterns that arise when reversing the computation graph (\secref{sec:scatter_gather}). Figure~\ref{fig:gradient_halide_overview} illustrates the workflow.

% Explain high-level strategy.
\subsection{High-level Strategy}
\label{sec:autodiff_halide}

We assume we want to compute the derivatives of some scalar $\Loss$, typically a cost function to be minimized.
Our system implements reverse-mode automatic differentiation, which computes the gradient with the same time complexity as the forward function (Chapter~\ref{sec:reverse_mode}).
We propagate the adjoints $\frac{\partial\Loss}{\partial g}$ to each function in the forward pipeline $g$, until we reach the inputs. The adjoints of the inputs are the components of the gradient.

Specifically, given a Halide program represented as a graph of Halide functions, we traverse the graph backward from the output and accumulate contributions to the adjoints using the chain rule. Halide function definitions are represented as expression trees, so within each function, we perform a similar backpropagation through the expression tree, propagating adjoints to all leaves.

A key difference between our algorithm and traditional automatic differentiation arises when an expression is a Halide function call. We need to construct a computation which accumulates adjoints onto the called function in the face of non-trivial data dependencies between the two functions. \secref{sec:halide_function_calls} describes this in detail.

We illustrate our algorithm on the example in \secref{sec:halide_background}, which performs gamma correction on an image and computes the $L^2$ distance between the output and some target image. To compute the gradients of the distance with respect to the input image and the gamma parameter, one would write:
\begin{lstlisting}[style=Halide]
  // Obtain gradients with respect to image and gamma parameters
  auto d_loss_d = propagate_adjoints(loss);
  Func d_loss_d_g = d_loss_d(g);
  Func d_loss_d_im = d_loss_d(im);
\end{lstlisting}

Throughout this chapter, we use the convention that prefixing a function's name with \lstinline{d_} refers to the gradient of that Halide function. We extend Halide with a key feature \lstinline{propagate_adjoints}. It takes a scalar Halide function and generates gradients in the form of new Halide functions for every Halide function, buffer, and real number parameter the output depends on. Our system can also be used as a component in other automatic differentiation systems that compute gradients. In this case, the user can specify a non-scalar Halide function and a buffer representing the adjoints of the function. Figure~\ref{fig:gradient_halide_overview} shows the computational graph for both the original and gradient computations.

\subsection{Differentiating Halide Function Calls}
\label{sec:halide_function_calls}

An important difference between automatic differentiation in Halide and traditional automatic differentiation, is that Halide functions are defined on multi-dimensional grids, therefore function calls and the elements on the grids can have non-trivial aggregate interactions.

Given each input-output pair of Halide functions, we synthesize a new Halide function definition that accumulates the adjoint of the output function onto the adjoint of the input. For performance, we want these new definitions to be as parallelizable as possible.

\subsubsection{Scatter-gather conversion}
\label{sec:scatter_gather}

\begin{figure}[!t]
\centering
\captionsetup[subfigure]{justification=centering}
  \begin{subfigure}[t]{0.32\linewidth}
    \includegraphics[width=\linewidth]{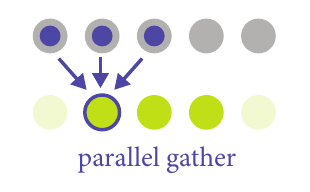}
    \caption{\label{fig:forward_1d_convolution} forward 1D convolution}
  \end{subfigure}
  \begin{subfigure}[t]{0.32\linewidth}
    \includegraphics[width=\linewidth]{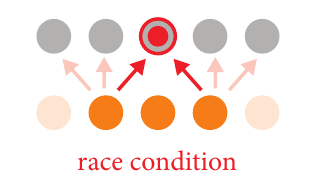}
    \caption{\label{fig:backward_general_scatter} backward general scatter}
  \end{subfigure}
  \begin{subfigure}[t]{0.32\linewidth}
    \includegraphics[width=\linewidth]{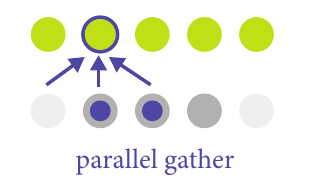}
    \caption{\label{fig:backward_with_our_gather_conversion} backward with \\ our gather conversion}
  \end{subfigure}
\caption{\textbf{Scatter-to-gather conversion.} Our compiler transform enables efficient, parallel code. In this example of a 1D 3-tap convolution, each dot represents a value in the input (resp. output) array. The forward computation (a) produces an output value from three inputs (the faded dots account for boundary conditions). This 3-tap reduction can easily be run in parallel over the output buffer (green dots). Computing the adjoint operator by simply reversing the dependency graph (b), that is by looping in parallel over the output nodes (orange), leads to race conditions since two inputs might need to write to the same location in the input's adjoint buffer (highlighted in red). This is a common issue with general scattering operations. Using our scatter-to-gather conversion, we convert this backward operation to a reduction over \lstinline{d_out} (the adjoint of a convolution is a correlation). In turn, this transformed computation is readily parallelized over \lstinline{d_out}'s domain (c).}
\label{fig:scatter_gather_convolution}
\end{figure}

\begin{lstfloat}[t]

\lstset{xleftmargin=2.5cm}
\begin{lstlisting}[style=Halide]
// We start with d_output, which contains the adjoint of output
// We propagate the derivatives from d_output to in and affine:
RDom ri(0, nci, 0, adjoints.channels());
d_in(x, y, ri.x, n) +=
  d_output(x, y, ri.y, n) * affine(x, y, ri.y * (nci + 1), n);
d_affine(x, y, ri.y*(nci+1)+ri.x, n) +=
  d_output(x, y, ri.y, n) * in(x, y, ri.x, n);
// Variable co is converted into a reduction variable rco.
RDom rco(0, adjoints.channels());
d_affine(x, y, rco*(nci+1)+nci, n) += d_output(x, y, rco, n);

// The derivatives are then propagated from affine to grid.
RDom rg(0, 2, 0, 2, 0, 2, 0, sigma_s, 0, sigma_s);
Expr inv_x = (x - rg[0]) * sigma_s + rg[3];
Expr inv_y = (y - rg[1]) * sigma_s + rg[4];
d_grid(x, y, fx + rg[2], c) +=
  d_affine(inv_x, inv_y, c, n) * d_tent;
// d_tent is tent with (x, y) replaced by (inv_x, inv_y).
// The scattering operation is transformed by solving
// x == inv_x/sigma_s+rt.x and y == inv_y/sigma_s+rt.y
// for inv_x and inv_y.

// Finally, and less obviously, affine also depends on guide.
RDom rgu(0, 2, 0, 2, 0, 2, adjoints.channels());
Expr wxy = abs(rgu[0] - wx) * abs(rgu[1] - wy);
Expr wz = select(rgu[2] - wz > 0.f, 1.f, -1.f);
d_guide(x, y, n) +=
  select(guide(x, y, n) >= 0.f && guide(x, y, n) <= 1.f,
    d_affine(x, y, rgu[3], c, n)*wxy*wz*grid.channels(), 0.f);
\end{lstlisting}

\caption{Derivatives generated by our algorithm for the bilateral slicing code in the left of \figref{fig:code_comparison}.}
\label{fig:bilateral_slice_grad}
\end{lstfloat}

Two cases require special care for correctness and efficiency. The first and most important case occurs when each output element reads and combines multiple input values. This happens for example in the simple convolution of Figure~\ref{fig:scatter_gather_convolution}(a). We call this pattern a \emph{gather} operation.

When computing gradients in reverse automatic differentiation, the natural reverse of this gather is a \emph{scatter} operation: each input writes to multiple elements of the output. Scattering operations, however, are not naturally parallelizable since they may lead to race conditions on write. For this reason, we want to convert scatters back to gathers whenever possible. We do this by shearing the iteration domain (e.g.~\cite{Lamport:1974:HMA}). To illustrate this transformation, consider the following code that convolves a 1D signal with a kernel, also illustrated in Figure~\ref{fig:scatter_gather_convolution}(a):
\begin{lstlisting}[style=Halide]
  Func output;
  output(x) = input(x - r.x) * kernel(r.x);
\end{lstlisting}
Assume that we are interested in propagating the gradient to \lstinline{input}. This is achieved by reversing the dependency graph between the input and output variables as shown in Figure~\ref{fig:scatter_gather_convolution}(b). In code, this transformation would yield:
\begin{lstlisting}[style=Halide]
  RDom ro;
  d_input(ro.y - ro.x) += d_output(ro.y) * kernel(ro.x);
\end{lstlisting}
where \lstinline{ro.x} iterates over the original \lstinline{r.x}, and \lstinline{ro.y} iterates over the domain of \lstinline{output}. For each argument in the calls to \lstinline{input}, we replace the pure variables (\lstinline{x} here) with reduction variables that iterate over the domain of the output (in this case \lstinline{ro.y}). \lstinline{r.x} is renamed to \lstinline{ro.x} so we can merge the reduction variables into a single reduction domain \lstinline{ro}.

This new update definition cannot be computed in parallel over \lstinline{ro.y} since multiple \lstinline{ro.y - ro.x} may write to the same memory location. A more efficient way to compute the update, illustrated in Figure~\ref{fig:scatter_gather_convolution}(c), is to rewrite the same computation as follows:
\begin{lstlisting}[style=Halide]
  d_output(x) = select(x >= a && x < b, d_output(x), 0.f);
  d_input(x) += d_output(x + r.x) * kernel(r.x);
\end{lstlisting}
where \lstinline{a} and \lstinline{b} are the bounds of \lstinline{output}.
By shearing the iteration domain with the variable substitution \lstinline{x = ro.y - ro.x}, we have made \lstinline{d_input} parallelizable over \lstinline{x}. Because Halide only iterates over rectangles, and the sheared iteration domain is no longer a rectangle, we add a zero-padding boundary condition to \lstinline{d_output}, and iterate over a conservative bounding box of the sheared domain: 

\begin{center}
\includegraphics[width=0.6\linewidth]{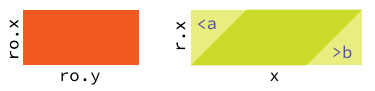}
\end{center}

We use Halide's equation-solving tools to deduce the variable substitution to apply. For each argument in a function call, we construct an equation e.g. $u = x - r_x$ and solve for $x$. Importantly, we solve for the smallest \textit{interval} of $x$ where the condition holds, since $x$ may map to multiple values. This may introduce new reduction variables, as in the following upsampling operation:
\begin{lstlisting}[style=Halide]
  output(x) = input(x/4);
\end{lstlisting}
Since \lstinline{x} is an integer, 4 values in \lstinline{input} are used to produce each value of \lstinline{output}. Accordingly, our converter will generate the following adjoint code:
\begin{lstlisting}[style=Halide]
  RDom r(0, 4); // loops from 0 to 3
  d_output(x) = d_input(4*x + r.x)
\end{lstlisting}

If any step of this procedure fails to find a solution, we fall back to a general scattering operation. It is still possible to parallelize general scatters using \emph{atomics}. We added atomic operations to Halide's GPU backend to handle this case. A general scatter with atomics usually remains significantly less efficient than our transformed code. For instance, the backward pass of a 2D convolution layer applied to a \(16 \times 16 \times 256 \times 256\) input takes 68 ms using atomics and 6 ms with our scatter-to-gather conversion.

Listing~\ref{fig:bilateral_slice_grad} shows some derivatives our system would generate for the bilateral slicing example in the left of Figure~\ref{fig:code_comparison}.

\subsubsection{Differentiating in-place updates}
\label{sec:partial_update}

The second case requiring special care arises when an update overwrites some variables of the function, introducing side effects. We categorize the in-place update further into two cases. In the first case the update statements do not reference the variables being overwritten (e.g. \lstinline{f(x) = 1.f}), and in the second case the overwritten variables are referenced (e.g. \lstinline{f(x) = 2 * f(x) + 1}).

Differentiating an update without self-reference is simpler. For example, consider the following forward code:
\begin{lstlisting}[style=Halide]
  g(x) = f(x);
  g(1) = 2.f; // update to f that overwrites a value
  h(x) = g(x);
\end{lstlisting}
When backpropagating the adjoints, we need to propagate correctly through the chain of update definitions. While \lstinline{h(x)} depends on \lstinline{f(x)} for most \lstinline{x} (via \lstinline{g(x)}), this is not true for \lstinline{x==1}. The update definition to \lstinline{g} hides the previous dependency on \lstinline{f(1)}. The corresponding gradient code is:
\begin{lstlisting}[style=Halide]
  d_g_update(x) = d_h(x); // Propagate to the first update
  d_g(x) = d_g_update(x); // Propagate to the initial definition
  d_g(1) = 0.f;           // Mask unwanted dependency
  d_f(x) = d_g(x);        // Propagate to f
\end{lstlisting}
In general, if we detect different update arguments between two consecutive function updates (in the example above, \lstinline{g(1)} is different from \lstinline{g(x)}), we mask the adjoint of the first update to zero using the update argument of the second update.

In the second case, when the update overwrites an intermediate value, and the intermediate value is required for the derivative, the situation is more complicated. For example:
\begin{lstlisting}[style=Halide]
  f(x) = g(x)
  f(x) = f(x) * f(x)
\end{lstlisting}
The gradient with respect to \lstinline{g} requires the overwritten \lstinline{f(x)}, making it impossible to backpropagate. Following is another example:
\begin{lstlisting}[style=Halide]
  f(x) = 0
  f(x) = 2 * f(x) + g(r.x)
\end{lstlisting}
In this case the reduction loop \lstinline{r.x} introduces a dependency between the adjoint of \lstinline{g(r.x)} and the intermediate \lstinline{f(x)}. On the other hand, if there is only one self-reference, and the adjoint to that self-reference is $1$, then we can differentiate as usual without special treatment:
\begin{lstlisting}[style=Halide]
  f(x) = 0
  f(x) = f(x) + g(r.x)
\end{lstlisting}
This is because all the intermediate \lstinline{f(x)} share the same adjoint.

For the first two examples, it is possible to rewrite the forward operation so that the update no longer overwrites intermediate, in a way similar to the stack we used for recording intermediate values in Chapter~\ref{sec:control_flow} and Pearlmutter and Siskind's lambda calculus approach~\cite{Pearlmutter:2008:RAF}. The first example can be rewritten as:
\begin{lstlisting}[style=Halide]
  f_(x, 0) = g(x)
  f_(x, 1) = f_(x, 0) * f_(x, 0)
  f(x) = f_(x, 1)
\end{lstlisting}
While the second example can be rewritten as:
\begin{lstlisting}[style=Halide]
  f_(x, 0) = 0
  f_(x, r.x + 1) = 2 * f_(x, r.x) + g(r.x)
  f(x) = f_(x, r.x.max() + 1)
\end{lstlisting}
It is possible for the compiler to do the rewrite automatically, but this transformation would change the original algorithm, making manual scheduling more difficult. We opt for more predictive behavior of the compiler. Therefore we detect the following two cases and return an error, asking the user to rewrite the function as above:
\begin{itemize}
  \item We check if the derivatives depend on a previous value, and if that particular value has been overwritten.
  \item For updates with reduction variables, unless the derivative of self-reference is $1$ or $0$, and there is at most one self-reference, we check if the overwritten derivative is used by others.
\end{itemize}

\subsection{Checkpointing}
\label{sec:checkpointing}

Reverse-mode automatic differentiation on complex pipelines must traditionally deal with a difficult trade-off. Memoizing values from the forward evaluation to be reused in the reverse pass saves compute, but costs memory. Even with unlimited memory, bandwidth is limited, so it can be more efficient to recompute values. In automatic differentiation systems, this trade-off is addressed with \emph{checkpointing}~\cite{Volin:1985:ACD}, which reduces memory usage by recomputing parts of the forward expressions. Fortunately, this is just a specific instance of the general recomputation-vs-memory trade-off already addressed by Halide's scheduling primitives.

For each function, we can decide whether to create an intermediate buffer for later reuse (the \lstinline{compute_root} construct), or recompute values at every call site (the \lstinline{compute_inline} construct). We can also compute these values at some intermediate granularity, i.e.,
by setting its computation somewhere in the loop nest of their consumers (the \lstinline{compute_at} construct).
Halide also allows checkpointing \emph{across} different Halide pipelines by using a global cache (the \lstinline{memoize} construct). This is useful when the forward pass and backward pass are in separately-compiled units.

As an example, consider the following 2D convolution implementation in Halide:
\begin{lstlisting}[style=Halide]
  RDom rk, rt;
  convolved(x, y) = 0.f;
  convolved(x, y) += in(x - rk.x, y - rk.y) * kernel(rk.x, rk.y);
  loss() = 0.f; // define an optimization objective
  loss() += pow(convolved(rt.x, rt.y) - target(rt.x, rt.y), 2.f);
  auto d = propagate_adjoints(loss);
  Func d_in = d(in);
\end{lstlisting}

We are interested in \lstinline{d_in}, the gradient of \lstinline{loss} with respect to \lstinline{in}. It is given by a cross correlation of \lstinline{2*(convolved-target)} with \lstinline{kernel}, where the cross correlation depends on the values of \lstinline{convolved}. Using the scheduling handles provided by Halide, we can easily decide whether to cache the values of \lstinline{convolved} for the gradient computation. For example, if we write:
\begin{lstlisting}
  convolved.compute_root();
\end{lstlisting}
the values of \lstinline{convolved} are computed once and will be fetched from memory when we need them for the derivative \lstinline{d_in}. On the other hand, if we write:
\begin{lstlisting}
  convolved.compute_inline();
\end{lstlisting}
the values of \lstinline{convolved} are computed on-the-fly and no buffer is allocated to store them. This can be advantageous when the convolution kernel is small (say \(2 \times 1\)) since this preserves memory locality, or when the pipeline is much longer and we cannot afford to store every intermediate buffer.

Halide provides scheduling primitives that are more general than binary checkpointing decisions. Fine-grained control over the schedule allows exploration of memory/recomputation trade-offs in the forward and gradient code. For instance, we can interleave the computation and storage of \lstinline{convolved} with the computation of another Halide function that consumes its value (in this case \lstinline{d_in}). The following code instructs Halide to compute and store a tile of \lstinline{convolved} for each \(32 \times 32\) tile of \lstinline{d_in} computed. This offers a potentially faster balance between computing all of \lstinline{convolved} before backpropagation, or recomputing each of its pixels on-demand:
\begin{lstlisting}
  d_in.compute_root().tile(x, y, xi, yi, 32, 32);
  convolved.compute_at(d_in, x);  // compute at each tile of d_in
\end{lstlisting}

We timed the three schedules above by computing \lstinline{d_in}. With multi-threading and vectorization on a CPU, on an image with size of \(2560 \times 1600\) and kernel size \(1 \times 5\), the \lstinline{compute_inline} schedule takes $5.6$ milliseconds while the \lstinline{compute_root} schedule takes $10.1$ milliseconds and the \lstinline{compute_at} schedule takes $9.7$ milliseconds. On the same image but with kernel size \(3 \times 5\), the \lstinline{compute_inline} schedule takes $66.2$ milliseconds while the \lstinline{compute_root} schedule takes $18.7$ milliseconds and the \lstinline{compute_at} schedule takes $12.3$ milliseconds.

\subsection{Automatic Scheduling}
\label{sec:autoscheduler}

Halide's built-in auto-scheduler~\cite{Mullapudi:2016:ASH} navigates performance trade-offs well for stencil pipelines, but struggles with patterns that arise when reversing their computational graph (\secref{sec:scatter_gather}). In particular, it does not try to optimize large reductions, like those needed to compute a scalar loss. It also does not generate GPU schedules for the current version of Halide \footnote{Mullapudi et al.'s work did include experiments on GPU and ARM, but as the Halide compiler has evolved, the original implementation was not able to consistently generate valid schedules.}.
Therefore we implemented a custom automatic scheduler for gradient pipelines.

Similar to Halide's built-in auto-scheduler, we ask the user to provide an estimate of the input and output buffer sizes. We then infer the extent of all the intermediate functions' domains.

Our automatic scheduler checkpoints (\lstinline{compute_root}) any stage that scatters or reduces, along with those called by more than one other function. We leave any other functions to be recomputed on-demand (\lstinline{compute_inline}). For the checkpointed functions, we tile the function domain and parallelize the computation over tiles when possible. Specifically, on CPUs, we split the function's domain into 2D tiles (\(16 \times 16\)) and launch CPU threads for each tile, vectorizing the innermost dimension inside a tile. On GPUs, we split the domain into 3D tiles (\(16 \times 16 \times 4\)). The tiles are mapped to GPU blocks, and elements within a tile to GPU threads. In both cases, we tile the first two (resp.\ three) dimensions of the function's domain that are large enough. We split the domain if its dimensionality is too low.

If the function's domain is not large enough for tiling, and the function performs a large associative reduction, we transform it into a parallel reduction using Halide's \lstinline{rfactor} scheduling primitive~\cite{Suriana:2017:PAR}. This allows us to factorize the reduction into a set of partial reductions which we compute in parallel and a final, serial reduction. Like before, we find the first two dimensions of the reduction domain which are large enough for tiling. We reduce the tiles in parallel over CPU threads (resp.\ GPU blocks). Within each 2D tile, we vectorize (resp.\ parallelize over GPU threads) the column-wise reductions. We also implemented a multi-level parallel reduction schedule but found it unnecessary in the applications presented. When compiling to GPUs, if both the function domain and the reduction domain are large enough for tiling, but the recursive update does not contain enough pure variables for parallelism, we parallelize the reduction using atomics.

To allow control over checkpointing, the automatic scheduler decisions can be overridden. We ask the user to provide optional lists of Halide functions they do or do not want to inline. We currently do not use \lstinline{compute_at} in our automatic scheduler.

\section{Applications and Results}

\newcommand{\deepbilateralcode}{\footnote{\url{https://github.com/mgharbi/hdrnet/blob/master/hdrnet/ops/bilateral_slice.cu.cc}}}
\newcommand{\flownetcode}{\footnote{FlowNet 2.0: \url{https://github.com/lmb-freiburg/flownet2/blob/master/src/caffe/layers/flow_warp_layer.cu}}}
\newcommand{\nvidiaflownet}{\footnote{Nvidia FlowNet 2.0: \url{https://github.com/NVIDIA/flownet2-pytorch}}}

We generate gradients for pipelines in three groups of applications (Figure~\ref{fig:gradient_halide_teaser}). First, we show that our system can be integrated into existing deep learning systems to more easily develop new custom operators. Second, we show that we can improve existing image processing pipelines by optimizing their internal parameters on a dataset of training images. Finally, we show how to use our derivatives to solve inverse imaging problems (i.e., optimizing for the image itself).
 
Unless otherwise specified, we use our automatic scheduler (\secref{sec:autoscheduler}) to schedule all the applications throughout the section (i.e., for both the forward code and the derivatives we generate). Therefore, our implementation only requires the programmer to specify the forward pass of the algorithm.

\subsection{Custom Neural Network Layers}
\label{sub:nnlayer}

\begin{table}[t]
  \centering
  \small
  \caption{\textbf{Custom neural network operators.} Performance of our approach for custom neural network operators. The runtime measures end-to-end latency for forward+backward evaluation. The spatial transformer transforms a batch of $4\times 16 \times512 \times 512$. The Flownet node warps a batch of $4\times 64 \times512 \times 512$ images with a 2D warping field. The BilateralSlice layer processes images with size $4\times 4 \times 1024 \times 1024$ and grid size $4\times 12 \times 64 \times 64$. Measurements were made on an Intel Core i7-3770K CPU @ 3.50GHz, with 16GB of RAM and an NVIDIA Titan X (Pascal) GPU with 12 GB of RAM\@.}
  \tabcolsep=0.11cm
\begin{tabulary}{1\linewidth}{l*{3}{c}}
operator        & SpatialTransformer & Flownet & BilateralSlice \\
\midrule
PyTorch (cpu)     & 1094  ~ms &  4240 ~ms & 19819 ~ms \\
ours (cpu)        & 461   ~ms &  2466 ~ms & 1957 ~ms  \\
\midrule
PyTorch (gpu)     & 11 ~ms   &  482 ~ms & 1440 ~ms \\
CNTK (gpu)        & 136 ~ms  &  404 ~ms & 270 ~ms \\
manual CUDA (gpu) &---&  181 ~ms & 430 ~ms \\
ours (gpu)        & 13 ~ms   &  178 ~ms & 64 ~ms \\
\end{tabulary}
\label{tab:performance}
\end{table}

% warped convolutions \cite{henriques2016warped}
% deformable convolution \cite{dai2017deformable}
% oriented response nets \cite{zhou2017oriented}
% active convolutions \cite{jeon2017active}
% polar transformer nets \cite{esteves2017polar}
% sparsity invariant nets \cite{uhrig2017sparsity}

The class of computations expressible with deep learning libraries such as Caffe~\cite{Jia:2014:CCA}, PyTorch~\cite{Paszke:2017:ADP}, TensorFlow~\cite{Abadi:2015:TLM}, or CNTK~\cite{Dong:2014:ICN} is growing increasingly rich. Nonetheless, it is still common for a practitioner to require a new, custom node tailored to their problem. For instance, TensorFlow offers a bilinear interpolation layer and a separable 2D convolution layer. However, even a simple extension of these operations to 3D would require implementing a new custom operator in C++ or CUDA to be linked with the main library. This can already be tedious and error-prone. Furthermore, while the forward algorithm is being developed, the gradient must be re-derived by hand and kept in sync with the forward operator. This makes experimentation and prototyping especially difficult. Finally, both the forward and backward implementations ought to be reasonably optimized so that a model can be trained in a finite amount of time to verify its design.

We implemented a PyTorch backend for Halide so that our derivatives can be plugged into PyTorch's autograd system. We used this backend to re-implement custom operators recently proposed in the literature: the transformation layer in the spatial transformer network~\cite{Jaderberg:2015:STN}, the warping layer in Flownet 2.0~\cite{Ilg:2017:FEO}, and the bilateral slicing layer in deep bilateral learning~\cite{Gharbi:2017:DBL}. The performance of our automatically scheduled code matches highly-optimized primitives written in CUDA, and is much faster than unoptimized code. We compare the runtime of our method to PyTorch, CNTK, and hand-written CUDA code in Table~\ref{tab:performance}.

\subsubsection{Spatial transformer network}

The spatial transformer network of Jaderberg~et~al.\cite{Jaderberg:2015:STN} applies an affine warp to an intermediate feature map of a neural network.

The function containing the forward Halide code is 31 lines long excluding comments, empty lines, and function declarations.
% The tensorflow code found here
% https://github.com/kevinzakka/spatial-transformer-network/blob/master/transformer.py
% is around 81 lines after removing garbage
%
% The PyTorch custom node found here
% https://github.com/fxia22/stn.pytorch
% is >1000 lines
Due to the popularity of this operator, deep learning frameworks have implemented specialized functions for the layer. The cuDNN library~\cite{Chetlur:2014:CEP} added its own implementation in version 5 (2016), a year after the original publication. It took another year for PyTorch to implement a wrapper around the cuDNN code. We compare our performance to PyTorch's \lstinline{grid_sample} and \lstinline{affine_grid} functions which use the cuDNN implementation on GPU\@. On \(512 \times 512\) images with 16 channels and a batch size of 4, our CPU code is around $2.3$ times faster than PyTorch's implementation, and our GPU code is around $20\%$ slower than the highly-optimized version implemented in cuDNN\@. Currently, Halide does not support texture sampling on GPU, which could be causing some of the slowdown. We also compare our performance to a CNTK implementation of spatial transformer using the \textit{gather} operation. Our GPU code is around 10 times faster than the CNTK implementation.

Having fixed functions such as \lstinline{affine_grid} can be problematic when users want to slightly modify their models and experiment with different ideas. For example, changing the interpolation scheme (e.g., bicubic or Lanczos instead of bilinear), or interpolating over more dimensions (e.g., transforming volume data) would require implementing a new custom operator. Using our system, these modifications only require minor code changes to the forward algorithm. Our system then generates the derivatives automatically, and our automatic scheduler provides performance without further effort.

\subsubsection{Warping layer}

FlowNet 2.0~\cite{Ilg:2017:FEO}, which targets optical flow applications, introduced a new 2D warping layer. Compared to the previous spatial transformer layer, this warping layer is a more general transform using a per-pixel warp-field instead of a parametric transformation.

The function containing the forward Halide code is 18 lines long. The original warping function was implemented as a custom node in Caffe. The authors had to write the forward and reverse code for both the CPU and GPU backends. In total it comprises more than 400 lines of code\flownetcode. While the custom node can handle 2D warps well, adapting it to higher-dimensional warps or semi-parametric warps would be challenging. Our system makes this much easier. In addition to PyTorch and CNTK, we also compare the performance of our GPU code with a highly-optimized reimplementation from NVIDIA\nvidiaflownet. The performance of our code is comparable to the highly-optimized CUDA code.

\subsubsection{Bilateral slicing layer}

Deep bilateral learning~\cite{Gharbi:2017:DBL} is a general, high-performance image processing architecture inspired by bilateral grid processing and local affine color transforms. It can be used to approximate complicated image processing pipelines with high throughput. The algorithm works by splatting a 2D image onto a 3D grid using a convolutional network. Each voxel of the grid contains an affine transformation matrix. A high-resolution guidance map is then used to \emph{slice} into the grid and produce a unique, interpolated, affine transform to apply to each input pixel. The original implementation in TensorFlow had to implement a custom node\deepbilateralcode~for the final slicing operation due to the lack of an efficient way to perform trilinear interpolation on the grid. This custom node also applies the affine transformation on the fly to avoid instantiating a high-resolution image containing all the affine parameters at each pixel. The reference custom node had around 300 lines of CUDA code excluding comments and empty lines. Using the recently introduced general scattering functionality, we can implement the same operation directly in PyTorch. Figure~\ref{fig:code_comparison} shows a comparison between our Halide code, reference CUDA code, and PyTorch code.

PyTorch and CNTK implementations are modestly more complex than our code. PyTorch is $20$ times slower while CNTK is 4 times slower on an $1024 \times 1024$ input with a grid size of $32 \times 32 \times 8$ and a batch size of $4$. CNTK is faster than PyTorch due to different implementation choices on the gather operations. The manual CUDA code aims for clarity more than performance, but is both more complicated and $6.7$ times slower than our code.

Gharbi~et~al.~\cite{Gharbi:2017:DBL} argue that training on high-resolution images is key to capturing the high-frequency features of the image processing algorithm being approximated. Both the PyTorch and CNTK code run out of memory on a $2048 \times 2048$ input with grid size $64 \times 64 \times 8$ on a Titan GPU with 12 GB of memory. This makes it almost impossible to experiment with high-resolution inputs. Our code is 13.7 times faster than the authors' reference implementation on this problem size.

\subsection{Parameter Optimization for Image Processing Pipelines}
\label{sub:param-opt}

\begin{figure}[t]
\centering
\captionsetup[subfigure]{justification=centering}
  \begin{subfigure}[t]{0.3\textwidth}
    \includegraphics[width=\textwidth]{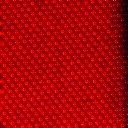}
    \caption*{AHD (19.6 dB)}
  \end{subfigure}
  \begin{subfigure}[t]{0.3\textwidth}
    \includegraphics[width=\textwidth]{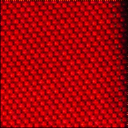}
    \caption*{ours, $8$ $5 \times 5$ filters (24.7 dB)}
  \end{subfigure}
  \begin{subfigure}[t]{0.3\textwidth}
    \includegraphics[width=\textwidth]{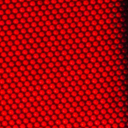}
    \caption*{reference}
  \end{subfigure}

  \begin{subfigure}[t]{0.3\textwidth}
    \includegraphics[width=\textwidth]{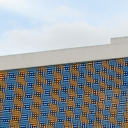}
    \caption*{AHD (19.7 dB)}
  \end{subfigure}
  \begin{subfigure}[t]{0.3\textwidth}
    \includegraphics[width=\textwidth]{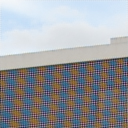}
    \caption*{ours, $8$ $5 \times 5$ filters (21.4 dB)}
  \end{subfigure}
  \begin{subfigure}[t]{0.3\textwidth}
    \includegraphics[width=\textwidth]{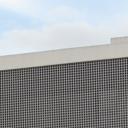}
    \caption*{reference}
  \end{subfigure}

  \caption{\label{fig:demosaic_comparison} \textbf{Tuning demosaicking algorithms.} We use our automatic gradients to relax the adaptive homogeneity-directed demosaicing (AHD) algorithm (a) by adding more filters to interpolate the green channel (8 instead of 2 here, with 5x5 footprint instead of 5x1). With this simple tweak, and by optimizing the filters using our automatically generated derivatives, we can obtain sharper images in difficult cases (b), first row. The small-footprint of this simple demosaicking method nevertheless inherits some of the limitations of AHD\@. In particular, it leads to artifacts in complex, moir\'e-prone patterns (second row). Images are taken from the deep demosaicking dataset~\protect\cite{Gharbi:2016:DJD}.}
\end{figure}

\begin{table}[t]
\centering
\small
\begin{tabulary} {1\linewidth} {l *{4}{c} r}
                                                   & kodak      & mcm        & vdp        & moir\'e  & time   \\
\midrule
  bilinear                                         & 32.9       & 32.5       & 25.2       & 27.6 & *127ms \\ 
  Adobe Camera Raw 9                               & 33.9       & 32.2       & 27.8       & 29.8 &---\\
  AHD~\cite{Hirakawa:2005:AHD}                     & 36.1       & 33.8       & 28.6       & 30.8 & *1618ms\\
  {\bf ours} (2 filters, 5x5)                      & 36.7       & 34.7       & 29.4       & 31.5 & 71ms \\
  {\bf ours} (9 filters, 5x5)                      & 36.8       & 35.2       & 29.8       & 31.7 & 177ms\\
  {\bf ours} (15 filters, 7x7)                     & 37.3       & 35.5       & 30.1       & 32.0 & 324ms \\
 Gharbi~\cite{Gharbi:2016:DJD}                     & 41.2       & 39.5       & 34.3       & 37.0 & 2932ms
\end{tabulary}
\caption{\textbf{Performance-accurarcy trade-offs.} Peak signal-to-noise ratio for several demosaicking techniques following the evaluation methodology of Gharbi et al.~\cite{Gharbi:2016:DJD} (higher is better). We implemented a version of AHD demosaicking algorithm~\cite{Hirakawa:2005:AHD} with our system. Despite the simplicity of our approach, by relaxing the algorithm's specifications (i.e.\ adding more filters on the green channel reconstruction with larger footprints) and re-optimizing the parameters, we achieve higher fidelity (over 1 dB better) for a similar computational cost. While our method does not rival state-of-the-art deep-learning-based techniques, it is significantly faster and opens up new avenues to optimize more parsimoniously parametrized algorithms tailored to the problem. (Timings reported for a 1 megapixel image. (*)Timing for these algorithms is from non-optimized MATLAB code.)}
\label{tab:demosaic}
\end{table}

\begin{figure}[t]
  \centering
  \captionsetup[subfigure]{justification=centering}

  \begin{subfigure}[t]{0.23\textwidth}
    \includegraphics[width=\textwidth]{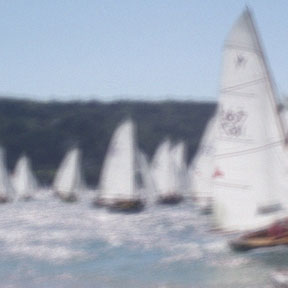}
    \caption*{blurred}
  \end{subfigure}
  \begin{subfigure}[t]{0.23\textwidth}
    \includegraphics[width=\textwidth]{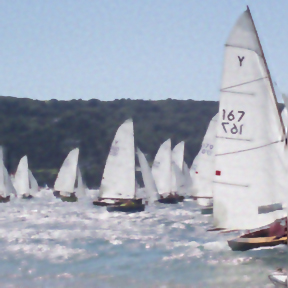}
    \caption*{Fortunato \\ (25.39 dB)}
  \end{subfigure}
  \begin{subfigure}[t]{0.23\textwidth}
    \includegraphics[width=\textwidth]{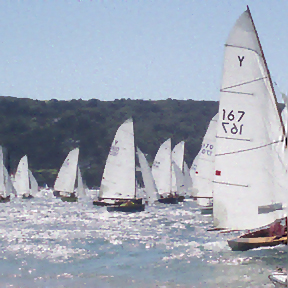}
    \caption*{ours \\ (27.37 dB)}
  \end{subfigure}
  \begin{subfigure}[t]{0.23\textwidth}
    \includegraphics[width=\textwidth]{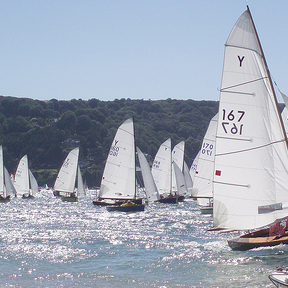}
    \caption*{reference}
  \end{subfigure}

  \begin{subfigure}[t]{0.23\textwidth}
    \includegraphics[width=\textwidth]{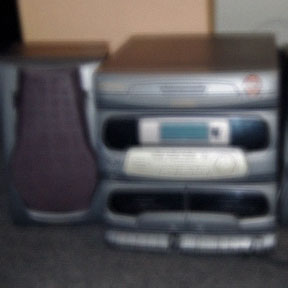}
    \caption*{blurred}
  \end{subfigure}
  \begin{subfigure}[t]{0.23\textwidth}
    \includegraphics[width=\textwidth]{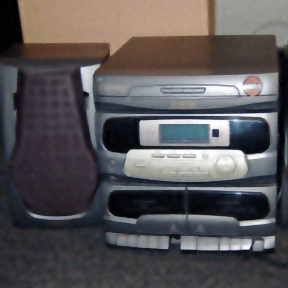}
    \caption*{Fortunato \\ (25.83 dB)}
  \end{subfigure}
  \begin{subfigure}[t]{0.22\textwidth}
    \includegraphics[width=\textwidth]{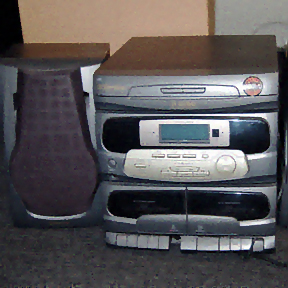}
    \caption*{ours \\ (27.86 dB)}
  \end{subfigure}
  \begin{subfigure}[t]{0.23\textwidth}
    \includegraphics[width=\textwidth]{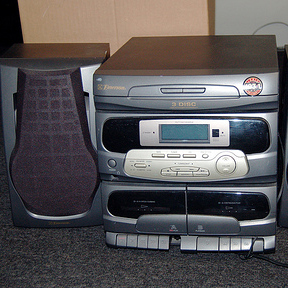}
    \caption*{reference}
  \end{subfigure}

  \caption{\textbf{Tuning deconvolution algorithms.} We use automatic gradients to enhance Fortunato and Oliveira's non-blind deconvolution algorithm~\cite{Fortunato:2014:FHN}. We use more iterations and automatically train the weights, thresholds and filtering parameters. We are able to get sharper results. On eight randomly selected-images we achieve an average PSNR of 29.57 dB. Using the original algorithm with its original parameters the PSNR is 28.51 dB. Image taken from ImageNet~\cite{Deng:2009:ILS}}
  \label{fig:deconv_comparison}
\end{figure}

Traditionally, when developing an image processing algorithm, a programmer manually tunes the parameters of their pipeline to make it work well on a small test set of images. When the number of parameters is large, manually determining these parameters becomes difficult.

In contrast, modern deep learning methods achieve impressive results by using a large number of parameters and many training images. We demonstrate that it is possible to apply a similar strategy to general image processing algorithms, by augmenting the algorithm with more parameters, and tuning these parameters through an offline training process. Our system provides the necessary gradients for this optimization. Users write the forward code in Halide, and then optimize the parameters of the code using training images.

We demonstrate this with an image demosaicking algorithm based on the adaptive homogeneity directed demosaicking~\cite{Hirakawa:2005:AHD} (AHD), and a non-blind image deconvolution algorithm based on sparse adaptive priors~\cite{Fortunato:2014:FHN}.

\subsubsection{Image demosaicking}

Demosaicking seeks to retrieve a full-color image from incomplete color samples captured through a color filter array, where each pixel only contains one out of three red, green and blue colors. Traditional demosaicking algorithms work well on most cases, but can exhibit structured aliasing artifacts such as \textit{zippering} and \textit{moir\'e} (Figure~\ref{fig:demosaic_comparison}). Recent methods using deep learning have achieved impressive results~\cite{Gharbi:2016:DJD}, however, the execution time is still an issue for practical usage. We relax the adaptive homogeneity-directed demosaicking algorithm (AHD)~\cite{Hirakawa:2005:AHD}, variations of which are the default algorithms in \emph{Adobe Camera Raw} and \emph{dcraw}. We increase the number of filters to interpolate the green channel. We also fine-tune the chrominance (red-blue) interpolation filters from the AHD reference. We experiment with different numbers of filters and filter sizes to explore the runtime versus accuracy trade-off. We optimized the filter weights on Gharbi~et~al.'s~\cite{Gharbi:2016:DJD} training dataset using the gradients provided by our system. The results are illustrated in Table~\ref{tab:demosaic}. With this simple modification, we obtain a significant 1 to 1.5 dB improvement on the more difficult datasets (\emph{moir\'e} and \emph{vdp}), depending on the number of filters used. We also obtain visually sharper images in many challenging cases, as shown in Figure~\ref{fig:demosaic_comparison}. 

With its limited footprint and filtering complexity, our optimized demosaicking still struggles on moir\'e-prone textures. Our system allows users to experiment with more complex ideas without having to implement the derivatives at each step. For instance, we were able to quickly experiment with (and ultimately discard) alternative algorithms (e.g.\ using filters that take the ratio between colors into account and 1D directional filters).

\subsubsection{Non-blind image deconvolution}

The task of non-blind image deconvolution is: given a point spread function and a blurry image, which is the result of a latent natural image convolved with the function, recover the underlying image. The problem is highly ill-posed, therefore the quality of the reconstruction heavily depends on the priors we place on the image. It is thus important to learn a good set of parameters for those priors.

We based our implementation on the sparse adaptive prior proposed by Fortunato and Oliveira~\cite{Fortunato:2014:FHN}. The original method works in a 2-stage fashion. In the first stage, they solve a conventional $L^2$ deconvolution using a set of discrete derivative filters as the prior. Then they use an edge-aware filter to clean up the noise in the image. In the second stage, another $L^2$ deconvolution is solved for large  discrete derivatives by matching the prior terms to the result of the first stage, masked by a smooth thresholding function.

We extend the method by increasing the number of stages (we use $4$ instead of $2$), and having a different set of filters for the priors for each stage. We optimize the weights of the prior filters, the smoothness parameters of the edge-aware filter (we use a bilateral grid), and the thresholding parameters in the smooth thresholding functions.

To demonstrate the ability of our system to handle nested derivatives, we implemented a generic conjugate gradient solver using a linear search algorithm based on Newton-Raphson to solve for the $L^2$ deconvolution. We write the conjugate gradient loop in PyTorch, but implement the gradient and vector-Hessian-vector product (required in the line search step) in Halide. We also implemented the bilateral grid filtering step in Halide. To optimize the parameters, we then differentiate through the gradients we used for the non-linear conjugate gradient algorithm. We train our method on ImageNet~\cite{Deng:2009:ILS} and use the point spread function generation scheme described in Kupyn~et~al.'s work~\cite{Kupyn:2017:DBM}. We initialize the parameters to the recommended parameters described in Fortunato and Oliveira's work. Figure~\ref{fig:deconv_comparison} shows the result.  

\subsection{Inverse Imaging Problems: Optimizing for the Image}
\label{sec:inverse_imaging}
\label{sec:burst_demosaick}

\begin{figure}[t]
  \centering
  \captionsetup[subfigure]{justification=centering}
  \begin{subfigure}[t]{0.3\textwidth}
    \includegraphics[width=\textwidth]{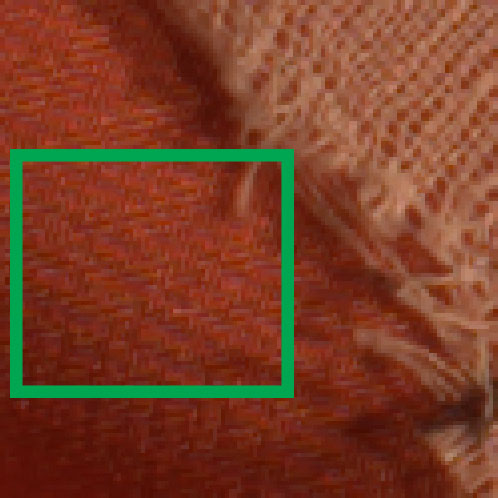}
  \end{subfigure}
  \begin{subfigure}[t]{0.3\textwidth}
    \includegraphics[width=\textwidth]{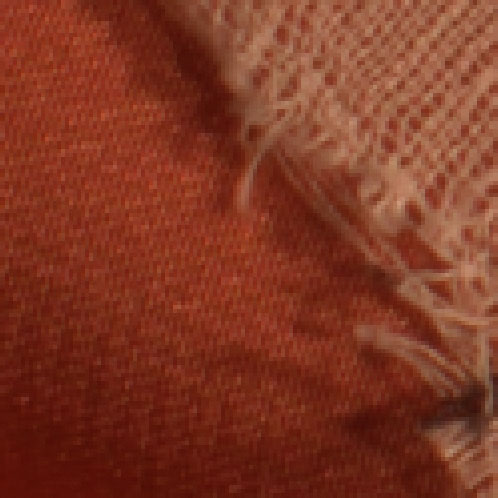}
  \end{subfigure}
  \begin{subfigure}[t]{0.3\textwidth}
    \includegraphics[width=\textwidth]{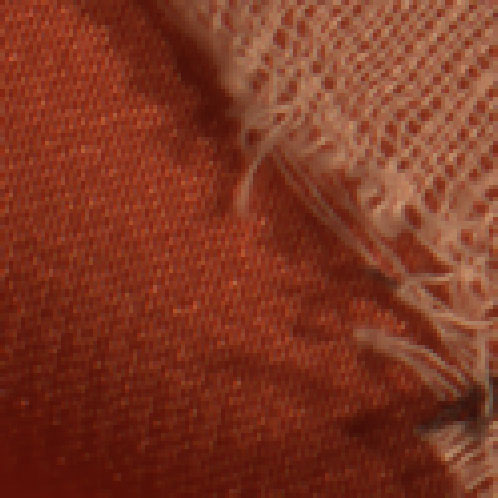}
  \end{subfigure}

  \begin{subfigure}[t]{0.3\textwidth}
    \includegraphics[width=\textwidth]{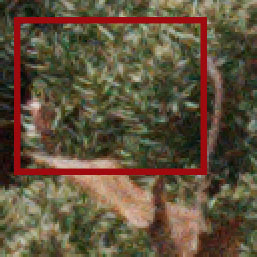}
    \caption{dcraw (AHD) \\ single frame}
  \end{subfigure}
  \begin{subfigure}[t]{0.3\textwidth}
    \includegraphics[width=\textwidth]{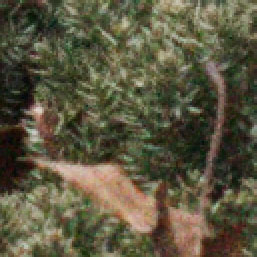}
    \caption{Gharbi et al.~\cite{Gharbi:2016:DJD} \\ single frame}
  \end{subfigure}
  \begin{subfigure}[t]{0.3\textwidth}
    \includegraphics[width=\textwidth]{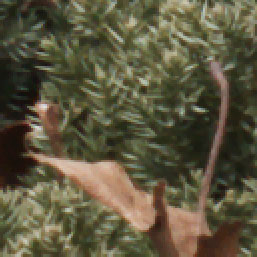}
    \caption{our output $R$}
  \end{subfigure}

  \caption{\label{fig:burst_demosaic} \textbf{Gradients for inverse problems.} Automatic gradients can be used for inverse problems such as high-resolution demosaicking from a  burst of images. The user only needs to implement the forward model. Bursts of RAW images are captured with a \emph{Nikon D810} camera then jointly aligned and demosaicked (13 and 23 images respectively, only showing crops). We initialize our reconstruction to a simple bilinear interpolation (not shown) and solve an inverse problem to recover both a set of homographies and a demosaicked image that matches the captured data when reprojected. Compared to the result of \emph{dcraw}'s AHD algorithm (a) and Gharbi~et~al.~\cite{Gharbi:2016:DJD} (b), our output (c) is much sharper, and shows less noise (red square) and color moir\'e (green square).}
\end{figure}

The derivatives produced by our automatic differentiation algorithm can be readily employed to solve inverse problems in computational photography. Using our system, users can quickly experiment with different forward models or different priors. We demonstrate this on a burst-demosaicking inverse pipeline.
 
Given $N$ misaligned Bayer RAW images, our goal is to reconstruct a full-color image as well as estimate the homography parameters that align our reconstruction to the input data. We do this by minimizing the following cost function:
\begin{equation} 
  \min_{R, H_i}\sum_{i=1}^{N} ||MH_iR - I_i||_2^2 + \lambda ||\nabla R||_1
  \label{eq:burst_demosaick}
\end{equation}
\noindent where $M$ decimates the color samples according to the Bayer mosaic pattern. The homographies $H_i$ align our reconstruction $R$ to the input data $I_i$.

Gradient descent can help us minimize the function locally, but Equation~\ref{eq:burst_demosaick} is highly non-convex, so a good initialization is critical. We initialize the $H_i$ using RANSAC~\cite{Fischler:1981:RSC} and SIFT-based features~\cite{Lowe:2004:DIF} in a pairwise fashion. We also initialize $R = I_0$. This part is implemented in OpenCV\footnote{OpenCV: https://github.com/opencv/opencv}. From this starting point, we jointly refine the alignment and our estimate of the full-color image by minimizing the loss function \eqref{eq:burst_demosaick}. Compared to any individual image $I_i$, our reconstruction is sharper, and does not suffer from color \textit{moir\'e} artifacts (Figure~\ref{fig:burst_demosaic}). We use the Adam gradient-descent optimizer~\cite{Kingma:2015:AMS} for $300$ iterations, setting the learning rate to $10^{-2}$ for $R$ and $10^{-4}$ for $H_i$. Our algorithm provides the gradient of the loss with respect to the reconstructed image $R$ and homographies $H_i$. We set $\lambda=10^{-3}$. For 13 $2048\times 2048$ images, computing the initial homographies takes $44.5s$, initializing the reconstruction $0.1s$. Minimizing the cost function takes $179.4s$ using the code generated by our automatic scheduler on a Titan X (Pascal) GPU\@.

\subsection{Non-image-processing Applications}
\label{sub:non_image_processing}

\begin{figure}[t]
  \centering
  \captionsetup[subfigure]{justification=centering}
  \begin{subfigure}[t]{0.49\textwidth}
    \includegraphics[width=\textwidth]{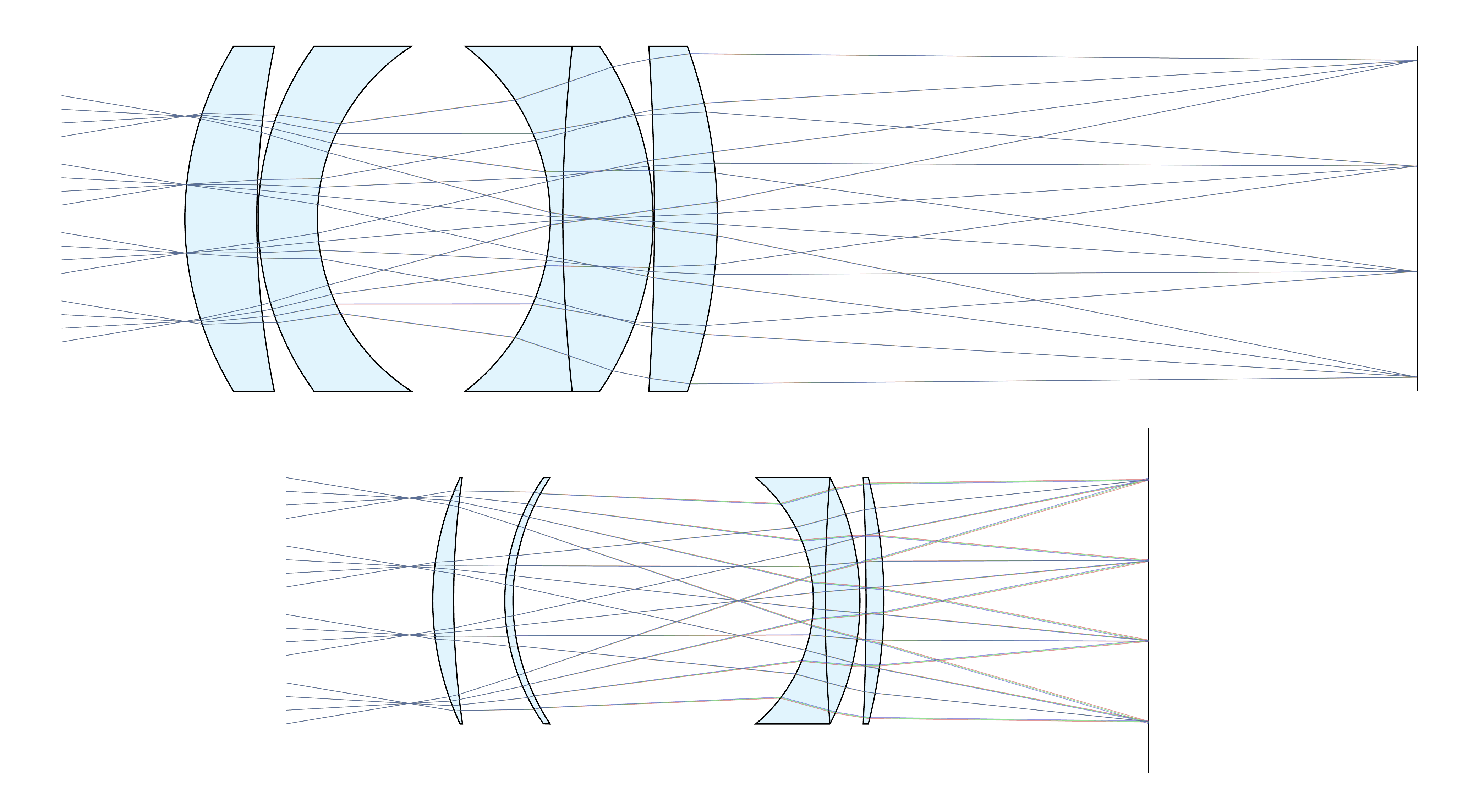}
    \caption{\label{fig:lens_optimization} lens optimization}
  \end{subfigure}
  \begin{subfigure}[t]{0.49\textwidth}
    \raisebox{0.3\height}{\includegraphics[width=\textwidth]{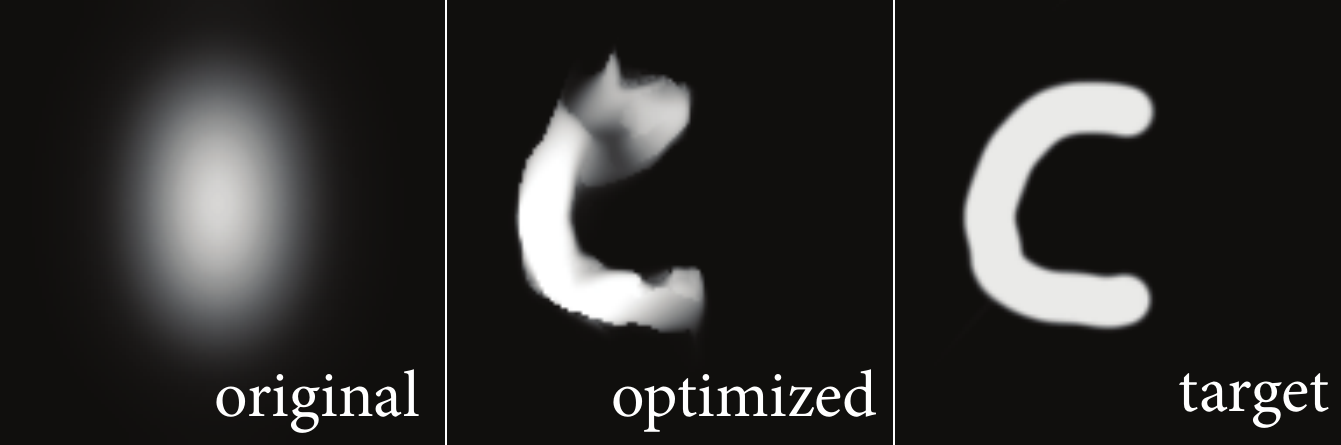}}
    \caption{\label{fig:fluid_simulation} inverse fluid simulation}
  \end{subfigure}
  \caption{\textbf{Non-image-processing applications.} Halide augmented with gradients is useful for a wider range of applications than just image processing and machine learning. (\subref{fig:lens_optimization}) By expressing a ray tracer for an optical system in Halide and taking derivatives of sharpness with respect to the lens parameters, we can reoptimize a classic Zeiss lens design~\cite{Lange:1957:GPO} (above) to be more compact (below) while maintaining as much sharpness as possible. (\subref{fig:fluid_simulation}) We can also optimize for fluid simulation, by taking a key frame from an original animation, and optimizing it to be as similar as possible to the target. We implement stable fluid~\cite{Stam:1999:SF} in Halide and optimize for the force field per frame, diffusion constant, viscosity, and time step size to make the last frame of the animation match the target.}
  \label{fig:non_image_processing}
\end{figure}

While we focus on image processing, Halide can express any feed-forward pipeline of arithmetic on multi-dimensional arrays (Figure~\ref{fig:non_image_processing}). There are numerous non-imaging applications in this class, and taking derivatives is useful for many of them. We implement two examples of this. First, we implemented a simple ray-tracer for a system of spherical lenses in Halide, and used our system to construct derivatives of the sharpness with respect to the lens positions and curvatures. In Figure~\ref{fig:lens_optimization}, we start from an existing Zeiss design~\cite{Lange:1957:GPO} and re-optimize it to be more compact while maintaining the field of view, F-number, and sharpness.

Secondly, we implement a classical grid-based fluid simulation algorithm~\cite{Stam:1999:SF} and use our system to differentiate the whole fluid simulation process. We implement a fluid control system (Figure~\ref{fig:fluid_simulation}). Given a target keyframe and an initial sequence of simulation images, we try to find a source force that ``bends'' the fluid to the desired image. We optimize for the force fields per frame, diffusion constant, viscosity, and time step size. This is not a novel application, but previously the derivatives were hand-derived~\cite{Treuille:2003:KCS, Mcnamara:2004:FCA}, while our system is capable of generating derivatives automatically, and adapt to new simulation algorithms.

\subsection{Future Work}
\label{sec:halide_future_work}

As these applications demonstrate, our system automatically delivers state of the art performance when computing the gradients of general image processing pipelines. We see three major directions for future work.

\paragraph{Higher-order derivatives and non-scalar outputs.}
Some optimization methods require derivatives of non-scalar outputs, the full Hessian matrix, or even higher-order derivatives~\cite{Girolami:2011:RML}. Our system supports repeated or nested application of differentiation. However, it only differentiates with respect to one scalar at a time. When the dimensionality of both the input and the output are high, there are automatic differentiation algorithms that are more efficient than both forward- and reverse-mode (Chapter~\ref{sec:beyond}). Incorporating these algorithms into our system, and developing better interfaces for non-scalar outputs and higher-order derivatives, will broaden the range of possible applications.

\paragraph{Better automatic scheduling.}
While it is possible to manually schedule the synthesized reverse computation, we found it challenging for non-trivial examples, and relied on our automatic scheduler entirely for this work. Its performance is good for gradient pipelines, but inspecting the generated code reveals plenty of room for further improvement. We consider the general Halide automatic scheduling problem still unsolved.

\paragraph{More general programming model.}
Halide assumes all operations are performed on a multi-dimensional grid. While this is a rather general programming model, there are many operations outside of image processing that are ill-suited for this model, such as sparse matrix multiplication or tree traversal. Generalizing Halide to handle differentiation of these operations, or developing new differentiable programming language to explore different trade-offs are both interesting directions.

\section{Conclusion}

Gradient-based optimization is revolutionizing many fields including image processing, but efficient computation of derivatives has so far been difficult, requiring one to either conform to limited building blocks or to error-prone manual derivation and challenging performance tuning. In contrast, our method can automatically generate high-performance gradient code for general image processing pipelines. Our method only requires the implementation of the operators in a language that is concise, easy to maintain, and portable. It then automatically derives the gradient code using reverse automatic differentiation. We have presented a new automatic performance tuner that handles the particular computation patterns exhibited by derivatives. Our code compiles to a variety of platforms such as x86, ARM, and GPUs, which is critical both for final deployment and for efficient training.

We have demonstrated that our work enables several types of applications, from custom neural network nodes, to the tuning of internal image processing parameters, to the solution of inverse problems. It dramatically simplifies the exploration of custom neural network nodes by automatically providing a level of performance that has so far been reserved to advanced CUDA programmers. It makes it easy to optimize internal weights and parameters for general image-processing pipelines, a step that few practitioners feel they can afford due to the cost of implementing gradients, which is especially true during the algorithmic exploration stages. Our system can also be used for inverse problems (which can even include unknown imaging parameters in addition to the unknown image). The user now only needs to worry about implementing the forward model. Each of the demonstrated applications was implemented initially in a few hours, and then evolved rapidly, with correct gradients and high-performance implementation automatically provided at each step by our method. We believe this will create new opportunities for rapid research and development in learning- and optimization-based imaging applications.

\chapter[Differentiable Monte Carlo Ray Tracing through Edge Sampling]{Differentiable Monte Carlo Ray Tracing \\ through Edge Sampling}
\label{chap:redner}

\begin{figure}[h]
  \centering
  \captionsetup[subfigure]{justification=centering}

  \begin{subfigure}[t]{0.16\textwidth}
  \includegraphics[width=0.99\linewidth]{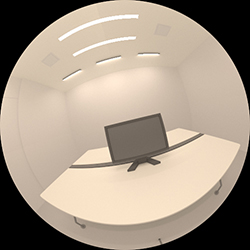} 
  \caption{initial guess}
  \label{fig:perception_lab_init}
  \end{subfigure}
  \begin{subfigure}[t]{0.16\textwidth}
  \includegraphics[width=0.99\linewidth]{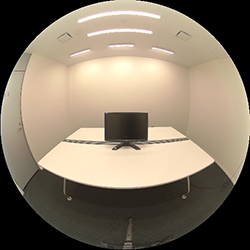}
  \caption{photograph}
  \label{fig:perception_lab_target}
  \end{subfigure}
  \begin{subfigure}[t]{0.16\textwidth}
  \includegraphics[width=0.99\linewidth]{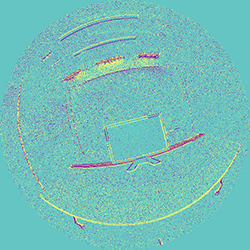}
  \caption{camera gradient}
  \label{fig:lab_gradient_camera}
  \end{subfigure}
  \begin{subfigure}[t]{0.16\textwidth}
  \includegraphics[width=0.99\linewidth]{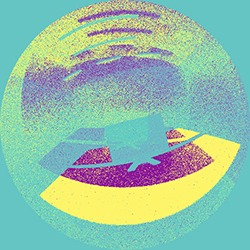}
  \caption{table albedo gradient}
  \label{fig:lab_gradient_table}
  \end{subfigure}
  \begin{subfigure}[t]{0.16\textwidth}
  \includegraphics[width=0.99\linewidth]{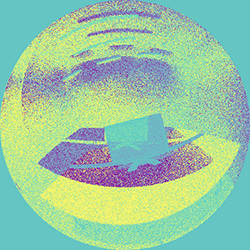}
  \caption{light gradient}
  \label{fig:lab_gradient_light}
  \end{subfigure}
  \begin{subfigure}[t]{0.16\textwidth}
  \includegraphics[width=0.99\linewidth]{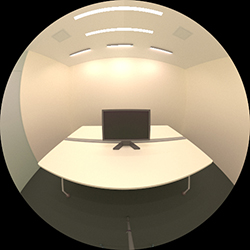}
  \caption{our fitted result}
  \label{fig:perception_lab_ours}
  \end{subfigure}
  \caption{\textbf{Differentiable Monte Carlo ray tracing.} We develop a general-purpose differentiable renderer that is capable of handling general light transport phenomena. Our method generates gradients with respect to scene parameters, such as camera pose (\subref{fig:lab_gradient_camera}), material parameters (\subref{fig:lab_gradient_table}), mesh vertex positions, and lighting parameters (\subref{fig:lab_gradient_light}), from a scalar loss computed from the output image. (\subref{fig:lab_gradient_camera}) shows the per-pixel gradient contribution of the $L^1$ difference with respect to the camera moving into the screen. (\subref{fig:lab_gradient_table}) shows the gradient with respect to the red channel of table albedo. (\subref{fig:lab_gradient_light}) shows the gradient with respect to the green channel of the intensity of one light source. As one of our applications, we use our gradient to perform an inverse rendering task by matching a real photograph (\subref{fig:perception_lab_target}) starting from an initial configuration (\subref{fig:perception_lab_init}) with a manual geometric recreation of the scene. The scene contains a fisheye camera with strong indirect illumination and non-Lambertian materials. We optimize for camera pose, material parameters, and light source intensity. Despite slight inaccuracies due to geometry mismatch and lens distortion, our method generates an image (\subref{fig:perception_lab_ours}) that almost matches the photo reference.}
  \label{fig:redner_teaser}
\end{figure}

The increasing importance of derivatives-based optimization creates a need for rendering algorithms that can be differentiated with respect to arbitrary input parameters, such as camera location and direction, scene geometry, lights, material appearance, or texture values. Unfortunately, the rendering integral includes visibility terms that are not differentiable at object boundaries. Whereas the final image function is usually differentiable once radiance has been integrated over pixel prefilters, light source areas, etc., the integrand of rendering algorithms is not. In particular, the derivative of the integrand has Dirac delta terms at occlusion boundaries that cannot be handled by traditional sampling strategies.

Previous work in differentiable rendering~\cite{Loper:2014:OAD, Kato:2018:N3M} has focused on fast, approximate solutions using simpler rendering models that only handle primary visibility, and ignore secondary effects such as shadows and indirect light. Analytical solutions exist for diffuse interreflection~\cite{Arvo:1994:IJP} but are difficult to generalize for arbitrary material models. The work by Ramamoorthi~et~al.~\cite{Ramamoorthi:2007:FAL} is an exception but it only differentiates with respect to image coordinates, whereas we want derivatives with respect to arbitrary scene parameters. Other previous work usually also relies on finite differences, with the usual limitation of these methods when the function is complex, namely that these methods work well for simple configurations but they do not propose a comprehensive solution to the full light transport equation.

In this chapter, we propose an algorithm that is, to the best of our knowledge, the first to compute derivatives of scalar functions over a physically-based rendered image with respect to arbitrary input parameters (camera, light materials, geometry, etc.). Our solution is stochastic and builds on Monte Carlo ray tracing~\cite{Kajiya:1986:RE}. For this, we introduce new techniques to explicitly sample edges of triangles in addition to the usual solid angle sampling of traditional approaches. This requires new spatial acceleration strategies and importance sampling to efficiently sample edges. Our method is general and can sample derivatives for arbitrary bounces of light transport. The running times we observed range from a second to a minute depending on the required precision, for an overhead of roughly $10\times - 20 \times$ compared to rendering an image alone.

We integrate our differentiable ray tracer with the automatic differentiation library PyTorch~\cite{Paszke:2017:ADP} for efficient integration with optimization and learning approaches. The scene geometry, lighting, camera, and materials are parameterized by PyTorch tensors, which enables a complex combination of 3D graphics, light transport, and neural networks. Backpropagation runs seamlessly across PyTorch and our renderer.

\section{Related Work}

\subsection{Inverse Graphics}

Inverse graphics techniques seek to find the scene parameters given observed images. Vision as inverse graphics has a long history in both computer graphics and vision (e.g.~\cite{Baumgart:1974:GMC,Yu:1999:IGI,Patow:2003:SIR}). Many techniques in inverse graphics use derivatives of the rendering process for inference.

Blanz and Vetter~\cite{Blanz:1999:MMS} optimized for the shape and texture of a face. Shacked and Lischinski~\cite{Shacked:2001:ALD} and Bousseau~et~al.~\cite{Bousseau:2011:OEM} optimized a perceptual metric for lighting design. Gkioulekas et~al.~\cite{Gkioulekas:2013:IVR, Gkioulekas:2016:ECI} focused on scattering parameters. Aittala~et~al.~\cite{Aittala:2013:PSC, Aittala:2015:TSC, Aittala:2016:RMN} inferred spatially varying material properties. Barron~et~al.~\cite{Barron:2015:SIR} proposed a solution to jointly optimize shape, illumination, and reflectance. Khungurn~et~al.~\cite{Khungurn:2015:MRF} and Zhao~et~al.~\cite{Zhao:2016:DSP} concentrated on optical properties of fabrics. All of the above approaches use gradients for solving the inverse problem, and had to develop a specialized solver to compute the gradient of the specific light transport scenarios they were interested in.

Loper and Black~\cite{Loper:2014:OAD} and Kato~et~al.~\cite{Kato:2018:N3M} proposed two general differentiable rendering pipelines. Rhodin~et~al.~\cite{Rhodin:2015:VSM} developed a differentiable volumetric ray caster. Liu~et~al.~\cite{Liu:2018:PSE} differentiated rendering process for inverse geometry editing. Athalye~et~al.~\cite{Athalye:2018:SRA}, Zeng~et~al.~\cite{Zeng:2017:AAB}, and Liu~et~al.~\cite{Liu:2019:BPN} all use differentiable rasterizers for adversarial example synthesis. All of them focus on performance and approximate the primary visibility gradients when there are multiple triangles inside a pixel, and assume Lambertian materials and do not compute shadows and global illumination. 

Recently, it is increasingly popular for deep learning methods to incorporate a \emph{differentiable rendering layer} in their architecture (e.g.~\cite{Liu:2017:MEU, Richardson:2017:LDF}). These rendering layers are usually special purpose and do not handle geometric discontinuities such as primary visibility and shadow.

To our knowledge, our method is the first that is able to differentiate through a full path tracer, while taking the geometric discontinuities into account.

\subsection{Derivatives in Rendering}

Analytical derivatives have been used for computing the footprint of light paths~\cite{Shinya:1987:PAP, Igehy:1999:TRD, Suykens:2001:PDA} and predicting the changes of specular light paths~\cite{Chen:2000:TAS, Jakob:2012:MEM, Kaplanyan:2014:NCR}. The derivatives are usually manually derived for the particular type of light paths the work focused on, making it difficult to generalize to arbitrary material models or lighting effects. Unlike these methods, we compute the gradients using a hybrid approach that mixes automatic differentiation and manually derived derivatives focusing on the discontinuous integrand.

Arvo~\cite{Arvo:1994:IJP} proposed an analytical method for computing the spatial gradients for irradiance. The method requires clipping of triangle meshes in order to correctly integrate the form factor, and does not scale well to scenes with large complexity. It is also difficult or impossible to compute closed-form integration for arbitrary materials.

Ramamoorthi~et~al.'s work on first order analysis of light transport~\cite{Ramamoorthi:2007:FAL} is highly related to our method. Their method is a special case of ours. Our derivation generalizes their method to differentiate with respect to any scene parameters. Furthermore, we handle primary visibility, secondary visibility, and global illumination.

Irradiance or radiance caching~\cite{Ward:1992:IG,Krivanek:2005:RCE,Jarosz:2012:TAA} numerically computes the gradient of interreflection with respect to spatial position and orientation of the receiver. To take discontinuities into account, these methods resort to stratified sampling. Unlike these methods, we estimate the gradient integral directly by automatic differentiation and edge sampling.

In Chapter~\ref{chap:h2mc}, we propose a variant of the Metropolis light transport~\cite{Veach:1997:MLT} algorithm by computing the Hessian of a light path contribution with respect to the path parameters using automatic differentiation. It focuses on second-derivatives for forward rendering and does not take geometric discontinuities into account. We will discuss more in Chapter~\ref{chap:h2mc}.

\section{Method}
\label{sec:redner_method}

\begin{figure}[t]
  \centering
  \captionsetup[subfigure]{justification=centering}

  \begin{subfigure}{0.33\textwidth}
  \includegraphics[width=0.99\linewidth]{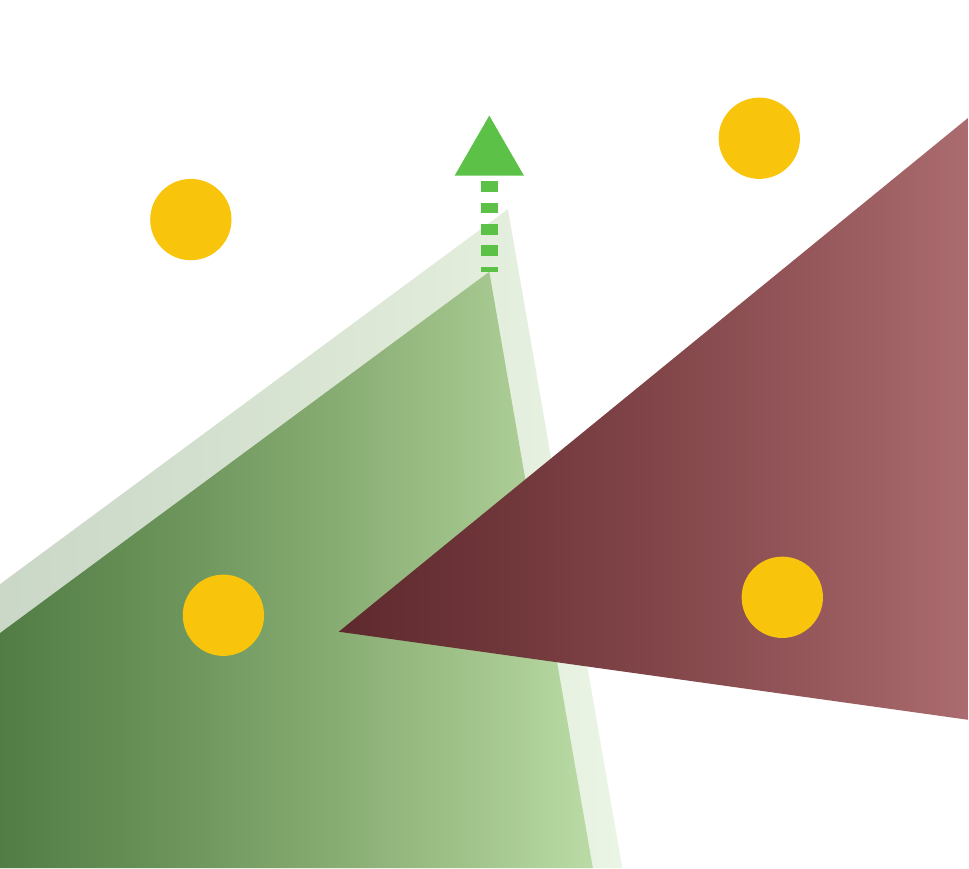} 
  \caption{area sampling}
  \label{fig:area}
  \end{subfigure}
  \begin{subfigure}{0.33\textwidth}
  \includegraphics[width=0.99\linewidth]{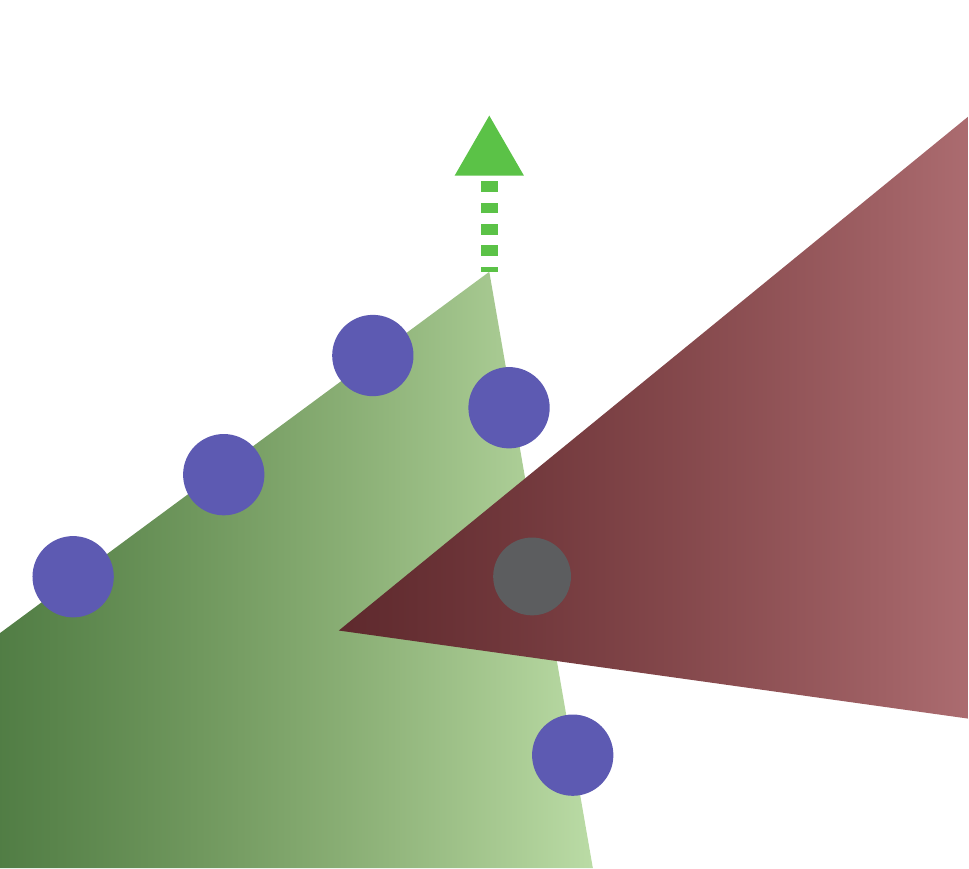}
  \caption{edge sampling}
  \label{fig:edge}
  \end{subfigure}
  \caption{\textbf{Area sampling v.s. edge sampling.} (\subref{fig:area}) The figure shows a pixel overlapped with two triangles. We are interested in computing the derivative of pixel color with respect to the green triangle moving up. Since the area covered by the green triangle increases, the final pixel color will contain more green area and less white background. Traditional area sampling (yellow samples) even instrumented with automatic differentiation does not account for the change in the covered area. (\subref{fig:edge}) In addition to traditional area sampling, we propose a novel edge sampling algorithm (blue samples) to sample the differential area on the edges. Our method computes unbiased gradients and correctly takes occlusion into account.}
  \label{fig:area_vs_edge}
\end{figure}

Our task is the following: given a 3D scene with a continuous parameter set $\Phi$ (including camera pose, scene geometry, material and lighting parameters), we generate an image using the path tracing algorithm~\cite{Kajiya:1986:RE}. Given a scalar function computed from the image (e.g. a loss function we want to optimize), our goal is to backpropagate the gradient of the scalar with respect to all scene parameters $\Phi$.

The pixel color is formalized as an integration over all light paths that pass through the pixel filter. We use Monte Carlo sampling to estimate both the integral and the gradient of the integral~\cite{Veach:1998:RMC, Pharr:2016:PBR}. However, since the integrand is discontinuous due to edges of geometry and occlusion, traditional area sampling does not correctly capture the changes due to camera parameters or triangle vertex movement (Figure~\ref{fig:area_vs_edge}~\subref{fig:area}). Mathematically, the gradient of the discontinuous integrand is a Dirac delta function, therefore traditional sampling has zero probability of capturing the Dirac deltas.

Our strategy for computing the gradient integral is to split it into smooth and discontinuous regions (Figure~\ref{fig:area_vs_edge}). For the smooth part of the integrand (e.g. Phong shading or trilinear texture reconstruction), we employ traditional area sampling with automatic differentiation. For the discontinuous part, we use a novel edge sampling method to capture the changes at boundaries. In this section, we first focus on primary visibility where we only integrate over the 2D screen-space domain (Chapter~\ref{sec:primary_visibility}). We then generalize the method to handle shadow and global illumination (Chapter~\ref{sec:secondary_visibility}).

We focus on triangle meshes and we assume the meshes have been preprocessed such that there is no interpenetration. We also assume no point light sources and no perfectly specular surfaces. We approximate these with area light sources and materials with very low roughness. We also focus on static scenes and leave integration over the time dimension for motion blur as future work.

\subsection{Primary Visibility}
\label{sec:primary_visibility}

\begin{figure}[t]
  \centering
  \captionsetup[subfigure]{justification=centering}

  \begin{subfigure}{0.33\textwidth}
  \includegraphics[width=0.99\linewidth]{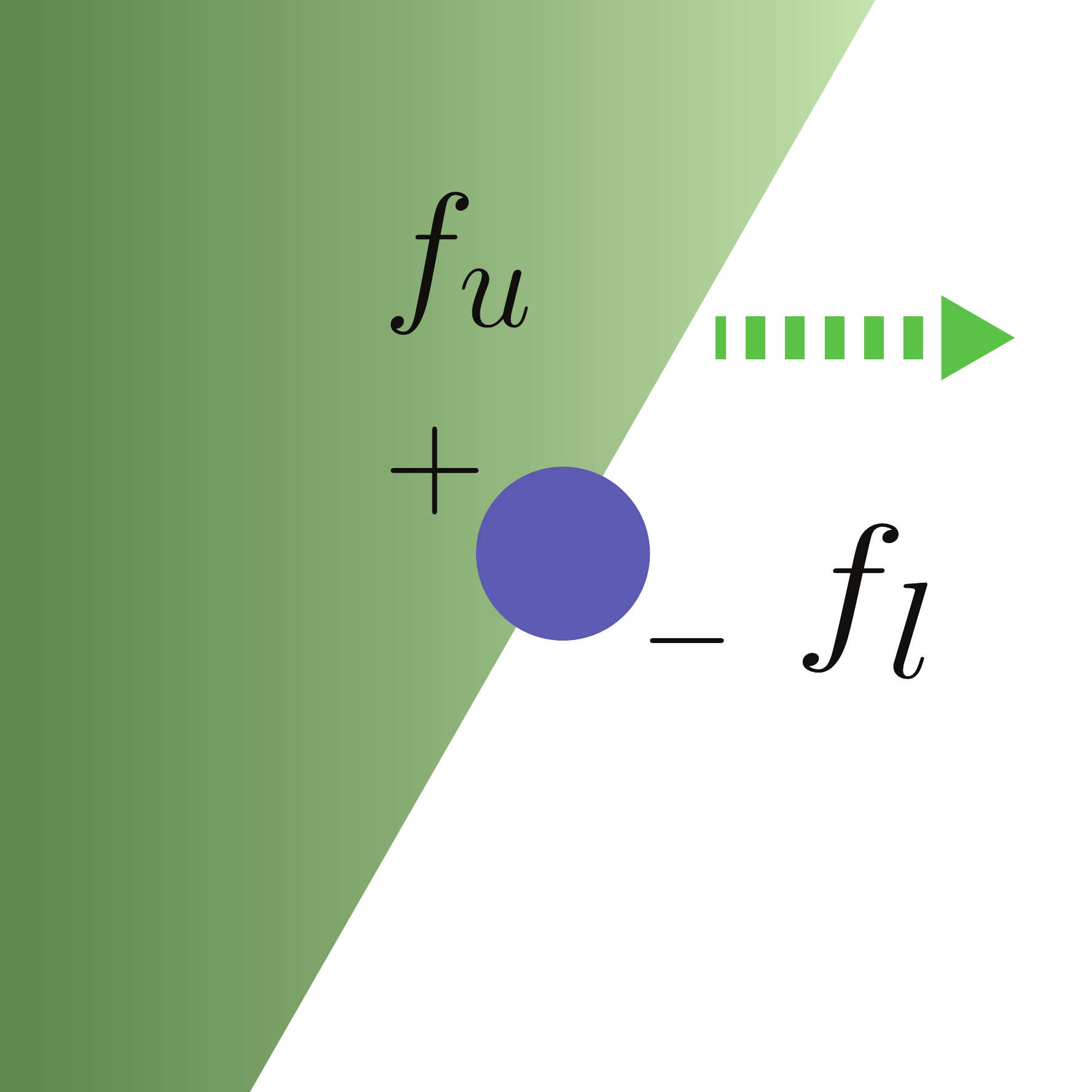} 
  \caption{half-spaces}
  \label{fig:edge_eps}
  \end{subfigure}
  \begin{subfigure}{0.33\textwidth}
  \includegraphics[width=0.99\linewidth]{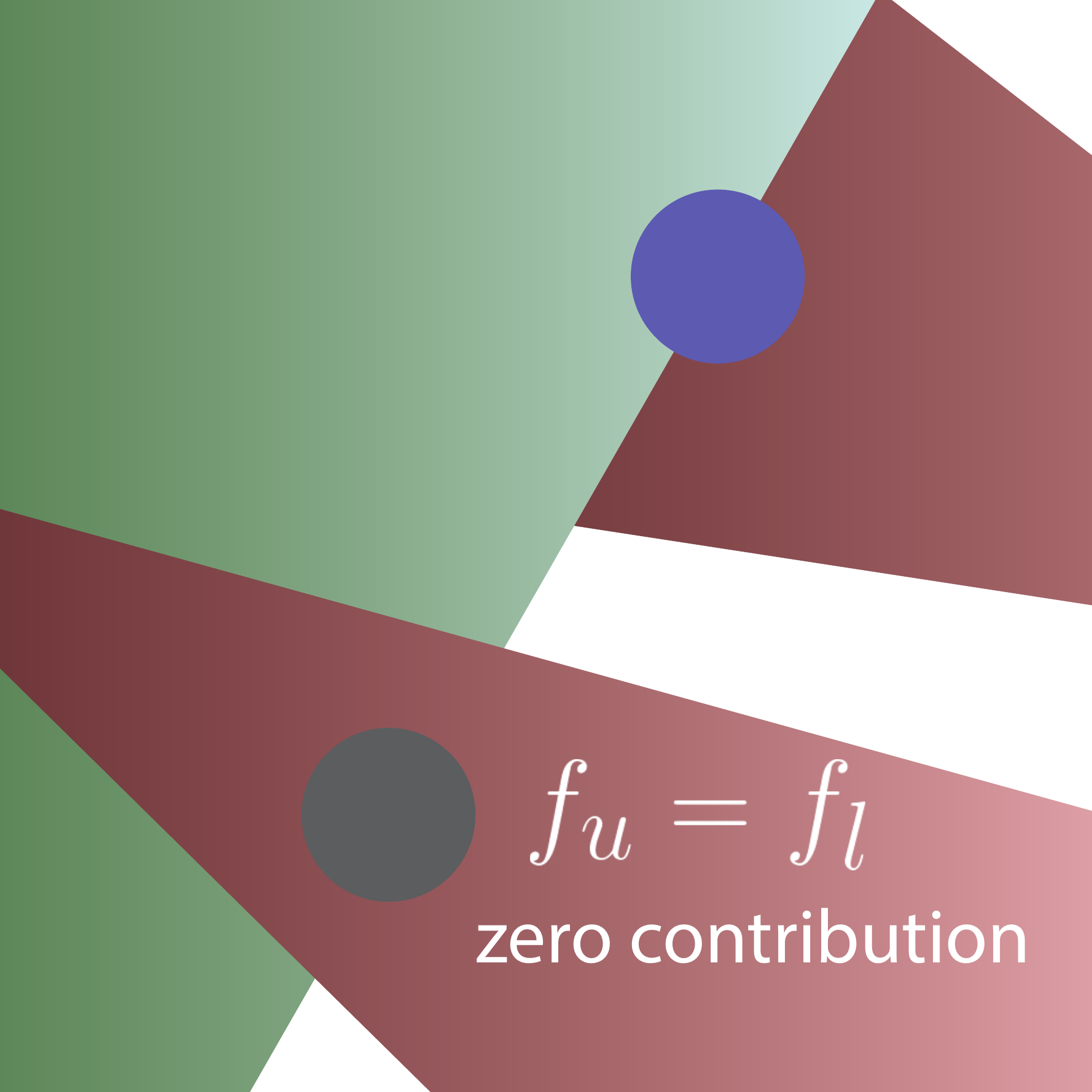}
  \caption{occlusion}
  \label{fig:edge_occlusion}
  \end{subfigure}
  \caption{\textbf{Edge sampling.} (\subref{fig:edge_eps}) An edge splits the space into two half-spaces $f_u$ and $f_l$. If the edge moves right, the green area increases while the white area decreases. We integrate over edges to compute gradients by taking into account the change in areas. To compute the integration, we sample a point (the blue point) on the edge and compute the difference between the half-spaces by computing the color on the two sides of the edge. (\subref{fig:edge_occlusion}) Our method handles occlusion correctly since an occluded sample will land on the continuous part of the path contribution function, thus having the exact same contribution on the two sides (for example, the grey sample has zero contribution to the gradient).}
  \label{fig:edge_sample}
\end{figure}

We start by focusing on the 2D pixel filter integral for each pixel that integrates over the pixel filter $k$ and the radiance $L$, where the radiance itself can be another integral that integrates over light sources or the hemisphere. We also focus on linear projective cameras for simplicity. We will generalize the method to handle discontinuities inside the radiance integral in Chapter~\ref{sec:secondary_visibility}. We will also generalize to nonlinear projections such as fisheye cameras in Chapter~\ref{sec:fisheye_cameras}. The pixel color $I$ can be written as:
\begin{equation}
I = \iint k(x, y) L(x, y) dxdy.
\label{eq:first}
\end{equation}

For notational convenience, we will combine the pixel filter and radiance and call them scene function $f(x, y) = k(x, y) L(x, y)$. We are interested in the gradients of the integral with respect to some parameters $\Phi$ in the scene function $f(x, y; \Phi)$, such as the position of a mesh vertex or camera pose:
\begin{equation}
\nabla I = \nabla \iint f(x, y; \Phi) dxdy.
\label{eq:gradient}
\end{equation}

The integral usually does not have a closed-form solution, especially when more complex effects such as non-Lambertian materials are involved. Therefore we rely on Monte Carlo integration to estimate the pixel value $I$:
\begin{equation}
I \approx \frac{1}{N}\sum_{i} f(x_i, y_i; \Phi),
\label{eq:gradient_monte_carlo}
\end{equation}
where $N$ is the number of samples we use for pixel $I$, and $x_i, y_i$ are the screen-space samples. Unfortunately, we cannot take the naive approach of applying the same Monte Carlo estimator for the gradient $\nabla I$, since the scene function $f$ is not necessarily differentiable with respect to the scene parameters (Figure~\ref{fig:area}).

A key observation is that all the discontinuities happen at triangle edges, since we assume no interpenetration. This allows us to explicitly integrate over the discontinuities. A 2D triangle edge splits the space into two half-spaces ($f_u$ and $f_l$ in Figure~\ref{fig:edge_sample}). We can model it as a Heaviside step function $\theta$:
\begin{equation}
\theta(\alpha(x, y))f_{u}(x, y) + \theta(-\alpha(x, y))f_{l}(x, y),
\label{eq:half_space}
\end{equation}
where $f_u$ represents the upper half-space, $f_l$ represents the lower half-space, and $\alpha$ defines the edge equation formed by the triangles.

For each edge with two endpoints $(a_x, a_y), (b_x, b_y)$, we can construct the edge equation by forming the line $\alpha(x, y) = Ax + By + C$, since we assume we are using a projective camera. If $\alpha(x, y) > 0$ then the point is in the upper half-space, and vice versa. For the two endpoints of the edge $\alpha(x, y) = 0$. Hence, by plugging in the two endpoints we obtain:
\begin{equation}
\alpha(x, y) = (a_y - b_y) x + (b_x - a_x) y + (a_x b_y - b_x a_y).
\label{eq:edge_equation}
\end{equation}

% Reviewer 1:
% At this point, I had a question about whether the method works for complex combinations of overlapping shapes, where occlusion edges lie on each other. This should be clarified in the paper and in the rebuttal. I think it might work, but it seems like it would fail when one triangle lies immediately on top of another, and so the second triangle starts becoming visible, but will have zero derivative since it is initially occluded. But this may be a measure 0 case. What about two occlusion edges of two triangles on top of each other not perfectly aligned; is there intersection point handled correctly, or is it a 0 measure set which is not consequential? 

We can rewrite the scene function $f$ as a summation of Heaviside step functions $\theta$ with edge equation $\alpha_i$ multiplied by an arbitrary function $f_i$:
\begin{equation}
\iint f(x, y) dxdy = \sum_i \iint \theta(\alpha_i(x, y))f_i(x, y)dxdy.
\label{eq:scene_function_step}
\end{equation}
$f_i$ itself can contain Heaviside step functions: for example, a triangle defines a multiplication of three Heaviside step functions. Occlusion can also be modeled by multiplying step functions from other edges closer to the camera. $f_i$ can even be an integral over light sources or the hemisphere. This is crucial for our later generalization to secondary visibility.

% Reviewer 1:
% The paper can do a better job of stressing that to differentiate visibility correctly, one needs to analytically differentiate the heaviside step function and integrate the delta, and only then numerically evaluate the resulting well-behaved integral. Efforts to differentiate numerically up front will fail because of the discontinuities and delta functions. This is a key point of the paper and should be stressed. However, if one does the analytical differentiation first, everything is actually well-behaved and can be analyzed. Right now, the paper just implicitly conveys this in equation 6, but the high-level intuition and insight is missed.

We want to analytically differentiate the Heaviside step function $\theta$ and explicitly integrate over its derivative -- the Dirac delta function $\delta$. To do this we first swap the gradient operator inside the integral, then we use the product rule to separate the integral into two:
\begin{equation}
\begin{aligned}
&\nabla \iint \theta(\alpha_i(x, y)) f_i(x, y) dxdy \\
= &\iint \delta(\alpha_i(x, y)) \nabla \alpha_i(x, y) f_i(x, y) dxdy + \iint \nabla f_i(x, y) \theta(\alpha_i(x, y)) dxdy.
\end{aligned}
\label{eq:redner_product_rule}
\end{equation}

Equation~\eqref{eq:redner_product_rule} shows that we can estimate the gradient using two Monte Carlo estimators. The first one estimates the integral over the edges of triangles containing the Dirac delta functions, and the second estimates the original pixel integral except the smooth function $f_i$ is replaced by its gradient, which can be computed through automatic differentiation.

To estimate the integral containing Dirac delta functions, we eliminate the Dirac function by performing a variable substitution to rewrite the first term containing the Dirac delta function to integrate over the edge, that is, over the regions where $\alpha_i(x, y) = 0$:
\begin{equation}
\begin{aligned}
\iint \delta(\alpha_i(x, y)) \nabla \alpha_i(x, y) f_i(x, y) dxdy 
= \int_{\alpha_i(x, y) = 0} 
\frac{\nabla\alpha_i(x, y)}{\left\| \nabla_{x, y}\alpha_i(x, y) \right\|} f_i(x, y)
d\sigma(x, y),
\end{aligned}
\label{eq:dirac_rewrite}
\end{equation}
where $\left\| \nabla_{x, y}\alpha_i(x, y) \right\|$ is the $L^2$ length of the gradient of the edge equations $\alpha_i$ with respect to $x, y$, which takes the Jacobian of the variable substitution into account. $\sigma(x, y)$ is the measure of the length on the edge~\cite{Hormander:1983:ALP}.

The gradients of the edge equations $\alpha_i$ are:
\begin{equation}
\begin{aligned}
&\left\| \nabla_{x, y} \alpha_i \right\| = \sqrt{(a_x - b_x)^2 + (a_y - b_y)^2} \\
&\frac{\partial \alpha_i}{\partial a_x} = b_y - y,\ 
 \frac{\partial \alpha_i}{\partial a_y} = x - b_x \\
&\frac{\partial \alpha_i}{\partial b_x} = y - a_y,\ 
 \frac{\partial \alpha_i}{\partial b_y} = a_x - x \\
&\frac{\partial \alpha_i}{\partial x} = a_y - b_y,\  
 \frac{\partial \alpha_i}{\partial y} = b_x - a_x.
\end{aligned}
\label{eq:redner_dirac_jacobian}
\end{equation}
As a byproduct of the derivation, we also obtain the screen space gradients $\frac{\partial}{\partial x}$ and $\frac{\partial}{\partial y}$, which can potentially facilitate adaptive sampling as shown in Ramamoorthi~et~al.'s first-order analysis~\cite{Ramamoorthi:2007:FAL}. We can obtain the gradient with respect to other parameters, such as camera parameters, 3D vertex positions, or vertex normals by propagating the derivatives from the projected triangle vertices using the chain rule:
\begin{equation}
\frac{\partial \alpha}{\partial p} =
	\sum_{k \in \{x, y\}}
	\frac{\partial \alpha}{\partial a_k} \frac{\partial a_k}{\partial p} +
	\frac{\partial \alpha}{\partial b_k} \frac{\partial b_k}{\partial p},
\end{equation}
where $p$ is the desired parameter.

We use Monte Carlo sampling to estimate the Dirac integral (Equation~\eqref{eq:dirac_rewrite}). Recall that a triangle edge defines two half-spaces (Equation~\eqref{eq:half_space}), therefore we need to compute the two values $f_l(x, y)$ and $f_u(x, y)$ on the edge (Figure~\ref{fig:edge_sample}). By combining Equation~\eqref{eq:half_space} and Equation~\eqref{eq:dirac_rewrite}, our Monte Carlo estimation of the Dirac integral for a single edge $E$ on a triangle can be written as:
\begin{equation}
\frac{1}{N}\sum_{j=1}^{N} 
\frac{\left\| E \right\| \nabla \alpha_i(x_j, y_j) (f_{u}(x_j, y_j) - f_{l}(x_j, y_j))} 
{P(E) \left\| \nabla_{x_j, y_j}\alpha_i(x_j, y_j) \right\|},
\label{eq:mc}
\end{equation}
where $\left\| E \right\|$ is the length of the edge and $P(E)$ is the probability of selecting edge $E$.
% where $p(x, y)$ is the probability density function of the edge sampling distribution.

In practice, we use a path tracer to evaluate the values on the two sides of an edge ($f_l(x, y)$ and $f_u(x, y)$). We trace two light paths from the screen space position $(x, y)$, and offset them by a small amount (say, $10^{-6}$). We use the same random number sequence for both light paths to minimize the variance through correlated sampling. Note that this is different from finite differences: we do not move the target parameter to acquire their derivatives. 

If we employ smooth shading, most of the triangle edges are in the continuous regions and the Dirac integral is zero for these edges since by definition of continuity $f_{u}(x, y) = f_{l}(x, y)$. Only the \textit{silhouette edges} (e.g.~\cite{Hertzmann:1999:INP}) have non-zero contribution to the gradients. We select the edges by projecting all triangle meshes to screen space and clip them against the camera frustum. We select one silhouette edge with probability proportional to the screen space lengths. We then uniformly pick a point on the selected edge. For thin-lens camera that produce depth-of-field effects, we use the center of the camera pupil as the basis of projection, and we conservatively test the silhouette using the four corners of the pupil bounding box.

\begin{figure}[t]
  \centering
  \includegraphics[width=0.5\linewidth]{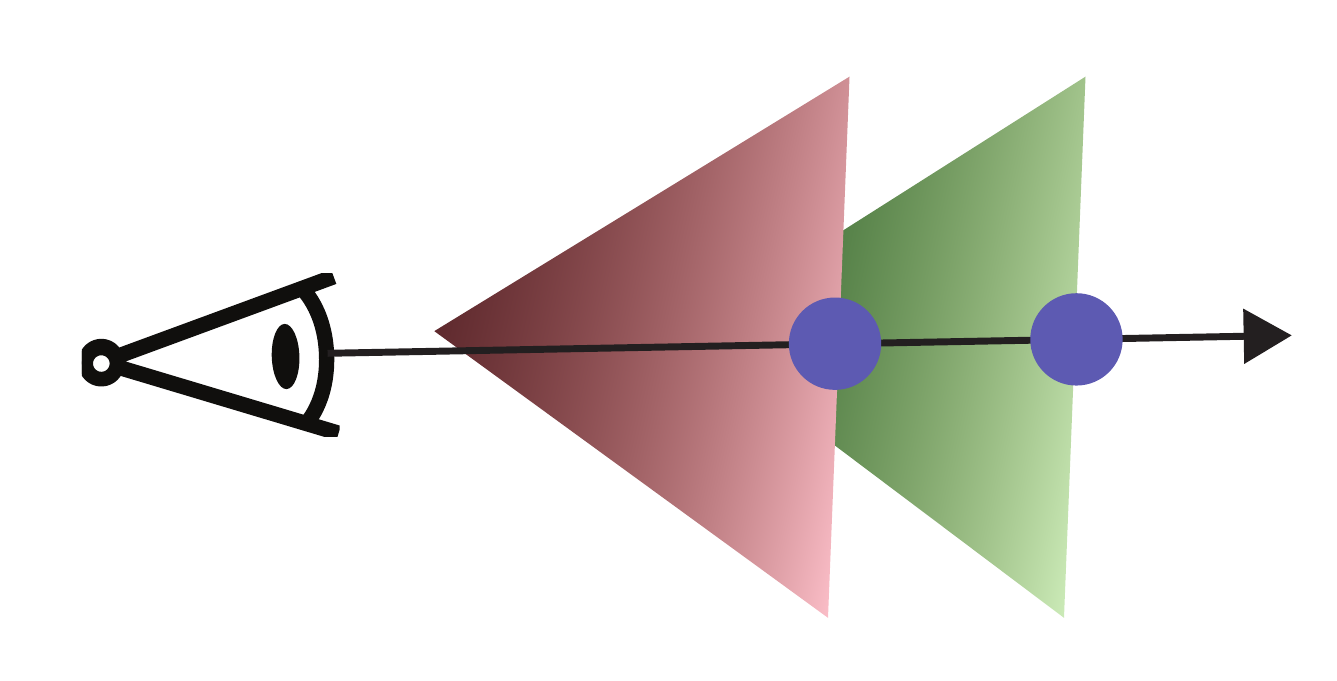} 
  \caption{\textbf{Parallel edges.} Our method almost always handles occlusion correctly. However, it can produce incorrect results in the pathological case where two edges are exactly parallel to the viewport. If we sampled the occluded parallel edge, we consider it occluded and wrongly ignore the color of the occluded triangle. If we sampled the edge that occludes other parallel edges, we do not take the occluded triangle color into consideration. Fortunately, this is a zero-measure case and it rarely happens in practice.}
  \label{fig:parallel_edge}
\end{figure}

% We compute them by picking two points along the edge normal from a small distance 
% $\epsilon$.
% Our method handles occlusion correctly since if the sample is blocked by another 
% surface, $(x, y)$ will always land on the continuous part of the contribution function 
% $f(x, y)$.
% Thus by definition of continuity $f_{u}(x, y) = f_{l}(x, y)$, such samples will have
% zero contribution to the gradients. Figure~\ref{fig:edge_sample} illustrates the process.
Our method handles occlusion correctly, since if the sample is blocked by another surface, $(x, y)$ will always land on the continuous part of the contribution function $f(x, y)$. Such samples will have zero contribution to the gradients. Figure~\ref{fig:edge_occlusion} illustrates the process. However, in the pathological case where two edges are exactly parallel to the viewport (Figure~\ref{fig:parallel_edge}), we can compute the wrong result. This is because occlusion is modeled as a multiplication between two Heaviside step functions in Equation~\eqref{eq:scene_function_step}, and if two step functions coincide exactly, we completely mask the occluded edge and wrongly ignore its contribution. In theory, this can be resolved by detecting parallel edges in the ray casting operation, and properly breaking even: If we sampled an edge occluded by a parallel edge, treat it as not occluded. On the other hand, if we sampled an edge that occludes other parallel edges, set the other half-space to the occluded triangle. However, since this is a zero-measure case and we never observe it in practice, we do not implement it at the moment.

To recap, we use two sampling strategies to estimate the gradient integral of pixel filter (Equation~\eqref{eq:gradient}): one for the discontinuous regions of the integrand (first term of Equation~\eqref{eq:redner_product_rule}), one for the continuous regions (second term of Equation~\eqref{eq:redner_product_rule}). To compute the gradient for discontinuous regions, we need to explicitly sample the edges. We compute the difference between two sides of the edges using Monte Carlo sampling (Equation~\eqref{eq:mc}).

\subsection{Secondary visibility}
\label{sec:secondary_visibility}

\begin{figure}[t]
  \centering
  \captionsetup[subfigure]{justification=centering}

  \begin{subfigure}{0.33\textwidth}
  \includegraphics[width=0.99\linewidth]{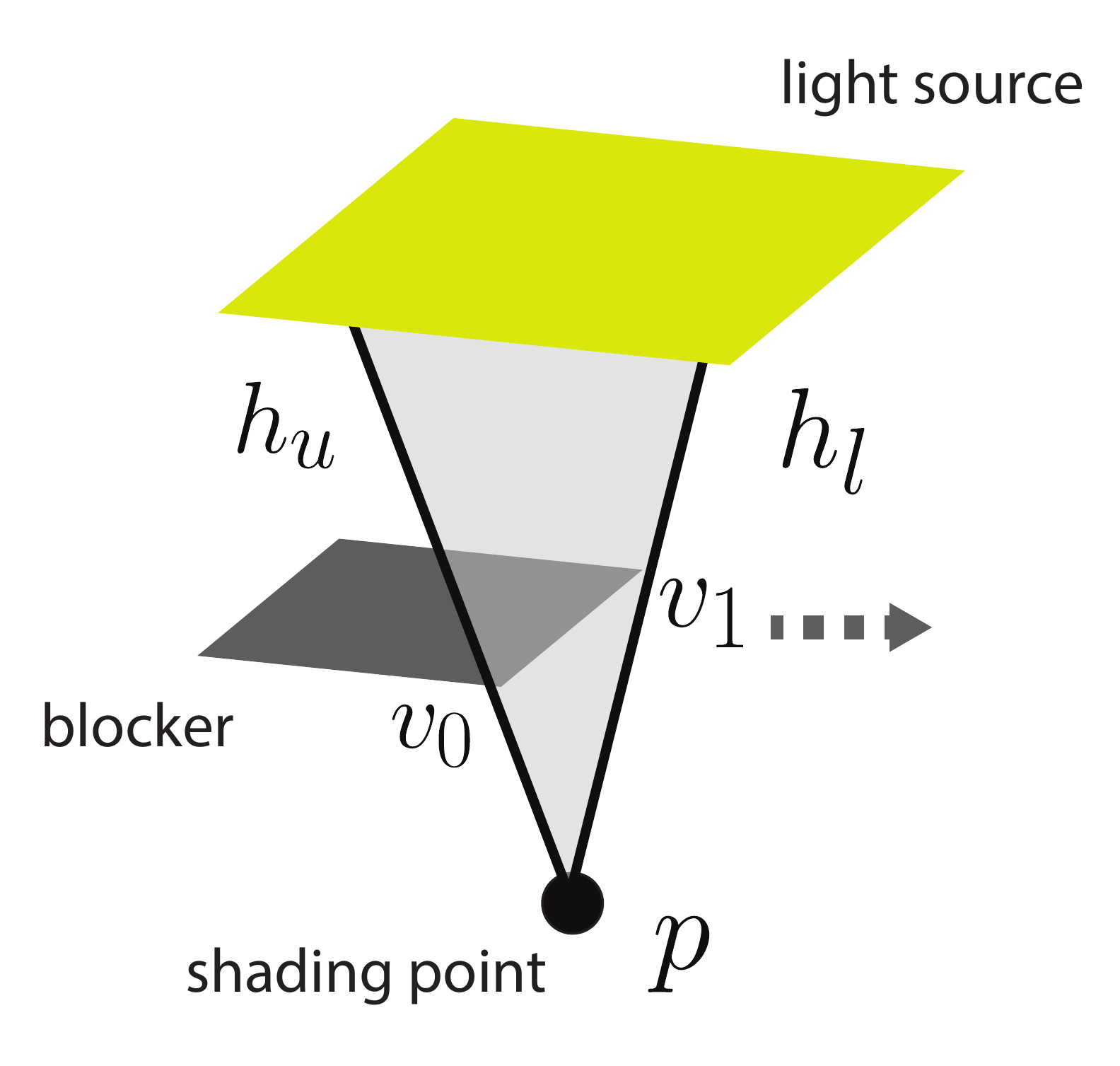} 
  \caption{secondary visibility}
  \label{fig:secondary_visibility}
  \end{subfigure}
  \begin{subfigure}{0.33\textwidth}
  \includegraphics[width=0.99\linewidth]{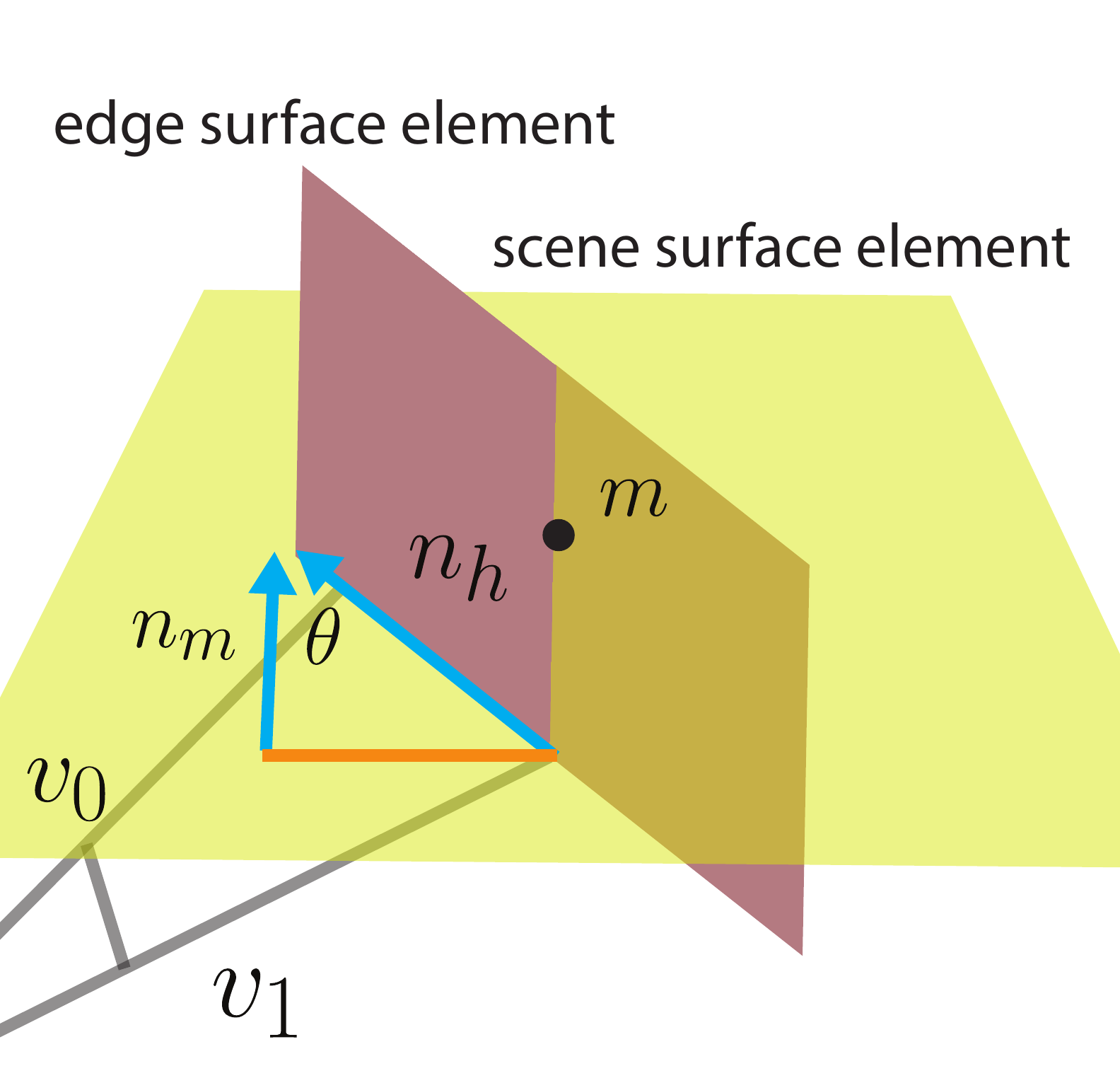} 
  \caption{width correction}
  \label{fig:width_correction}
  \end{subfigure}
  \caption{(\subref{fig:secondary_visibility}) Our method can be easily generalized to handle shadow and global illumination. Similar to the primary visibility case (Figure~\ref{fig:edge_sample}), a geometry edge $(v_0, v_1)$ and the shading point $p$ splits the 3D space into two half-spaces $f_u$ and $f_l$ and introduces a discontinuity. Assuming the blocker is moving right, we integrate over the edge to compute the difference. By doing so we take account of the increase in blocker area and the decrease in light source area looking from the shading point. The integration over the edge is defined on the intersection between the scene manifold and the plane formed by the shading point and the edge (the semi-transparent triangle). (\subref{fig:width_correction}) The orientation of the infinitesimal width of the edge differs from the scene surface element the edge intersects with. During the integration, we need to project the scene surface element width onto the edge surface element. The ratio of the widths between the two is determined by $\frac{1}{\sin \theta}$, which is one over the length of the cross product between the normal of the edge plane and the scene surface.}
  \label{fig:blocker_half_spaces}
\end{figure}

Our method can be easily generalized to handle effects such as shadow and global illumination by integrating over the 3D scene. Figure~\ref{fig:blocker_half_spaces} illustrates the idea.

We focus on a single shading point $p$ since we can propagate the derivatives back to screen space and camera parameters using Equation~\eqref{eq:redner_product_rule}. Given the shading point, the shading equation involves integration over all points $m$ on the scene manifold $\mathcal{M}$:
\begin{equation}
g(p) = \int_{\mathcal{M}} h(p, m) dA(m),
\end{equation}
where $A$ is the area measure of point $m$, and $h$ is the scene function including material response, geometric factor, incoming radiance, and visibility. Note that $g(p)$ can itself be part of the pixel integrand $f(x, y)$ in the previous section (Equation~\eqref{eq:first}). Therefore we can propagate the gradient of $g(p)$ using the chain rule or automatic differentiation with Equation~\eqref{eq:redner_product_rule}.

Similar to the primary visibility case, an edge $(v_0, v_1)$ in 3D introduces a step function into the scene function $h$:
\begin{equation}
\theta(\alpha(p, m))h_{u}(p, m) + \theta(-\alpha(p, m))h_{l}(p, m).
\label{eq:3d_half_space}
\end{equation}
We can derive the edge function $\alpha(m)$ by forming a plane using the shading point $p$ and the two points on the edge. The sign of the dot product between the vector $m - p$ and the plane normal determines the two half-spaces. Therefore, the edge equation $\alpha(m)$ can be defined by
\begin{equation}
\alpha(p, m) = (m - p) \cdot \left(v_0 - p\right) \times \left(v_1 - p\right).
\end{equation}

To compute the gradients, we analogously apply the derivation used for primary visibility, using the 3D version of Equation~\eqref{eq:redner_product_rule} and Equation~\eqref{eq:dirac_rewrite} with $x, y$ replaced by $p, m$. The edge integral integrating over the line on the scene surface, analogous to Equation~\eqref{eq:dirac_rewrite} is:
\begin{equation}
\begin{aligned}
& \int_{\alpha(p, m) = 0}
\frac{\nabla\alpha(p, m)}{\left\| \nabla_{m}\alpha(p, m) \right\|} h(p, m)
\frac{1}{\left\| n_m \times n_h \right\|} d\sigma'(m) \\
& n_h = \frac{\left(v_0 - p\right) \times \left(v_1 - p\right)}{\left\| \left(v_0 - p\right) \times \left(v_1 - p\right) \right\|},
\end{aligned}
\label{eq:dirac_3d}
\end{equation}
where $n_m$ is the surface normal on point $m$. There are two crucial differences between the 3D edge integral (Equation~\eqref{eq:dirac_3d}) and the previous screen space edge integral (Equation~\eqref{eq:dirac_rewrite}). First, while the measure of the screen space edge integral $\sigma(x, y)$ coincides with the unit length of the 2D edge, the measure of the 3D edge integral $\sigma'(m)$ is the length of projection of a point on the edge from the shading point $p$ to a point $m$ on the scene manifold (the semi-transparent triangle in Figure~\ref{fig:secondary_visibility} illustrates the projection). Second, there is an extra area correction term $\left\| n_m \times n_h \right\|$, since we need to project the scene surface element onto the infinitesimal width of the edge (Figure~\ref{fig:width_correction}).

To integrate the 3D edge integral using Monte Carlo sampling we substitute the variable again from the point $m$ on the surface to the line parameter $t$ on the edge $v_0 + t(v_1 - v_0)$:
\begin{equation}
\int_{0}^{1}
\frac{\nabla\alpha(p, m(t))}{\left\| \nabla_{m}\alpha(p, m(t)) \right\|} h(p, m(t))
\frac{\left\| J_m(t) \right\|}{\left\| n_m \times n_h \right\|} dt,
\label{eq:dirac_3d_edge}
\end{equation}
where the Jacobian $J_m(t)$ is a 3D vector describing the projection of edge $(v_0, v_1)$ onto the scene manifold with respect to the line parameter. We derive the Jacobian in Appendix~\ref{sec:secondary_measure_derivation}.

The derivatives for $\alpha(p, m)$ needed to compute the edge integral are:
\begin{equation}
\begin{aligned}
&{v_0}' = v_0 - p \text{, } {v_1}' = v_1 - p \text{, } m' = m - p \\
&\left\| \nabla_{m} \alpha(p, m) \right\| = \left\| {v_0}' \times {v_1}' \right\| \\
&\nabla_{v_0} \alpha(p, m) = {v_1}' \times m' \\
&\nabla_{v_1} \alpha(p, m) = m' \times {v_0}' \\
&\nabla_{p} \alpha(p, m) = {v_0}' \times {v_1}' + m' \times {v_1}' + {v_0}' \times m'.
\end{aligned}
\label{eq:alpha_3d}
\end{equation}

Efficient Monte Carlo sampling of secondary edges is more involved. Unlike primary visibility where the viewpoint does not change much, shading point $p$ can be anywhere in the scene. The consequence is that we need a more sophisticated data structure to prune the edges with zero contribution. Chapter~\ref{sec:redner_implementaiton} describes the process for importance sampling edges.

\subsection{Cameras with Non-linear Projections}
\label{sec:fisheye_cameras}

\begin{figure}[t]
  \centering
  \includegraphics[width=0.45\linewidth]{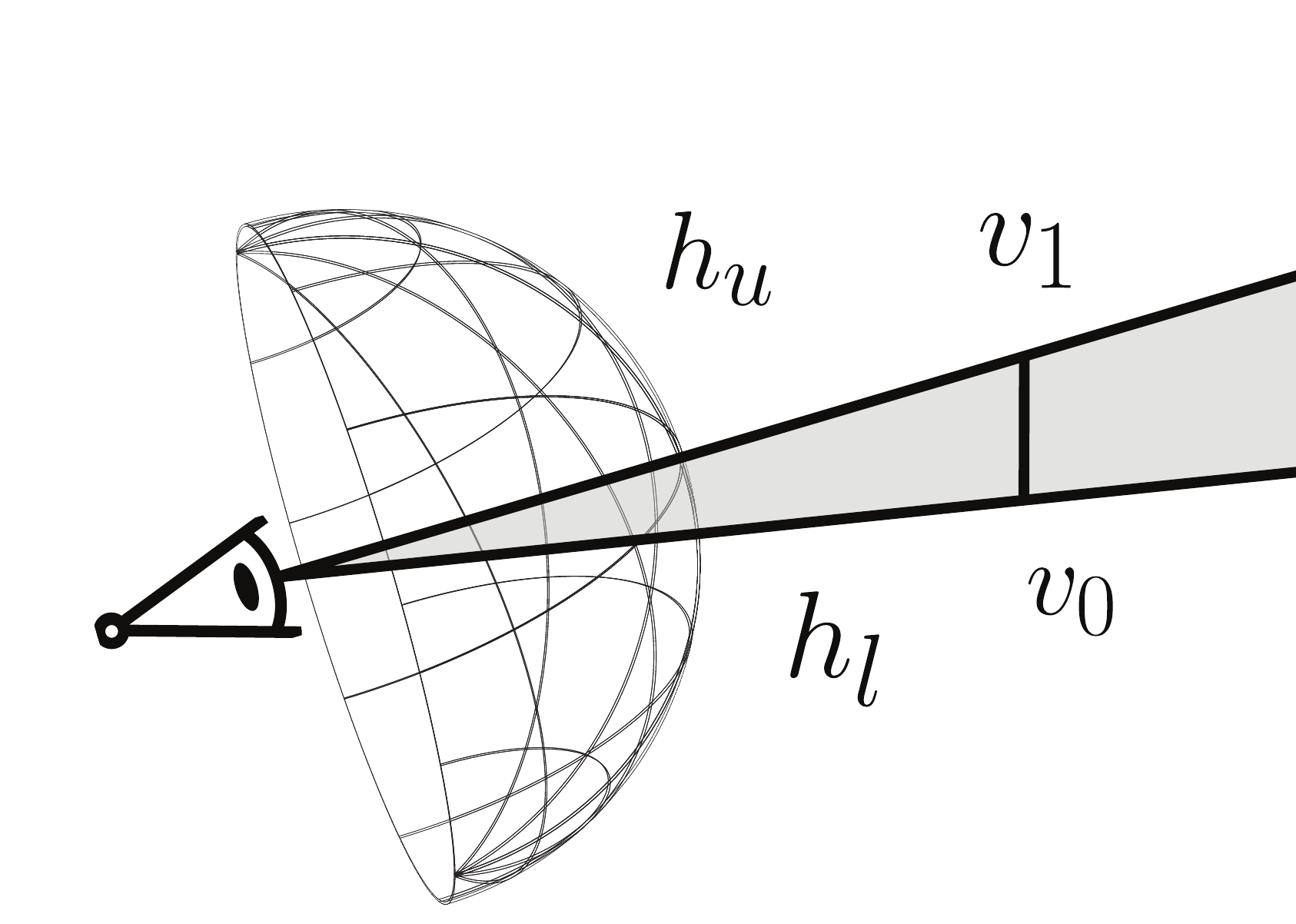} 
  \caption{\textbf{Cameras with non-linear projection.} Our method can also be extended to work with cameras with non-linear projection, such as a fisheye camera as shown in this figure. Gradients of these camera models cannot be computed directly using the 2D edge sampling described in Section~\ref{sec:primary_visibility} and Figure~\ref{fig:edge_sample}, because the projection of edges are not straight lines. Instead of sampling the projected edge in screen space, we directly sample the edge ($v_0$, $v_1$) in the 3D space, and use the 3D step function as in the secondary visibility case to split the spaces into two half-spaces $h_u$ and $h_l$ (Section~\ref{sec:secondary_visibility} and Figure~\ref{fig:blocker_half_spaces}).}
  \label{fig:fisheye}
\end{figure}

In Chapter~\ref{sec:primary_visibility}, we desribed how we compute gradients for projective cameras. In this subsection, we discuss generalization of the camera model to handle non-linear projections. These cameras need a different treatment because we assumed we could obtain a closed-form of the edge equations in screen-space after projection. For projective cameras, a line in 3D is still a line after projection to screen space. For non-linear projections such as fisheye cameras, this assumption does not hold. 

Fortunately, as Figure~\ref{fig:fisheye} illustrated, we can reuse the 3D extention we developed in the previous subsection for secondary visibility. For a camera location at position $p$ and a 3D edge $(v_0, v_1)$ in camera space, we can form the exact same edge equation $\alpha(p, m)$ as formulated in Equation~\ref{eq:3d_half_space}. The gradient then can be computed the same way as in the previous section (Equation~\eqref{eq:dirac_3d_edge} and ~\eqref{eq:alpha_3d}).

\subsection{Relation to Reynolds transport theorem and shape optimization}
\label{sec:redner_reynolds}

While we derive the gradients caused by discontinuities in the integrand using Dirac delta functions, it is also possible to represent the discontinuities using the integral boundary. Instead of using Heaviside functions to represent the shape boundaries, we can integrate over the domain on one side of the edge. Therefore, in the primary visibility case, Equation~\eqref{eq:scene_function_step} can be rewritten as:
\begin{equation}
\nabla \iint \theta(\alpha_i(x, y)) f_i(x, y) dxdy = \nabla \iint_{\Omega_i\left(\Theta\right)} f_i(x, y) dxdy,
\end{equation}
where $\Omega_i\left(\Theta\right)$ represents the domain where the Heaviside function $\theta(\alpha_i(x, y)) = 1$. Critically, the domain depends on the scene parameters $\Theta$. Reynolds transport theorem~\cite{Reynolds:1903:SU}, commonly used in fluid mechanics, addresses this specific case:
\begin{equation}
\begin{aligned}
&\nabla_t \iint_{\Omega_i\left(\Theta\right)} f_i(x, y) dxdy = \\
&\iint_{\Omega_i\left(\Theta\right)} \nabla f_i(x, y) dxdy +
\iint_{\partial \Omega_i\left(\Theta\right)} \left(v \cdot n\right) f_i(x(s), y(s)) ds,
\end{aligned}
\end{equation}
where $\partial \Omega_i\left(\Theta\right)$ is the boundary of the domain, $v$ is the \emph{velocity} at the boundary of the domain, and $n$ is the normal vector of the boundary. This mirrors Equation~\eqref{eq:redner_product_rule} and also separates the gradient integral into a continuous part and a discontinuous part. If we choose to differentiate the $x$ component, the velocity would be $(1, 0)$ and the normal before normalization would be $\left(a_y - b_y, b_x - a_x\right)$, which matches the results in Equation~\eqref{eq:redner_dirac_jacobian}. Similar derivation also appears in shape optimization~\cite{Sokolowski:1992:ISO}, where the goal is to find the optimal shape boundary that minimizes certain cost integral, using gradient-based optimization. Shape optimization has been applied in computer graphics in the context of diffusion curve optimization~\cite{Zhao:2018:IDC}.

\begin{figure}[t]
    \centering
    \captionsetup[subfigure]{justification=centering}
    \setlength{\tabcolsep}{1pt}
    \begin{tabular}{cccccc}
        &
        uniform &
        w/o hierarchy &
        w/o LTC &
        w/o NEE &
        all \\
        &
        &
        ours &
        ours &
        ours &
        ours \\
        \includegraphics[width = 0.16\textwidth]{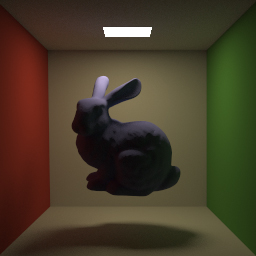} &
        \includegraphics[width = 0.16\textwidth]{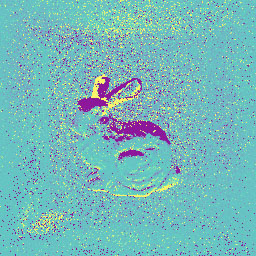} &
        \includegraphics[width = 0.16\textwidth]{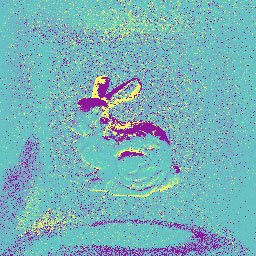} &
        \includegraphics[width = 0.16\textwidth]{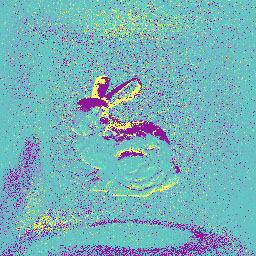} &
        \includegraphics[width = 0.16\textwidth]{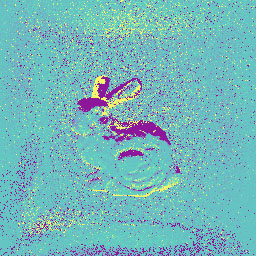} &
        \includegraphics[width = 0.16\textwidth]{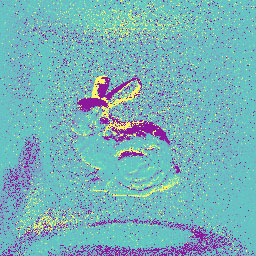}
        \\
        bunny & 144 spp & 84 spp & 72 spp & 103 spp & 72 spp \\
        \includegraphics[width = 0.16\textwidth]{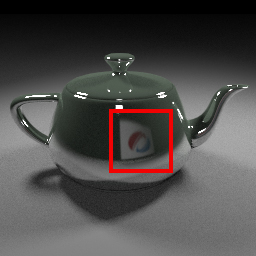} &
        \includegraphics[width = 0.16\textwidth]{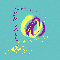} &
        \includegraphics[width = 0.16\textwidth]{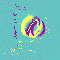} &
        \includegraphics[width = 0.16\textwidth]{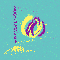} &
        \includegraphics[width = 0.16\textwidth]{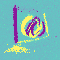} &
        \includegraphics[width = 0.16\textwidth]{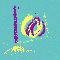}
        \\
        teapot & 375 spp & 194 spp & 140 spp & 179 spp & 135 spp
    \end{tabular}
  \caption{Equal time (24 seconds) comparison between sampling with and without our importance sampling methods. We tested our algorithm on scenes with soft shadow, global illumination, and specular reflection. We show the per-pixel derivatives of average color with respect to the bunny moving up in the top row, and the derivatives with respect to the reflected plane with the SIGGRAPH logo moving right in the second row. For the second row we only show the gradients in the red inset. \emph{Uniform} indicates uniformly picking an edge based on length and uniformly picking a point on the edge, and is inefficient at sampling important contributions. We selectively turn off the three optimizations we introduce for importance sampling: the hierarchical edge selection (Chapter~\ref{sec:edge_tree}), the linearly transformed cosines (LTC) based importance sampling for a single edge (Chapter~\ref{sec:edge_importance_sampling}), and the sampling based on the next event estimation (NEE) rays' intersections with edge billboards (Chapter~\ref{sec:edge_nee}). Both the hierarchical edge selection and the LTC sampling are important for picking important or silhouette edges (see the teapot). The NEE intersection is important for shadow caused by relatively small light sources (see the bunny).}
  \label{fig:importance_sampling}
\end{figure}
\section{Importance Sampling the Edges}
\label{sec:redner_implementaiton}

Our edge sampling method described in Chapter~\ref{sec:redner_method} requires us to sample an edge from hundreds of thousands, or even millions of triangles in the scene. The problem is two-fold: we need to sample an edge and then sample a point on the edge efficiently. Typically only a tiny fraction of these edges contribute to the gradients, since most of the edges are not silhouette (e.g.~\cite{Hertzmann:1999:INP, Benard:2018:LD3}), and some of them are shadow blockers which can have significant contributions. Naive sampling methods fail to select important edges (Figure~\ref{fig:importance_sampling}). Even if the number of edges is small, it is often the case that only a small region on an edge has non-zero contributions, especially when there exist highly-specular materials.

As mentioned in Chapter~\ref{sec:primary_visibility}, the case for primary visibility is easier since the viewpoint is the camera. We project all edges onto the screen in a preprocessing pass, and test whether they are silhouettes with respect to the camera position. We sample an edge based on the distance of two points on the screen and uniformly sample in screen space. For secondary visibility, the problem is much more complicated. The viewpoint can be anywhere in the scene, and we need to take the material response between the viewpoint and the point on the edge into account.

In this section, we describe a scalable edge sampling implementation given arbitrary viewpoint. We introduce three optimizations: a hierarchical data structure for selecting important edges, an importance sampling scheme for selecting a point on a single edge, and a method that extracts shadow blocker edges using next event estimation rays. Our solution is inspired by previous methods for sampling many light sources using hierarchical data structures (e.g.~\cite{Paquette:1998:HVD, Walter:2005:LSA, Estevez:2018:ISM}), efficient data structures for selecting silhouette edges~\cite{Sander:2000:SC, Hertzmann:2000:ISS, Olson:2006:SEH}, and the more recent closed-form solution for linear light sources~\cite{Heitz:2016:RPS, Heitz:2017:LSL}.

\subsection{Edge selection}
\label{sec:edge_tree}
Given a shading point, our first task is to importance sample one or more triangle edges. There are several factors to take into account when selecting the edges: whether the edge is a silhouette, the geometric foreshortening factor inversely proportional to the distance to the edge, the material response between the shading point and the point on the edge, and the radiance incoming from the edge direction (e.g. whether it hits a light source or not). We address the incoming radiance in Chapter~\ref{sec:edge_nee} and the rest in this subsection.

% We propose two different methods for sampling edges. The first one based on importance resampling~\cite{Rubin:1987:CPD, Talbot:2005:IRG} is simpler to implement, has less overhead, and works well in many common cases. The second one based on traversing an edge hierarchy is more sophisticated. It is more costly but can be more useful in difficult scenes.

% Our first method has two stages. The first stage select $64$ edges from a global importance function, and the second stage resample one edge from the selected ones based on their importance to the shading point. As observed by previous works~\cite{Markosian:1997:RNR, Mcguire:2004:OSS}, the probability of an edge being a silhouette of a random looking direction is proportional to the exterior dihedral angle. Therefore, in the first stage, we build a global importance function using the length of the edge multiplied by the exterior dihedral angle. In the second stage, we set the weight of non-silhouette edges to zero, and take the geometric factor and material response into consideration by treating the edge as a constant linear light source, and integrate over the edge using Linearly Transformed Cosine Distribution~\cite{Heitz:2016:RPS, Heitz:2017:LSL} (see more discussion in the next subsection). We use this method for our GPU implementation. The downside of this method is that the first stage can be inefficient at sampling edges that has strong contributions.

Our method involves building hierarchies of edges and probabilistically pruning out unimportance ones during traversal. We follow Olson and Zhang's Hough space approach~\cite{Olson:2006:SEH} and build two hierarchies. The first contains the triangle edges that are silhouettes with respect to the camera position. The second contains the remaining edges. For the first set of edges, we build a 3D bounding volume hierarchy using the 3D positions of the two endpoints of an edge. For the second set of edges, we build a 6D bounding volume hierarchy using the two endpoint positions and the two planes associated with the two faces of an edge, transformed into the Hough space. For quick rejections of non-silhouette edges, we form a sphere between the shading point and the camera position (the \emph{v-sphere}~\cite{Olson:2006:SEH}), and test the intersection of the sphere and the bounding box of the planes in Hough space. We build the hierarchy parallelly by building a radix tree on top of the sorted Morton codes of the bounding box centroids~\cite{Lauterbach:2009:FBC, Karras:2012:MPC}. We also optimize for the surface area heuristics cost~\cite{Macdonald:1990:HRT} using the treelet approach proposed by Karras and Aila~\cite{Karras:2013:FPC}.

We traverse the hierarchies to sample edges. For better stratification, we use a scheme similar to the Gaussian kd-tree~\cite{Adams:2009:GKF}. We start with a number of samples at the root (say, 16). During the traversal, for each node in the hierarchy, we distribute the samples to the two children proportional to an importance estimation of the contributions, similar to the lightcuts algorithm~\cite{Walter:2005:LSA}. We estimate the importance using the total length of edges, multiplied by the inverse distance to the center of the bounding box, times an upper bound estimation of the BRDF, by fitting the BRDF to a Linearly Transformed Cosine Distribution~\cite{Heitz:2016:RPS}. We use the linearly transformed cosines for BRDF upper bound estimation since the linear transformation preserves the mode of the distribution, therefore we linearly transform the bounding box and compute the upper bound of the cosine distribution using Walter et al.'s method~\cite{Walter:2005:LSA}. We set the importance to zero if the node does not contain any silhouette. We set the importance of both children to one if the shading point is inside the bounding box of their parent. Finally, we select one edge using weighted reservoir sampling~\cite{Chao:1982:GPU}, where the weight is the importance of the leaf node.

\subsection{Importance sampling on an edge}
\label{sec:edge_importance_sampling}

Oftentimes with a highly-specular BRDF, only a small portion of the edges have significant contributions. We employ the recent technique on integrating linear light sources over the linearly transformed cosines~\cite{Heitz:2017:LSL}. Heitz and Hill's work provides a closed-form solution of the integral between a point and a linear light source, weighted by BRDF and geometric foreshortening. We numerically invert the integrated cumulative distribution function using Newton's method for importance sampling. We precompute a table of fitted linearly transformed cosines for our BRDFs.

\subsection{Next event estimation for edges}
\label{sec:edge_nee}

The techniques above do not address the case of a small fraction of edges blocking a relatively small light sources. These edges cause sharp boundaries that have large gradients and are difficult to sample even with the edge hierarchy. We propose a method to address this by reusing the next event estimation samples we used in the forward pass. We transform all edges into billboards with finite width facing the next event estimation ray. We then collect all the edge billboards that intersect with the ray. We sample one of them based on the importance of the edge described in Chapter~\ref{sec:edge_tree}. We project the intersection of the ray and the billboard onto the nearest point on the edge, and the point on the edge is our sample. We set the width of the edge billboards to the $L^2$ sum of two percent of the mean absolute deviation of the triangle vertex positions over each coordinate.

We evaluate our importance sampling method using equal-time comparisons on GPU and show the results in Figure~\ref{fig:importance_sampling}. We compare against the baseline approach of uniformly sampling all edges by length. We also perform ablation study by selectively turning off the three optimizations. The baseline approach is not able to efficiently sample rare events such as shadows cast by a small light source or highly-specular reflection of edges, while our importance sampling generates images with much lower variance.

\section{Results}
\label{sec:redner_results}

\begin{figure}[t]
    \setlength{\tabcolsep}{2pt}
    \centering
    \begin{tabular}{cccccc}
        \begin{overpic}[width=0.16\textwidth]{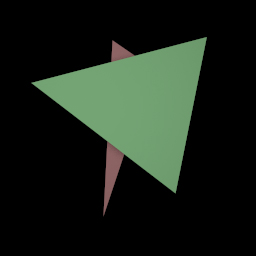}
            \put (6, 5) {\textcolor{white}{\footnotesize initial guess}}
        \end{overpic} &
        \includegraphics[width = 0.16\textwidth]{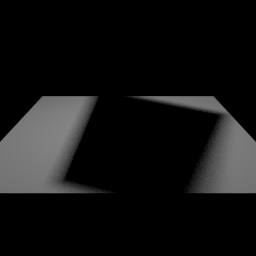} &
        \includegraphics[width = 0.16\textwidth]{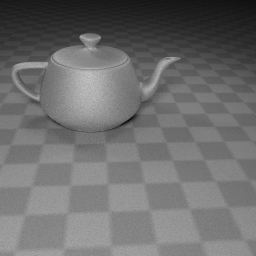} &
        \includegraphics[width = 0.16\textwidth]{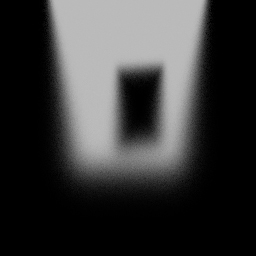} &
        \includegraphics[width = 0.16\textwidth]{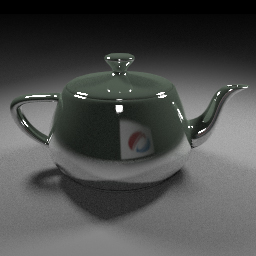} &
        \includegraphics[width = 0.16\textwidth]{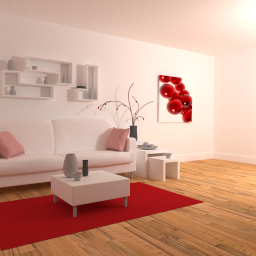} \\
        \begin{overpic}[width=0.16\textwidth]{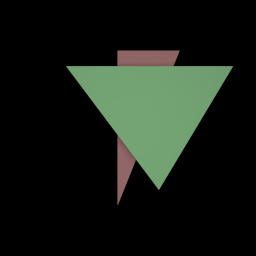}
            \put (6, 5) {\textcolor{white}{\footnotesize target}}
        \end{overpic} &
        \includegraphics[width = 0.16\textwidth]{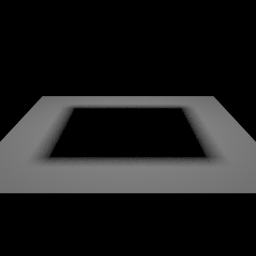} &
        \includegraphics[width = 0.16\textwidth]{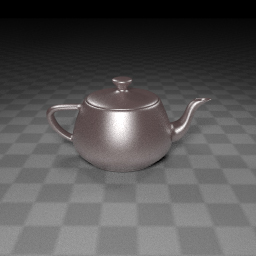} &
        \includegraphics[width = 0.16\textwidth]{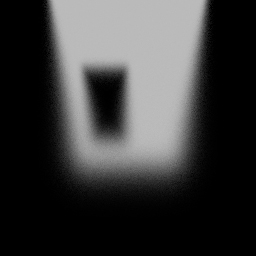} &
        \includegraphics[width = 0.16\textwidth]{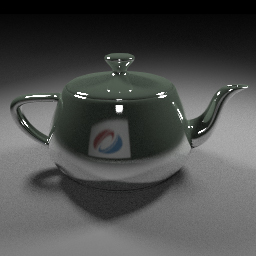} &
        \includegraphics[width = 0.16\textwidth]{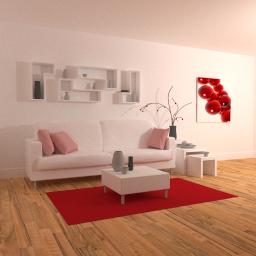} \\
        \begin{overpic}[width=0.16\textwidth]{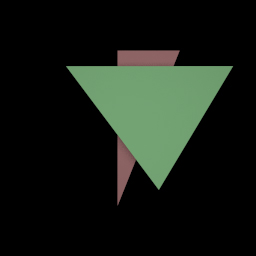}
            \put (6, 5) {\textcolor{white}{\footnotesize optimized}}
        \end{overpic} &
        \includegraphics[width = 0.16\textwidth]{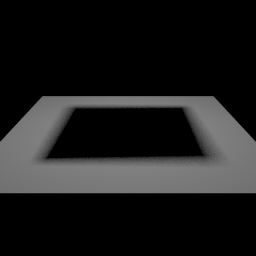} &
        \includegraphics[width = 0.16\textwidth]{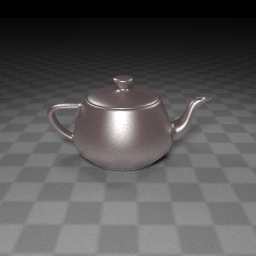} &
        \includegraphics[width = 0.16\textwidth]{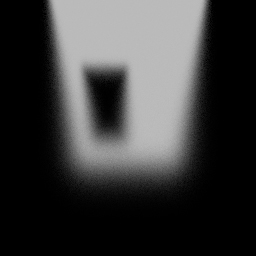} &
        \includegraphics[width = 0.16\textwidth]{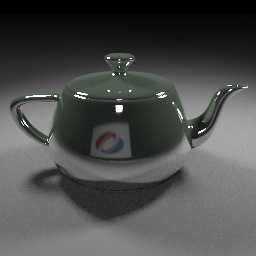} &
        \includegraphics[width = 0.16\textwidth]{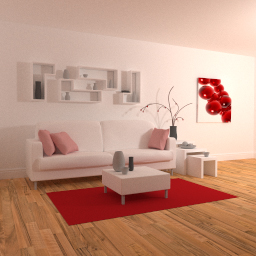} \\
        {\footnotesize (a) \makecell{primary \\ occlusion}} &
        {\footnotesize (b) shadow} &
        {\footnotesize (c) \makecell{camera \& \\ glossy}} &
        {\footnotesize (d) \makecell{glossy \\ receiver}} &
        {\footnotesize (e) \makecell{near \\ specular}} &
        {\footnotesize (f) \makecell{global \\ illumination}}
    \end{tabular}
	\caption{We verify our renderer by matching a range of synthetic scenes with different light transport configurations. For each scene, we start from an initial parameter (first row) and attempt to set scene parameters so that the rendering matches the target (second row) using gradient-based optimization. Each scene is intended to test a different aspect of the renderer. (a) optimizes triangle positions under the presence of occlusion. (b) optimizes blocker position for shadow. (c) optimizes camera pose and material parameters over textured and glossy surfaces. (d) optimizes the blocker position where the shadow receiver is highly glossy. (e) optimizes an almost specular reflection of a plane behind the camera; the free parameter is the plane position. (f) optimizes camera pose under the presence of global illumination and soft shadow. Our method is able to generate gradients for these scenes and to optimize the parameters correctly, resulting in minimal difference between the optimized result (final row) and target (second row). All the scenes are rendered with 4 samples per pixel during optimization. The final renderings are produced with 625 samples per pixel, except for (f) we use 4096 samples. We encourage the reader to refer to the project page (https://people.csail.mit.edu/tzumao/diffrt/) for videos and more scenes.}
	\label{fig:verification}
\end{figure}
\begin{figure}[t]
    \centering
    \captionsetup[subfigure]{justification=centering}
    \setlength{\tabcolsep}{2pt}
    \begin{tabular}{ccc}
        \begin{overpic}[width=0.17\textwidth]{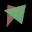}
            \put (6, 5) {\textcolor{white}{\footnotesize image}}
        \end{overpic} &
        \includegraphics[width = 0.17\textwidth]{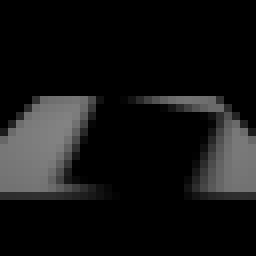} &
        \includegraphics[width = 0.17\textwidth]{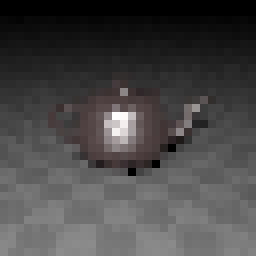} \\
        \begin{overpic}[width=0.17\textwidth]{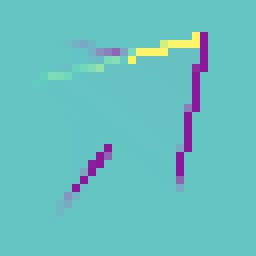}
            \put (6, 5) {\textcolor{white}{\footnotesize finite differences}}
        \end{overpic} &
        \includegraphics[width = 0.17\textwidth]{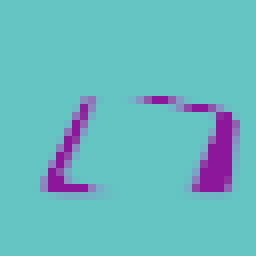} &
        \includegraphics[width = 0.17\textwidth]{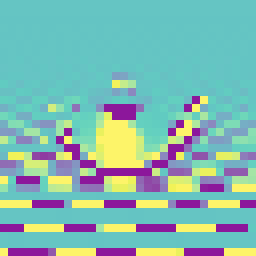} \\
        \begin{overpic}[width=0.17\textwidth]{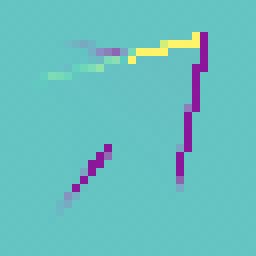}
            \put (6, 5) {\textcolor{white}{\footnotesize ours}}
        \end{overpic} &
        \includegraphics[width = 0.17\textwidth]{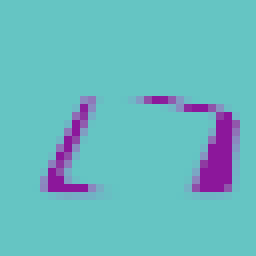} &
        \includegraphics[width = 0.17\textwidth]{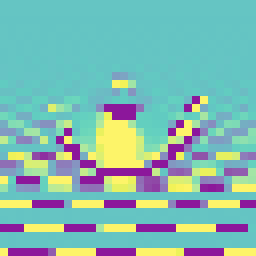} \\
        (a) triangles & (b) shadow & (c) teapot
    \end{tabular}
  \caption{We compare with central finite differences by rendering the scenes in Figure~\ref{fig:verification} at $32 \times 32$. The scenes are slightly adjusted to make the per-pixel gradient look clearer in the image. The derivatives are with respect to (a) each rightmost vertex of the two triangles moving left (b) the shadow blocker moving up (c) the camera moving into the screen. Our derivatives match the finite differences within an error of $1\%$ relative to the $L^1$ norm of the gradients. Finite differences usually take two or three orders of magnitude more samples to reach the same error. For our method, we use $16$ thousand samples per pixel for the scene with two triangles and $32$ thousand samples per pixel for the other two scenes. For finite differences, we use $1$ million samples per pixel for the triangles scene and $10$ million samples per pixel for the rest.}
  \label{fig:finite_difference}
\end{figure}
\begin{figure}[t]
  \centering
  \captionsetup[subfigure]{justification=centering}

  \begin{subfigure}{0.24\textwidth}
  \includegraphics[width=0.99\linewidth]{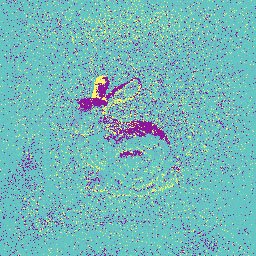} 
  \caption{4 spp}
  \label{fig:bunny_dx_1spp}
  \end{subfigure}
  \begin{subfigure}{0.24\textwidth}
  \includegraphics[width=0.99\linewidth]{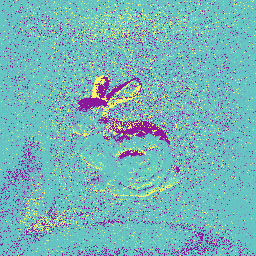}
  \caption{16 spp}
  \label{fig:bunny_dx_16spp}
  \end{subfigure}
  \begin{subfigure}{0.24\textwidth}
  \includegraphics[width=0.99\linewidth]{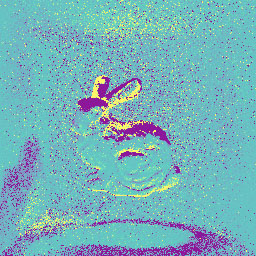}
  \caption{128 spp}
  \label{fig:bunny_dx_128spp}
  \end{subfigure}
  \begin{subfigure}{0.24\textwidth}
  \includegraphics[width=0.99\linewidth]{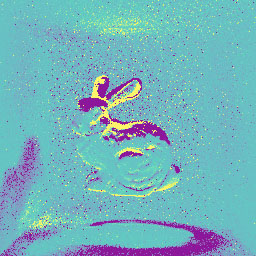}
  \caption{1024 spp}
  \label{fig:bunny_dx_1024spp}
  \end{subfigure}
  \caption{We visualize the per-pixel gradient contribution generated by our method over different numbers of samples per pixel. We take the bunny scene from Figure~\ref{fig:importance_sampling}. The gradient is the average of color with respect to the bunny moving right. The $1024$ samples per pixel image took around $5$ minutes to compute on a Pascal GPU. In practice we usually use 4 samples per pixel for inverse rendering.}
  \label{fig:convergence}
\end{figure}

We implement our method in a stand-alone C++/CUDA renderer with an interface to the automatic differentiation library PyTorch~\cite{Paszke:2017:ADP}. To use our system, the user constructs their scenes using lists of PyTorch tensors. For example, triangle vertices and indices are represented by floating point and integer tensors. Our renderer in the forward pass outputs an image which is also a PyTorch tensor. The user can then compute a scalar loss on the output image and obtain the gradient by backpropagating to the scene parameters.

Our renderer is structured similarly to a wavefront path tracer~\cite{Laine:2013:MCH}. We trace one path per pixel at a time for all pixels. In the forward pass we store the path vertex information for each bounce (i.e. we checkpoint at each path vertex). We manually backpropagate both the edge gradients and smooth gradients in Equation~\eqref{eq:redner_product_rule} by traversing the light path backward (using in principle introduced in Chapter~\ref{chap:autodiff}). We use Embree~\cite{Wald:2014:EKF} and OptiX prime~\cite{Parker:2010:OGP} for our ray casting operations. The renderer supports pinhole or thinlens camera with planar and equiangular spherical projection, Lambertian and Blinn-Phong BRDFs with Schlick approximation for Fresnel reflection, trilinear reconstruction of textures for diffuse and specular reflectance and roughness, area light sources with triangle meshes and environment maps.

\subsection{Verification of the method}

We tested our method on several synthetic scenes covering a variety of effects, including occlusion, non-Lambertian materials, and global illumination. Figure~\ref{fig:verification} shows the scenes. We start from an initial parameter, and optimize the parameters to minimize the $L^2$ difference between the rendered image and target image using gradients generated by our method (except for the living room scene in Figure~\ref{fig:verification} (f) where we optimize for the $L^2$ difference between the Gaussian pyramids of the rendered image and target image). Our PyTorch interface allows us to apply their in-stock optimizers, and backpropagate to all scene parameters easily. We use the Adam~\cite{Kingma:2015:AMS} algorithm for optimization. The number of parameters ranges from $6$ to $30$. The experiment shows that our renderer is able to generate correct gradients for the optimizer to infer the scenes. It also shows that we are able to handle many different light transport scenarios, including cases where a triangle vertex is blocked but we still need to optimize it into the correct position, optimization of blocker position when we only see the shadow, joint optimization of camera and material parameters, pose estimation in presence of global illumination, optimizing blockers occluding highly-glossy reflection, and inverting near specular reflection. See the supplementary materials for more results.

We also compare our method to central finite differences on a lower resolution version of the synthetic scenes in Figure~\ref{fig:finite_difference}. Our derivatives match the finite difference within an error of $1\%$ relative to the $L^1$ norm of the gradients. The comparison is roughly equal quality. We increase the number of samples for the finite differences until the error is low enough. In general finite differences require a small step size to measure the visibility gradient correctly, thus they usually take two or three orders of magnitude more samples to reach the same error as our result. In addition, finite differences do not scale with the number of parameters, making them impractical for most optimization tasks.

Figure~\ref{fig:convergence} demonstrates the convergence of our method by visualizing the gradients of the bunny scene in Figure~\ref{fig:importance_sampling} over different numbers of samples per pixel. We show the gradients of the average of pixel colors with respect to the bunny moving right on the screen. Generating the near-converged $1024$ samples per pixel image takes around 5 minutes on a Pascal GPU. In practice we don't render converged images for optimization. We utilize stochastic gradient descent and render a low sample count image (usually $4$).

\subsection{Comparison with previous differentiable renderers}
\begin{figure}[t]
  \centering
  \captionsetup[subfigure]{justification=centering}

  \begin{subfigure}{0.2\textwidth}
  \includegraphics[width=0.99\linewidth]{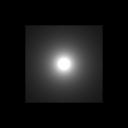} 
  \caption{planar scene}
  \label{fig:planar}
  \end{subfigure}
  \begin{subfigure}{0.2\textwidth}
  \includegraphics[width=0.99\linewidth]{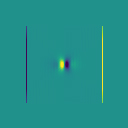}
  \caption{OpenDR}
  \label{fig:opendr}
  \end{subfigure}
  \begin{subfigure}{0.2\textwidth}
  \includegraphics[width=0.99\linewidth]{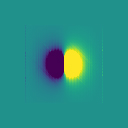}
  \caption{Neural}
  \label{fig:neural_renderer}
  \end{subfigure}
  \begin{subfigure}{0.2\textwidth}
  \includegraphics[width=0.99\linewidth]{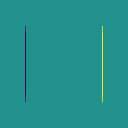}
  \caption{ours}
  \label{fig:ours}
  \end{subfigure}
  \caption{(\subref{fig:planar}) A plane lit by a point light close to the plane. We are interested in the derivative of the image with respect to the plane moving right. Since the point light stays static, the derivatives should be zero except for the boundary. (\subref{fig:opendr}) (\subref{fig:neural_renderer}) Previous work uses color buffer differences to approximate the derivatives, making them unable to take large variation between pixels into account and output non zero derivatives at the center. (\subref{fig:ours}) Our method outputs the correct derivatives.}
  \label{fig:redner_comparison}
\end{figure}

In this subsection we compare our method with two previously proposed differentiable renderers: OpenDR~\cite{Loper:2014:OAD} and Neural 3D Mesh renderer~\cite{Kato:2018:N3M}. Both previous methods focus on speed and approximate the gradients even under Lambertian materials with unshadowed direct lighting. In contrast, our method outputs consistent gradients and supports arbitrary non-Dirac materials, shadow, and global illumination, as shown in Figure~\ref{fig:verification}.

Both OpenDR and the Neural 3D Mesh renderer follow the approach of first rendering into a color buffer using a traditional rasterizer with z-buffer. They then approximate the derivatives with respect to screen-space triangle vertex positions using the rendered color buffer. OpenDR performs a screen-space filtering approach based on a brightness constancy assumption~\cite{Jones:1996:MML}. The shape of the filter is determined by boundary detection using triangle ID. For the horizontal derivatives of a pixel neighboring an occlusion boundary on the left, they use the kernel $\left[0,-1,1\right]$. For pixels that are not neighboring any boundaries, or are intersecting with boundaries, or are neighboring more than one occlusion boundary, they use the kernel $\frac{1}{2}\left[-1,0,1\right]$. The Neural 3D Mesh renderer performs an extra edge rasterization pass of the triangle edges and accumulates the derivatives by computing the difference between the color difference on the color buffer around the edge. The derivative responses are modified by applying a smooth falloff.

Both previous differentiable renderers output incorrect gradients in the case where there is brightness variation between pixels due to lighting. Figure~\ref{fig:redner_comparison} shows an example of a plane lit by a point light with inverse squared distance falloff. We ask the two renderers and ours to compute the derivatives of the pixel color with respect to the plane moving right. Since the light source does not move, the illumination on the plane remains static and the derivatives should be zero except for the boundaries of the plane. Since both previous renderers use the differences between color buffer pixels to approximate derivatives, they incorrectly take the illumination variation as the changes that would happen if the plane moves right, and output non-zero derivatives around the highlights. On the other hand, since we sample \emph{on} the edges, our method correctly outputs zero derivatives for continuous regions.

OpenDR's point light does not have distance falloff and the Neural 3D mesh renderer does not support point lights so we modified their renderers. Our renderer does not support pure point lights so we use a small planar area light to approximate a point light. We also tessellate the plane into $256 \times 256$ grids as both previous renderers use Gouraud shading.

\subsection{Differentiable geometry buffer/AOV extension}
\label{sec:redner_g_buffer}
\begin{figure}[t]
  \centering
  \captionsetup[subfigure]{justification=centering}

  \begin{subfigure}{0.24\textwidth}
  \includegraphics[width=0.99\linewidth]{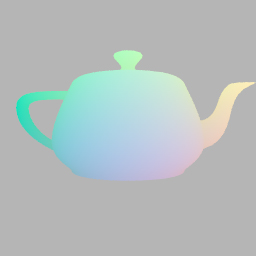} 
  \caption{3D position}
  \label{fig:dr_pos}
  \end{subfigure}
  \begin{subfigure}{0.24\textwidth}
  \includegraphics[width=0.99\linewidth]{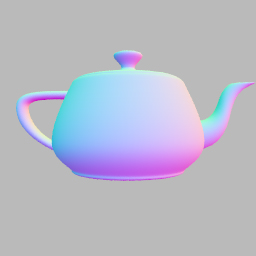} 
  \caption{normal}
  \label{fig:dr_normal}
  \end{subfigure}
  \begin{subfigure}{0.24\textwidth}
  \includegraphics[width=0.99\linewidth]{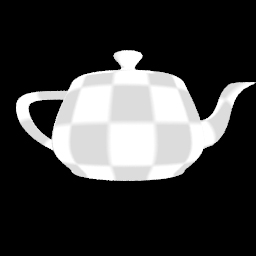} 
  \caption{albedo}
  \label{fig:dr_albedo}
  \end{subfigure}
  \begin{subfigure}{0.24\textwidth}
  \includegraphics[width=0.99\linewidth]{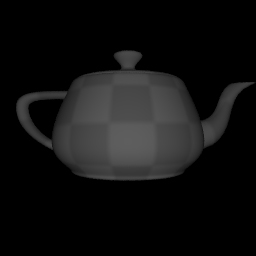} 
  \caption{deferred shading}
  \label{fig:dr_color}
  \end{subfigure}

  \caption{\textbf{Geometry buffer rendering.} Our method can also output correct gradients of (\subref{fig:dr_pos}) 3D position (\subref{fig:dr_normal}) normal, (\subref{fig:dr_albedo}). This enables vision applications such as matching RGB-D signals. The buffers can also be combined with deferred shading techniques to produce final rendering (\subref{fig:dr_color}), for high-performance rendering.}
  \label{fig:deferred_shading}
\end{figure}

Our method naturally generalizes to arbitrary shading functions, and can be used for generating geometry or AOV (arbitrary output variable) buffers, such as 3D position, normal, or material parameters. Our primary edge sampling (Chapter~\ref{sec:primary_visibility}) backpropagates the derivatives of these auxiliary buffers correctly. This can be useful for computer vision applications (e.g. matching RGB-D signal), and also enables us to apply deferred shading techniques~\cite{Deering:1988:TPN} to speed up rendering. Only coherent primary rays need to be traced in this case. Figure~\ref{fig:deferred_shading} shows the geometry buffers generated by our renderer.

\subsection{Inverse rendering application}

We apply our method on an inverse rendering task for fitting camera pose, material parameters, and light source intensity. Figure~\ref{fig:redner_teaser} shows the result. We take the scene photo and geometry data from the thesis work of Jones~\cite{Jones:2017:VID}, where the scene was used for validating daylight simulation. The scene contains strong indirect illumination and has non-Lambertian materials. We assign most of the materials to white except for plastic or metal-like objects, and choose an arbitrary camera pose as an initial guess. There are in total $177$ parameters for this scene. We then use gradient-based optimizer Adam and the gradients generated by our method to find the correct camera pose and material/lighting parameters. In order to avoid getting stuck in local minima, we perform the optimization in a multi-scale fashion, starting from $64 \times 64$ and linearly increasing to the final resolution $512 \times 512$ through $8$ stages. For each scale we use an $L^1$ loss and perform $50$ iterations. We exclude the light source in the loss function by setting the weights of pixels with radiance larger than $5$ to $0$.

\subsection{3D adversarial example}

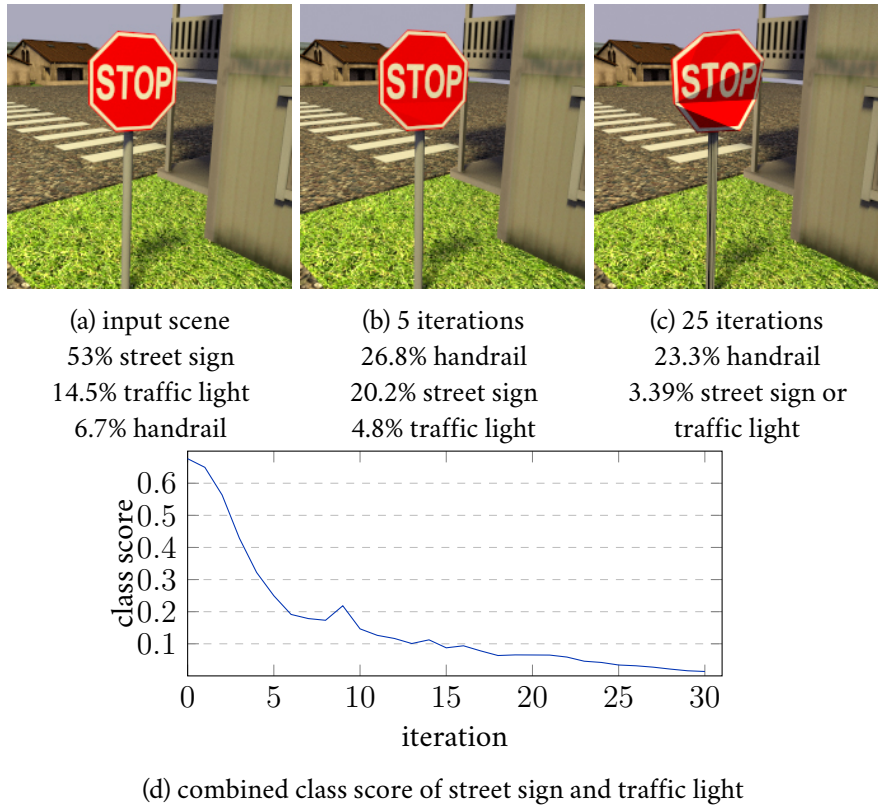
\begin{figure}[t]
  \centering
  \captionsetup[subfigure]{justification=centering}

  \begin{subfigure}{0.25\textwidth}
  \includegraphics[width=0.99\linewidth]{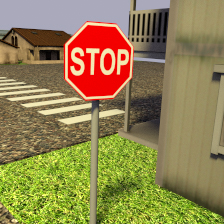} 
  \caption{input scene \\ 53\% street sign \\ 14.5\% traffic light \\ 6.7\% handrail}
  \label{fig:adv_init}
  \end{subfigure}
  \begin{subfigure}{0.25\textwidth}
  \includegraphics[width=0.99\linewidth]{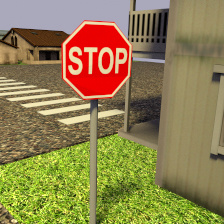}
  \caption{5 iterations \\ 26.8\% handrail \\ 20.2\% street sign \\ 4.8\% traffic light}
  \label{fig:adv_5_iter}
  \end{subfigure}
  \begin{subfigure}{0.25\textwidth}
  \includegraphics[width=0.99\linewidth]{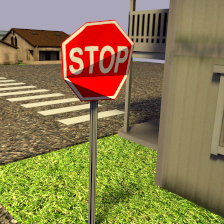}
  \caption{25 iterations \\ 23.3\% handrail \\ 3.39\% street sign or \\ traffic light}
  \label{fig:adv_25_iter}
  \end{subfigure}
  \begin{subfigure}{0.6\textwidth}
    \begin{tikzpicture}
    \begin{axis}[
        width=0.95\textwidth,
        height=0.5\textwidth,
        title={},
        xlabel={iteration},
        ylabel={class score},
        xmin=0, xmax=31,
        ymin=0, ymax=0.7,
        xtick={0,5,10,15,20,25,30},
        ytick={0.1,0.2,0.3,0.4,0.5,0.6},
        y label style={at={(axis description cs:0.05,0.5)}},
        ymajorgrids=true,
        grid style=dashed,
    ]
     
    \addplot[
        color=blue,
        mark=circle,
        ]
        coordinates {
        (0,0.6764)
        (1,0.6495)
        (2,0.5634)
        (3,0.4290)
        (4,0.3222)
        (5,0.2496)
        (6,0.1914)
        (7,0.1785)
        (8,0.1732)
        (9,0.2184)
        (10,0.1465)
        (11,0.1263)
        (12,0.1165)
        (13,0.1008)
        (14,0.1123)
        (15,0.087454)
        (16,0.093748)
        (17,0.077811)
        (18,0.063452)
        (19,0.065389)
        (20,0.065149)
        (21,0.064822)
        (22,0.058651)
        (23,0.045775)
        (24,0.041838)
        (25,0.033912)
        (26,0.031515)
        (27,0.027214)
        (28,0.021186)
        (29,0.015953)
        (30,0.013708)
        };
    \end{axis}
    \end{tikzpicture}
  \caption{combined class score of street sign and traffic light}
  \label{fig:class_score}
  \end{subfigure}
  \caption{Our method can be used for finding 3D scenes as adversarial examples for neural networks. We use the gradient generated by our method to optimize for the geometry of the stop sign, camera pose, light intensity and direction to minimize the class scores of street sign and traffic light classes. After 5 iterations the network classifies the stop sign as a handrail, and after 25 iterations both street sign and traffic light are out of the top 5 predictions. In (\subref{fig:class_score}) we plot the sum of street sign and traffic light class scores as a function of iteration. As we optimize scene parameters such as the stop sign shape, gradient descent tries to find the geometry that minimizes the class scores, thus we see decreasing of the score.}
  \label{fig:adversarial}
\end{figure}

Recently, it has been shown that gradient-based optimization can also be used for finding adversarial examples for neural networks (e.g. ~\cite{Szegedy:2014:IPN, Goodfellow:2015:EHA}) for analysis or mining training data. The idea is to take an image that was originally labelled correctly by a neural network classifier, and use backpropagation to find an image that minimizes the network's output with respect to the correct output. Our system can be used for mining adversarial examples of 3D scenes since it provides the ability to backpropagate from image to scene parameters. Similar ideas have been explored~\cite{Athalye:2018:SRA, Zeng:2017:AAB, Liu:2019:BPN}, but we use a more general renderer.

We demonstrate this in Figure~\ref{fig:adversarial}. We show a stop sign classified correctly as a street sign by the VGG16 classifier~\cite{Simonyan:2014:VDC}. We then optimize for $2256$ parameters including camera pose, light intensity, sun position, global translation, rotation, and vertex displacement of the stop sign. We perform stochastic gradient descent to minimize the network's output of the classes street sign and traffic light, using $256$ samples per pixel. After 5 iterations the network starts to output handrail as the most probable class. After 23 iterations both the street sign class and traffic light class are out of the top-5 predictions and the sum of the two has less than $5\%$ probability.

We do not claim this as a robust way to break or to attack neural networks, since the CG scene we use has different statistics compared to real world images. Nevertheless this demonstrates that our gradient can be used for finding interesting scene configurations and can be potentially used for mining training data.

\subsection{Limitations}
\label{sec:future_work}

\emph{Performance.}
Our current GPU implementation takes a few hundred milliseconds to generate a small resolution image (say $256 \times 256$) with a small number of samples (say $4$). Note though that when using stochastic gradient descent it is usually not necessary to use high sample counts.

We have found that, depending on the type of scene, the bottleneck can be at the edge sampling phase or during automatic differentiation of the light paths, when we need to perform large reductions for the pixels hitting the same object (Chapter~\ref{chap:gradient_halide}). Developing better sampling algorithms such as incorporating bidirectional path tracing~\cite{Lafortune:1993:BPT} or photon mapping~\cite{Jensen:1996:GIP} could be an interesting avenue of future work. In particular, sampling the Dirac delta introduced by the edges is related to photon beams~\cite{Jarosz:2011:CTV}. Developing better compiler techniques for optimizing automatic differentiation code is also an important task. While we achieved promising results in Chapter~\ref{chap:gradient_halide}, the programming model focused on image processing and is unsuitable for many tasks in rendering such as tree traversal.

\emph{Other light transport phenomena.}
We assume static scenes with no participating media. Differentiating motion blur requires sampling on 4D edges with an extra time dimension. Combining our method with Gkioulekas~et~al.'s work~\cite{Gkioulekas:2013:IVR} for handling participating media is left as future work. 

\emph{Interpenetrating geometries and parallel edges.}
Dealing with the derivatives of interpenetration of triangles requires a mesh splitting process and its derivatives. Interpeneration can happen if the mesh is generated by some simulation process. As shown in Figure~\ref{fig:parallel_edge}, our method also does not handle the case where two edges are perfectly aligned as seen from the center of projection (camera or shadow ray origin). However, these are zero-measure sets in path space, and as long as the two edges are not perfectly aligned to the viewport, we will be able to converge to the correct solution.

\emph{Shader discontinuities.}
We assume our BRDF models and shaders are differentiable and do not handle discontinuities in the shaders. We handle textures correctly by differentiating through the smooth reconstruction, and many widely-used reflection models such as GGX~\cite{Walter:2007:MMR} (with Smith masking) or Disney's principled BRDF~\cite{Burley:2012:PBS} are differentiable. However, we do not handle the discontinuities at total internal reflection and some other BRDFs relying on discrete operations, such as the discrete stochastic microfacet model of Jakob et al.~\cite{Jakob:2014:DSM}. Extremely high frequency textures also require prefiltering to have low variance on both the rendered images and the gradients. Compiler techniques for band-limiting BRDFs can be applied to mitigate the shader discontinuity issue~\cite{Yang:2018:APS}.

\section{Conclusion}

We have introduced a differentiable Monte Carlo ray tracing algorithm that is capable of generating correct and unbiased gradients with respect to arbitrary input parameters such as scene geometry, camera, lights and materials. For this, we have introduced a novel edge sampling algorithm to take the geometric discontinuities into consideration, and derived the appropriate measure conversion. For increased efficiency, we use a new discrete sampling method to focus on relevant edges as well as continuous edge importance sampling. We believe this method and the software that we release will have an impact in inverse rendering and deep learning.

\begin{subappendices}

\section{Derivation of the 3D edge Jacobian}
\label{sec:secondary_measure_derivation}
We derive the Jacobian $J_m(t)$ in Equation~\ref{eq:dirac_3d_edge}. 
The goal is to compute the derivatives of point $m(t)$ with respect to the line 
parameter $t$. The relation between $m(t)$ and $t$ is described by a ray-plane 
intersection. That is, we are intersecting a plane at point $m$ with normal $n_m$ 
with a ray of origin $p$ and unnormalized direction $\omega(t)$:
\begin{equation}
\begin{aligned}
& \omega(t) = v_0 + (v_1 - v_0) t - p \\
& \tau(t) = \frac{(m - p) \cdot n_m}{\omega(t) \cdot n_m} \\
& m(t) = \tau(t) \omega(t).
\end{aligned}
\end{equation}
We can then derive the derivative $J_m(t) = \frac{\partial m(t)}{\partial t}$ as:
\begin{equation}
J_m(t) = \tau(t) \left( \left( v_1 - v_0 \right) - \omega(t) \frac{\left( v_1 - v_0 \right) \cdot n_m}{\omega(t) \cdot n_m} \right)
\end{equation}

\end{subappendices}
\chapter[Hessian-Hamiltonian Monte Carlo Rendering]{Hessian-Hamiltonian Monte Carlo \\ Rendering}
\label{chap:h2mc}

\begin{figure}[h]
	\setlength{\tabcolsep}{1pt}
	\def\arraystretch{0.5}
	\begin{tabular}{cccccccc}	
	\multicolumn{3}{c}{	
		\multirow{2}{*}[0.55in] { 
			\includegraphics[width=0.42\linewidth]{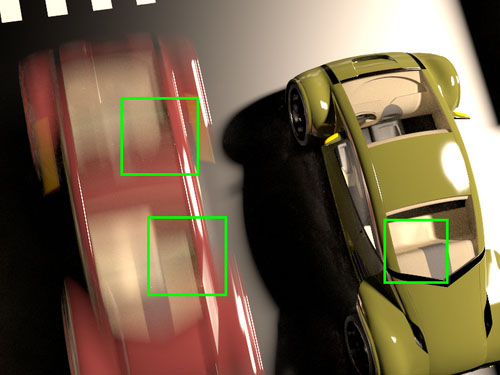}
		}
	} &
	\includegraphics[width=0.1\linewidth]{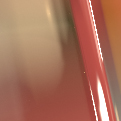} &
	\includegraphics[width=0.1\linewidth]{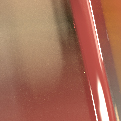} &
	\includegraphics[width=0.1\linewidth]{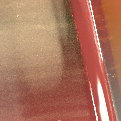} &
	\includegraphics[width=0.1\linewidth]{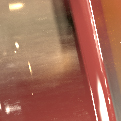} &
	\includegraphics[width=0.1\linewidth]{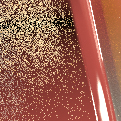} \\
	\multicolumn{3}{c}{} &	
	\includegraphics[width=0.1\linewidth]{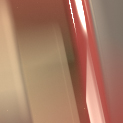} &
	\includegraphics[width=0.1\linewidth]{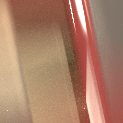} &
	\includegraphics[width=0.1\linewidth]{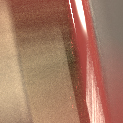}	&
	\includegraphics[width=0.1\linewidth]{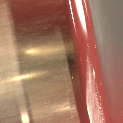} &
	\includegraphics[width=0.1\linewidth]{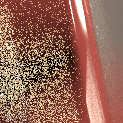} \\
	\multicolumn{3}{c}{} &	
	\includegraphics[width=0.1\linewidth]{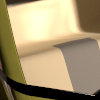} &
	\includegraphics[width=0.1\linewidth]{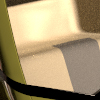} &
	\includegraphics[width=0.1\linewidth]{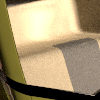} &
	\includegraphics[width=0.1\linewidth]{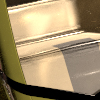} &
	\includegraphics[width=0.1\linewidth]{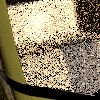} \\
	\multicolumn{3}{c}{OURS} & REF & OURS & MEMLT & MMLT & BDPT
	\end{tabular}
	\caption{\textsc{Cars}: Equal-time (20 minutes) comparison on the \textit{cars} scene, with a static car and a moving car lit by an area light. The direct lighting is computed separately. The interior of the car is enclosed by near-specular glass windows, which gives rise to specular-diffuse-specular paths that are challenging to sample. The three insets show the renderings of our method (H\textsuperscript{2}MC), Manifold Exploration Metropolis Light Transport (MEMLT)~\cite{Jakob:2012:MEM}, Multiplexed Metropolis Light Transport (MMLT)~\cite{Hachisuka:2014:MML}, and Bidirectional Path Tracing (BDPT)~\cite{Veach:1994:BEL}. The reference (REF) is rendered by our method in roughly 15 hours. BDPT cannot efficiently sample the sparse contribution function and suffers from severe noise. MMLT tends to get trapped in the hard-to-find features and produces correlated noise. MEMLT specializes in finding difficult static specular paths, but does not consider the anisotropy in the time domain, resulting in ghosting artifacts. Our method can efficiently resolve the hard-to-find caustics light paths like the specialized method, and is more general so that it can resolve moving caustic paths in the window by capturing the correlation between the time domain and path space.}
	\label{fig:comp_cars}
\end{figure}

In the previous chapters, we focused on \emph{inverse} applications, where we try to find a set of parameters or inputs satisfying certain outputs. In this chapter, we discuss something slightly different, in particular we show that derivatives can also be useful for accelerating \emph{forward} rendering.

\begin{figure}[t]
	\centering
	\captionsetup[subfigure]{justification=centering}
		\begin{minipage}[t]{0.395\linewidth}
			\centering
			\begin{subfigure}[t]{0.49\textwidth}
				\includegraphics[width=1.0\linewidth]{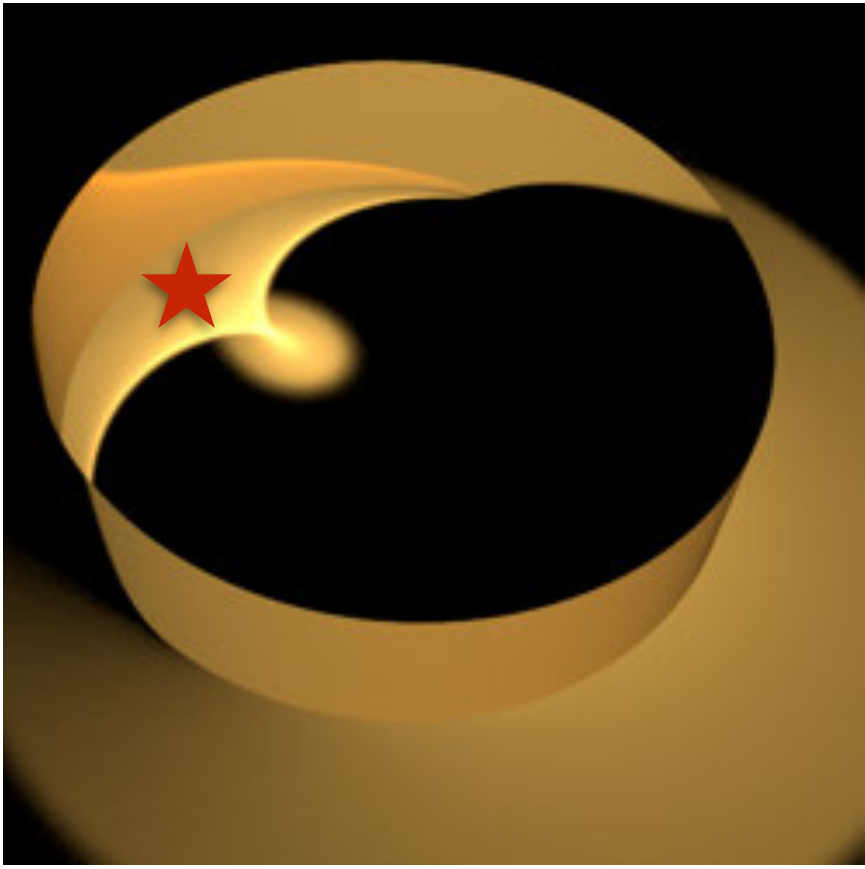}
				\caption{\footnotesize ring}
			\end{subfigure}
			\begin{subfigure}[t]{0.49\textwidth}
				\includegraphics[width=1.0\linewidth]{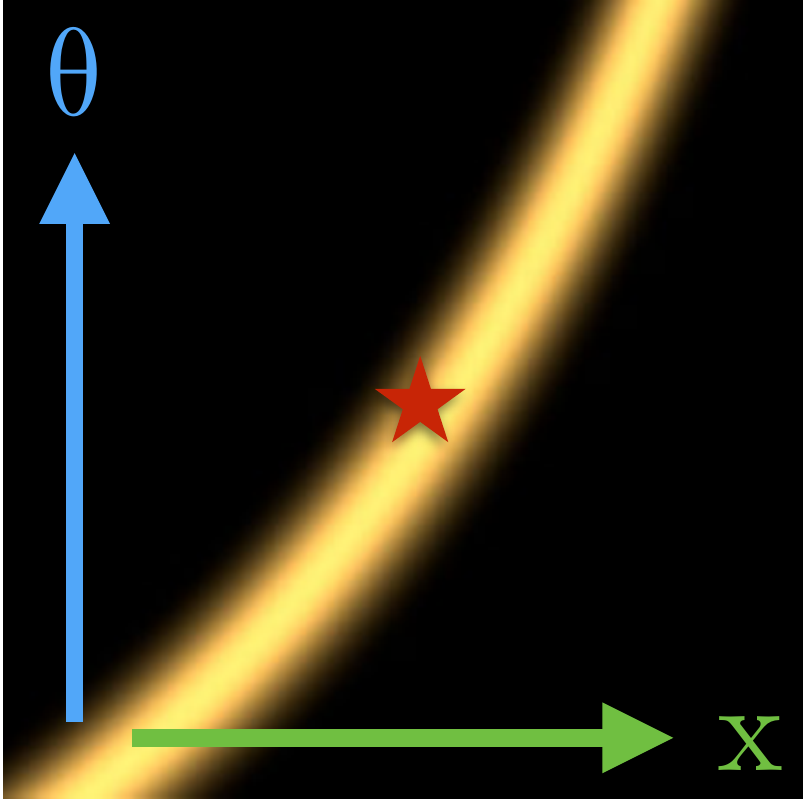}
				\caption{\footnotesize path space slice}
			\end{subfigure}
			\begin{subfigure}[t]{1.0\textwidth}
				\vspace{1em}
				\includegraphics[width=0.75\linewidth]{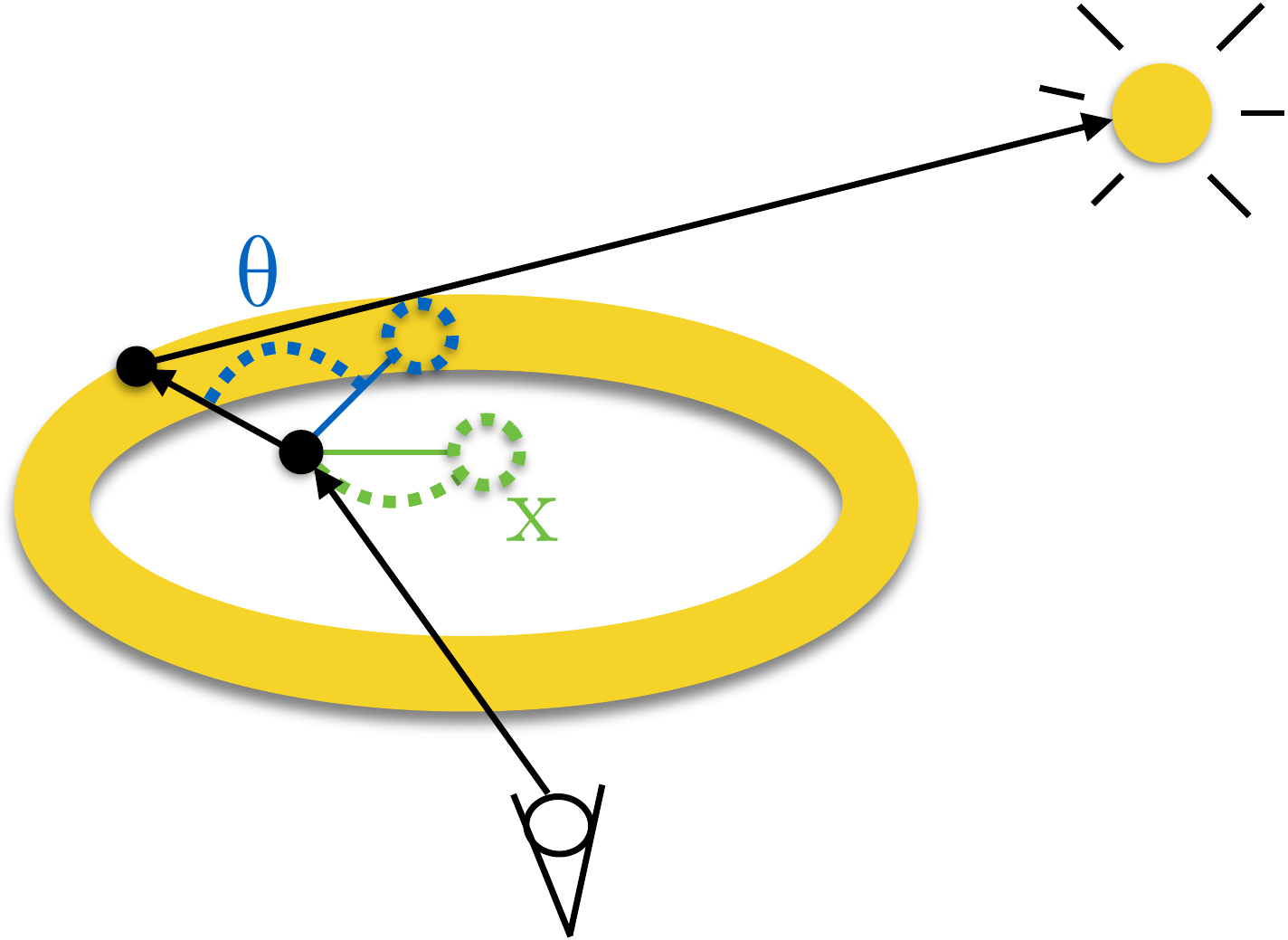}
				\caption{\footnotesize scene configuration}
			\end{subfigure}
		\end{minipage}		
	\hspace*{-0.15cm}
	\raisebox{1pt}{
		\begin{minipage}[t]{0.59\linewidth}		
			\centering
			\begin{subfigure}[t]{0.32\textwidth}
				\begin{overpic}[width=1.0\linewidth]{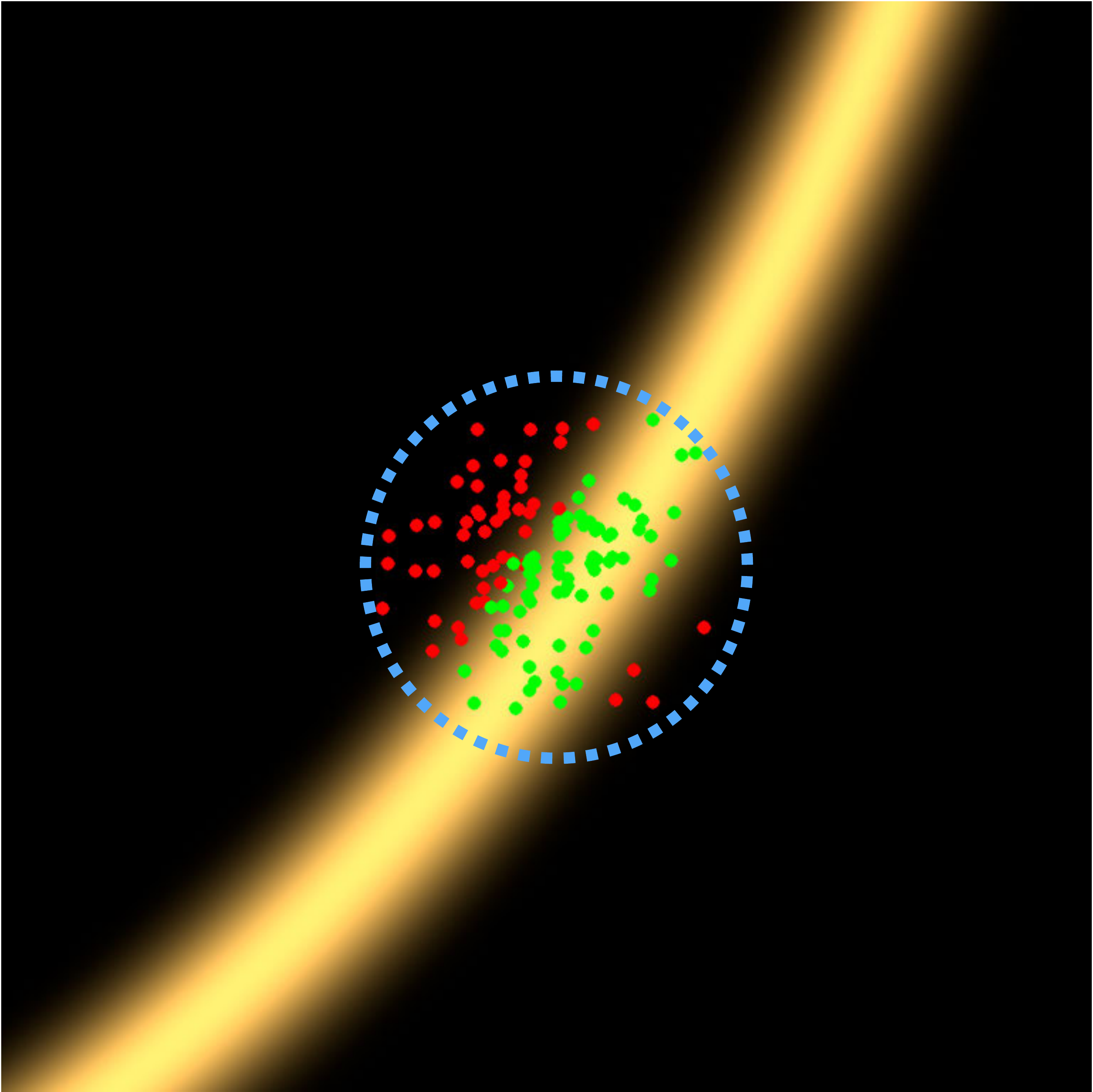}
					\put (6, 80) {\parbox{1in}{\textcolor{white}{\scriptsize green: accept \\ red: reject}}}
				\end{overpic}
				\caption{\footnotesize traditional \\ $128$ proposals \\ (acc. rate $59.37\%$)}
			\end{subfigure}
			\begin{subfigure}[t]{0.32\textwidth}
				\includegraphics[width=1.0\linewidth]{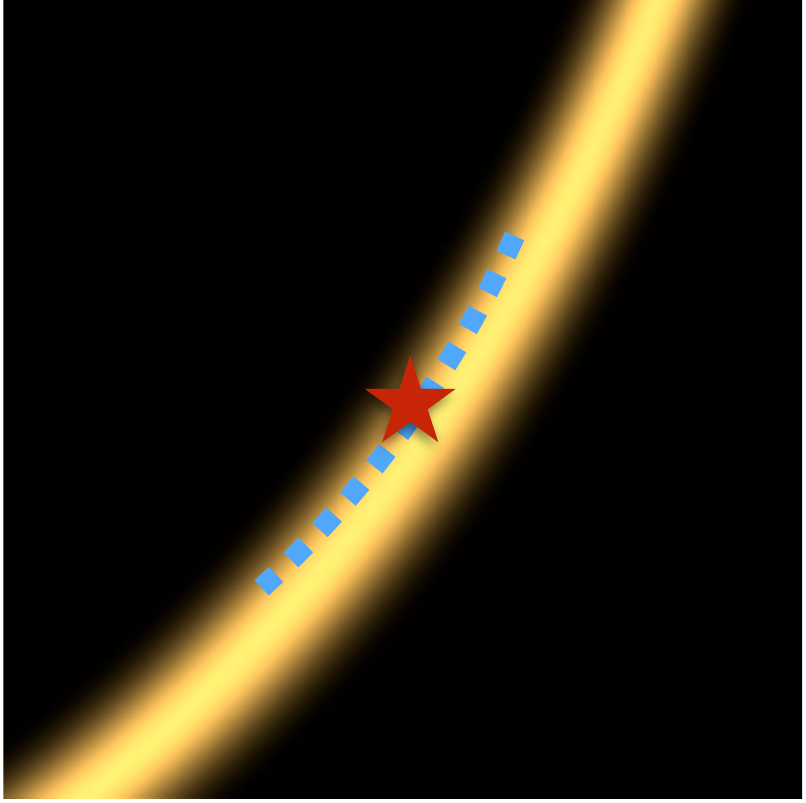}
				\caption{\label{fig:ring_me} \footnotesize ME~\cite{Jakob:2012:MEM} \\ (schematic view)}
			\end{subfigure}
			\begin{subfigure}[t]{0.32\textwidth}
				\includegraphics[width=1.0\linewidth]{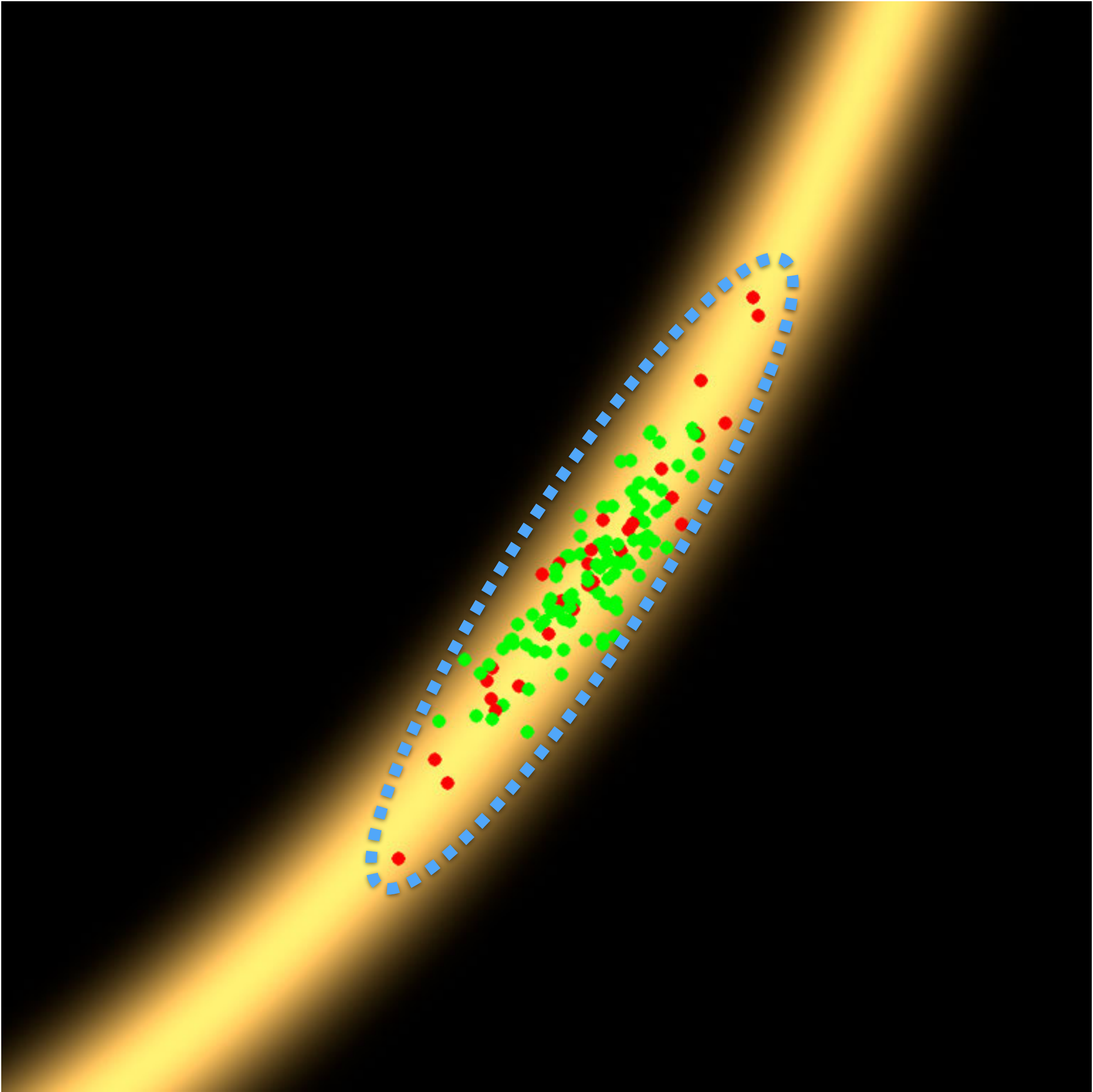}
				\caption{\footnotesize ours \\ $128$ proposals \\ (accept rate $75.58\%$)}
			\end{subfigure}

			\begin{subfigure}[t]{0.32\textwidth}
				\includegraphics[width=1.0\linewidth]{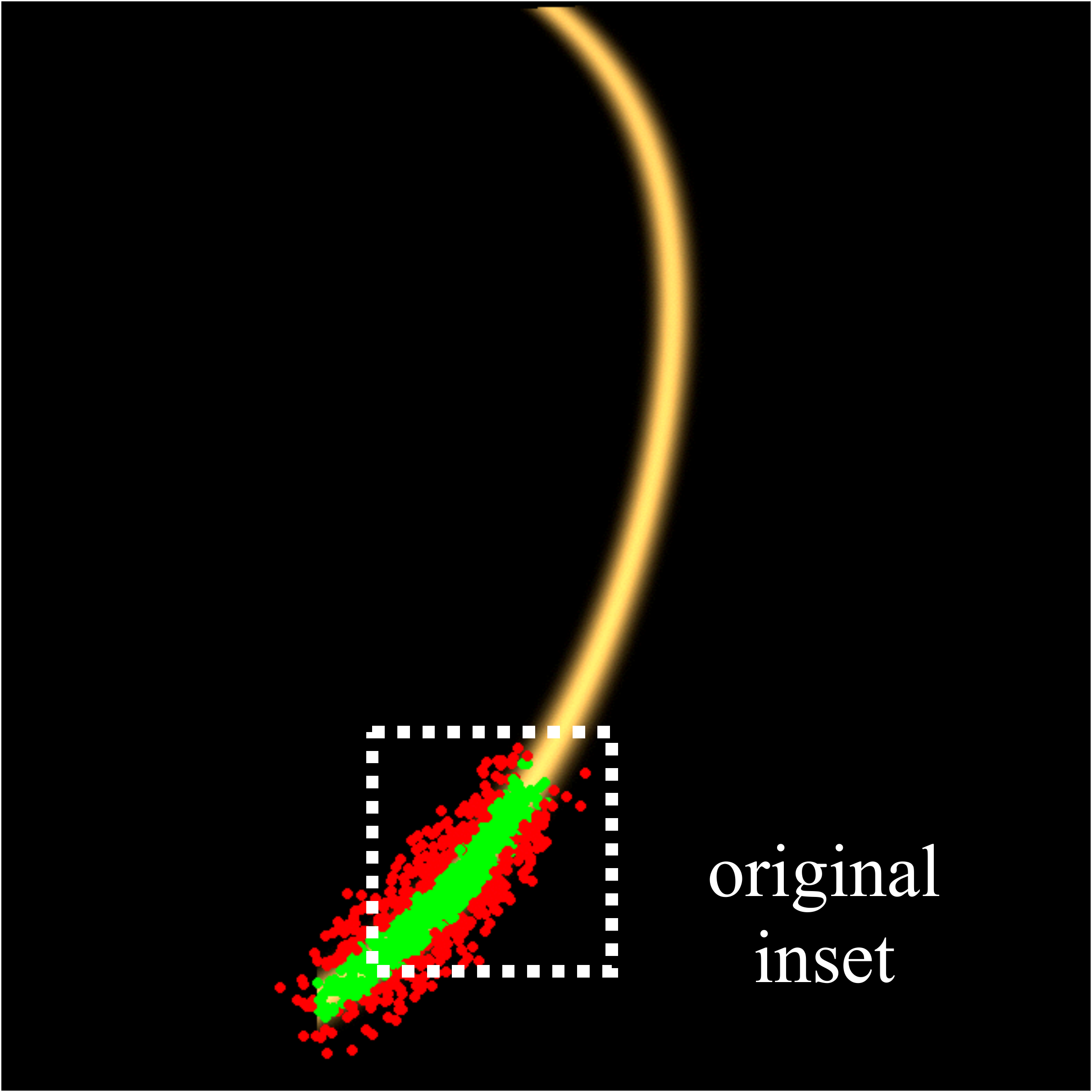}
				\caption{\footnotesize traditional \\ $1024$ MCMC states \\ (acc. rate $57.52\%$)}
			\end{subfigure}
			\begin{subfigure}[t]{0.32\textwidth}
				\includegraphics[width=1.0\linewidth]{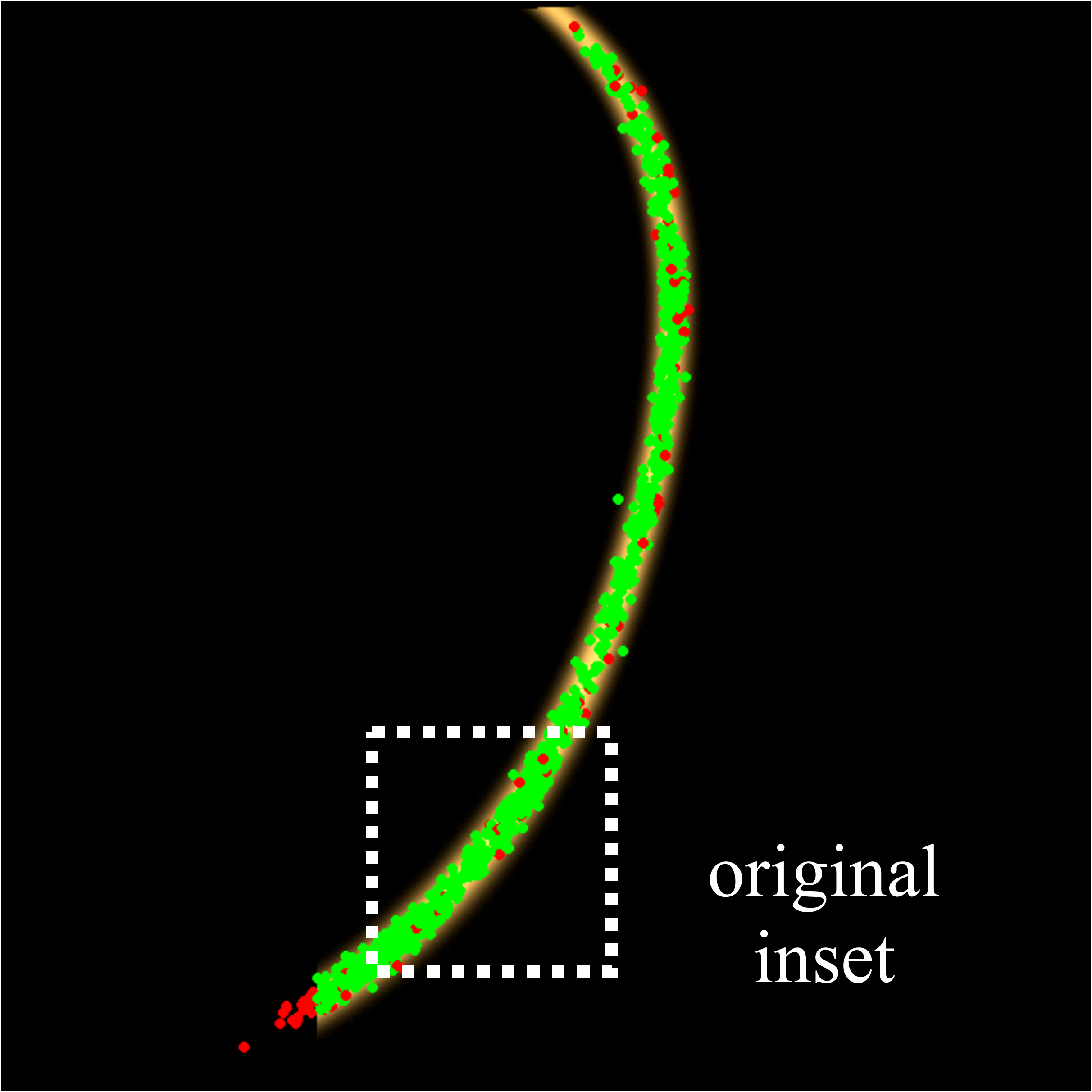}
				\caption{\footnotesize ours \\ $1024$ MCMC states \\ (accept rate $92.99\%$)}
			\end{subfigure}
			\begin{subfigure}[t]{0.32\textwidth}
				\includegraphics[width=1.0\linewidth]{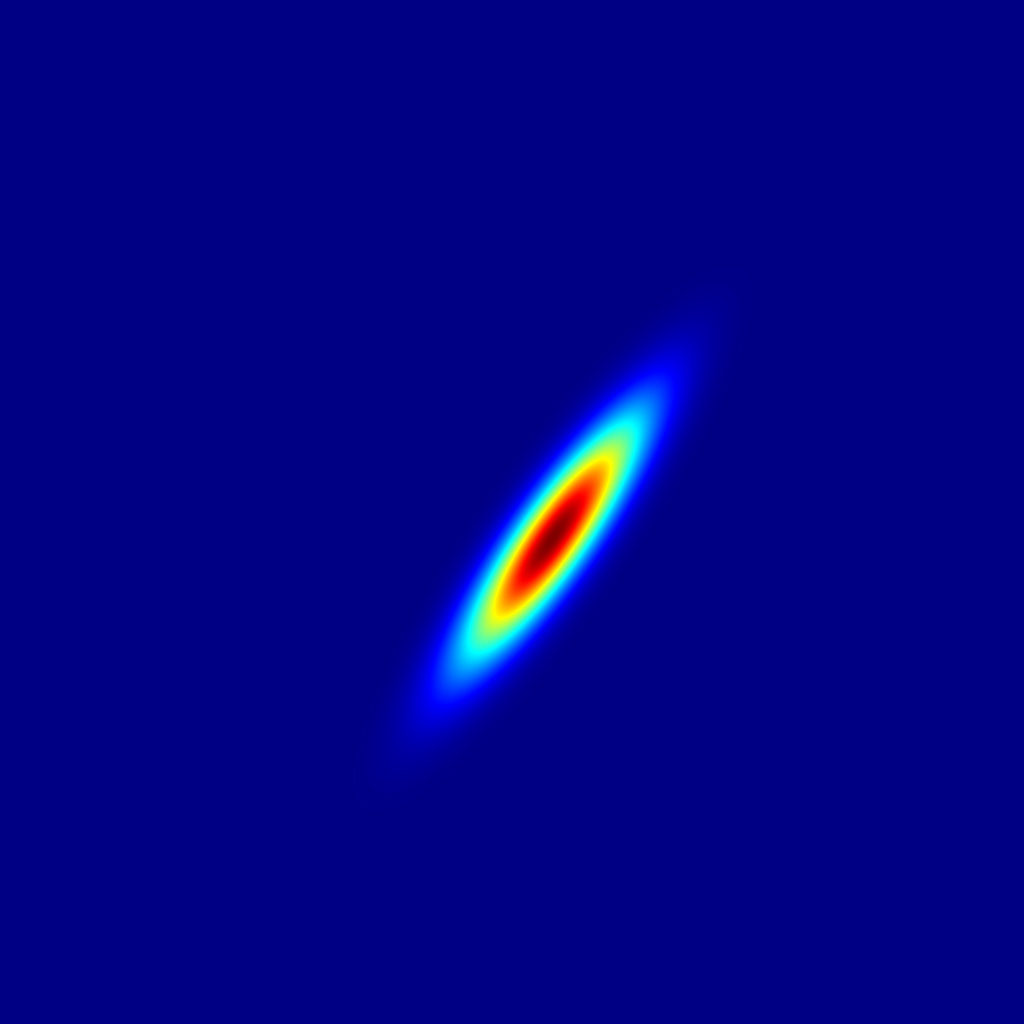}
				\caption{\footnotesize our proposal \\ is a closed-form \\ Gaussian}
			\end{subfigure}
		\end{minipage}	
	}

	\caption{\textsc{Ring example}: (a) A motivating example showing the caustics caused by a highly-glossy gold ring, lit by a distant point light. (b) A slice of the two-bounce indirect light field around the red star, where $x$ represents one of the dimensions in screen-space, and $\theta$ represents one of the dimensions along the BRDF sampling direction (the configuration is shown in (c)). The path contribution is sparse, and most of the contributions are zero. (d) The green/red dots represent the accepted/rejected proposal samples of a traditional MCMC rendering algorithm~\cite{Kelemen:2002:SRM}, which uses isotropic mutation that makes the sampling inefficient. (e) We also show the schematic of Manifold Exploration (ME)~\cite{Jakob:2012:MEM}, which only travels on the tangent of a lower dimensional space. (f) Our approach builds a Gaussian approximation around the neighborhood, enabling us to efficiently traverse the target function locally. Some samples are rejected due to the adaptivity, but it still results in a higher acceptance rate. (g)(h) We show the zoomed out slice (the positions of the original insets (d) and (e) are at the bottom of the images) with dots now representing the samples obtained by simulating the Markov chain for $1024$ states; our method explores the space more thoroughly. We also show the false color visualization of the Gaussian approximation in (i), which takes the width of the function into consideration.}
	\label{fig:ring_example}
\end{figure}

Light transport phenomena such as caustics, multiple-bounce glossy transport and motion blur often concentrate high contributions in a narrow volume within the high-dimensional sample space. While efficient methods exist for local importance sampling of individual scattering events, their combined effect on path throughput is intricate and hard to sample, leading to noisy images. Figure~\ref{fig:ring_example} shows a  caustic caused by a glossy gold ring. The integrand (Figure \ref{fig:ring_example} (b)) is sparse: for points on the floor ($x$), only a few incident directions ($\theta$) contribute radiance through reflection. Even in this simple scene,  sparsity makes standard numerical integration methods inefficient. The region of high-contribution is continuous,  but highly anisotropic, and the anisotropy varies over the integrand. In this chapter, we present a general solution by extending Metropolis Light Transport~\cite{Veach:1997:MLT} (a Markov Chain Monte Carlo sampler) to exploit the local structure of the path contribution function over its entire high-dimensional domain.

Adapting to the local anisotropic behavior of the integrand has been a long-standing challenge in rendering. Previous work has focused on model-based characterizations of anisotropy that are tied to specific effects (specular transfer, motion, etc.) \cite{Jakob:2012:MEM,Belcour:2013:CTF,Kaplanyan:2014:NCR}, and combining them is not easy. Closest to our work is Manifold Exploration \cite{Jakob:2012:MEM} and Half-vector Space Light Transport~\cite{Kaplanyan:2014:NCR,Hanika:2015:IHV} which use assumptions about the mirror direction and specular reflection to derive major directions of anisotropy (Figure \ref{fig:ring_me}), and walk along a lower-dimensional manifold. In contrast, we seek for a general solution that can characterize the ``thickness'' of the manifold in all directions, avoiding case-specific manual derivations.

The adaptation boils down to two main problems: 1) characterizing the anisotropy using local information and 2) sampling according to the derived information. We solve 1) by characterizing the local throughput using its derivatives. Since the gradient provides weak directional information, we also use the second derivative, the Hessian matrix. Whereas the gradient points only into the direction of the strongest increase, the Hessian additionally captures the correlation between coordinates. While the Hessian has been used before in rendering, e.g. \cite{Holzschuch:1998:EEB,Schwarzhaupt:2012:PHB}, its manual derivation is tedious and has usually been restricted to specific transport phenomena such as diffuse-only. In contrast, we use automatic differentiation (Chapter~\ref{chap:autodiff}) which allows us to handle general effects.

While the Hessian captures anisotropy well, the second problem of sampling remains: it is not possible to directly sample from the resulting quadratic approximation because it does not define a proper distribution and grows to infinity. Instead, we start from Hamiltonian Monte Carlo ~\cite{Duane:1987:HMC}, a Markov chain Monte Carlo sampling algorithm that proposes new sample locations by simulating the dynamics of a particle that starts at the current sample with a random initial velocity, briefly mentioned in Chapter~\ref{sec:hmc_background}. The particle evolves under gravity in a landscape composed of the contribution function flipped upside down so that the particle is attracted to high contribution areas (low height) by gravity. Crucially, we do not apply Hamiltonian Monte Carlo directly: this would be too expensive, because it would require numerical integration to generate just a single sample, and each integration time step would involve costly ray tracing, shading, and derivatives which do not directly contribute to the image. In practice, up to a hundred time steps per sample may be needed~\cite{Neal:2010:MUH}. Instead, we apply a modified version of HMC that results in closed-form integration. As we show in the paper, running Hamiltonian dynamics on a second-order function with a Gaussian distribution of initial momentums leads to a Gaussian distribution of final positions, and it results in a standard Metropolis-Hastings sampling. While traditional Metropolis sampling also uses a Gaussian distribution of proposals, it is usually isotropic and is centered on the current sample. In contrast, our Gaussian proposal is anisotropic, conforms to the shape of the contribution function, and is centered towards higher values according to the local gradient and Hessian. 

While our method can be used with arbitrary path parametrization, a carefully designed parametrization can lead to better mixing of the Markov chain. We propose a modified parameterization of the path space based on the \emph{primary sample space} proposed by Kelemen et al.~\cite{Kelemen:2002:SRM}. The modified parameterization reduces the correlation between the dimensions (Figure~\ref{fig:comp_param}).

Our method is general thanks to the use of the second-order Taylor expansion and automatic differentiation. In particular, it can be easily extended to time for motion blur effects, so that we are able to resolve the correlation between path-space and time for a light path that contains a moving caustic in a window (Figure~\ref{fig:comp_cars}). We focus on surface rendering in this paper, though conceptually our general approach could be extended to handle a variety of other phenomena such as BSSRDFs or participating media.

\section{Related Work}

Our work is closely related to the rendering algorithms that build upon MCMC sampling and the methods that utilize derivatives to drive the sampling process.

\paragraph{Metropolis Light Transport}
In light transport simulation, we need to compute the path integral~\cite{Veach:1998:RMC} $I_j$ for each pixel $j$:
\begin{equation}
I_j = \int_{\Omega}
	{h_j \left(\bm{x}\right) 
	 f\left(\bm{x}\right) 
	 d\mu \left(\bm{x}\right)},
\end{equation}
where $\Omega$ is \textit{path space}, which contains all the light paths, $h_j$ is the camera response function for pixel $j$, $f\left(\bm{x}\right)$ is the path contribution function~\cite{Veach:1998:RMC}, and $\mu\left(\bm{x}\right)$ is the area density of path $\bm{x}$.

Veach and Guibas~\cite{Veach:1997:MLT} apply the Markov Chain Monte Carlo sampling method (see Chapter~\ref{sec:sampling} for a brief review) by generating a sequence of Markov chain samples $\bm{x}_i$. A new proposal sample is \emph{mutated} from the previous sample, and probabilistically accepted or rejected. Specifically, given a sample $\bm{x}_{i-1}$, and a target function $f^{*}\left(\bm{x}\right)$, which is commonly set to the luminance of $f\left(\bm{x}\right)$, we first generate a \textit{proposal} sample $\bm{x}'$ with the transition probability $Q\left(\bm{x}_{i-1} \rightarrow \bm{x}'\right)$, and set the next sample $\bm{x}_{i}$ as follows:
\begin{equation}
\bm{x}_{i} = \begin{cases}
\bm{x}' & \mbox{with probability } a\left(\bm{x}_{i-1} \rightarrow \bm{x}'\right) \\
\bm{x}_{i-1} & \mbox{otherwise,}
\end{cases}
\label{eq:metropolis_hastings}
\end{equation}
where the acceptance probability $a$ is defined as
\begin{equation}
a\left(\bm{x}_{i-1} \rightarrow \bm{x}'\right) = \min\left(1, \frac{f^{*}\left(\bm{x}'\right) Q\left(\bm{x}' \rightarrow \bm{x}_{i-1}\right)}{f^{*}\left(\bm{x}_{i-1}\right) Q\left(\bm{x}_{i-1} \rightarrow \bm{x}'\right)}\right)
\label{eq:acceptance_prob}
\end{equation}
This satisfies the \textit{detailed balance} condition. That is, for any light paths $\bm{x}$ and $\bm{y}$, we have
\begin{equation}
f^{*}\left(\bm{x}\right) Q\left(\bm{x} \rightarrow \bm{y}\right) a\left(\bm{x} \rightarrow \bm{y}\right) = f^{*}\left(\bm{y}\right) Q\left(\bm{y} \rightarrow \bm{x}\right) a\left(\bm{y} \rightarrow \bm{x}\right).
\label{eq:detailed_balance2}
\end{equation}

As mentioned in Chapter~\ref{sec:sampling}, if a transition function satisfies the detailed balance condition, and if there is a strict positive probability to sample all light paths with non-zero contribution (\emph{ergodicity}), it will converge to a distribution proportional to the target function $f^{*}(\bm{x})$. Veach and Guibas then approximated the path integral $I_j$ at pixel $j$ using the weighted average of the Markov chain samples:
\begin{equation}
I_j = \frac{b}{N}\sum_{i=1}^{N}\frac{h_j\left(\bm{x}_i\right)f\left(\bm{x}\right)}{f^{*}\left(\bm{x}\right)},
\end{equation}
where $b$ is a normalization constant, which is the average of $f^{*}\left(\bm{x}\right)$ over the image.\footnote{Since we need to estimate the normalizing constant $b$, this particular use of Markov chain is only useful when we are interested in multiple integrals, where their integrals are correlated.} Originally Veach and Guibas designed several specialized mutation strategies to address different lighting scenarios. Each strategy has a different asymmetric probability distribution, which introduces a significant challenge to implement all the strategies correctly. To simplify the algorithm, Kelemen et al.~\cite{Kelemen:2002:SRM} proposed to mutate the state in the \emph{random number space}, which makes the mutation agnostic to the particular lighting effect. Later, Jacopo~\cite{Pantaleoni:2017:CML}, Otsu et al.~\cite{Bitterli:2017:RJM} and Bitter et al.~\cite{Bitterli:2017:RJM} propose methods to combine the two mutation strategies by finding the random number that generates a certain light path produced by Veach and Guibas's mutation. Unfortunately, both the mutation strategies proposed by Veach and Guibas and Kelemen et al. do not respect the complex local structure in the sampling domain, which makes them inefficient in some difficult cases.

Metropolis Light Transport has been extended in several aspects. Cline et al.~\cite{Cline:2005:ERP} proposed the Energy Redistribution Path Tracing technique by running many short Markov chains. Lai et al.~\cite{Lai:2007:PIR} adapted mutations with different parameters using Population Monte Carlo. Kitaoka et al.~\cite{Kitaoka:2009:REL} introduced replica exchange, or parallel tempering~\cite{Swendsen:1986:RMC}, that exchanges states between multiple Markov chains to avoid getting stuck at local modes. All these methods require some form of local mutation strategies. We introduce a new local sampling strategy that adapts to the local structure of the function. Lai et al.~\cite{Lai:2009:PAR} proposed a temporal mutation strategy based on object-space transformation.  Unlike their method, which requires a specially designed mutation, we treat the time dimension the same as the other dimensions, and handle the correlation between coordinates using second derivatives.

Jakob and Marschner~\cite{Jakob:2012:MEM}, Kaplanyan et al.~\cite{Kaplanyan:2014:NCR}, and Hanika et al.~\cite{Hanika:2015:IHV} use the first derivatives of the half-vectors of a specular light path to guide the MCMC sampling. These methods apply a form of Newton-iteration to sample new light paths satisfying certain constraints. While they improve the sampling efficiency of glossy and specular surfaces significantly, their methods can sometimes be inefficient on small, highly-curved surfaces, because of their first-order approximation. In addition, they only account for a subset of terms in the path contribution function, ignoring important effects such as the Fresnel reflection or light source emission profiles. In contrast, we utilize second-order derivatives and do not assume any particular effect. For example, we are able to render difficult moving caustics (Figure~\ref{fig:comp_cars}), where their methods would suffer from ghosting artifacts

Hachisuka et al.~\cite{Hachisuka:2014:MML} proposed Multiplexed Metropolis Light Transport that combines Kelemen et al's mutation startegy with multiple importance sampling~\cite{Veach:1995:OCS}. Their method is orthogonal to our algorithm, and we build our bidirectional path tracer based on their approach.

\paragraph{Derivatives in rendering}

Shinya et al.~\cite{Shinya:1987:PAP} used a second-order power series along with paraxial approximation to approximate the neighborhood of a ray. Irradiance caching techniques~\cite{Ward:1988:RTS,Ward:1992:IG,Schwarzhaupt:2012:PHB} compute the gradients and the Hessians of the irradiance with respect to the screen coordinates for sparse interpolation for diffuse or low-glossy surfaces. Ray differentials~\cite{Igehy:1999:TRD} and path differentials~\cite{Suykens:2001:PDA} compute the footprint of the light paths for texture filtering using first derivatives. Chen and Arvo~\cite{Chen:2000:TAS} use first and second-order derivatives of the specular light paths for sparse interpolation. Path gradients~\cite{Suykens:2001:PDA} are used for hierarchical radiosity applications, where the gradients of the paths are hand-derived. Ramamoorthi et al.~\cite{Ramamoorthi:2007:FAL} performed a first-order analysis for the direct illumination light field. Gradient-domain rendering approaches, e.g.~\cite{Lehtinen:2013:GDM,Kettunen:2015:GPT}, sample in the gradient domain to exploit the sparsity of gradients in image space. They use finite differences of the path on the image coordinates, whereas our method uses analytical derivatives of all dimensions. While finite differences could capture the discontinuities of the signal, they are more expensive to generate and do not scale well with dimensionality.

Our usage of derivatives differs from previous works in several respects. First, we use automatic differentiation to compute the derivatives, which means that we do not assume any particular effect. This enables our method to handle various combinations of lighting scenarios. Second, we take the derivatives with respect to all the sampling dimensions, so we can capture the high-dimensional structure of the light path.  Third, we take both the first and the second derivatives. The Hessians enable us to take the correlation between the dimensions of the sampling domain into account. Finally, we apply the derivatives in the MCMC sampling context.

\section{Hamiltonian Monte Carlo}

\begin{figure}[t]
	\centering
	\captionsetup[subfigure]{justification=centering}	
	\begin{minipage}[t]{0.29\linewidth}
		\begin{subfigure}[t]{\textwidth}
			\includegraphics[width=1.0\linewidth]{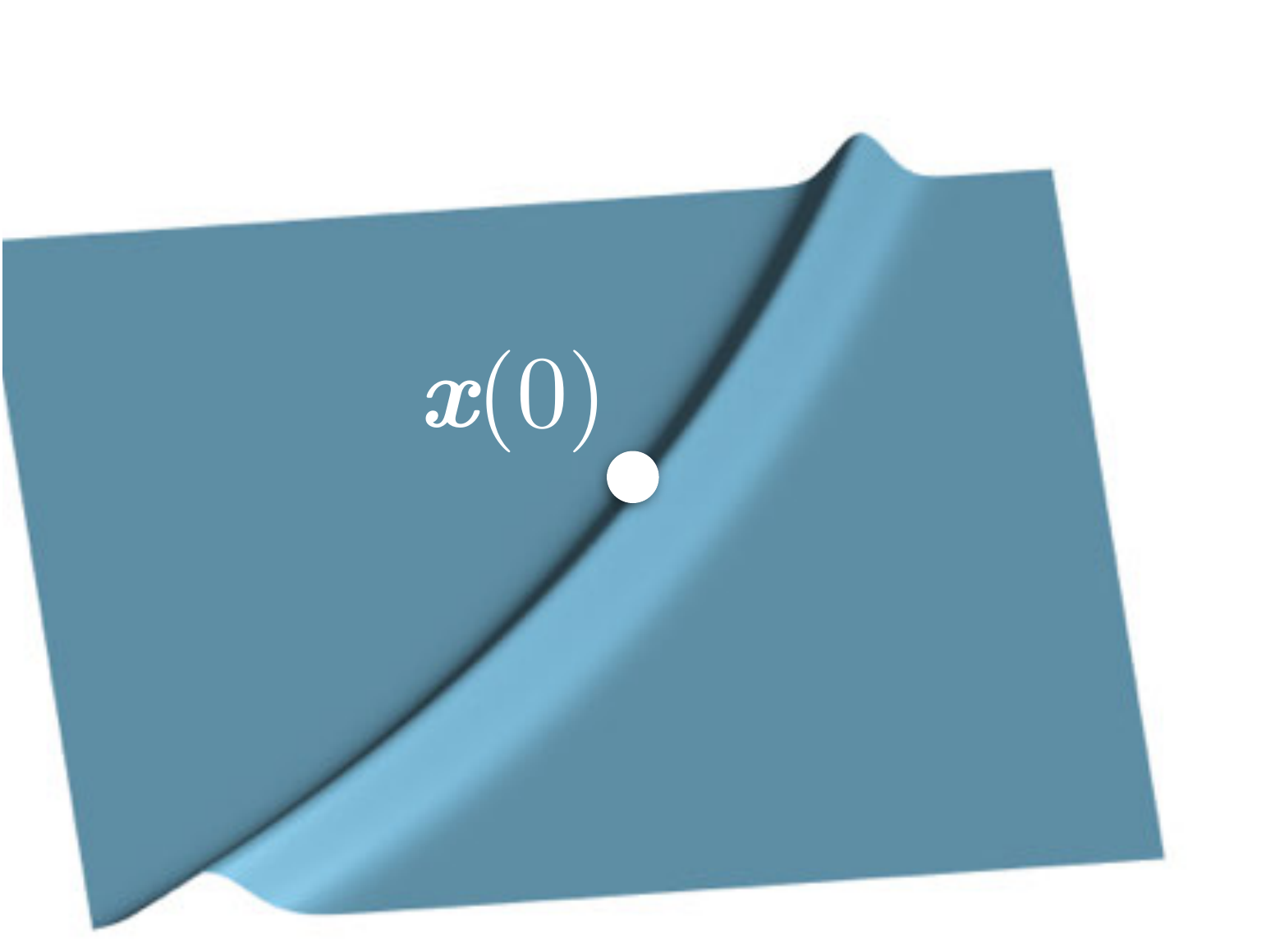} 
			\caption{\footnotesize given current sample $\bm{x}(0)$ and target function}
		\end{subfigure}
	\end{minipage}	
	\raisebox{40pt}{$\xrightarrow{-\log}$}
	\begin{minipage}[t]{0.29\linewidth}
		\begin{subfigure}[t]{\textwidth}
			\includegraphics[width=1.0\linewidth]{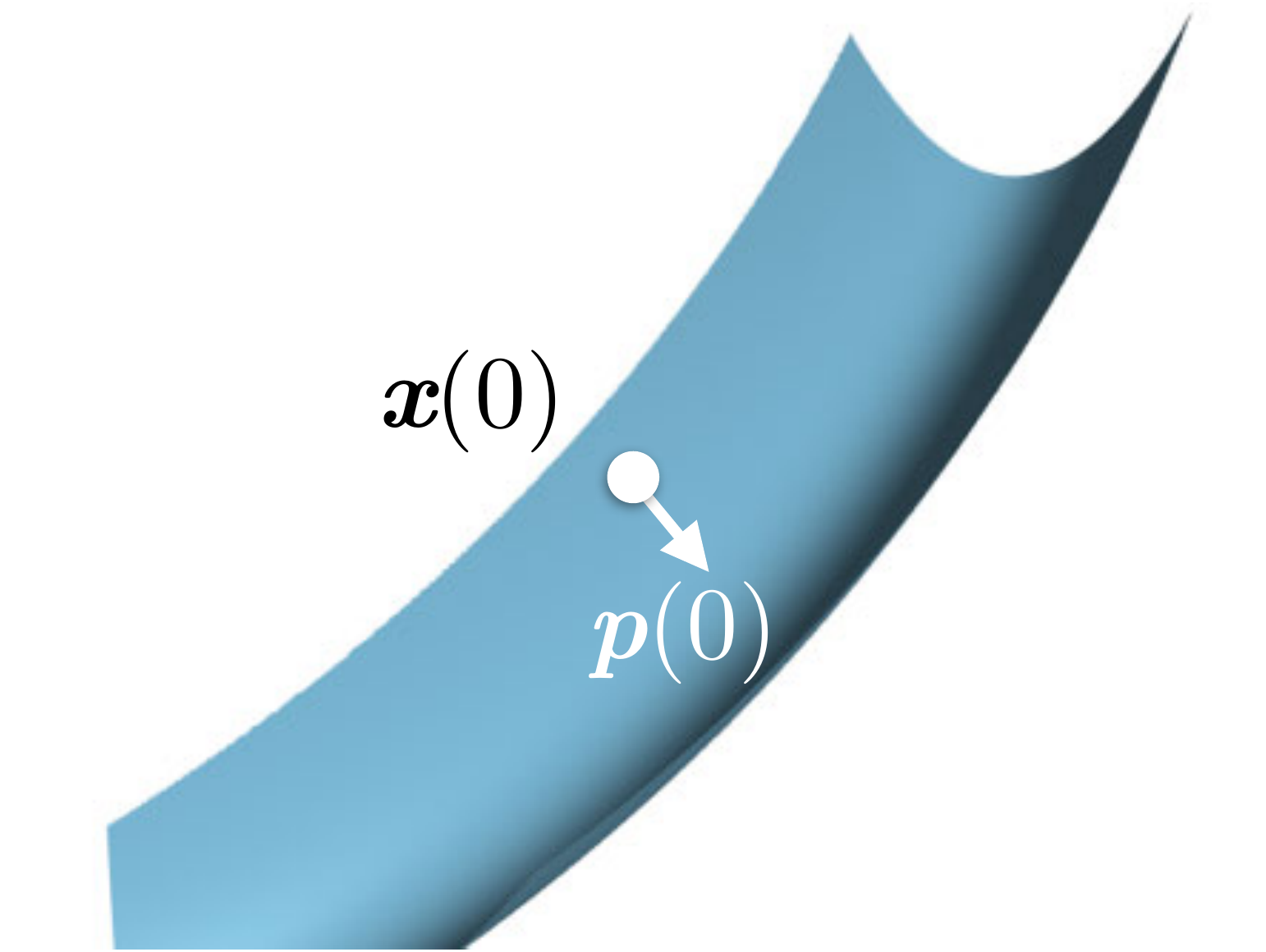} 
			\caption{\footnotesize flip function so that gravity pulls the particle, sample initial momentum $\bm{p}(0)$}
		\end{subfigure}
	\end{minipage}
	\raisebox{40pt}{$\xrightarrow{\quad\;\,}$}
	\begin{minipage}[t]{0.29\linewidth}
		\begin{subfigure}[t]{\textwidth}
			\includegraphics[width=1.0\linewidth]{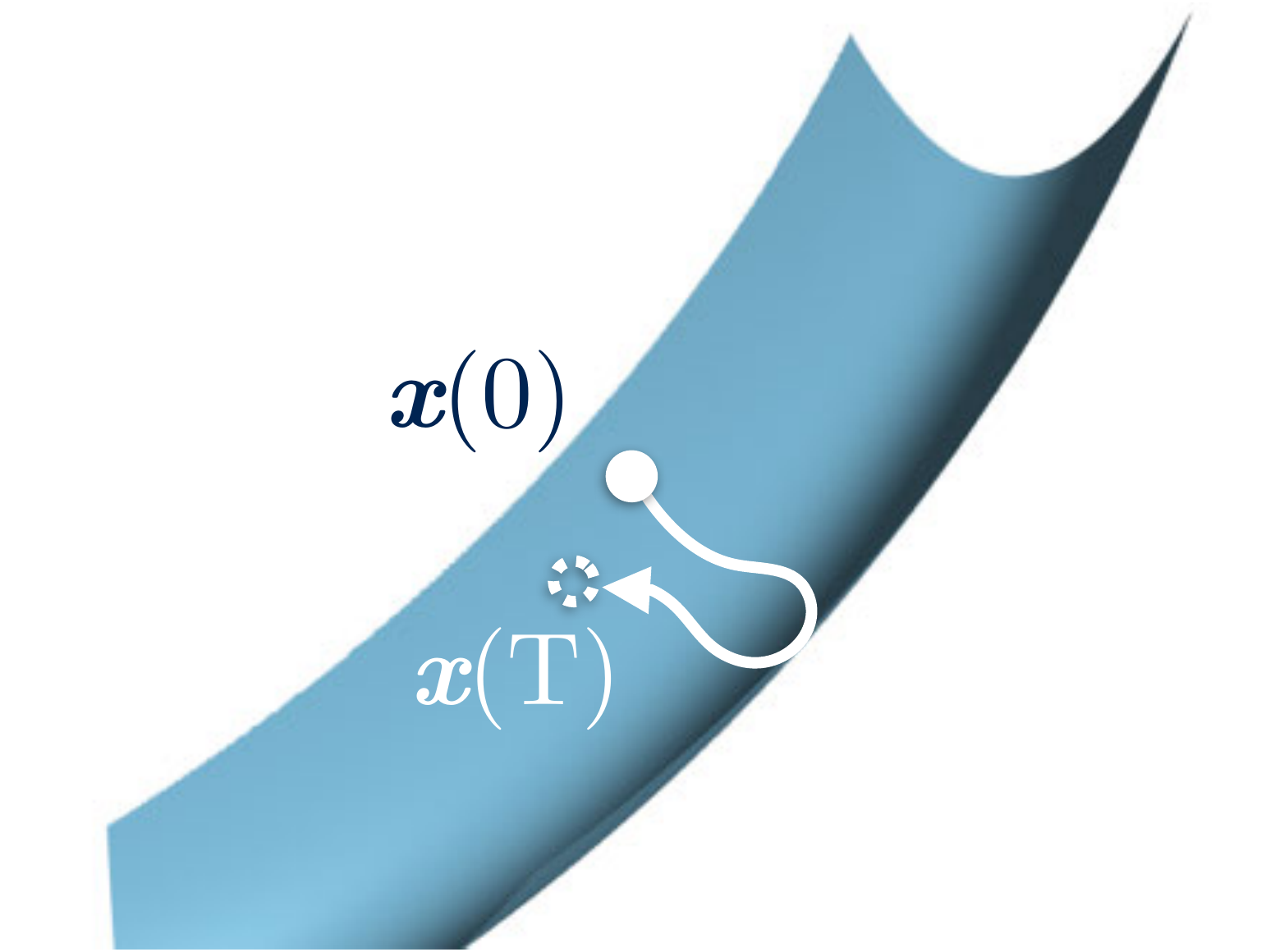} 
			\caption{\footnotesize simulate Hamiltonian dynamics for time $T$ to obtain proposal at $\bm{x}(T)$}
		\end{subfigure}
	\end{minipage}	
	
	\caption{\textsc{Hamiltonian Monte Carlo}: Given the current sample position $\bm{x}(0)$ and a target function (the 2D slice from Figure~\ref{fig:ring_example}), a physical analogy of Hamiltonian Monte Carlo is: (a) first it takes the logarithm of the target function and flips it upside down so that ``gravity'' pulls towards high contribution areas. (b) Then it gives the current sample an initial momentum $\bm{p}(0)$ and (c) lets the point move for some time $T$ with respect to the geometry of the flipped function.}
	\label{fig:hmc_heightfield}
\end{figure}
\begin{figure}[t]
	\centering
	\captionsetup[subfigure]{justification=centering}
	\begin{minipage}[t]{0.22\linewidth}
		\begin{subfigure}[t]{\textwidth}
			\includegraphics[width=1.0\linewidth]{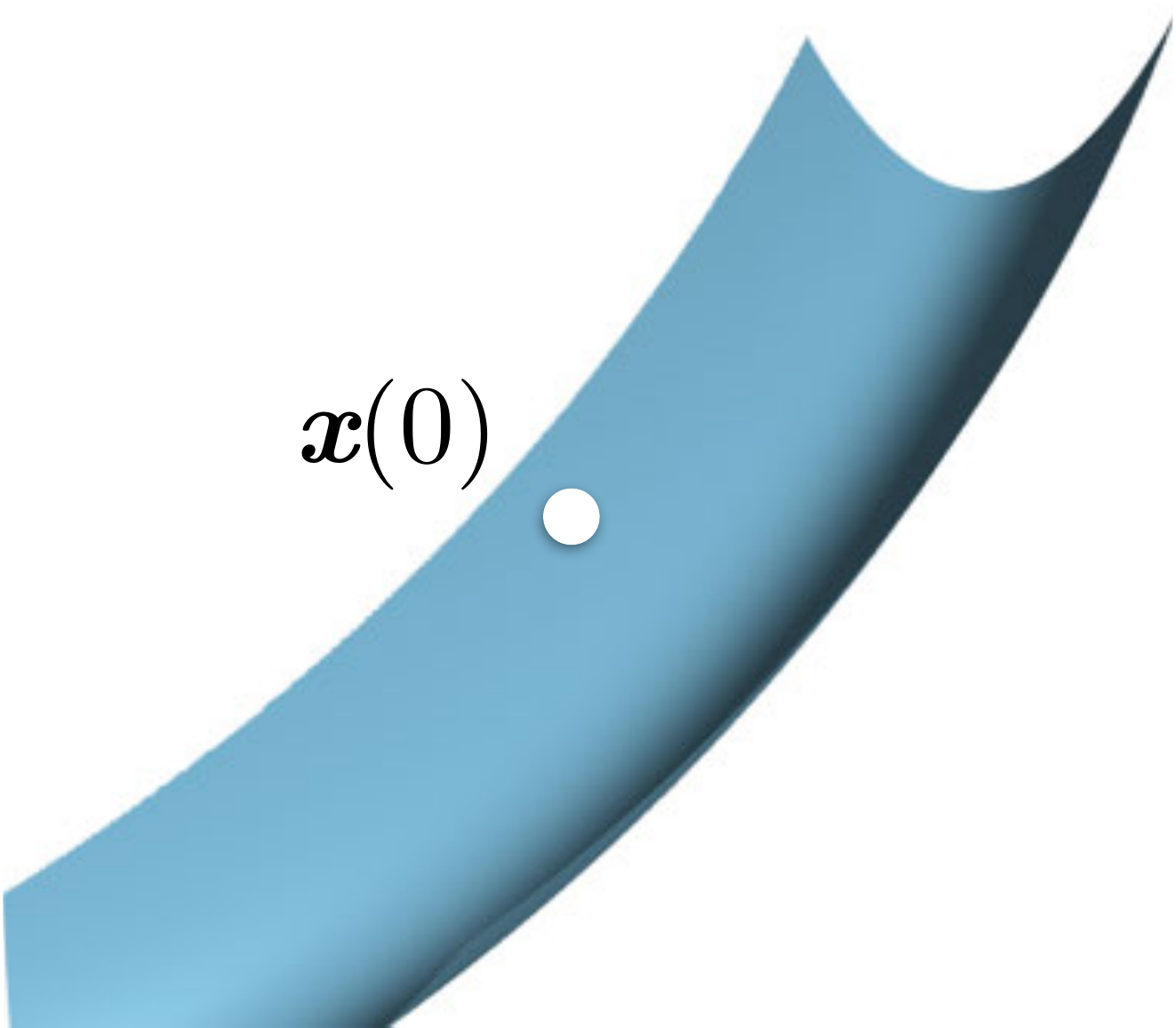} 
			\caption{\footnotesize original target and sample position}
		\end{subfigure}
	\end{minipage}
	\raisebox{40pt}{$\rightarrow$}
	\begin{minipage}[t]{0.22\linewidth}
		\begin{subfigure}[t]{\textwidth}
			\includegraphics[width=1.0\linewidth]{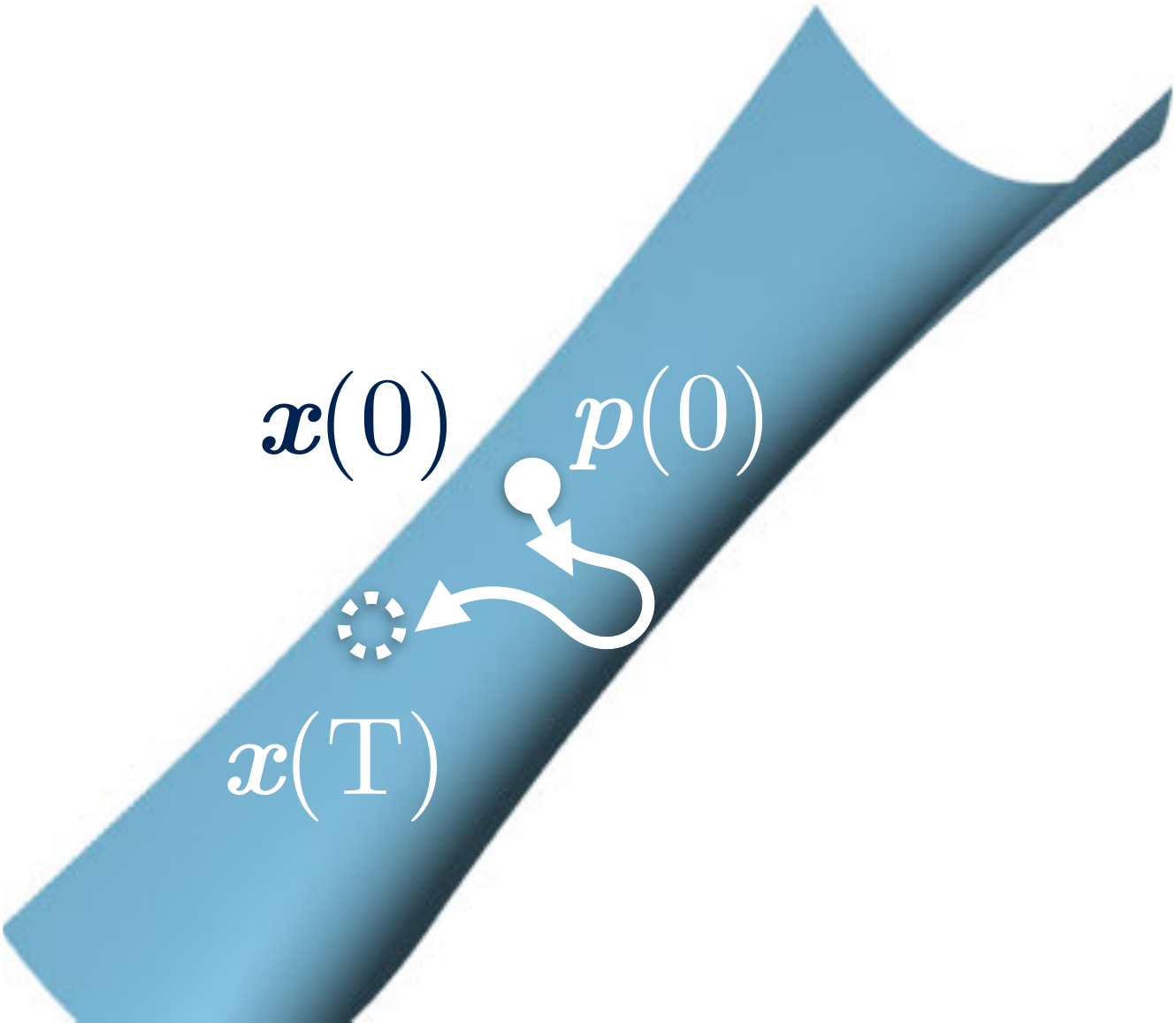} 
			\caption{\footnotesize simulate dynamics on local quadratic approximation}
		\end{subfigure}
	\end{minipage}
	\raisebox{40pt}{$\rightarrow$}
	\begin{minipage}[t]{0.22\linewidth}
		\begin{subfigure}[t]{\textwidth}
			\includegraphics[width=1.0\linewidth]{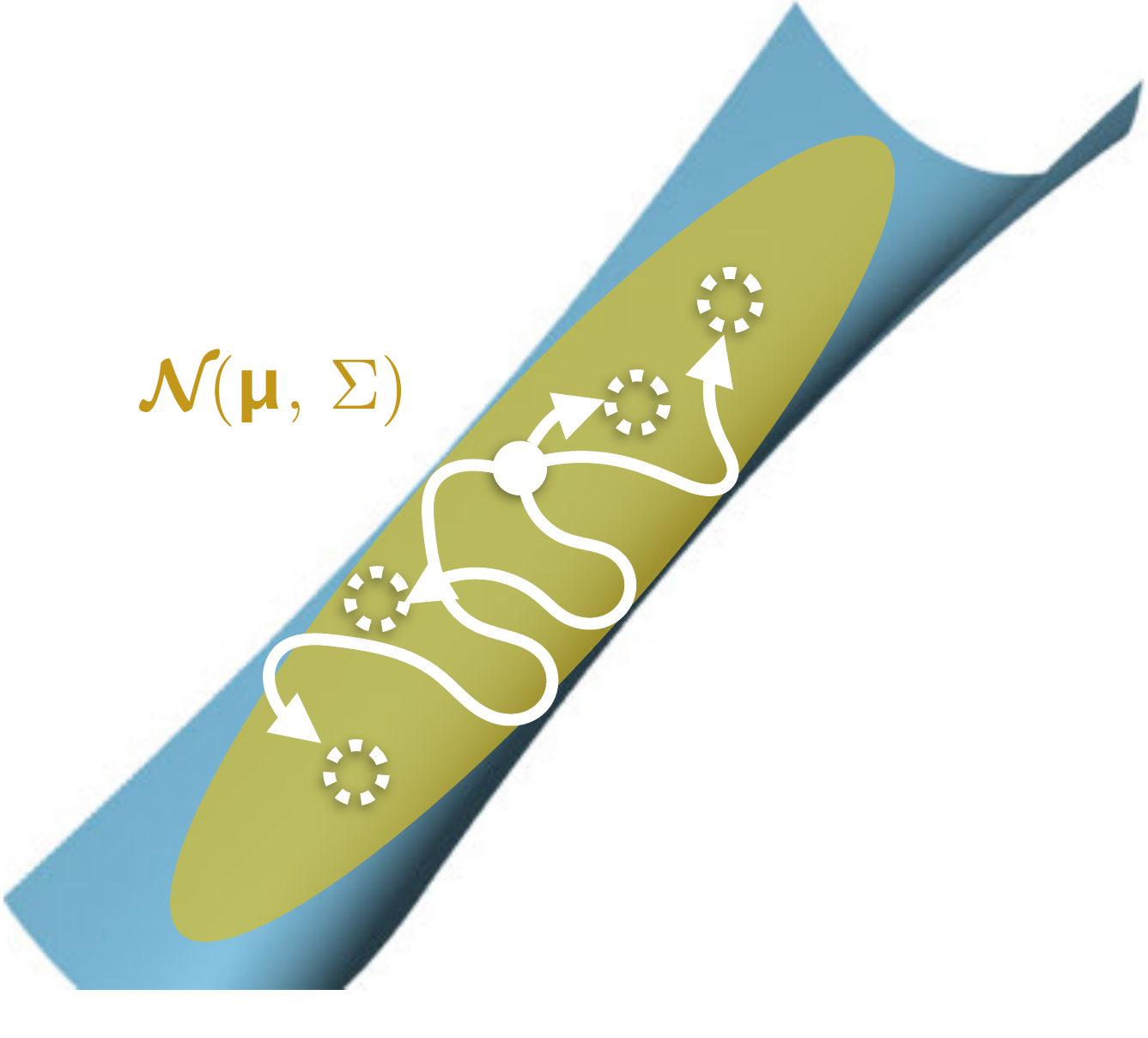} 
			\caption{\footnotesize proposals result in a Gaussian distribution}
		\end{subfigure}
	\end{minipage}
	\raisebox{40pt}{$\rightarrow$}
	\begin{minipage}[t]{0.22\linewidth}
		\begin{subfigure}[t]{\textwidth}
			\includegraphics[width=1.0\linewidth]{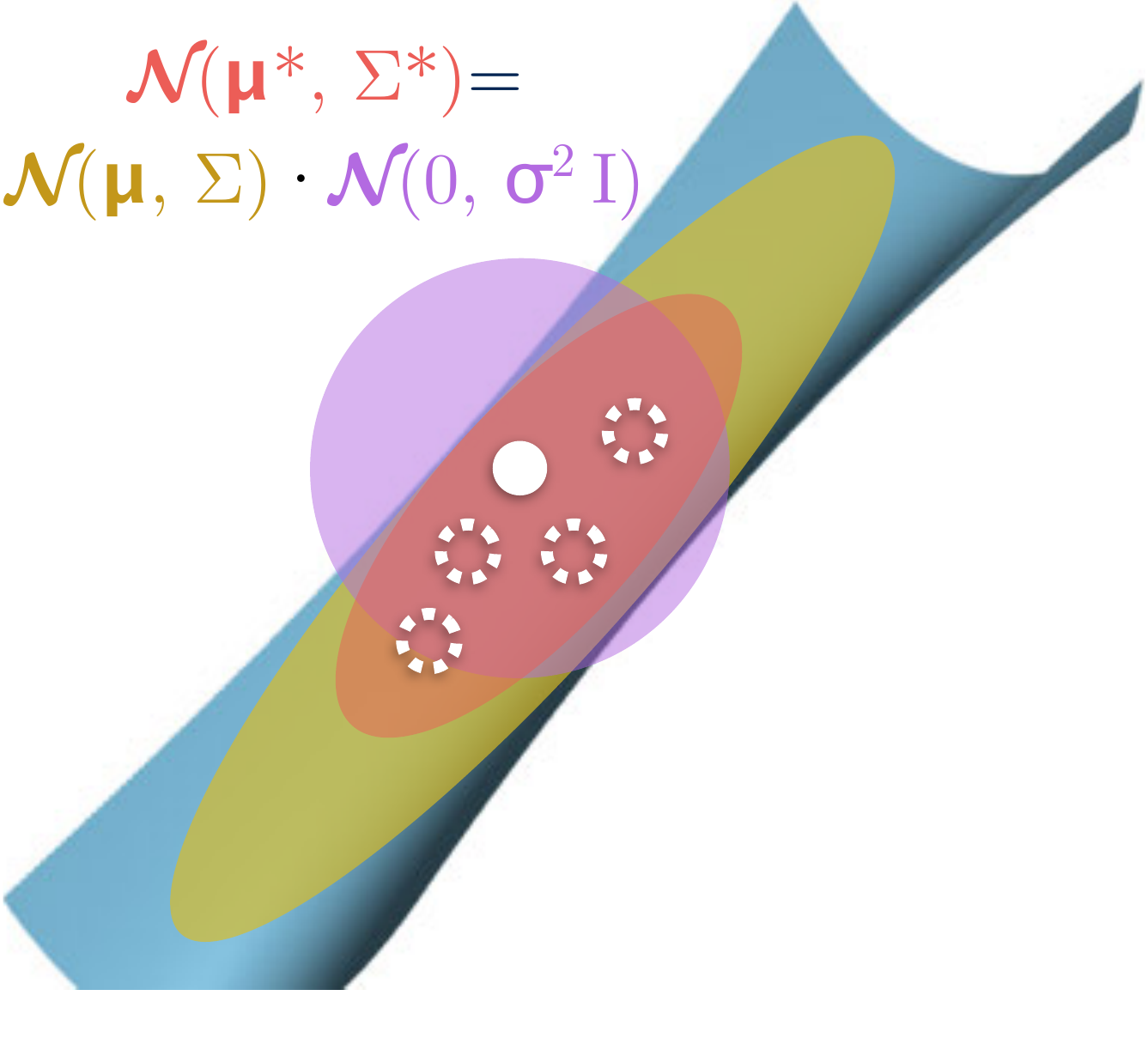} 
			\caption{\footnotesize multiply with a prior to limit variance, draw from red}
		\end{subfigure}
	\end{minipage}

	\caption{\textsc{Hessian-Hamiltonian Monte Carlo}: (a) Given the original function and sample position $x\left(0\right)$, (b) we approximate the costly Hamiltonian dynamics simulation by first constructing a local quadratic approximation at the current sample position $\bm{x}\left(0\right)$. (c) Different initial momentum $\bm{p}\left(0\right)$ results in different proposal positions $\bm{x}\left(T\right)$, which makes $\bm{x}\left(T\right)$ a random variable. We show that trajectories with a Gaussian distribution for initial momentum result in a Gaussian distribution (the yellow shaded area) for final destination. (d) Finally, we multiply the Gaussian with a prior (the purple shaded area) to prevent proposals from going too far when the second derivative is low. The resulting sample proposal distribution is shown in red, and we draw our proposals from the resulting distribution.}
	\label{fig:qhmc}
\end{figure}

We review the Hamiltonian Monte Carlo~\cite{Duane:1987:HMC} method; also see Neal's introduction article~\cite{Neal:2010:MUH} for a more thorough description and survey. Hamiltonian Monte Carlo generates the new proposal samples from the current sample by simulating Hamiltonian dynamics driven by the landscape of the target function. 

Markov chain Monte Carlo methods generate a sequence of samples $\bm{x}_i$, whose distribution converges to a distribution proportional to a specific target function $f^{*}\left(\bm{x}_i\right)$, by forming a Markov chain of the sample sequence. For the sake of notational simplicity, we denote the target function as $f\left(\bm{x}\right)$. At iteration $i+1$, a new proposal sample is drawn from a distribution based on $\bm{x}_i$. Then the proposal sample is probabilistically accepted or rejected. If accepted, it forms the new state of the Markov chain. From now on, we assume that the samples $\bm{x}_i$ lie in a hypercube of $\left[0,1\right]^N$, similar to the primary sample space~\cite{Kelemen:2002:SRM}. Operating directly in path-space~\cite{Veach:1998:RMC} is more challenging due to its definition as a cross product of lower-dimensional manifolds.

Figure~\ref{fig:hmc_heightfield} gives an illustration of Hamiltonian Monte Carlo where, in a nutshell, state is modified by giving the current sample a random initial velocity (or more precisely, a momentum), and simulating its motion under gravity. The target function first needs to be ``flipped'' so that high contribution regions correspond to a lower height and samples are attracted there by gravity (Figure~\ref{fig:hmc_heightfield} (b)).  The particle is given an initial momentum, typically drawn from a Gaussian, and its motion is simulated in the height field given by the flipped contribution function for a fixed amount of time.  Acceptance rules are then applied, although if the integrator preserves energy, samples are always accepted.  This approach helps the samples stay in the high contribution region (low height in the flipped function) because of the effect of gravity.

\paragraph{Hamiltonian dynamics} Formally, Hamiltonian dynamics is a system of differential equations defined on the Hamiltonian energy $E$:
\begin{align}
\begin{split}
\frac{\partial \bm{x}}{\partial t} &= \frac{\partial E}{\partial \bm{p}} \\
\frac{\partial \bm{p}}{\partial t} &= -\frac{\partial E}{\partial \bm{x}}.
\end{split}
\label{eq:Hamiltonian_equation}
\end{align}
The auxiliary momentum variable $\bm{p}$ is introduced to drive the sampling of position $\bm{x}$, and $\bm{p}$ has the same number of dimensions as $\bm{x}$. The Hamiltonian energy $E\left(\bm{x}, \bm{p}\right)$ is a composite of the potential energy $U\left(\bm{x}\right)=-\log{f\left(\bm{x}\right)}$ and the kinetic energy $K\left(\bm{p}\right)=\frac{1}{2}\bm{p}^TA\bm{p}$. The potential energy is defined in the logarithmic domain to better capture the dynamic range of the target functions:
\begin{equation}
E\left(\bm{x}, \bm{p}\right) = U\left(\bm{x}\right) + K\left(\bm{p}\right) = -\log{f\left(\bm{x}\right)} + \frac{1}{2}\bm{p}^TA\bm{p},
\label{eq:Hamiltonian_energy}
\end{equation}
where $A$ is a user-defined ``inverse mass matrix'', which represents the inverse of the mass of the particle. Typically, it is set to a scalar $\frac{1}{m}$ times an identity matrix, where $m$ is the mass, but in our work we will use a full matrix (discussed in Section~\ref{sec:setting_param}). The negative of the function $\log{f\left(x\right)}$ is taken to enable high contribution regions to have low potential energy (as shown in Figure~\ref{fig:hmc_heightfield}).

We substitute the definition of the Hamiltonian energy (Equation~\eqref{eq:Hamiltonian_energy}) into the Hamiltonian equation (Equation~\eqref{eq:Hamiltonian_equation}), and obtain:
\begin{align}
\begin{split}
\frac{\partial \bm{x}}{\partial t} &= A\bm{p} \\
\frac{\partial \bm{p}}{\partial t} &= \frac{\partial \log{f\left(\bm{x}\right)}}{\partial \bm{x}}.
\end{split}
\label{eq:Hamiltonian_equation_2}
\end{align}
Equation~\eqref{eq:Hamiltonian_equation_2} defines a trajectory of position $\bm{x}$ and momentum $\bm{p}$ over time $t$. Intuitively, if the momentum at time $t$ is high, we will make a large jump from the current position $\bm{x}\left(t\right)$, and if the derivatives of the target function at $\bm{x}\left(t\right)$ are low, the increment to the momentum will be small. Hamiltonian Monte Carlo is highly adaptive to the local structure of the target function.

\paragraph{Markov chain Monte Carlo with Hamiltonian dynamics} To apply Hamiltonian dynamics in the context of Markov chain Monte Carlo, we first take the exponent of the negative Hamiltonian energy:
\begin{equation}
\exp{\left(-E\left(\bm{x}, \bm{p}\right)\right)} = f\left(\bm{x}\right) \exp\left(-\frac{1}{2}\bm{p}^TA\bm{p}\right) \eqqcolon f\left(\bm{x}) \phi(\bm{p}\right).
\label{eq:exp_hamil_energy}
\end{equation}
$\exp\left(-\frac{1}{2}\bm{p}^TA\bm{p}\right)$, which we denote as $\phi\left(\bm{p}\right)$, is proportional to the PDF of a zero-mean Gaussian with covariance $A^{-1}$.  To generate a new proposal position, we pick a zero-mean Gaussian distributed momentum $\bm{p}\left(0\right)$ with covariance $A^{-1}$ and a fixed time $T$, and simulate the Hamiltonian dynamics to obtain the position at $\bm{x}\left(T\right)$. 

The proposal position is probabilistically accepted with the probability $a$, where
\begin{align}
\begin{split}
& a\left( 
	\left( \bm{x} \left( 0 \right), 
		   \bm{p} \left( 0 \right) 
		   \right) \rightarrow 
	\left( \bm{x} \left( T \right), 
	       \bm{p} \left( T \right) 
	       \right) \right) \\ 
& \quad = \min\left(
	\frac{\exp \left(-E \left(\bm{x} \left(T \right), 
	                          \bm{p} \left(T \right) \right) \right)}
	     {\exp \left(-E \left(\bm{x} \left(0 \right), 
	                          \bm{p} \left(0 \right) \right) \right)}
	, 1 \right) \\
& \quad = \min\left(
	\frac{   f \left(\bm{x} \left( T \right) \right)
		  \phi \left(\bm{p} \left(T \right) \right)}
	     {   f \left(\bm{x} \left( 0 \right) \right)
	      \phi \left(\bm{p} \left(0 \right) \right)}, 1 \right).
\end{split}
\label{eq:hmc_accept_prob}
\end{align}
Intuitively, the acceptance rule resembles the Metropolis-Hastings rule (Equations~\eqref{eq:metropolis_hastings} and~\eqref{eq:acceptance_prob}), where the transition probability $T$ is substituted with the (unnormalized) PDF of the momentum Gaussian $\phi$. Furthermore, if the Hamiltonian dynamics is simulated perfectly, $\exp\left(-E\left(\bm{x}, \bm{p}\right)\right)$ is a constant throughout the simulation because of energy conservation, and the acceptance probability is $1$.  

\paragraph{Properties of Hamiltonian dynamics} More formally, given a fixed time $T$, the Hamiltonian equation creates a mapping $M$ between $\left(\bm{x} \left(0 \right), \bm{p}\left(0 \right)\right)$ and $\left(\bm{x}\left(T\right), \bm{p}\left(T\right)\right)$. Neal~\cite{Neal:2010:MUH} showed that this mapping has several important properties:
\begin{enumerate}
\item The mapping is time-reversible: if we flip the momentum at time $T$ and use $\left(\bm{x}\left(T\right), -\bm{p}\left(T\right)\right)$ as the input to $M$, the output of the mapping would be $\left(\bm{x}\left(0\right), -\bm{p}\left(0\right)\right)$. That is, if we simulate the Hamiltonian dynamics in a backward manner from the end point, it will go back to the starting point.
\item The mapping preserves the volume: If we apply the mapping for a region $R_{0}$ of points $\left(\bm{x}\left(0 \right), \bm{p}\left(0 \right)\right)$, and map them to another region $R_{T}$, the volumes of the two regions in the position-momentum space remain the same (known as Liouville's theorem).
\item The mapping preserves energy: the Hamiltonian energy $E$ (Equation~\eqref{eq:Hamiltonian_energy}) remains the same after the mapping.
\end{enumerate}
The first property is crucial for the detailed balance condition (Equation~\eqref{eq:detailed_balance2}) to hold, since it ensures that the mapping is one-to-one. The second property ensures that we do not need to account for the Jacobian of the mapping in the Metropolis acceptance rule. The energy preservation property shows that the probability of acceptance is in fact $1$ since $E\left(\bm{x}\left(0\right), \bm{p}\left(0\right)\right) = E\left(\bm{x}\left(T\right), \bm{p}\left(T\right)\right)$. Recently, it has been shown~\cite{Sohl-Dickstein:2014:HMC} that it is also possible to design transition rules for Hamiltonian Monte Carlo to converge without satisfying the detailed balance condition.

Unfortunately, Equation~\eqref{eq:Hamiltonian_equation_2} does not have a known analytical solution for an arbitrary target function. It is usually required to integrate the differential equation using numerical integrators such as leapfrog integrators. These integrators maintain the time-reversibility and volume-preservation, but do not preserve energy. The Hamiltonian dynamics are approximated and the acceptance probability is no longer $1$. Furthermore, numerical integrators are expensive for light transport simulation because each step involves costly ray tracing operations and derivative computations of the shader.

\paragraph{Discussion.} As discussed in Chapter~\ref{sec:sampling}, Langevin Monte Carlo~\cite{Roberts:1996:ECL} is a one-step approximation to Hamiltonian Monte Carlo. Its proposal distribution is isotropic, except that the mean of the proposal distribution is shifted by the first derivatives (gradient) times a user-specified constant. Our method is also a one-step approximation, but the proposal distribution of our method adapts to the anisotropy of the signal, because we utilize the second derivatives. It is possible to precondition Langevin Monte Carlo using a positive-definite mass matrix, such as the Fisher information matrix~\cite{Girolami:2011:RML}. However, it remains unclear how to relate the Hessian matrix to the positive-definite mass matrix. Betancourt~\cite{Betancourt:2013:GMR} proposed the \textsc{SoftAbs} metric that removes the sign of the eigenvalues of the Hessian matrix using a smooth mapping. In contrast, we treat positive and negative eigenvalues differently by directly simulating Hamiltonian dynamics on the quadratic landscape. 

\section{Hessian-Hamiltonian Monte Carlo}

Figure~\ref{fig:qhmc} illustrates our sampling algorithm. We compute the second order Taylor expansion (local quadratic approximation) of the logarithm of the target function first, where the gradient and the Hessian are computed using automatic differentiation.  The quadratic function does not define a proper distribution, since it might grow to infinity, which prevents us from directly importance sampling it.  Hamiltonian dynamics enables us to sample from this quadratic function to obtain the proposal position, since it works on any continuous function.

The Hamiltonian dynamics have an analytical solution in the case of a quadratic function. However, we cannot use the acceptance rule in standard Hamiltonian Monte Carlo (Equation~\eqref{eq:hmc_accept_prob}) to compute the acceptance probability. It would break time-reversibility, since each light path would have a different associated quadratic function. Fortunately, we can derive from the analytical solution that the distribution of a proposal, given a Gaussian momentum, is a Gaussian distribution (Figure~\ref{fig:qhmc} (c)).  Therefore, we associate each light path with a Gaussian distribution derived from the quadratic function and Hamiltonian dynamics, and it is possible to compute the acceptance probability using the Metropolis-Hastings rule (Equation~\eqref{eq:acceptance_prob}). Finally, we multiply the analytical Gaussian with a prior Gaussian distribution to place a limit on its variance (Figure~\ref{fig:qhmc} (d)), so that the proposals do not go too far away where the second order approximation can be inaccurate.

\paragraph{Approximating Hamiltonian dynamics}

We first show how to derive the closed-form solution to the differential equations for Hamiltonian dynamics (Equation~\eqref{eq:Hamiltonian_equation_2}), given an initial momentum and position. Then, we will show how to infer the Gaussian distribution of proposals.  We start from a second-order approximation of $\log{f}$. For the sake of simplicity and without loss of generality, in the following we assume the current position $\bm{x}(0)$ is at the origin. Any small offset $\bm{x}$ from the origin can be approximated by:
\begin{equation}
\log{f\left(\bm{x}\right)} \approx \frac{1}{2}\bm{x}^T H\bm{x} + G^T \bm{x} + \log{f\left(0 \right)},
\label{eq:second_order_approx}
\end{equation}
where $H$ is the Hessian matrix and $G$ is the gradient vector at $\log f\left(\bm{0}\right)$. If we substitute this approximation into the Hamiltonian equation~(Equation \eqref{eq:Hamiltonian_equation_2}) using $\frac{\partial \log{f\left(\bm{x}\right)}}{\partial \bm{x}} \approx H\bm{x} + G$ and combine the two differential equations, we get:
\begin{equation}
\frac{\partial^2 \bm{x}\left(t\right)}{\partial t^2} = AH\bm{x}\left(t \right) + AG.
\label{eq:approx_Hamiltonian_equation}
\end{equation}
The above equation is a standard second-order differential equation system, and has an analytical solution. We start from the one-dimensional case, then generalize it to higher dimensions. Assuming $x$ is a one-dimensional variable, if we let ${\alpha=AH, \beta=AG}$, an analytical solution is:
\begin{equation}
x\left(t\right) = \begin{cases}
	c_{1} \exp\left(\sqrt{\alpha} t\right) + c_{2} \exp\left(-\sqrt{\alpha} t\right) - \frac{\beta}{\alpha} &\mbox{if } \alpha > 0 \\
	c_{1} \cos\left(\sqrt{-\alpha} t\right) + c_{2} \sin\left(\sqrt{-\alpha} t\right) - \frac{\beta}{\alpha} &\mbox{if } \alpha < 0 \\
	c_{1} t + c_{2} + \frac{\beta t^2}{2} &\mbox{if } \alpha = 0,
\end{cases}
\label{eq:analytical_solution}
\end{equation}
which can be verified by plugging the solution back into the equation.  The constant multipliers $c_{1}$, $c_{2}$ can be obtained by plugging in the initial condition $x(0)=0$, $x^{\prime}\left(0\right)=Ap\left(0 \right)$ (where $p\left(0 \right)$ is sampled from the Gaussian distribution $\phi$ defined in Equation~\eqref{eq:exp_hamil_energy}) into the original Hamiltonian equation (Equation~\eqref{eq:Hamiltonian_equation_2}). Specifically, the constants are:
\begin{align}
\begin{split}
c_{1} = \begin{cases}
	\frac{1}{2}\left(\frac{\beta}{\alpha} + \frac{\hat{p}\left(0 \right)}{\sqrt{\alpha}} \right) &\mbox{if } \alpha > 0 \\
	\frac{\beta}{\alpha} &\mbox{if } \alpha < 0 \\
	\hat{p}\left(0 \right) &\mbox{if } \alpha = 0
\end{cases} \text{, }
c_{2} = \begin{cases}
	\frac{1}{2}\left(\frac{\beta}{\alpha} - \frac{\hat{p}\left(0\right)}{\sqrt{\alpha}}\right) &\mbox{if } \alpha > 0 \\
	\frac{\hat{p}\left(0 \right)}{\sqrt{-\alpha}} &\mbox{if } \alpha < 0 \\
	0 &\mbox{if } \alpha = 0,
\end{cases}
\end{split}
\label{eq:analytical_solution_2}
\end{align}
where we denote $\hat{p}\left(0 \right) = Ap\left(0 \right)$ for clarity.

To illustrate, since the inverse mass $A$ is required to be positive, if the second derivative $H$ is strictly negative, we consider the $\alpha < 0$ case and the trajectory $x\left(t\right)$ becomes a linear combination of a cosine curve and a sine curve, which oscillates in the ridges of the flipped function.  On the other hand, if the second derivative is strictly positive, then the trajectory climbs straight up the hill and goes to infinity as $t$ increases.

If $x$ is an $N$-dimensional vector instead, the general solution of this differential equation system becomes a linear combination of the eigenvectors $\bm{e}_i$ of the matrix $AH$:
\begin{equation}
\bm{x}\left(t\right) = \sum_{i=1}^{N} x_{i}\left(t \right) \bm{e}_i,
\label{eq:linear_combination}
\end{equation}
where the coefficient $x_{i}\left(t \right)$ is similar to the one-dimensional case (Equation~\eqref{eq:analytical_solution}), but with $\alpha$ substituted with matrix $AH$'s $i$-th eigenvalue $\lambda_{i}$, and $\beta$ and $\hat{p}\left(0 \right)$ substituted with the projection of the vector $AG$ and $\hat{\bm{p}}\left(0 \right)$ on the $i$-th eigenvector $\bm{e}_i$, respectively. Again, we can obtain the constant multipliers as in the one-dimensional case by plugging in the initial conditions.

\paragraph{A Gaussian equivalent to the approximation}

We have derived an analytical trajectory for a fixed initial momentum. However, having the analytical trajectory is not enough. Recall that Hamiltonian Monte Carlo starts by generating a Gaussian distributed momentum $\bm{p}\left(0 \right) \sim \mathcal{N}\left(0, A^{-1}\right)$, and generates a new position proposal $x\left(T \right)$ at a fixed time $T$. Unfortunately, a direct application of the analytical solution to Hamiltonian Monte Carlo using the original acceptance rule (Equation~\eqref{eq:hmc_accept_prob}) is infeasible. The gradient and Hessian generally would be different at the proposal position, and the time-reversibility would be violated.

An observation from the analytical solution (Equations~\eqref{eq:analytical_solution} and~\eqref{eq:analytical_solution_2}) reveals that the Hamiltonian dynamics are actually linear mappings from the Gaussian distributed variable $\bm{p}\left(0 \right)$ to the new position $\bm{x}\left(T \right)$ if we have $t=T$ fixed. This means that $\bm{x}\left(T \right)$ is also Gaussian distributed since Gaussian variables are closed under linear transformations. Therefore, we can generate $\bm{x}\left(T \right)$ using \emph{a single Gaussian distribution}. Furthermore, the probability density of the Gaussian can be used as the transition probability $T$ to compute the Metropolis-Hastings acceptance probability (Equation~\eqref{eq:acceptance_prob}).

Now we will show why the mapping is linear and how to derive the covariance and the mean of $\bm{x}\left(T \right)$. Again we start from the one-dimensional case. If we plug the multipliers $c_1$ and $c_2$ (Equation~\eqref{eq:analytical_solution_2}) into the analytical solution (Equation~\eqref{eq:analytical_solution}) and rearrange the terms in one-dimensional $x\left(T \right)$, we have:
\begin{align}
\begin{split}
x\left(T\right) &= \begin{cases}
	\left(\frac{\exp\left(\sqrt{\alpha}T \right) - \exp\left(-\sqrt{\alpha}T\right)}{2\sqrt{\alpha}}\right) \hat{p}\left(0 \right) \\ 
	\quad + \frac{\beta}{2\alpha}\left(\exp\left(\sqrt{\alpha}T\right) + \exp\left(-\sqrt{\alpha}T\right) - 1\right) &\mbox{if } \alpha > 0 \\
	\frac{1}{\sqrt{-\alpha}} \sin\left(\sqrt{-\alpha}T\right) \hat{p}\left(0\right) \\ 
	\quad + \frac{\beta}{\alpha}\left(\cos\left(\sqrt{-\alpha}T\right) - 1\right) &\mbox{if } \alpha < 0 \\
	T \hat{p}\left(0\right) + \frac{\beta{T}^2}{2} &\mbox{if } \alpha = 0
\end{cases} \\
&= s \hat{p}\left(0\right) + o \\
&= s A p\left(0\right) + o,
\end{split} 
\label{eq:Hamiltonian_traj_rearrange}
\end{align}
which is a linear function of $p\left(0\right)$ and we denote the scaling coefficient as $s$ and the offset coefficient as $o$.

For the $N$-dimensional case, since $\bm{x}\left(T\right)$ is a linear combination of $x_{i}\left(T\right)$ (Equation~\eqref{eq:linear_combination}), it is still a linear transform.  Moreover, if we write $x_{i}\left(T\right) = s_{i} \cdot \hat{p}_{i}\left(0 \right) + o_{i}$, where $\hat{p}_{i}\left(0 \right)$ is the projection of $\hat{\bm{p}}\left(0 \right)$ on the $i$-th eigenvector $\bm{e}_{i}$ of the matrix $AH$, we can write out the linear transformation in matrix form:
\begin{equation}
\bm{x}\left(T \right) = S A \bm{p}\left(0 \right) + \bm{o},
\end{equation}
where the matrix $S$ and the vector $\bm{o}$ can be obtained from the eigenvectors $\bm{e}_{i}$, and the coefficients $s_{i}$ and $o_{i}$:
\begin{equation}
S = \sum_{i=1}^{N} s_{i} \bm{e}_{i}, \enskip \bm{o} = \sum_{i=1}^{N} o_{i} \bm{e}_{i}.
\label{eq:vector_scale_offsets}
\end{equation}
Recall that $\bm{p}\left(0 \right)$ is a zero-mean Gaussian variable with covariance $A^{-1}$. Therefore the covariance matrix $\Sigma$ and the mean $\bm{\mu}$ of the Gaussian random variable $\bm{x}\left(T \right)$ are 
\begin{equation}
\Sigma = \left(SA\right)A^{-1}\left(SA\right)^T = SAS^T, \enskip \bm{\mu} = \bm{o}.
\label{eq:before_prior}
\end{equation}

\paragraph{Multiplying with prior Gaussian}
\begin{figure}[t]
	\centering
	\captionsetup[subfigure]{justification=centering}	
	\begin{minipage}[t]{0.32\linewidth}
		\begin{subfigure}[t]{\textwidth}
			\includegraphics[width=1.0\linewidth]{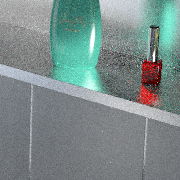} 
			\caption{\footnotesize $\sigma^2 = 0.028$ \\ accept rate $28.96\%$}
		\end{subfigure}
	\end{minipage}
	\begin{minipage}[t]{0.32\linewidth}
		\begin{subfigure}[t]{\textwidth}
			\includegraphics[width=1.0\linewidth]{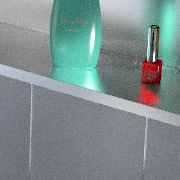} 
			\caption{\footnotesize $\sigma^2 = 0.007$ \\ accept rate $54.02\%$}
		\end{subfigure}
	\end{minipage}		
	\begin{minipage}[t]{0.32\linewidth}
		\begin{subfigure}[t]{\textwidth}
			\includegraphics[width=1.0\linewidth]{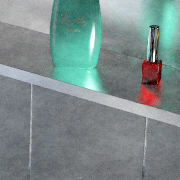} 
			\caption{\footnotesize $\sigma^2 = 0.001$ \\ accept rate $82.11\%$}
		\end{subfigure}
	\end{minipage}	
	\caption{We show the effect of the prior Gaussian parameter $\sigma^2$ using an inset from the \textsc{bathroom} scene (Figure~\ref{fig:comp_bathroom}). (a) High $\sigma^2$ results in low acceptance rate and a noisy image. (c) Low $\sigma^2$ results in high acceptance rate, but produces correlated noise. (b) We choose a $\sigma^2$ so that the acceptance rate falls in the ranges from $50\%-70\%$.}
	\label{fig:comp_prior}	
\vspace*{-.1in}
\end{figure}

In practice, our second order approximation (Equation~\eqref{eq:second_order_approx}) can be inaccurate when the proposal is far from the current state, or if there are discontinuities such as visibility changes. To compensate for this, we introduce a prior Gaussian distribution with zero mean and isotropic variance using a user specified constant $\sigma^2$, and multiply the PDF of it with the PDF of the Gaussian random variable $\bm{x}\left(T\right)$, to effectively place a limit on the maximum variance (which corresponds to the movement of the path in path space).  

Another way to think about the prior is that it acts as a regularization term that penalizes high variance. If $\sigma^2$ is high, then the change of the light path would be large, and the acceptance rate would be lower.  On the other hand, if $\sigma^2$ is low, then the change of the light path is small, and the acceptance rate would be higher.  We show the effects of different $\sigma^2$ in Figure~\ref{fig:comp_prior}.  In our current implementation we manually set $\sigma^2$ to achieve a certain acceptance rate ($50\%$ to $70\%$), but it may be possible to automatically adjust the parameter using adaptive MCMC~\cite{Andrieu:2008:TAM}. The final mean $\bm{\mu}^*$ and covariance $\Sigma^*$ are
\begin{equation}
\Sigma^* = \left(\Sigma^{-1} + \frac{1}{\sigma^2}\right)^{-1}, \enskip \bm{\mu}^* = \Sigma^*\Sigma^{-1}\bm{o}.
\label{eq:final_gaussian}
\end{equation}

\paragraph{Computing acceptance probability}

In order to apply the Metropolis-Hastings rule (Equation~\eqref{eq:metropolis_hastings}) given a current position $\bm{x}$, we generate a new proposal position $\bm{y}$ from a Gaussian variable with mean $\bm{\mu}_{x}^*$ and covariance matrix $\Sigma_{x}^*$ computed using Equation~\eqref{eq:final_gaussian}.  Then we compute the mean $\bm{\mu}_{y}^*$ and covariance matrix $\Sigma_{y}^*$ at the proposal position.  The acceptance probability (Equation~\eqref{eq:acceptance_prob}) is computed using the density of Gaussians:
\begin{align}
\begin{split}
a\left(\bm{x} \rightarrow \bm{y}\right) &= \min\left(1, \frac{f\left(\bm{y}\right) Q\left(\bm{y} \rightarrow \bm{x}\right)}{f\left(\bm{x}\right) Q\left(\bm{x} \rightarrow \bm{y}\right)}\right) \\
&= \min\left(1, \frac{f\left(\bm{y}\right) \Phi_{\bm{y}}\left(\bm{x} - \bm{y}\right)}{f\left(\bm{x}\right) \Phi_{\bm{x}}\left(\bm{y} - \bm{x}\right)}\right),
\end{split}
\end{align}
where $\Phi_{\bm{x}}\left(\bm{y} - \bm{x}\right)$ is the Gaussian PDF with covariance $\Sigma_{x}^*$ and mean $\bm{\mu}_{x}^*$ computed at $\bm{x}$ (Equation~\eqref{eq:final_gaussian}).  Specifically, if we define $\bm{z} = \bm{y} - \bm{x}$, it is:
\begin{equation}
\Phi_{\bm{x}}\left(\bm{z}\right) = \left(2\pi\right)^{-\frac{N}{2}} \left| \Sigma^*_{x} \right|^{\frac{1}{2}} \exp\left(-\frac{1}{2}\left(\bm{z} - \bm{\mu}_{x}^*\right)^T {\Sigma_{x}^{*}}^{-1} \left(\bm{z} - \bm{\mu}_{x}^* \right)\right).
\end{equation}
$\Phi_{\bm{y}}\left(-\bm{z}\right)$ is defined similarly with covariance and mean computed at $\bm{y}$.

\paragraph{Setting parameters $A$ and $T$}
\label{sec:setting_param}

A remaining question is how to choose the inverse mass matrix $A$ and simulation time $T$. Previous work in Hamiltonian Monte Carlo suggests setting $A$ to the covariance of the target function~\cite{Neal:2010:MUH,Girolami:2011:RML}. As an example, consider a target function $f\left(\mathbf{x}\right)$ that is a Gaussian distribution with covariance $\Sigma_f$. If we ignore multiplication with the prior, setting $A$ to the covariance of the target function, and setting $T=\pi/2$ will result in a Gaussian mutation (Equation~\eqref{eq:before_prior}) that precisely matches the target function. \footnote{In this case, the Hessian $H$ for $\log f$ will simply be $-\Sigma_f^{-1}$, and $\alpha = A H$ will be a negative identity matrix.  Therefore, we consider the $\alpha < 0$ case in Equation~\eqref{eq:Hamiltonian_traj_rearrange}, where $s = 1$ for $T = \pi/2$, and $S$ is the identity matrix.  Therefore, the covariance matrix $\Sigma$ from Equation~\eqref{eq:before_prior} is given simply by $A$, leading to a Gaussian distribution with covariance $\Sigma_f$, which is exactly the target function. This justifies setting $A$ to the covariance of the target function, and setting $T=\pi/2$.}

In general, the target function need not be a Gaussian and a global covariance $\Sigma_f$ may not be sufficient to describe the function.  We approximate the covariance locally using the fact that we have the Hessian $H$ of the log of the function.  If the target function is a Gaussian, the negative inverse of the Hessian $-H^{-1}$ would exactly be the covariance of the target function.  It would be tempting to directly set $A$ to $-H^{-1}$, but the covariance matrix of a Gaussian distribution is required to be positive semidefinite (all eigenvalues need to be positive), and $-H^{-1}$ is not necessarily positive definite in general.  We approximate the local covariance of the function by substituting the eigenvalues in $-H^{-1}$ by their absolute values, and set $A$ to the approximated local covariance:
\begin{equation}
A = \sum_{i=1}^{N} \begin{cases}
        \frac{1}{\left| \lambda^{H}_{i} \right|}\bm{e}^{H}_{i} & \mbox{if }
\lambda^{H}_{i} \neq 0 \\
        0 & \mbox{otherwise},
\end{cases}
\end{equation}
where $\bm{e}^{H}_{i}$ and $\lambda^{H}_{i}$ are the $i$-th eigenvector and eigenvalue of $H$. Finally, we set $T$ to $\frac{\pi}{2}$, as in the Gaussian example above.

The construction of $A$ and $T$ also simplifies the implementation, since $A$ and $H$ share the same set of eigenvectors and $A$'s eigenvalues are the inverse of the absolute value of $H$'s eigenvalues or zero.  The eigenvalues $\lambda_i$ of matrix $AH$ would then be either $-1$, $1$, or $0$, depending on the sign of the eigenvalue of $H$.  The magnitudes of the eigenvalues in the Hessian $H$ (and hence $A$) are still taken into consideration when sampling from the momentum using the inverse mass matrix $A$. We show the pseudo-code of our algorithm in Appendix~\ref{sec:h2mc_appendix}, which outputs the final mean $\bm{\mu}^*$ and covariance $\Sigma^*$ given the gradient $G$, Hessian $H$, and prior $\sigma^2$.

\begin{figure}[t]
	\centering
	\captionsetup[subfigure]{justification=centering}
	\framebox{
		\begin{minipage}[t]{0.18\linewidth}
			\begin{subfigure}[t]{\textwidth}
				\includegraphics[width=1.0\linewidth]{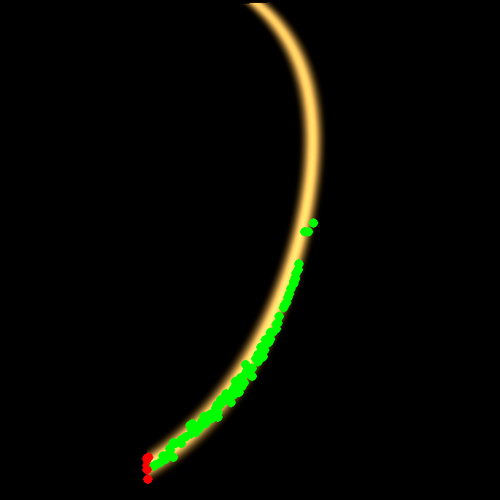}
				\caption{\footnotesize HMC \\ $128$ proposals \\ no Markov chain \\ accept rate $94.53\%$ \\ $100$ steps \\ $12929$ function evaluations}
			\end{subfigure}
		\end{minipage}
		\begin{minipage}[t]{0.18\linewidth}
			\begin{subfigure}[t]{\textwidth}
				\includegraphics[width=1.0\linewidth]{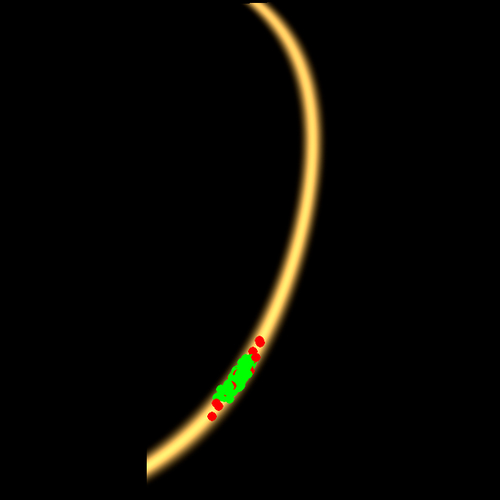}
				\caption{\footnotesize ours \\ $128$ proposals \\ no Markov chain \\ accept rate $75.78\%$ \\ $1$ step \\ $129$ function evaluations \\ ($\sim 100$ times fewer)}
			\end{subfigure}
		\end{minipage}
	}
	\framebox{
		\begin{minipage}[t]{0.18\linewidth}
			\begin{subfigure}[t]{\textwidth}
				\includegraphics[width=1.0\linewidth]{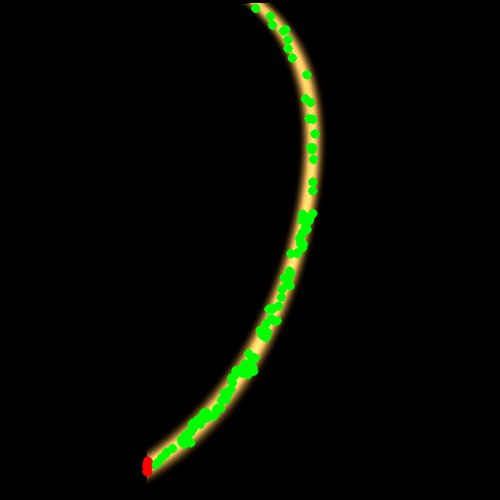}
				\caption{\footnotesize HMC \\ $128$ states \\ accept rate $92.97\%$ \\ $100$ steps \\ $12929$ function evaluations}
			\end{subfigure}
		\end{minipage}
		\begin{minipage}[t]{0.18\linewidth}
			\begin{subfigure}[t]{\textwidth}
				\includegraphics[width=1.0\linewidth]{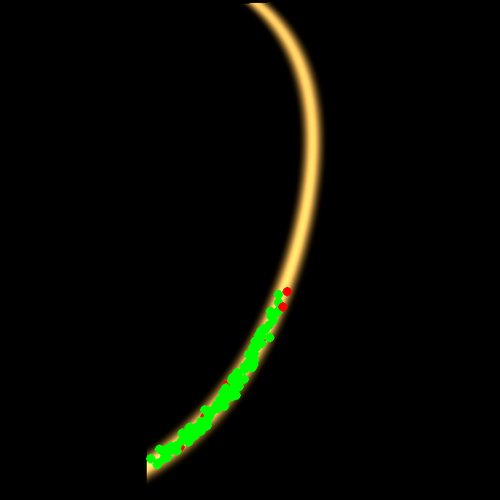}
				\caption{\footnotesize ours \\ $128$ states \\ accept rate $94.53\%$ \\ $1$ step \\ $129$ function evaluations \\ ($\sim 100$ times fewer)}
			\end{subfigure}
		\end{minipage}
		\begin{minipage}[t]{0.18\linewidth}
			\begin{subfigure}[t]{\textwidth}
				\includegraphics[width=1.0\linewidth]{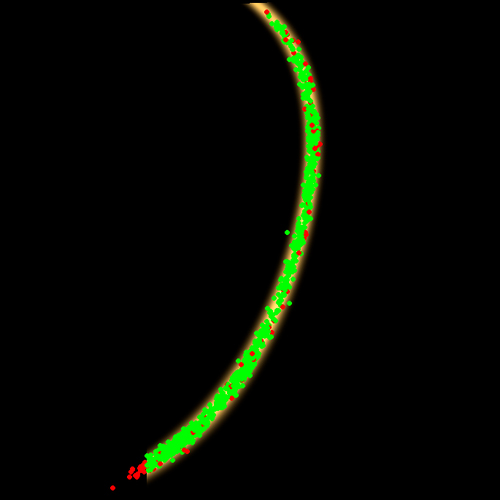}
				\caption{\footnotesize ours \\ $1024$ states \\ accept rate $92.29\%$ \\ $1$ step \\ $1025$ function evaluations \\ ($\sim 12.5$ times fewer)}
			\end{subfigure}
		\end{minipage}
	}
	\caption{We compare the sample distribution of the original Hamiltonian Monte Carlo (HMC) method and our method using the zoomed out slices from Figures~\ref{fig:ring_example} (g) and (h). The left box shows the ``proposals'' drawn from the current sample position, without running the Markov chain, and the right box shows the actual Markov chain states.  While the original HMC is able to generate proposals with high acceptance probability, and over longer trajectories (compare (a) to (b) and (c) to (d)), each proposal in the original HMC requires many steps to compute ($100$ steps in this case), and each step involves costly ray tracing, shading, and derivative computation. Our method achieves much better space coverage using a single step (as opposed to the $100$ steps in the original HMC), and requires an order of magnitude fewer function evaluations (e).}
	\label{fig:comp_hmc_h2mc}
\end{figure}

Finally, we compare the proposals and the Markov chain of original Hamiltonian Monte Carlo with a leapfrog integrator, and our Hessian-HMC method in Figure~\ref{fig:comp_hmc_h2mc}, using a 2D slice in the \textsc{ring} scene (Figure~\ref{fig:ring_example}). We use a step size of $0.0005$ with $100$ steps for the leapfrog numerical integrator in Hamiltonian Monte Carlo, and we set the prior Gaussian $\sigma^2=0.01$ for our method. The target acceptance rate is set higher because the dimensionality of the function is low~\cite{Neal:2010:MUH}. Although original Hamiltonian Monte Carlo is able to use longer trajectories to explore the space more thoroughly with the same number of samples, a single sample in Hamiltonian Monte Carlo requires $100$ steps of ray tracing, shading and derivatives computation. Choosing a bigger step size or smaller step number for HMC may result in energy loss or inferior space exploration efficiency, and this parameter of original Hamiltonian Monte Carlo is notoriously hard to tune.  Our H\textsuperscript{2}MC method can explore the space better using an order of magnitude fewer function evaluations (Figure~\ref{fig:comp_hmc_h2mc}(e)).

\section{Implementation}
\label{sec:h2mc_implementation}
\begin{figure}[t]
	\centering
	\captionsetup[subfigure]{justification=centering}
	\begin{minipage}[t]{0.24\linewidth}
		\begin{subfigure}[t]{\textwidth}
			\includegraphics[width=1.0\linewidth]{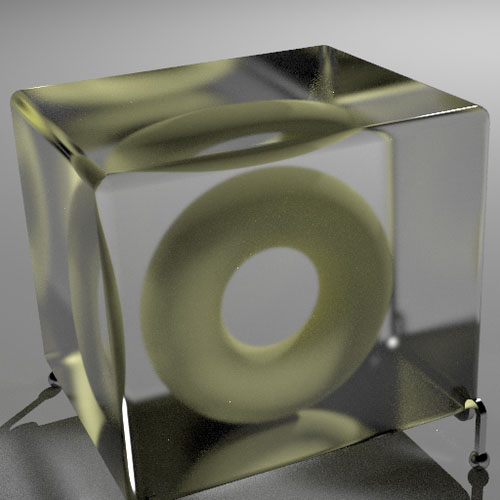} 
			\caption{\textsc{torus}}
		\end{subfigure}
	\end{minipage}		
	\begin{minipage}[t]{0.24\linewidth}		
		\begin{subfigure}[t]{\textwidth}
			\includegraphics[width=1.0\linewidth]{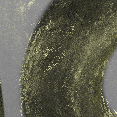} 
			\caption{Kelemen et al.~\cite{Kelemen:2002:SRM}, \\ isotropic proposal}
		\end{subfigure}
	\end{minipage}		
	\begin{minipage}[t]{0.24\linewidth}
		\begin{subfigure}[t]{\textwidth}
			\includegraphics[width=1.0\linewidth]{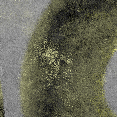} 
			\caption{Ours, \\ isotropic proposal}
		\end{subfigure}
	\end{minipage}		
	\begin{minipage}[t]{0.24\linewidth}
		\begin{subfigure}[t]{\textwidth}
			\includegraphics[width=1.0\linewidth]{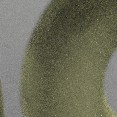} 
			\caption{Ours, \\ H\textsuperscript{2}MC proposal}
		\end{subfigure}
	\end{minipage}	
	\caption{We compare our new parameterization to Kelemen et al.'s parameterization on the \textsc{torus} scene, with a diffuse torus inside a glossy glass cube lit by a point light. The left image is computed using 5000 samples per pixel using our method, and the three insets are computed with 256 samples per pixel. The original parameterization incurs correlation between screen space and the outgoing sample directions on the glass and the torus, creating streaks on the torus.  Our new parameterization greatly reduces this correlation.  Our Hessian-Hamiltonian proposal further improves the sampling efficiency dramatically.}
	\label{fig:comp_param}
\end{figure}

We implement our method in a stand-alone renderer with the Embree ray tracing engine~\cite{Wald:2014:EKF}. We implement an embedded automatic differentiation compiler in C++ that overloads all the common functions and operators, and compiles the gradient and Hessian of the path throughput function into an ispc~\cite{Pharr:2012:ISC} kernel. The renderer supports the Phong BRDF, the microfacet refraction model~\cite{Walter:2007:MMR}, point and area light sources, environment maps, and linear object motion. Each sample in the Markov chain represents a single light path that connects the light to the camera. As in most previous Markov chain Monte Carlo rendering methods, we employ multiple mutation strategies to better cover different types of light paths. Specifically, we adopt three different types of mutation strategies: a multiplexed \emph{large step} mutation~\cite{Kelemen:2002:SRM,Hachisuka:2014:MML}, a novel modified \emph{small step} perturbation, and a lens perturbation. The large step mutation is responsible for making large jumps between different disconnected components of light paths, the small step perturbation is responsible for making a small change to all dimensions of the function, and finally the lens perturbation changes only part of the light path to alleviate difficult visibility issues.  We apply the H\textsuperscript{2}MC sampling on the small step and the lens perturbation to explore the local structure of the path throughput function.  In the rest of this section, we address some technical details of the implementation.

\paragraph{Multiplexed large step mutation} To ensure the ergodicity of the Markov chain, that is, to ensure we have a strictly positive probability to sample all light paths with non-zero contribution, we include a large step mutation to generate a proposal light path that is completely independent of the current sample. Our large step mutation is a hybrid between Multiplexed Metropolis Light Transport (MMLT)~\cite{Hachisuka:2014:MML} and Kelemen et al.'s mutation. Specifically, like MMLT, the state of our Markov chain only represents one of the $N^2$ pairs connections (where in Kelemen style the state would be the sum of all connections). However, instead of choosing path length and subpath length a priori as in MMLT, we sample all pairs of connections of a bidirectional path tracer, and probabilistically pick one based on their contributions weighted by multiple importance sampling (similar to Multiple-try Metropolis~\cite{Liu:2000:MML} and importance resampling~\cite{Rubin:1987:CPD}). Comparing to MMLT, this has the benefit of stratification, since we always sample all path and subpath lengths instead of randomly choosing one. Comparing to Kelemen's mutation, during the other local perturbations, we only keep a single subpath in a bidirectional path tracer, therefore we benefit from multiple importance sampling just like MMLT. We notice that the acceptance rate of large steps significantly increases in difficult scenes compared to MMLT when using our approach.

\paragraph{H\textsuperscript{2} small step perturbation}  We adopt a modified version of the \emph{small step} perturbation~\cite{Kelemen:2002:SRM} as the main component to explore the path throughput function locally. In Kelemen et al.'s work, the light paths are represented as the random numbers that are used to generate them. The perturbation is done by making small changes to the random numbers, and results in a new light path.  We make two modifications to the parameterization.

First, we classify the surfaces into specular and non-specular by applying a user-defined threshold on the roughness. If the surface is near-specular, the outgoing directions are parameterized using the random numbers. On the other hand, if the surface is non-specular, the outgoing directions are parameterized using the global directions expressed in absolute spherical coordinates. We found that this change improves sampling efficiency because the correlation between the dimensions is reduced. Kelemen's parameterization handles specular surfaces well, because importance sampling captures the peak of the target function well. On the other hand, the local parameterization introduces extra correlation between dimensions, because the outgoing direction depends on the normal of the surface, and the normal depends on the previous outgoing direction. The parameterization change is beneficial because H\textsuperscript{2}MC is invariant to linear parametrization changes, while the parameterization change is non-linear. We show a comparison of the original parameterization with the new one in Figure~\ref{fig:comp_param}.

Second, if the light path hits a light source without next event estimation (that is, no explicit connection is made), we substitute the parameterization of the last outgoing direction, to the position on the light source, so that the perturbation is more likely to hit the light source (a limited form of Reversible Jump Markov Chain Monte Carlo~\cite{Pantaleoni:2017:CML, Otsu:2017:FSS, Bitterli:2017:RJM}). We assume a pinhole camera in our implementation, but the second change can also apply in the case when the light path starts from the light source and hits the camera lens without explicit connection.  The new parameterization represents the sample position $\bm{x}$ in the H\textsuperscript{2}MC sampling. 

The time dimension is treated the same as other dimensions.  The generality of H\textsuperscript{2}MC sampling makes it agnostic to the underlying representation. This enables us to detect the correlation between time and other dimensions, which was not considered in previous Markov chain Monte Carlo rendering methods.  

\paragraph{H\textsuperscript{2} lens perturbation} Consider light paths involving small and flat surfaces. If we mutate the whole path, chances are high that we will miss the surfaces and result in zero contribution. A better strategy for these light paths is to perturb only a subset of the full path, and keep the rest of the vertices fixed.  We implement the lens perturbation in the original Metropolis Light Transport algorithm~\cite{Veach:1997:MLT}, which mutates only the lens subpath. For lens perturbation, the sample position $\bm{x}$ in the H\textsuperscript{2}MC sampling is the two dimensional image coordinate.

\begin{figure}[t]
	\centering
	\setlength{\tabcolsep}{1pt}
	\begin{tabular}{ccccc}
	\multicolumn{5}{c}{	
		\includegraphics[width=0.97\linewidth]{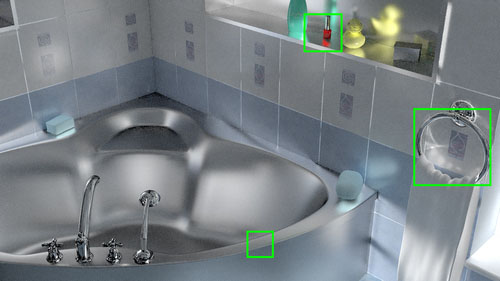}	
	} \\
	\includegraphics[width=0.19\linewidth]{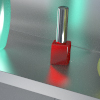} &
	\includegraphics[width=0.19\linewidth]{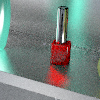} &
	\includegraphics[width=0.19\linewidth]{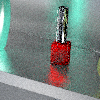} &
	\includegraphics[width=0.19\linewidth]{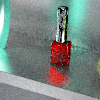} &
	\includegraphics[width=0.19\linewidth]{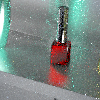} \\
	\includegraphics[width=0.19\linewidth]{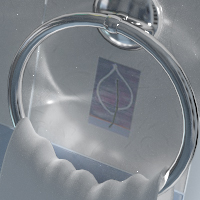} &
	\includegraphics[width=0.19\linewidth]{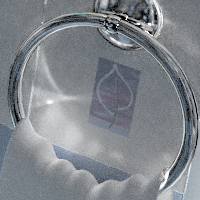} &
	\includegraphics[width=0.19\linewidth]{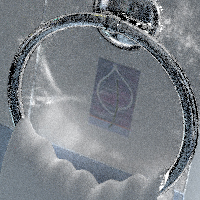} &
	\includegraphics[width=0.19\linewidth]{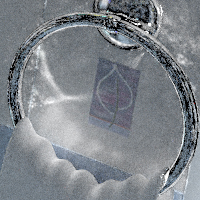} &
	\includegraphics[width=0.19\linewidth]{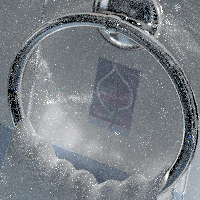} \\
	\includegraphics[width=0.19\linewidth]{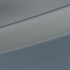} &
	\includegraphics[width=0.19\linewidth]{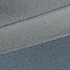} &
	\includegraphics[width=0.19\linewidth]{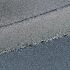}	&
	\includegraphics[width=0.19\linewidth]{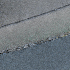} &
	\includegraphics[width=0.19\linewidth]{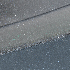} \\
	REF & H\textsuperscript{2}MC & MEMLT & HSLT & MMLT
	\end{tabular}
	\caption{\textsc{Bathroom}: An equal-time (10 minutes) comparison on the \emph{bathroom} scene with multiple glossy reflections lit by a distant area light. The top image is generated by our method in 10 minutes.  Our method achieves less noisy results on highly curved glossy surfaces and the caustics because we can adapt to the curvatures of the surfaces using second-order derivatives.}
	\label{fig:comp_bathroom}
\end{figure}
\begin{figure}[t]
	\begin{center}
		\includegraphics[width=0.92\linewidth]{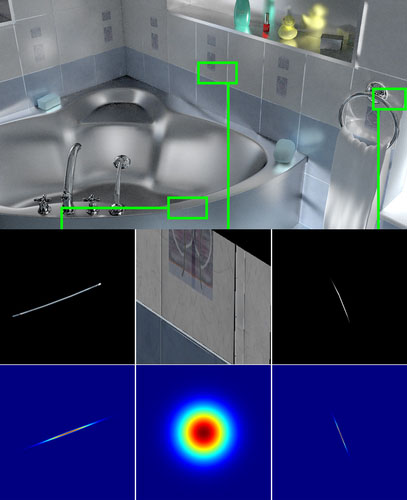}
	\end{center}
	\caption{We visualize the screen space slice of the contribution of three different light paths and our Gaussian approximation $\mathcal{N}\left(\bm{\mu}^*, \Sigma^*\right)$ in the \textsc{bathroom} scene.  The center row of the insets shows the contribution of perturbing the light path in the screen space.  The left column shows a 4 bounce glossy reflection light path, the center column shows a 3 bounce diffuse reflection light path, and the right column shows a 3 bounce caustic light path caused by the metal towel ring.  Glossy/specular transport results in sparse and anisotropic contributions, which are hard to sample using isotropic mutations. The bottom row shows our Gaussian approximation projected onto the screen space. The approximation matches the sharp contribution function and falls back to isotropic sampling when the contribution is smooth.  Note that our method is anisotropic in all sampling dimensions, and we only show the screen space slices for visualization purposes. }
	\label{fig:screen_space_slice}
\end{figure}

\begin{figure}[t]
	\centering
	\setlength{\tabcolsep}{1pt}
	\begin{tabular}{ccccc}
	\multicolumn{5}{c}{	
		\includegraphics[width=0.97\linewidth]{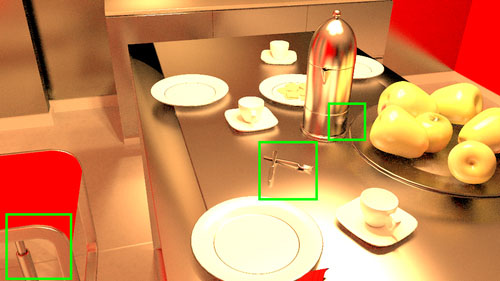}	
	} \\
	\includegraphics[width=0.19\linewidth]{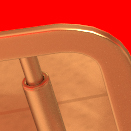} &
	\includegraphics[width=0.19\linewidth]{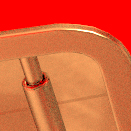} &
	\includegraphics[width=0.19\linewidth]{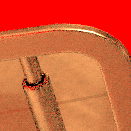} &
	\includegraphics[width=0.19\linewidth]{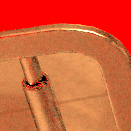} &
	\includegraphics[width=0.19\linewidth]{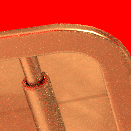} \\
	\includegraphics[width=0.19\linewidth]{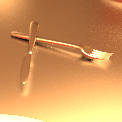} &
	\includegraphics[width=0.19\linewidth]{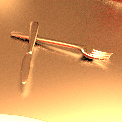} &
	\includegraphics[width=0.19\linewidth]{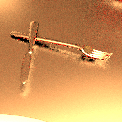} &
	\includegraphics[width=0.19\linewidth]{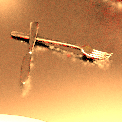} &
	\includegraphics[width=0.19\linewidth]{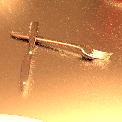} \\
	\includegraphics[width=0.19\linewidth]{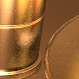} &
	\includegraphics[width=0.19\linewidth]{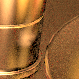} &
	\includegraphics[width=0.19\linewidth]{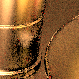} &
	\includegraphics[width=0.19\linewidth]{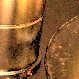} &
	\includegraphics[width=0.19\linewidth]{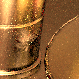} \\
	REF & H\textsuperscript{2}MC & MEMLT & HSLT & MMLT
	\end{tabular}
	\caption{\textsc{Kitchen}: An equal-time (1 hour) comparison on the \textit{kitchen} scene with complex material and geometry configuration lit by four area lights right above the table.  The top image is generated by our method in an hour.  This is a challenging scene and the reference rendered by PSSMLT is still slightly noisy after 2 days of computation on a 64 core machine.  Our method excels at following the small features of the image such as the fork and the knife on the table, or the edges on the chair.  It is also good at following the multiple glossy reflections on the highly curved surfaces such as the reflection on the flask.}
	\label{fig:comp_kitchen}
\end{figure}

\begin{figure}[t]
\centering
\setlength{\tabcolsep}{1pt}
\begin{tabular}{cccc}
\raisebox{20pt}{
\multirow{12}{*}{
	\includegraphics[width=0.47\linewidth]{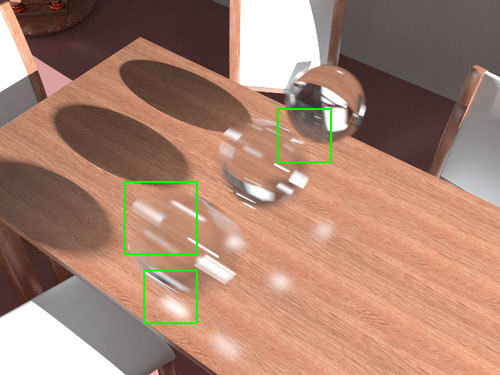}	
}
} &
\includegraphics[width=0.16\linewidth]{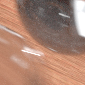} &
\includegraphics[width=0.16\linewidth]{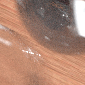} &
\includegraphics[width=0.16\linewidth]{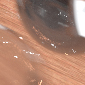} \\
& 
\includegraphics[width=0.16\linewidth]{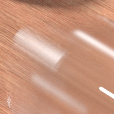} &
\includegraphics[width=0.16\linewidth]{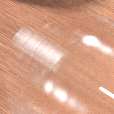} &
\includegraphics[width=0.16\linewidth]{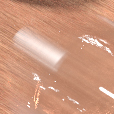} \\
&
\includegraphics[width=0.16\linewidth]{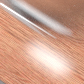} &
\includegraphics[width=0.16\linewidth]{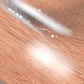} &
\includegraphics[width=0.16\linewidth]{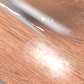} \\
& H\textsuperscript{2}MC & MEMLT & MMLT
\end{tabular}
\caption{\textsc{Balls}: 30 minute rendering of the \textit{balls} scene, which consists of three moving near-specular glass balls lit by a point light. The left image is generated by our method in 30 minutes.  The moving balls show complex patterns with a combination of reflection from the room and the resulting caustics on the table.  Neither MMLT nor MEMLT are able to efficiently resolve the moving features within the given time budget.  Our method is able to closely follow the specular highlights and caustics in the glass because it detects the correlation between the time domain and path-space.}
\label{fig:comp_balls}
\end{figure}

\section{Results and Discussion}

\begin{table}[t]
    \centering
    \begin{tabular}[t]{|c|r|r|r|r|}
    \hline
                      & $H^2MC$ & MMLT & MEMLT & HSLT \\
    \hline
    \textsc{Bathroom} & 610 & 1288 & 600 & 331 \\
    \hline
    \textsc{Kitchen} & 5169 & 12453 & 4749 & 3319 \\
    \hline
    \textsc{Balls} & 2943 & 8554 & 2961 & N/A \\
    \hline
    \textsc{Cars} & 1576 & 5361 & 1422 & N/A \\
    \hline
    \end{tabular}
    \captionsetup{justification=centering}
    \caption{Sample count per pixel of each method for the equal-time comparisons.}
    \label{table:sample_count}
\end{table}

We compare against three other MCMC rendering methods: Multiplexed MLT (MMLT)~\cite{Hachisuka:2014:MML}, Manifold Exploration MLT (MEMLT)~\cite{Jakob:2012:MEM}, and the improved Half-vector Space Light Transport (HSLT)~\cite{Hanika:2015:IHV}. MMLT is a general rendering algorithm that does not assume any particular lighting effect, but its isotropic mutation makes it inefficient on difficult light paths such as highly-glossy transports.  We compare to MMLT to show the efficiency of the anisotropic proposal sampling. MEMLT and HSLT are two rendering algorithms dedicated to specular and glossy transport by using first-order derivatives of the half-vectors. They can efficiently resolve difficult specular light paths, but often produce noisy results on highly-curved surfaces. Furthermore, since they assume a specific lighting scenario, they cannot resolve difficult moving caustics, and usually result in ghosting artifacts (Figures~\ref{fig:comp_cars} and~\ref{fig:comp_balls}). We did not compare to HSLT on the scenes with motion blur because their implementation does not allow it. We render four scenes -- \textsc{Bathroom} ($1280 \times 720$), \textsc{Kitchen} ($1024 \times 576$), \textsc{Balls} ($768 \times 576$), \textsc{Cars} ($768 \times 576$) -- with different lighting, material, and geometry configurations (Figure~\ref{fig:comp_cars} and Figures~\ref{fig:comp_bathroom} to~\ref{fig:comp_balls}).

For MMLT we use our own implementation, for MEMLT and HSLT we use the implementation in the Mitsuba~\cite{Jakob:2010:Mitsuba} renderer. HSLT is used with the lens perturbation because in our experiments it results in better images. The comparisons are equal-time using an Intel Core i7-4770 at 3.40GHz using 4 cores. The maximum path length is set to $7$. References are rendered using the PSSMLT~\cite{Kelemen:2002:SRM} implementation in Mitsuba and rendered for 2-3 days on a 64 core machine, except that the reference for the \textsc{Cars} is rendered using our method for roughly 15 hours on the 4 core machine (PSSMLT did not converge in 2-3 days computation). We show the sample count per pixel of each method in each scene in Table~\ref{table:sample_count}.  In general our method is 2-3.5 times slower per sample than MMLT because of the derivatives and Gaussian computation, and is about the same speed as MEMLT. HSLT is slower than MEMLT because it works on a higher-dimensional manifold.

\paragraph{Bathroom} Figure~\ref{fig:comp_bathroom} shows an equal-time (10 minutes) comparison on the \textit{bathroom} scene with multiple glossy-to-glossy transports lit by a distant area light.  For this particular scene, only indirect illumination is shown to highlight the differences between the algorithms.  MMLT generates noisy results because of their isotropic mutation distribution.  MEMLT and HSLT do generally well, but produce noisy results on high curvature surfaces because they use a first-order approximation on the surface. Our method is able to capture the local structure of the function and generates accurate results. 

To demonstrate the anisotropic proposal distribution of our method, we visualize the screen space slice of the contribution of some light paths and the slice of our Gaussian approximation in Figure~\ref{fig:screen_space_slice}. Our method is able to adapt to the sparse and sharp path contribution function, and fall back to isotropic sampling when the contribution function is smooth. MEMLT and HSLT often fail to capture small screen space features, because they isotropically sample some dimensions first, and such sampling often misses the feature.  Note that our method adapts to all dimensions, and we only show the screen space slice for the sake of visualization.

\paragraph{Kitchen} Figure~\ref{fig:comp_kitchen} shows an equal-time (1 hour) comparison on the \textit{kitchen} scene with complex materials and a difficult geometry configuration lit by four area lights close to the table. This is a challenging scene and the reference rendered by PSSMLT is still slightly noisy after 2 days of computation on a 64 core machine. MMLT produces spiky noise because some glossy-to-glossy light paths have small and high-contribution regions. MEMLT and HSLT generate noisy results on small and highly curved surfaces.  Our method is able to follow the small image features closely, producing smoother results.

In general, light paths involving highly curved surfaces can be troublesome for MEMLT and HSLT, which only use first derivatives. Both of them need to start from an initial subpath, then iteratively converge to the new light path on the manifold. The light paths involving curved surfaces often have narrow contribution areas, and are highly non-linear. It is likely that the initial subpath will miss the highlight entirely, making it impossible to converge to a new light path.  Even if the initial subpath hits the highlight, it could take many iterations to converge due to the non-linearity. In contrast, the second derivatives along with the anisotropic Gaussian mutation enable us to generate the proposal path directly with respect to the local shape of the function, avoiding the convergence issue.

\paragraph{Balls} Figure~\ref{fig:comp_balls} shows a 30 minute rendering of the \textit{balls} scene, which consists of three moving near-specular glass balls lit by a point light.  MMLT is unable to resolve the difficult specular-diffuse-specular paths inside the moving balls and the caustics on the table.  While MEMLT excels at resolving the specular light paths given the time fixed, it relies on seeding to sample the time dimension, which causes the ghosting artifacts on the balls. Our method is able to capture correlation between the time and the path-space, so that it can efficiently sample the difficult moving caustics and specular highlights.

\paragraph{Cars.} Figure~\ref{fig:comp_cars} shows a 20 minute rendering of the \textit{cars} scene, with a static car and a moving car lit by an area light. This is a challenging scene because of the hard-to-find specular-diffuse-specular (SDS) light paths between the car interior and the near-specular window.  MMLT has a hard time finding the specular light paths, and is often trapped in local modes, producing streaks on the image.  MEMLT is able to resolve the static SDS paths more efficiently, but produces ghosting artifacts since it does not move in the time dimension.  Our method moves in all dimensions and generates smooth results.

\subsection{Limitations and Future Work}
\label{sec:h2mc_limitation}

Integrating our method into an existing renderer requires some work, because we need to automatically differentiate the shaders. However, once automatic differentiation has been set up, it is easier to integrate other distributed effects such as motion blur. Automatic differentiation could also be helpful for the shaders/integrators that require the derivatives of the light path (e.g. ray differentials). The production renderer Arnold~\cite{Kulla:2018:SPI} uses forward-mode automatic differentiation to compute ray differentials.

As with most Markov-chain Monte Carlo rendering algorithms, high frequency visibility changes can significantly lower the efficiency. Our Gaussian prior reduces this effect but tiny geometry can still cause problems. In addition to visibility changes, there can also be some pathological cases where the path contribution function is extremely noisy. For example, multiple-bounce reflections involving glossy surfaces with high frequency displacement maps. In these cases the derivatives become unreliable, and our method might start to produce correlated noise or have low acceptance rate. Combining our method with the visibility gradient introduced in Chapter~\ref{chap:redner}, or the recent cone fitting approach~\cite{Otsu:2018:GML} could be interesting future work. Proper prefiltering of geometry and texture is also important for the derivatives to be well-behaved (e.g.~\cite{Loubet:2017:HML}).

We also observe that light transport integration involves both global and local exploration challenges. We need to globally find high-contribution regions, and then locally sample them despite their narrowness. Our method improves local sampling, but it still needs seed paths that are globally reasonably well distributed. Combining recent data-driven methods, e.g., ~\cite{Vorba:2014:OLP, Reibold:2018:SGS}, with local perturbation is one possible research direction.

Finally, since the derivatives and covariance computation incurs extra overhead, for relatively simple scenes and BSDFs where ray casting is cheap and isotropic mutation is sufficient, the adaptiveness of our method may not be worth the cost.

\section{Conclusion}

We presented a novel Hessian-based Hamiltonian Monte Carlo method and applied it to light transport simulation. By introducing Hamiltonian dynamics, we are able to sample from the local quadratic representation that does not define a distribution. Our method can capture the local correlation of the path throughput function, making it suitable for rendering difficult lighting scenarios such as the combination of glossy-to-glossy transport and motion blur. We anticipate that the method's generality will make it possible to render a wider variety of effects such as retroreflective materials, spectral effects, and participating media.
\begin{subappendices}

\section{Pseudo-code for \texorpdfstring{H\textsuperscript{2}MC}{H2MC}}
\label{sec:h2mc_appendix}
Given the gradient $G$ and the Hessian $H$ of the log target function $\log f(x)$, and a user parameter $\sigma^{2}$, our method outputs an anisotropic Gaussian distribution $\Sigma^*, \bm{\mu}^*$. The following page shows the pseudo code of this procedure. Note that we simplify the algorithm using the fact that the inverse mass matrix $A$ and $H$ have the same set of eigenvectors.

\newpage

\begin{spacing}{0.75}
\begin{algorithmic}[1]
	\Procedure{H2MC}{$G,H,\sigma^2$}\Comment{gradient, Hessian, and prior}
	\State $N \gets $ dimension of the target function
	\State $T = \frac{\pi}{2}$ \Comment{Simulation time}    	
	\For {$i \leftarrow 1, N$} \Comment{Eigendecomposition of $H$}
    	\State $\bm{e}^{H}_{i} \gets i$-th eigenvector of $H$
    	\State ${\lambda}^{H}_{i} \gets i$-th eigenvalue of $H$
    \EndFor	           
    \State $A = 0_{N \times N}$ \Comment{Initialize with zero matrix}
    \For {$i \leftarrow 1, N$} \Comment{Construction of $A$}
    	\State $\bm{e}^{A}_{i} \gets \bm{e}^{H}_{i}$
    	\If{$\left| {\lambda}^{H}_{i} \right| > \epsilon$} \Comment{$\epsilon$ is set to a small number.}
   	 		\State ${\lambda}^{A}_{i} \gets \frac{1}{\left| {\lambda}^{H}_{i} \right|}$
   	 	\Else
   	 		\State ${\lambda}^{A}_{i} \gets 0$
    	\EndIf
    	\State $A = A + {\lambda}^{A}_{i} \bm{e}^{A}_{i}$
    \EndFor	           
    \For {$i \leftarrow 1, N$} \Comment{Eigendecomposition of the matrix $AH$}	  	    	
    	\State $\bm{e}_{i} \gets \bm{e}^{H}_{i}$
    	\If{$\left| {\lambda}^{H}_{i} \right| > \epsilon$} 
    		\State ${\lambda}_{i} \gets \frac{{\lambda}^{H}_{i}}{\left| {\lambda}^{H}_{i} \right|}$ 
    		\Comment{${\lambda}^{A}_{i}=\frac{1}{\left| {\lambda}^{H}_{i} \right|}$}
    	\Else
    		\State ${\lambda}_{i} \gets 0$
    	\EndIf
    \EndFor	    
    \State $S = 0_{N \times N}$
    \State $\bm{o} = 0_{N \times 1}$
    \For {$i \leftarrow 1, N$} \Comment{Scales and offsets (Equation~\eqref{eq:Hamiltonian_traj_rearrange})}
    	\State $\alpha \gets \lambda_{i}$ \Comment{$AH$'s $i$-th eigenvalue}
    	\State $\beta \gets {\lambda}^{A}_{i} G^T \bm{e}_i$ \Comment{Projection of $AG$ on $\bm{e}_i$}
    	\If{$\lambda_{i} > 0$} 
    		\State $s_i \gets \frac{\exp\left(\sqrt{\alpha} T\right) - \exp\left(-\sqrt{\alpha} T\right)}{2 \sqrt{\alpha}}$
    		\State $o_i \gets \frac{\beta}{2 \alpha} \left(\exp\left(\sqrt{\alpha} T\right) + \exp\left(-\sqrt{\alpha} T \right) - 1 \right)$
    	\ElsIf{$\lambda_{i} < 0$}
    		\State $s_i \gets \frac{1}{\sqrt{-\alpha}} \sin\left(\sqrt{-\alpha}T\right)$
    		\State $o_i \gets \frac{\beta}{\alpha}\left(\cos\left(\sqrt{-\alpha}T\right) - 1 \right)$
    	\Else
    		\State $s_i \gets T$
    		\State $o_i \gets -\frac{\beta T^2}{2}$	
    	\EndIf
    	\State $S = S + s_i \bm{e}_i$ \Comment{Equation~\eqref{eq:vector_scale_offsets}}
    	\State $\bm{o} = \bm{o} + o_i \bm{e}_i$
    \EndFor
    \State $\Sigma = SAS^T$ \Comment{Equation~\eqref{eq:before_prior}}
    \State $\bm{\mu} = \bm{o}$ 
    \State $\Sigma^* = \left(\Sigma^{-1} + \frac{1}{\sigma^2}\right)^{-1}$ \Comment{Prior multiplication (Equation~\eqref{eq:final_gaussian})}
    \State $\bm{\mu}^* = \Sigma'\Sigma \bm{\mu}$
    \State \textbf{return} $\Sigma^*, \bm{\mu}^*$
    \EndProcedure
\end{algorithmic}
\end{spacing}

\end{subappendices}
\sethead[\textit{\chaptertitle}]% even left
  []% even centre
  [\thepage]% even right
  {\thepage}% odd left
  {}% odd centre
  {\textit{\chaptertitle}}% odd right

\chapter{Conclusion and Future Vision}

This dissertation introduced three very different tools related to computing and applying derivatives for computer graphics, image processing, and deep learning applications. Applying derivatives is desirable, but also challenging. At the system level, accounting for parallelism and locality, while preserving expressiveness of a programming language, is not trivial. At the algorithm level, resolving discontinuities, making use of first- or higher-order derivatives often require domain-specific knowledge. The three tools we introduced tackled these challenges in different ways, and are important first steps toward making all programs differentiable.

In the future, my vision is that the distinction between deep learning and traditional algorithms will become even blurrier than they are at the moment. Neural network architectures will become more and more sophisticated, and traditional methods will become more data-driven. For problems where data is limited, it is useful to apply our prior knowledge by formulating a forward model. A key to bridge the gap between deep learning and traditional methods is the generality of automatic differentiation, and the challenges that arose when developing the methods in this dissertation will repeatedly appear when trying to differentiate other programs.

I envision the following future research directions to be important:

\paragraph{Fully differentiable computer graphics}
While we have been successful at differentiating physically-based rendering (Chapter~\ref{chap:redner} and ~\ref{chap:h2mc}) and a small fluid simulation example (Chapter~\ref{chap:gradient_halide}), these are only subsets of the whole field of computer graphics. It would be desirable to make the whole 3D modeling, simulation, and rendering pipeline differentiable. As shown throughout this dissertation, having fully differentiable pipelines enables data-driven training and inverse inference. As computer graphics models the world with tremendous detail and principled simulation, it would have enormous use in real-world applications.

\paragraph{Automatic differentiation compilers for other computation}
While we can automatically generate efficient gradient code for image processing and deep learning operators (Chapter~\ref{chap:gradient_halide}), the programming model is still relatively limited compared to a fully general compiler. This is necessary to achieve high performance while writing concise code. However, it is important to cover other computation for automatic differentiation, such as tree traversal, sparse or graph data structures, sorting, etc. As we did in Chapter~\ref{chap:gradient_halide}, we need to properly hint the compiler about the computation patterns involved in order to reason about the computation and generate efficient code. I argue that we need more domain specific languages for high performance automatic differentiation.

\paragraph{Generalizing differentiation}
Derivatives are only one possibility for representing the neighborhood of a point in a function. For example, the Fourier transform and wavelets also characterize the smoothness of a function, with non-local information. While using derivatives are better than treating the program as a black box, it is natural to question whether derivatives are the most useful information we can retrieve from a program. After all, automatic differentiation is merely a program transformation. Finding other transformations of programs that benefit the tasks we are interested in is an interesting direction to pursue. For example, is it possible to find a transformation that is an approximation to the derivatives, but cheaper to compute, or more robust to high frequency changes of the function? Mathematics-wise, generalizations of derivatives have been proposed. Discrete calculus studies functions with integer inputs and real number outputs. Subgradients are commonly used in optimization theory. Systems-wise, program synthesis techniques~\cite{Lezama:2008:PSS} have been proven to be useful for finding program transformations.

\paragraph{Local minimum, overparametrization, and prefiltering}
When dealing with function landscapes that are noisy and bumpy, derivative-based methods are known to be unstable. Deep learning does not seem to suffer from this, most likely due to their large amount of parameters (e.g.~\cite{Frankle:2018:LTH, Kawaguchi:2019:EAB}). Figuring out how to overparametrize traditional algorithms is the key to truly fuse the two domains. Prefiltering in signal processing could also play an important role: if we can smooth out a function \emph{before} we sample it, it will make the derivatives much more well-behaved. Yang and Barnes~\cite{Yang:2018:APS} hinted on how to do this automatically for shader programs in computer graphics, but generalizing their approach to arbitrarily high-dimensional functions requires future research.

\appendix
%% This defines the bibliography file (main.bib) and the bibliography style.
%% If you want to create a bibliography file by hand, change the contents of
%% this file to a `thebibliography' environment.  For more information 
%% see section 4.3 of the LaTeX manual.
\begin{singlespace}
\bibliography{phdthesis}
\bibliographystyle{plain}
\end{singlespace}

\end{document}